\pdfoutput=1
\documentclass[b5paper, 10pt, twoside]{scrbook} 

\areaset[16mm]{12.3cm}{19.5cm} 

\usepackage{amsmath}  
\usepackage{graphicx} 
\usepackage[dutch,english]{babel} 
\usepackage[T1]{fontenc}
\usepackage[adobe-utopia]{mathdesign}
\usepackage[scaled=.92]{helvet}
\usepackage{textcomp} 

\usepackage[sectionbib]{chapterbib}         
\usepackage[numbers,sort&compress]{natbib}  
\bibpunct{[}{]}{,}{n}{}{,}                  

\newcommand{\includetwographics}[2]{
\begin{minipage}{0.49\textwidth}
\begin{flushleft}
\includegraphics[width=0.96\textwidth]{#1}
\end{flushleft}
\end{minipage}
\begin{minipage}{0.49\textwidth}
\begin{flushright}
\includegraphics[width=0.96\textwidth]{#2}
\end{flushright}
\end{minipage}}

\newcommand{\comment}[1]{}

\input sectiontitles

\usepackage{fancyhdr}

\fancypagestyle{plain} {%
    \fancyhf{}
}

\renewcommand{\chaptermark}[1]{\markboth{#1}{}}

\fancyheadoffset[RO]{25pt} \fancyheadoffset[LE]{25pt}

\newcommand{\avg}[1]{\left\langle#1\right\rangle}
\newcommand{\vect}[1]{\mathbf{#1}}
\newcommand{\bopt}{{\beta}}
\newcommand{\bucket}{C_{4,2}}
\newcommand{\mum}[1]{#1\,\text{\textmu m}}
\newcommand{\mm}[1]{#1\,\text{mm}}
\newcommand{\cm}[1]{#1\,\text{cm}}
\newcommand{\nm}[1]{#1\,\text{nm}}
\newcommand{\ms}[1]{#1\,\text{ms}}
\newcommand{\us}[1]{#1\,\text{\textmu s}}
\newcommand{\gram}[1]{#1\,\text{g}}
\newcommand{\ml}[1]{#1\,\text{ml}}
\newcommand{\ul}[1]{#1\,\text{\textmu l}}
\newcommand{\Fourier}[1]{\mathcal{F}\left\{#1\right\}}
\newcommand{\vr}{\vect{r}}
\newcommand{\optim}[1]{\widetilde{#1}}
\newcommand{\trb}{\vr_{\bopt}}
\newcommand{\rb}{\vr_b}
\newcommand{\ra}{\vr_a}
\newcommand{\ain}{{a'}}
\newcommand{\ralpha}{\vr_{\ain}}
\newcommand{\rk}{\vr_k}
\newcommand{\Dr}{\Delta\vr}
\newcommand{\qperp}{\vect{q}_\perp}
\newcommand{\Iin}{I_\text{in}}
\newcommand{\Iavg}{I_0}
\newcommand{\Ptot}{P_\text{tot}}
\newcommand{\Pin}{P_\text{in}}
\newcommand{\tPtot}{\optim{P}_\text{tot}}
\newcommand{\Ttot}{T_\text{tot}}
\newcommand{\tTtot}{\optim{T}_\text{tot}}
\newcommand{\Tcontrol}{T_\text{c}}
\newcommand{\NA}{\text{NA}}
\newcommand{\gamavg}{\avg{|\gamma|^2}}
\newcommand{\Neff}{N_\text{eff}}
\newcommand{\Meff}{M_\text{eff}}
\renewcommand{\Im}{\text{Im}}
\renewcommand{\Re}{\text{Re}}
\newcommand{\elltr}{\ell_\text{tr}}
\newcommand{\subfig}[1]{\textbf{#1})}
\newcommand{\subfigref}[1]{\textbf{#1}}
\newcommand{\ud}{\,\text{d}}
\renewcommand{\iint}{\int\!\!\!\!\!\int}
\renewcommand{\iiint}{\int\!\!\!\!\!\int\!\!\!\!\!\int}
\newcommand{\myfussy}{\fussy
\vfuzz1pt 
\hfuzz1pt 
}

\newenvironment{abstract}
{\begin{center}\vskip -10pt\begin{minipage}[t]{0.9\textwidth}\small\textbf{Abstract:\ }}
{\end{minipage}\vskip 20pt\end{center}}

\newcounter{sectionapp}
\newenvironment{sectionappendix}
{%
  \setcounter{sectionapp}{1}
  
}%
{
  
}

\usepackage{ifpdf}
\ifpdf
    \usepackage[pdftex, a4paper=false, b5paper=true,
        pdftitle={Controlling the propagation of light in disordered scattering media},
        pdfauthor={I. M. Vellekoop},
        pdfsubject={PhD thesis, University of Twente 2008},
        pdfkeywords={Scattering, Inverse diffusion, Channel demixing, Diffusion, Speckle},
        bookmarks,
        breaklinks=true,
        colorlinks=true,
        linkcolor=black, 
        anchorcolor=black, 
        citecolor=black, 
        bookmarksnumbered=true,
        pdfstartview=FitV,
        bookmarksopen=false]{hyperref}

\else
    \usepackage[dvips, bookmarks,
        colorlinks=false,
        pdftitle={Controlling the propagation of light in disordered scattering media},
        pdfauthor={I. M. Vellekoop},
        pdfsubject={PhD thesis, University of Twente 2008},
        pdfkeywords={Scattering, Inverse diffusion, Channel demixing, Diffusion, Speckle},
        bookmarks,
        breaklinks=true,
        colorlinks=true,
        linkcolor=black, 
        anchorcolor=black, 
        citecolor=black,
        bookmarksnumbered=true
        ]{hyperref}
\fi

\begin{document}

\newcommand{\wideimage}{0.9\textwidth}

\newcommand{\smallimage}{0.5\textwidth}

\newcommand{\tinyimage}{0.3\textwidth}

\newcommand{\mediumimage}{0.75\textwidth}

\pdfbookmark[0]{{Title page}}{title} 
%
%
\thispagestyle{empty}   
\vspace*{30pt}           %

\begin{center}
    \huge CONTROLLING THE PROPAGATION\\ OF LIGHT IN DISORDERED SCATTERING MEDIA
\end{center}

%
%
\newpage
\thispagestyle{empty}%
\vspace*{20pt}

\noindent Promotiecommissie\\

\noindent
\begin{tabular}{@{}p{3.5cm}p{6cm}}
    Promotor            & prof. dr. A. Lagendijk\\
    Assistent Promotor  & dr. A. P. Mosk \\[\baselineskip]

    Overige leden   & prof. dr. D. Lohse\\
                    & prof. dr. J. P. Woerdman\\
                    & prof. dr. W. L. Vos\\
                    & prof. dr. A. C. Boccara\\[2\baselineskip]

    Paranimfen      & H. E. Holland\\
                    & M. Pil
\end{tabular}

\vfill

\begin{center}
The work described in this thesis is part of the research program of the\\
`Stichting voor Fundamenteel Onderzoek der Materie (FOM)',\\
which is financially supported by the\\
`Nederlandse Organisatie voor Wetenschappelijk Onderzoek' (NWO)'.

\vspace{.5\baselineskip}

This work was carried out at the\\
\emph{
Complex Photonic Systems Group,\\
Department of Science and Technology\\
and MESA$^+$ Institute for Nanotechnology,\\
University of Twente, P.O. Box 217,\\
7500 AE Enschede, The Netherlands.\\
} \vspace{.5\baselineskip}

\end{center}

\vspace{1.25\baselineskip}
\noindent This thesis can be downloaded from \\
{http:{//}www.wavesincomplexmedia.com.}\\[0.25\baselineskip]
\noindent ISBN: 978-90-365-2663-0

%
%
\newpage
\thispagestyle{empty}

\vspace*{30pt}

\begin{center}
    \huge CONTROLLING THE PROPAGATION\\ OF LIGHT IN DISORDERED SCATTERING MEDIA
\end{center}

\vspace{80pt}

\begin{center}
    \Large
    PROEFSCHRIFT\\[\baselineskip]
    \normalsize

    ter verkrijging van\\
    de graad van doctor aan de Universiteit Twente,\\
    op gezag van de rector magnificus,\\
    prof.~dr.~W.H.M.~Zijm,\\
    volgens besluit van het College voor Promoties\\
    in het openbaar te verdedigen\\
    op donderdag 24 april 2008 om 16.45 uur\\[5\baselineskip]

    door\\[\baselineskip]

    {\Large Ivo Micha Vellekoop}\\

    \vspace{\baselineskip} geboren op 11 november 1977\\
    te 's-Gravenhage
\end{center}

%
%
\newpage
\thispagestyle{empty}

\vspace*{20pt}

\noindent Dit proefschrift is goedgekeurd door:

\vspace{\baselineskip}

\noindent prof. dr. A. Lagendijk en dr. A. P. Mosk

%
%
\newpage
\thispagestyle{empty}

\vspace*{3cm}
\begin{center}
\large\textit{``... no more substance than a pattern formed by frost\\
that a simple rise in temperature would reduce to nothing.''}\\[\baselineskip]
\end{center}

\begin{center}
\large\textit{``... pas plus de consistance qu'un motif form\'e par le givre\\
qu'un simple redoux suffit \`a an\'eantir.''}\\[\baselineskip]
\end{center}

\begin{flushright}
\large - Michel Houellebecq,\\La Possibilit\'e d'une \^ile\\
\end{flushright}

%
%
\newpage
\thispagestyle{empty} \vspace*{6cm}

\renewcommand\arraystretch{1.5}

\myfussy 

\newpage
\pdfbookmark[0]{Contents}{contents} 
\tableofcontents

\chapter{Introduction\label{cha:introduction}}

\noindent Devices that use or produce light play an important role
in modern life. Among the numerous daily applications of light are
displays, telecommunication, data storage, sensors. In industry,
medicine and scientific research, optical techniques are also
absolutely indispensable. Light is used for imaging and microscopy,
but also for detecting and treating diseases, analyzing chemical
compounds and investigating living cells on a molecular level.

\begin{figure}
\centering
  \includegraphics[width=\wideimage]{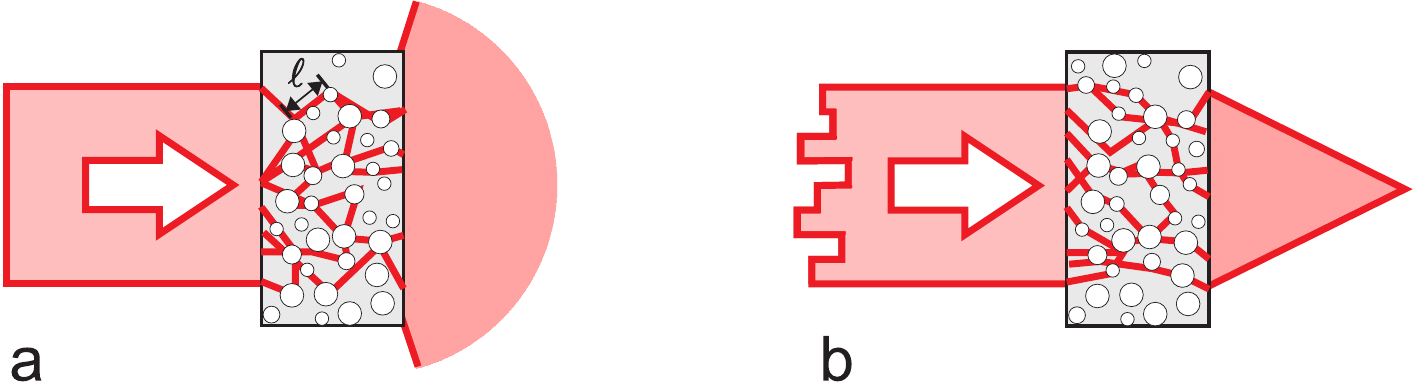}\\
  \caption{Principle of opaque lensing. \subfig{a} A plane wave impinges
  on an opaque scattering object. In the object, light performs a random walk with a typical distance
  given by the mean free path for light ($\ell$). The little light that makes it trough is
  scattered in all directions. \subfig{b} The incident wave is shaped to
  match the scattering in the object. The
  opaque object focuses the shaped wave to any desired point, thereby
  acting as an `opaque lens'.}\label{fig:intro-principle}
\end{figure}
In a transparent medium, like glass or air, light propagates along a
straight line. However, as we all know from daily experience, it is
impossible to see through, for instance, white paint or the shell of
an egg. Such opaque materials have a microscopic structure that
makes it impossible for light to go straight through.
Figure.~\ref{fig:intro-principle}a shows schematically what happens
when a beam of light impinges on a white opaque object: collisions
with tiny particles causes the light to spread out and lose all
directionality. This process is called diffusion of light; it is
comparable to the irreversible diffusion process that makes a drop
of ink in a glass of water spread out evenly.

Scattering and diffusion of light are huge limitations for optical imaging, but also severely hinder
telecommunication, spectroscopy, and other optical techniques. In the last few decades, a tremendous
effort was put in developing imaging methods that work in strongly scattering media.\cite{Sebbah2001} That
research has brought forward important new imaging methods like optical coherence
to\-mog\-ra\-phy\cite{Huang1991}, diffusion to\-mog\-ra\-phy\cite{Yodh1995} and laser speckle
velocimetry\cite{Briers2001}.

In this thesis, a new approach is taken to tackle the problem of
scattering. We develop a wavefront shaping technique to steer light
through opaque objects. When we shape the wavefront so that it
exactly matches the scattering properties of the object, the object
focuses light to a point (see Fig.~\ref{fig:intro-principle}b). The
term `opaque lens' was introduced in a news item\cite{Schewe2007}
about our research to describe this focusing behavior. Using our
wavefront shaping method, we steered light through opaque objects,
focused it inside and even projected simple images through the
objects.

In this chapter, we first explain in general terms how we shape the
wavefront to control the propagation of light. Then, we relate our
work to other experimental fields. In Section~\ref{sec:intro-tools},
we give a brief introduction to the concepts and tools that are used
for analyzing light propagation in opaque objects. We end this
introductory chapter by giving an overview of this thesis.

\section{Opaque lenses\label{sec:opaque-lenses-definition}}
\noindent The particles in a strongly scattering medium are smaller
than the wavelength of light. Therefore, the wave nature of light
needs to be taken into account and light propagation cannot be
described in terms of light rays. Diffusion of waves is
fundamentally different from diffusion of particles since waves
exhibit interference. We first briefly explain how a wave propagates
through a disordered medium. Then, we show how interference was used
in our experiments to make an `opaque lens' focus light.

\begin{figure}[t]
\centering
  \includegraphics[width=\wideimage]{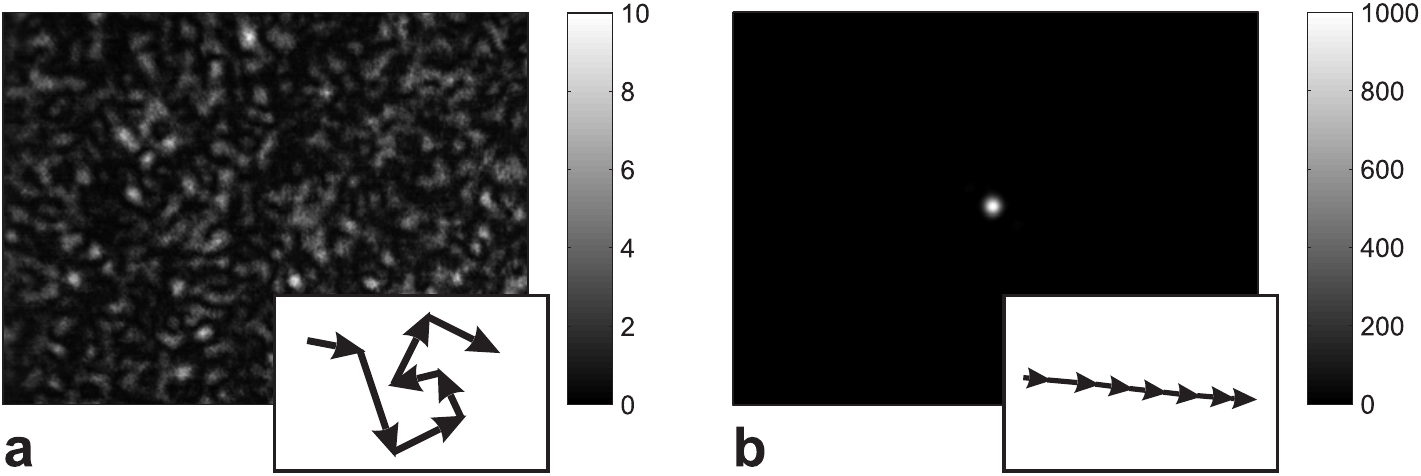}\\
  \caption{Interference in an opaque lens. \subfig{a} Transmitted
intensity of an unshaped incident beam. Scattered light forms a
random interference pattern known as laser speckle. Inset (phasor
plot), at each point many waves interfere randomly, resulting in a
low overall intensity. \subfig{b} Transmitted intensity of a
wavefront that is optimized for focusing at a single point. The
intensity in the focus is approximately 1000 times as high as the
average intensity in \subfigref{a}. Inset, in the focus all waves
are in phase.}\label{fig:speckle-focus-phasors}
\end{figure}

Wave propagation in a disordered medium can be made insightful with
the help of the Huygens
principle\cite{Huygens1690}\nocite{Shapiro1989}. When an incident
beam of light hits a small particle in the object, part of the light
is scattered and forms a spherical wave moving away from the
particle. In turn, this spherical wave hits other particles, giving
rise to more and more waves. Light propagation in a disordered
scattering medium is extremely complex; light is typically scattered
hundreds or thousands of times before it reaches the other side of
the sample. Figure~\ref{fig:speckle-focus-phasors}a shows the
transmitted intensity for a sample that is illuminated with laser.
The complicated random pattern is the result of the interference of
very many different waves; this pattern is known as laser speckle.

We now illuminate the opaque object with thousands of light beams,
instead of one. Each of this beams forms a different random speckle
pattern when it is scattered by the object. The total field at a
given point behind the sample is the sum of the speckle patterns of
all incident beams. Since the object is disordered, each beam
contributes to the field with a random phase. Therefore, the
contributions from different beams mostly cancel each other (see the
inset in Fig.~\ref{fig:speckle-focus-phasors}a for a graphical
representation of this statement).

Next, a spatial light modulator is used to delay each of the beams
with respect to the others and thereby shape the wavefront of the
incident light. The modulator is a liquid crystal on silicon display
(LCoS). These displays have been developed in the last decennium for
use in commercial video projectors. A computer controlled algorithm
optimizes the intensity in a single target point. It does this by
changing the phase of the beams one by one, until their speckle
patterns are all in phase in the target. In
Fig.~\ref{fig:speckle-focus-phasors}b we show the transmission after
a successful optimization. All contributions interfere
constructively in the target and the intensity increases
dramatically (by a factor of thousand in this case). The opaque
object now sharply focuses light to a single point. The shaped
wavefront uniquely matches the scattering object like a key matches
a lock. When the microscopic scatterers in the object have a
different position or orientation, a completely different wavefront
is required. Therefore, this method only works when the scatterers
do not move during the optimization.

\section{Relation with earlier work}
\noindent Our work is inspired by an article of
Dorokhov\cite{Dorokhov1984} about electron scattering in a metal
wire. Dorokhov predicted that there exist electron wave functions
that are fully transmitted through the wire, regardless of the
length of the wire. Scattering of light in a disordered medium is,
in many ways, analogous to scattering of electrons in a wire. The
optical equivalent of this prediction is that it should be possible
to construct a shaped wave of which all 100\% of the intensity is
transmitted through an otherwise opaque object. We wanted to
construct this wavefront. The work on opaque lenses was initially
performed as a first step towards achieving full transmission.
Because opaque lenses turned out to be a fascinating research
subject on their own, we first explored that field for a while.
Finally, we we used our optical analogue of a disordered electronic
system to confirm Dorokhov's hypothesis.

In our opaque lens experiments, we `borrowed' a lot of ideas from
time-reversal experiments with ultrasound and microwaves. In
pioneering work by Fink et al. (see Ref.~\citealt{Fink1999} for a
review) a short pulse impinged on a strongly scattering system. The
amplitude of the scattered wave was recorded and played back
reversed in time. Thanks to time reversal symmetry, the wave focused
back to the original source. In a series of beautiful experiments,
it was shown that time-reversal can be used to focus light through a
disordered medium\cite{Fink1989}, break the diffraction
limit\cite{Derode1995, Lerosey2007}, and improve communication by
using disorder\cite{Lerosey2007}.

In this thesis we show experiments that use our wavefront shaping method to obtain similar result with
light. Our approach has two fundamental advantages over time-reversal methods. First of all, no time
reversal symmetry is required. And secondly, for time reversal one needs to have a source at the target
focus. With our method, it is sufficient to only have a detector in the target focus. This difference
allows us to focus waves \emph{inside} a scattering medium, as is demonstrated in
Chapter~\ref{cha:focusing-inside}.

An optical method that is related to time-reversal is optical phase
conjugation\cite{Fisher1983}. Using a non-linear crystal and high
power lasers, it is possible to reverse both the direction and the
phase of a speckle pattern to project the light back to its original
source. Like with time-reversal, this method requires a source at
the point where we want the waves to focus.

Our experimental apparatus is similar to the setups used in adaptive
op\-tics\cite{Tyson1998}. Adaptive optics is a technique for
correcting aberrations in lenses or other transparent media, such as
a turbulent atmosphere\cite{Roddier1997} or the human
eye\cite{JOSA-retinal-imaging}. Adaptive optics works for situations
where light propagates along rays. Diffraction and interference
effects are not taken into account. In opaque scattering media,
however, there are no rays of light that one can steer; light
propagation is dominated by diffraction and interference. Therefore,
our method uses \emph{interference} instead of ray optics to steer
light.

\section{Mathematical tools for analyzing complex system\label{sec:intro-tools}}
\noindent The propagation of light is described perfectly well by
Maxwell's equations, so why would we need anything else? The problem
is that these equations are so hard to solve that exact results can
only be found for very simple geometries. Even for a simple sphere
geometry the result is not a closed expression, but a complicated
sum of Bessel functions.\cite{Mie1908}\nocite{Hulst1957} Obviously,
this approach cannot be scaled to a system containing billions of
spheres, let alone irregularly shaped grains. Even worse, we do not
even know the exact positions and orientations of the scatterers in
a sample to begin with.

Although this seems to be a hopeless situation, it is possible to capture the overall characteristics of
the system using statistical and physical tools. In this section, we introduce the most important
statistical tools that are used in this thesis: ensemble averaging, correlation functions and probability
density functions. Then, we highlight the powerful physical concept of diffusion. These tools are
applicable to all complex systems, whether it is light propagation in a disordered medium, or the dynamics
of a complex biological, chemical or economical system.

\subsection{Ensemble averaging}
\noindent Every sample scatters light in a unique way. Even if all
macroscopic properties (layer composition and thickness, porosity,
etcetera) are the same, the microscopic structure of two samples
will be completely different. Therefore, it is impossible to predict
the exact scattering properties of a specific sample.

Instead, one calculates average quantities for a whole ensemble of samples. For example, we could
calculate the optical field averaged over all conceivable samples that consist of a $\mum{10}$-thick layer
of zinc oxide pigment. We write this quantity as $\avg{E}$, where $E$ is the field and $\avg{\cdot}$
denotes averaging over the ensemble of all possible samples of a given type.

In an experiment it is, most of the time, not needed to average over an ensemble of samples. Instead, one
can average over the response of different portions of the sample. When both averaging methods are
equivalent, the system is called ergodic. We assume ergodicity for all our samples.

\subsection{Correlation functions}
\noindent The majority of the research in random scattering is
concerned with calculating or measuring correlations of some kind. A
correlation function relates the value of a quantity at one
coordinate to the value of that quantity at a different
coordinate\footnote{Here `coordinate' can denote position, angle,
frequency or any other independent variable. It is also very common
to have correlation functions involving four or more coordinates.}.
For instance, the position correlation function of the electrical
field is defined as
\begin{equation}
C_E(\vr_1, \vr_2) \equiv \avg{E^*(\vr_1)E(\vr_2)},
\end{equation}
where $^*$ denotes the complex conjugate. When two quantities are statistically independent, they can be
averaged separately. For instance, when two points $\vr_1$ and $\vr_2$ are so far apart that the fields at
both points are uncorrelated we can write (note the essential difference in the placement of the brackets)
\begin{equation}
C_E(\vr_1, \vr_2) = \avg{E^*(\vr_1)}\avg{E(\vr_2)} \qquad \text{for
$\vr_1$ far from $\vr_2$.}\label{eq:cE-uncorr}
\end{equation}
Since $E$ oscillates rapidly around $0$, it quickly averages out. Inside a diffusive medium
$\avg{E}\approx 0$, and Eq.~\eqref{eq:cE-uncorr} vanishes. However, in general the decomposition in
Eq.~\eqref{eq:cE-uncorr} cannot be made. Especially when $\vr_1=\vr_2$, the correlation function cannot be
separated
\begin{equation}
C_E(\vr, \vr) = \avg{E^*(\vr) E (\vr)} \neq \avg{E^*(\vr)}\avg{E(\vr)}.\label{eq:cE-corr}
\end{equation}
We adopt the convention that the intensity $I$ (unit $\mathrm{W}\mathrm{m}^{-2}$) is defined as
$I\equiv|E|^2$ (see e.g. \cite{Rossum1999}). Using this convention, $C_E(\vr, \vr)$ equals the average
intensity. Since the intensity is always positive, its average will not vanish and $C_E(\vr, \vr)$ is
positive.

\subsection{Probability density functions}
\noindent The probability density function (pdf) gives the
probability that a variable has a certain value. An important pdf is
that of the intensity $I$ of a speckle\cite{Goodman2000}
\begin{equation}
p(I) = \frac{1}{\avg{I}}\exp\left(-\frac{I}{\avg{I}}\right)\label{eq:exponential-pdf},
\end{equation}
Eq.~\eqref{eq:exponential-pdf} tells us that the most likely intensity in a speckle pattern is zero. The
chance of finding bright speckles decreases exponentially with intensity of that speckle. A typical
speckle pattern with this distribution is shown in Fig.~\ref{fig:speckle-focus-phasors}a.

A different pdf that is used extensively in this thesis describes
the field of a speckle. The joint probability density of the real
part of the field ($E_r$) and the imaginary part of the field
($E_i$) is given by a so-called circular Gaussian distribution
\begin{equation}
p(E_i, E_r) = \frac{1}{\pi \avg{I}} \exp\left( -\frac{|E_i|^2+|E_r|^2}{\avg{I}}
\right)\label{eq:gaussian-pdf}.
\end{equation}
The Gaussian distribution is very common and always arises when many
uncorrelated random variables with a finite variance are added. This
important statistical fact is known as the Central Limit Theorem. In
the case of a speckle pattern, the field in a single speckle is the
sum of the contributions from a large number of light paths. When
these light paths are independent, the field has a Gaussian
distribution\footnote{The assumption of independent paths is not
completely true. Tiny deviations from Gaussian statistics have been
observed experimentally.\cite{Boer1994} In
Chapter~\ref{cha:dorokhov-experiment} we describe an experiment in
which we observed very large effects of correlations between
paths.}.

\subsection{The diffusion equation\label{sec:intro-diffusion}}
\noindent The average propagation of intensity in a disordered
medium can be described very well with a diffusion equation. For
continuous wave illumination, the stationary diffusion
equation\cite{Carslaw1959, Chandrasekhar1960} applies
\begin{equation}
     - D\nabla^2 I(\vr; t) = S(\vr; t)\label{eq:diffusion},
\end{equation}
where $D$ is the diffusion constant for light, $S$ is the source of diffuse light, and $\nabla^2$ is the
Laplace operator.

We now solve the diffusion equation to find the intensity
distribution in a sample. All our samples are effectively infinitely
large in the $x$ and $y$ directions and have a finite thickness $L$
in the $z$ direction. For such a geometry, it is convenient to work
in Fourier transformed coordinates $\qperp \equiv (q_x, q_y)$ for
the traversal coordinates $x$ and $y$. In these coordinates,
Eq.~\eqref{eq:diffusion} transforms to
\begin{equation}
     q_\perp^2 I(\qperp, z) - \frac{\partial^2 I(\qperp, z)}{\partial z^2}
     = \frac{S(\qperp, z)}{D}\label{eq:diffusion-static-slab},
\end{equation}
with $q_\perp\equiv |\qperp|$. We use the Dirichlet boundary
conditions $I(-z_{e1}) = 0$, and $I(L+z_{e2})=0$ to describe the
interfaces of the sample. These boundary conditions give a much
simpler and more insightful result than the slightly more accurate
mixed boundary conditions.\cite{noteBC}\nocite{Vellekoop2005} Here,
$L$ is the thickness of the sample and $z_{e1}$ and $z_{e2}$ are the
so called extrapolation lengths that account for reflection at the
front and back surface of the medium, respectively. These
extrapolation lengths depend on the effective refractive index of
the sample and the refractive index of its
surroundings\cite{Lagendijk1989, Zhu1991, Vera1996}. When a slab of
diffusive material is illuminated by an external source, the
incident light can be described by a diffuse source at a depth of
one mean free path $\ell$.\cite{Akkermans1986} Then, the solution to
Eq.~\eqref{eq:diffusion-static-slab} is (see e.g.
Ref.~\cite{Molen2007a})
\begin{equation}
I(\qperp, z) =
    \begin{cases}
    S(\qperp)\frac{\displaystyle\sinh{(q_\perp [L_e-z-z_{e1}])}\sinh{(q_\perp [\ell+z_{e1}])}}{\displaystyle D q_\perp \sinh{(q_\perp
    L_e)}}& z \geq z_0\\
    \;\\
    S(\qperp)\frac{\displaystyle\sinh{(q_\perp [L_e-\ell-z_{e1}])}\sinh{(q_\perp [z+z_{e1}])}}{\displaystyle D q_\perp \sinh{(q_\perp L_e)}}& z < z_0\\
    \end{cases},
\label{eq:intro-diffusion-equation-solution}
\end{equation}
\begin{figure}
\centering
  \includegraphics[width=\wideimage]{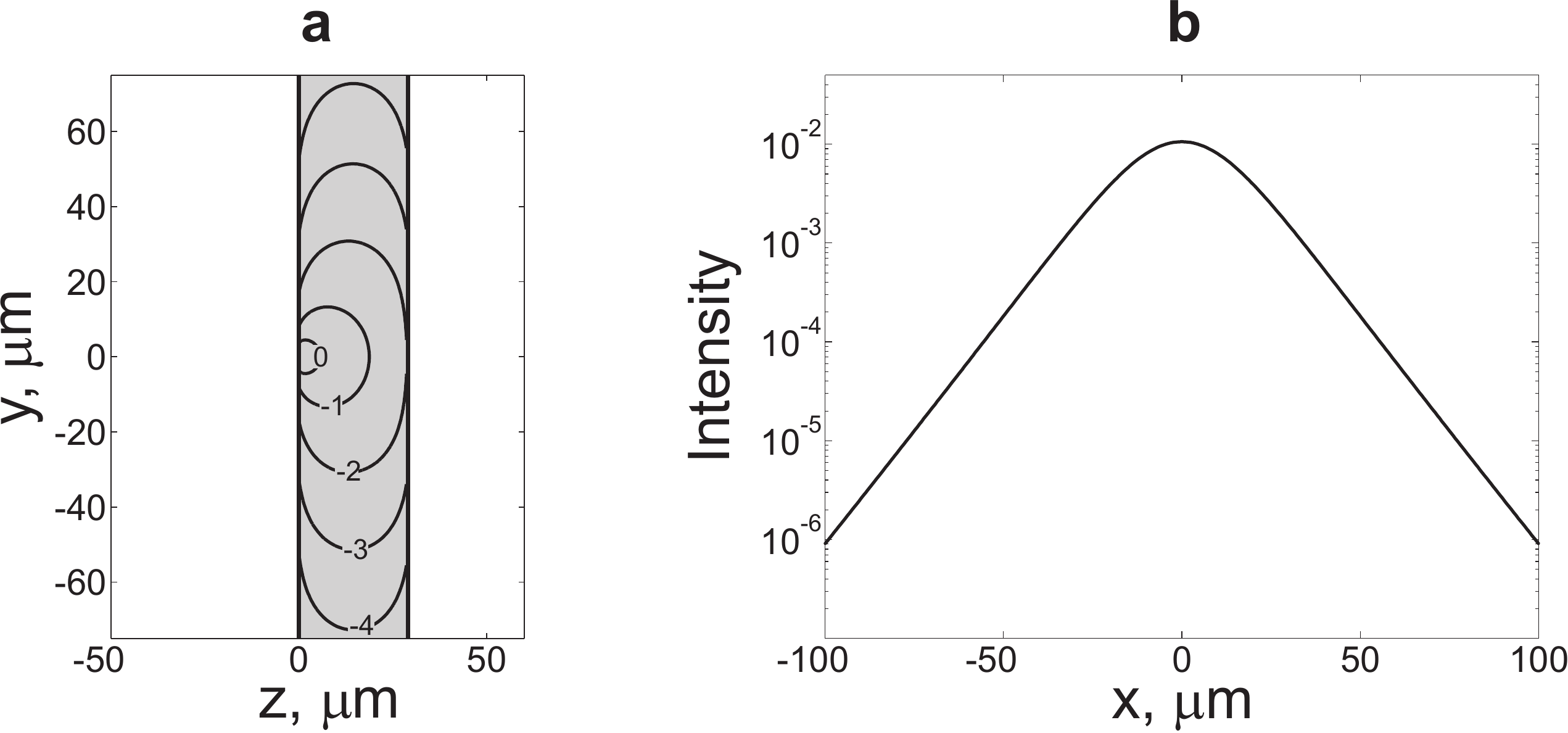}\\
  \caption{Average distribution of diffuse light in a slab (shaded rectangle of thickness $\mum{30}$) of strongly scattering material.
  \subfig{a} Contour plot of the energy density distribution inside the sample. The source is placed at a
  depth of $z=\ell=\mum{0.72}$ and $x=y=0$. The numbers on the equal-density curves indicate $\log^{10}(I)$.
  \subfig{b} Logarithmic plot of the intensity profile at the back of the sample (at $z=\mum{30}$).
  }\label{fig:contour-profile}
\end{figure}

\noindent with $L_e \equiv L + z_{e1} + z_{e2}$. A numerical Fourier
transform gives $I$ in real space coordinates. The intensity
distribution in a diffusive slab is shown in
Fig.~\ref{fig:contour-profile}. The intensity is maximal close to
the source and spreads out over the medium. Far away from the
source, the intensity decays exponentially with distance. From an
expansion of Eq.~\eqref{eq:intro-diffusion-equation-solution} in
small variable $q_\perp$, we find that the intensity decreases
exponentially with a decay length of $L_e/\sqrt{6}$.

\section{Outline of this thesis}
This thesis describes experiments and theory on controlling the
propagation of light in opaque objects. The theoretical framework
was developed in parallel with the experiments. Therefore, most of
the theory is presented together with the experimental results. The
simple notation that was used in the first experiments did not
suffice to describe the more advanced experiments and a more
powerful notation was developed along with the theory. Therefore
there are slight differences in the notation in the different
chapters. Care has bee taken that all chapters are self-contained
and can be read and understood separately.

In Chapter~\ref{cha:setup} we describe the experimental apparatus,
the samples and the control program that were used in the
experiments. Special attention is given to a novel wavefront
modulation method that we developed.

In Chapter~\ref{cha:focusing-through} the first experimental results
of opaque lensing are presented. We also explain the optimization
algorithm and calculate the maximally achievable intensity of the
focus. The focusing resolution of opaque lenses is studied
quantitatively in Chapter~\ref{cha:diffraction-limit}. It appears
that an opaque lens focuses light as sharply as the best possible
lens. In Chapter~\ref{cha:focusing-inside} the concept of opaque
lenses is extended to focus light inside an opaque object. In
Chapter~\ref{cha:communication} we demonstrate that disorder can
improve optical communication. In Chapter~\ref{cha:algorithms}
dynamic algorithms for use with non-stationary samples are
investigated.

In Chapter~\ref{cha:dorokhov-theory} a transport theory for optimized wavefronts is developed. We show
that wavefront shaping significantly increases the total transmission through opaque objects. The
experimental observation of this effect is presented in Chapter~\ref{cha:dorokhov-experiment}.

Finally, in Chapter~\ref{cha:summary}, we summarize our findings and
give an outlook of the many possible applications of our work.

\enlargethispage*{0.5\baselineskip}
\bibliography{../../bibliography,endnotes}
\bibliographystyle{Ivo_sty}

\setcounter{chapter}{1}

\chapter{Experimental
apparatus\label{cha:setup}}
\noindent In this chapter we discuss the experimental apparatus that
we built to control the propagation of light in disordered media. We
discuss the considerations that played a role in designing the
experiment and give special attention to a new wavefront shaping
method that we developed for our experiments. The setup was modified
for each of the individual experiments that are described in this
thesis. Here we discuss the common elements of the design and
explain what specific modifications were made.

\begin{figure}
\centering
  \includegraphics[width=\mediumimage]{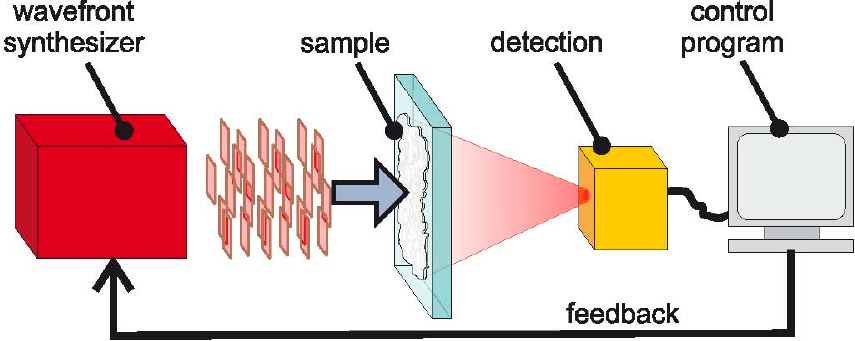}\\
  \caption{Block scheme of the experiment. A wavefront synthesizer generates a shaped monochromatic wavefront.
  The light is scattered by a strongly scattering sample. A detector defines a target position for the
  optimization procedure and provides a feedback signal. A computer analyzes the signal and
  reprograms the phase modulator until the light optimally focuses on the target.}\label{fig:block-scheme}
\end{figure}

A general diagram of the experiments is shown in
Fig.~\ref{fig:block-scheme}. A strongly scattering sample is
illuminated by a wavefront synthesizer that is able to construct a
spatially modulated beam. The light that is scattered by the sample
is collected by a detection system. This detection system provides
feedback to a computer algorithm that controls the wavefront
synthesizer.

The most complex part of the apparatus is the wavefront synthesizer.
It contains a liquid crystal display (LCD) that spatially modulates
incident light. In Section~\ref{sec:wavefront-synthesizer} we
analyze the characteristics of the LCD and explain how it was used
as a phase modulator. We also developed a new method for using the
LCD to spatially modulate both amplitude and phase of the light.
This method was published in Ref.~\citealt{Putten2008thesis}. In
Section~\ref{sec:detection} the detection system is described. Here
we also discuss how the detection is synchronized with the wavefront
synthesizer. We comment on the overall stability issues of the setup
in Section~\ref{sec:stability}. Then, in Section~\ref{sec:samples},
we describe which types of samples were used and introduce a new
fabrication method that we developed for making strongly scattering
samples. The structure of the control program that coordinates the
experiment is discussed in Section~\ref{sec:control-program}.
Finally, in Section~\ref{sec:outlook} we give an outlook of what can
be expected in future experiments.

\section{Wavefront synthesizer\label{sec:wavefront-synthesizer}}
\noindent Computer controlled wavefront shaping is a very versatile technique that is used in many fields
of optics. In adaptive optics\cite{Tyson1998}, for example, deformable mirrors or other spatial light
modulators are used to correct aberrations in a variety of optical systems. Another area that relies on
computer controlled light modulators is that of digital holography. Topics in digital holography include
holographic data storage\cite{Psaltis1998}, 3D display technology\cite{Travis1997}, and holographic image
processing\cite{Davis2001}.

Liquid crystal displays are among the most popular types of light modulators because of their high optical
efficiency, the high number of degrees of freedom and their wide availability. In our experiment, light is
modulated with a twisted nematic (TN) liquid crystal display\footnote{We used two Holoeye LC-R 2500
systems. The LCDs have a size of $\mm{19.6}\times\mm{14.6}$. One system was customized by the manufacturer
to control two LCDs with one control box. That system was used for the experiments in
Chapter~\ref{cha:dorokhov-experiment}, where we needed control over both polarizations.}. This LCD can
modulate light at a refresh rate of $60$ or $72$~Hz at a resolution of $1024\times768$ pixels. In
Section~\ref{sec:lcd-principle} we discuss the operating principle of a TN LCD. LCDs are designed to
modulate light intensity. To use the LCD as a phase modulator, a thorough characterization of the display
is required. This characterization procedure is described in Section~\ref{sec:lcd-characterization}. In
Section~\ref{sec:phase-mostly}, we describe a common method to use TN LCDs as phase modulators. This
method was used to perform the experiments that are described in Chapters~\ref{cha:focusing-through} and
\ref{cha:algorithms}. For the other experiments in this thesis, we developed a new modulation method that
has several advantages over existing methods. This novel modulation technique is introduced in
Section~\ref{sec:amplitude-phase} and experimentally demonstrated in
Section~\ref{sec:amplitude-phase-experiment}. The switching behavior of the display is described in
Section~\ref{sec:lcd-transient}.

\subsection{Principle of a twisted nematic liquid crystal display\label{sec:lcd-principle}}
\begin{figure}
\centering
  \includegraphics[width=\smallimage]{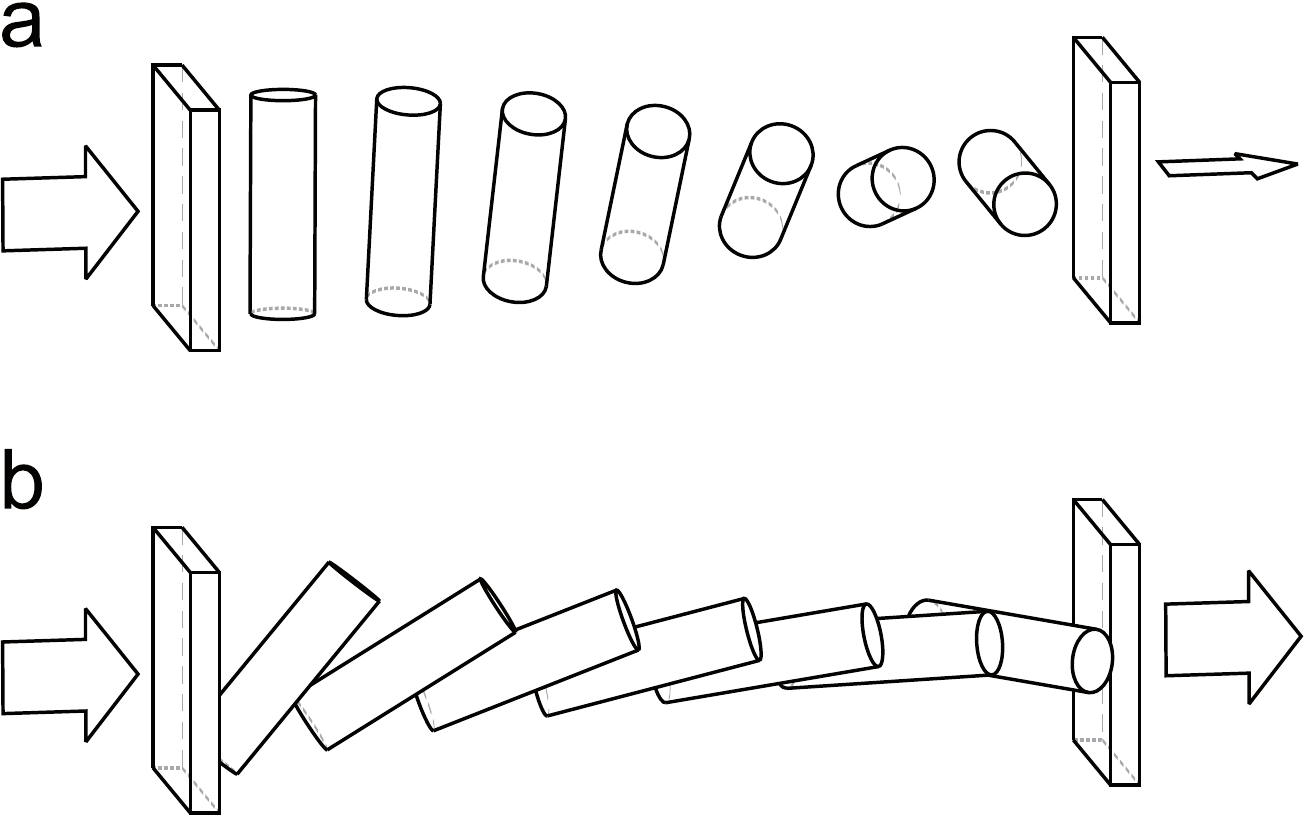}\\
  \caption{Operating principle of a transmissive TN LCD. \subfig{a} No voltage is applied
  and the rod-like liquid crystal molecules are ordered in a helix.
  The polarization of incident light follows the twist. \subfig{b} A voltage
  is applied between the two electrodes. The rods orient in the direction
  of the field and the twist disappears. The polarization of the incident light
  is not rotated.}\label{fig:TN-LCD}
\end{figure}
\noindent The operating principle of LCDs is based on the birefringence of rod-like organic molecules. In
the twisted nematic phase, these rods are ordered in a helix as is shown in Fig~\ref{fig:TN-LCD}a. A
transmissive LCD is typically sandwiched between two crossed polarizers. The $90^\circ$ helical structure
of the liquid crystal rotates the angle of polarization of the incident light so that the light passes the
second polarizer. When a voltage is applied over the liquid crystal cell, the molecules align with the
electrical field, as is shown in Fig~\ref{fig:TN-LCD}b. Now the optical axis of the molecules is parallel
to the direction of light propagation and the polarization of the incident light is not rotated. As a
result, the light is blocked by the second polarizer and the pixel is dark.

The LCDs in our experiments are TN liquid crystal on silicon (LCoS)
displays. These reflective displays are designed to be used with a
polarizing beam splitter or with crossed polarizers in an off axis
configuration. The thickness of the liquid crystal layer and the
twist angle are carefully engineered to optimize brightness,
contrast and response time for projecting video.

The operation principle of a reflective LCD is more complicated than
that of a transmissive LCD (see e.g. Ref.~\citealt{Wu2001}). When no
voltage is applied, the polarization of the linearly polarized
incident light follows the helix. At the back surface of the LCD the
light is reflected and, on its way back it is rotated back to its
original polarization. When a voltage is applied, the helix is
distorted (see Fig.~\ref{fig:TN-LCD}b). Now, linearly polarized
light cannot completely follow the twist anymore and becomes
elliptically polarized. In the `on' state, the light is exactly
circularly polarized at the back surface of the LCD. Reflecting
circularly polarized light changes its handedness. Therefore, on its
way back, the polarization is not rotated back to its original
polarization, but to the orthogonal polarization state. In the
crossed polarizer configuration, the pixel now is reflecting.

\subsection{Liquid crystal display characterization\label{sec:lcd-characterization}}
The optical characteristics of a single pixel of the LCD can be
described by its Jones matrix $J$. The Jones matrix relates the
horizontal and vertical polarization components of the incident
field (denoted as, respectively $E^\text{in}_H$ and $E^\text{in}_V$)
to the components of the outgoing field

\begin{equation}
\begin{bmatrix}
E^\text{out}_H\\
E^\text{out}_V
\end{bmatrix}
=J
\begin{bmatrix}
E^\text{in}_H\\
E^\text{in}_V
\end{bmatrix}.
\end{equation}\label{eq:Jones-def}

\noindent Elliptic and circular polarization are described by
complex values of $E_H$ and $E_V$. In general, the elements of the
Jones matrix are also complex numbers. A pixel of the LCD is fully
characterized by measuring $J(V)$ for all voltage settings $V$.

\begin{figure}
\centering
  \includegraphics[width=\textwidth]{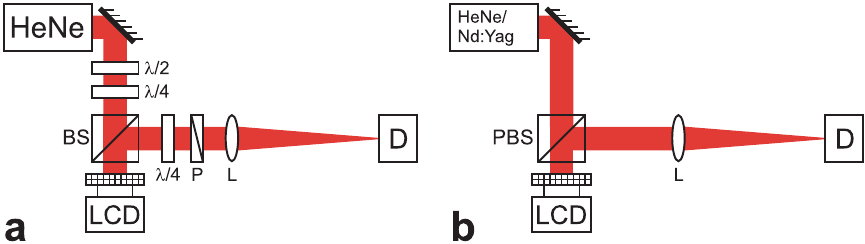}\\
  \caption{Setups used for characterizing the LCD. \subfig{a} Setup for full
  characterization. $\lambda/2$, half waveplate; $\lambda/4$, quarter waveplate;
  BS, non-polarizing 50\% beam splitter; P, polarizer; L, $\cm{60}$ lens;
  D, detector. The light source was a $\nm{632.8}$ HeNe laser.
  \subfig{b} Setup for partial characterization. PBS, polarizing
  beam splitter. Partial characterization was performed at wavelengths of
  $\nm{632.8}$ (HeNe laser) and $\nm{532}$ (Nb:YAG laser).}\label{fig:lcd-characterization}
\end{figure}

The setup in Fig.~\ref{fig:lcd-characterization}a was used to measure the Jones matrix of the LCD. A laser
illuminates a disk with a diameter of approximately $\mm{3}$ in the center of the LCD. With two
waveplates, any polarization of incident light can be generated. A third waveplate and a polarizer are
used to analyze the modulated light in any desired polarization basis. The modulated light is focused on a
detector. To obtain, for instance, the $J_{12}$ component, the polarization optics were rotated so that
the incident light was vertically polarized and the reflected horizontally polarized light was analyzed.
We later used the simpler setup that is shown in Fig.~\ref{fig:lcd-characterization}b. That setup has no
polarization optics that need to be rotated and, therefore, is less sensitive to alignment inaccuracies.
With the simplified setup only the $J_{12}$ component can be measured. For the modulation scheme that is
discussed in Section~\ref{sec:amplitude-phase}, such a partial characterization of the LCD is sufficient.

In both setups, $J_{12}$ was measured using a diffractive technique
that is comparable to the method described in
Refs.~\citealt{Zhang1994} and \citealt{Remenyi2003} but only
required detection of the $0^\text{th}$ order diffraction mode. We
assumed that all pixels of the LCD have the same Jones matrix. First
the absolute value of $J_{12}(V)$ was obtained by varying the
voltage over all pixels of the LCD simultaneously. Then, we
programmed the LCD with a binary grating with a duty cycle of 50\%
(a so called Ronchi grating). The voltage over the notches of the
grating was kept constant at $V_0$ while the voltage over the rules
was varied. The intensity in the $0^\text{th}$ diffraction order
responds as
\begin{equation}
\frac{I(V,V_0)}{|E^\text{in}_H|^2} =  |J_{12}(V_0)|^2 + |J_{12}(V)|^2 + 2 |J_{12}(V_0)||J_{12}(V)|\cos
\left[\arg J_{12}(V) - \arg J_{12}(V_0)\right],\label{eq:I-LCD-characterization}
\end{equation}

\begin{figure}
\centering
  \includegraphics[width=0.6\textwidth]{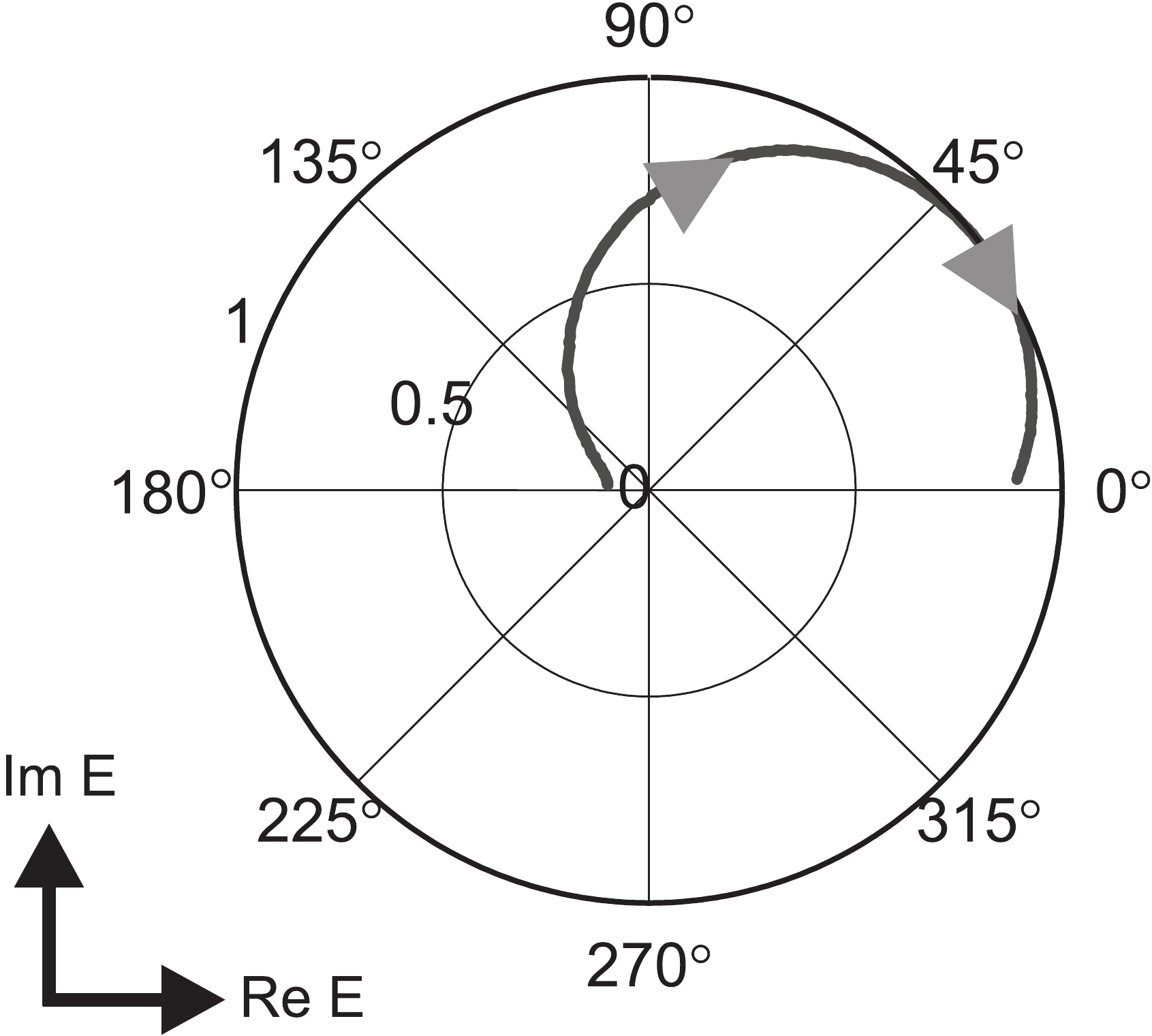}\\
  \caption{Polar plot of the modulator output amplitude
  with vertical polarization in, horizontal polarization out ($J_{12}$), at a wavelength of $\nm{532}$. The modulation voltage
  increases in the direction of the gray arrows.
  }\label{fig:results-lcd-modulation-curve}
\end{figure}

\noindent which gives us, in principle, enough information to obtain
the phase of $J_{12}(V)$ up to an overall phase offset. In practice,
however, the inversion of Eq.~\eqref{eq:I-LCD-characterization} is
very sensitive to noise for some values of $V$. We solved this
problem by measuring $I(V,V_0)$ for different values of $V_0$ (0\%,
25\%, 50\%, 75\% and 100\% of the maximum voltage). Then we combined
the data, using only measurements for which the inversion of
Eq.~\eqref{eq:I-LCD-characterization} was stable. The result of
these measurements is shown in
Fig.~\ref{fig:results-lcd-modulation-curve}. We find that the phase
of the reflected light changes from $\pi$ to $0$ with increasing
voltage. The amplitude increases from $0$ to a maximum at the `on'
state (around a phase of $35^\circ$) and then decreases slightly.

To obtain all four elements of the Jones matrix, the measurement was
repeated for each element of $J$. Moreover, the measurements were
also performed in a rotated basis with left and right hand
circularly polarized light. These extra measurements resolved the
relative phase of the different components of the $J$ matrix.

All in all, measuring the full Jones matrix of an LCD is cumbersome.
A complicating factor is the fact that all polarization states
(except for horizontal or vertical polarization) are changed by
reflecting off a coated mirror or passing through a beam splitter.
Theoretical analysis is complicated by the fact that in thin LCDs
surface effects start to play a role. Also, the total amount of
light that was reflected by the LCD was found to depend on the
voltage, which means that the Jones matrix is not unitary.
Therefore, many of the theoretical models that are used to describe
LCDs\cite{Aso1997, Davis1998, Fernandez-Pousa2000} can only give an
approximate result.

\subsection{Liquid crystal phase-mostly modulation\label{sec:phase-mostly}}
%
TN LCoS displays for intensity modulation typically achieve a maximum phase retardation of around $\pi$
for horizontally or vertically polarized light. Nevertheless, by choosing a suitable combination of
incident polarization and analyzer orientation, it is often possible to find an operating mode where the
phase retardation is $2\pi$ while the amplitude modulation is relatively low. This mode of operation is
called `phase-mostly' modulation and has been the subject of extensive experimental and theoretical
research\cite{Konforti1988, Lien1990, Zhang1994, Yamauchi1995, Bergeron1995, Dou1996, Davis1998,
Fernandez-Pousa2000, Davis2002, Remenyi2003, Chavali2007}.

\begin{figure}
\centering
  \includegraphics[width=\smallimage]{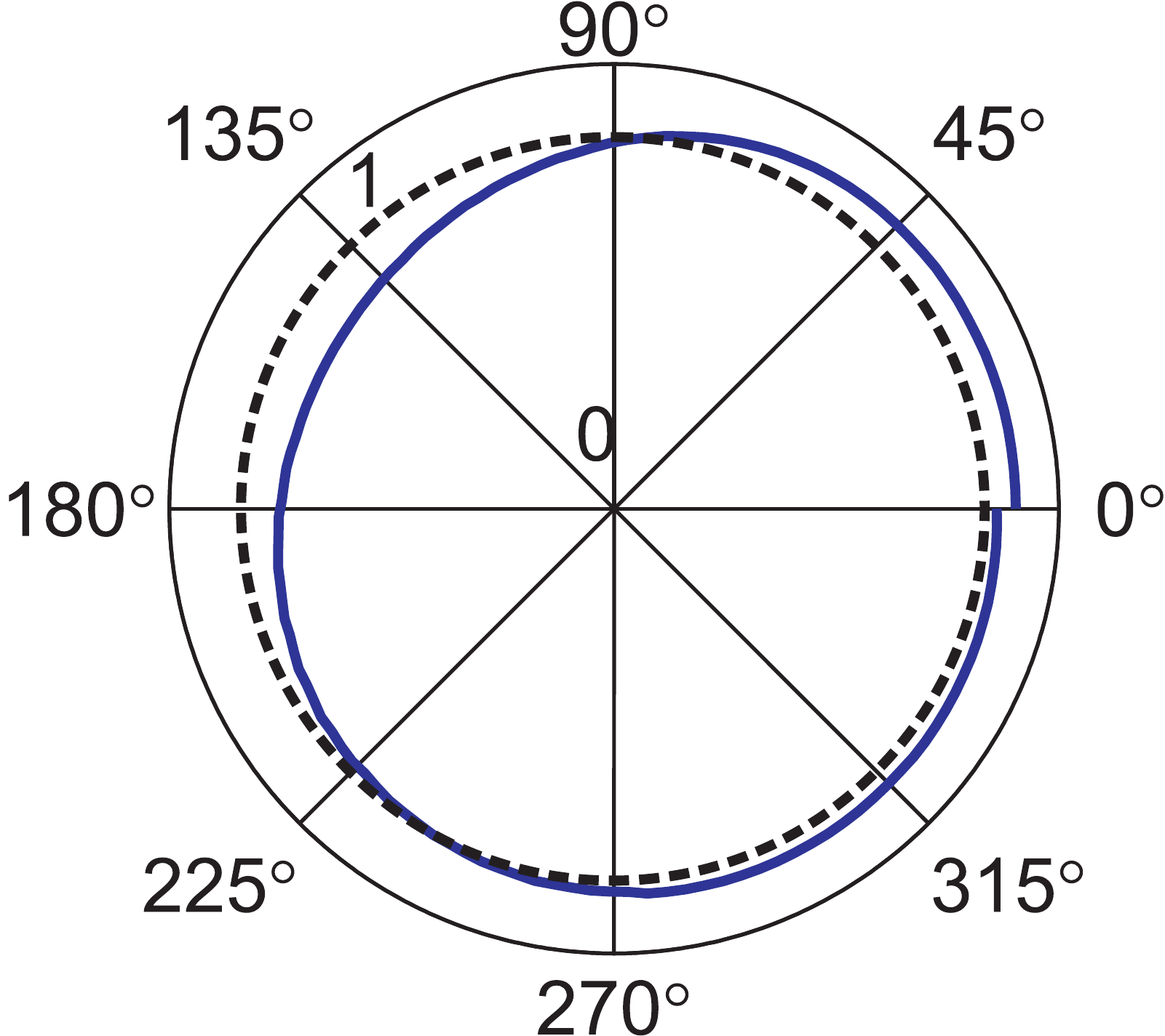}\\
  \caption{Polar plot of the modulator output amplitude. Dashed circle, perfect phase only modulation; solid curve, measured response
  in the configuration for optimal phase-mostly modulation at a wavelength of $\nm{632.8}$.}\label{fig:simulation-optimal-modulation-curve}
\end{figure}

The optimal configuration of the polarizers is different for each specific series of LCDs and also depends
on the wavelength of the light. In general, the optimal combination of polarization states is
elliptical.\cite{Davis2002} We found the optimal polarizations for a wavelength of $\nm{632.8}$ with a
brute force optimization method. This method used the measured Jones matrix of the LCD to numerically try
all configurations. In the optimal configuration the transmission is low and the system is very sensitive
to variations in the angles of the waveplates. The measured modulation curve (see
Fig.~\ref{fig:simulation-optimal-modulation-curve}) has an intensity variation of $21\%$ of the mean
value. This configuration was used successfully in the wavefront shaping experiments that are described in
Chapters~\ref{cha:focusing-through} and \ref{cha:algorithms}. In these experiments, a reference detector
compensated for the intensity variations.

\subsection[Decoupled amplitude and phase
modulation]{Decoupled amplitude and phase modulation\footnote{This section and the following section are
based on the article \emph{Spatial amplitude and phase modulation using commercial twisted nematic LCDs},
E. G. van Putten, I. M. Vellekoop, and A. P. Mosk, accepted for publication in Applied Optics
(2008).}}\label{sec:amplitude-phase}

\noindent In this section, we describe a novel modulation scheme
that uses a TN LCD in combination with a spatial filter to achieve
individual control over the phase and the amplitude of the light.
The method has four major advantages over the scheme described in
the previous section. Firstly, it allows separate control over
amplitude and phase of the wavefront. Secondly, the residual
amplitude-phase cross-modulation that occurred with the phase-mostly
modulation scheme is almost completely eliminated. Thirdly, the LCD
operates in a simple horizonal-in vertical-out configuration, this
saves components and characterization time, and makes the setup less
sensitive to align. And finally, the experimental setup can be
easily extended to control both polarizations.

Since the introduction of liquid crystals, many techniques have been developed to achieve combined
amplitude and phase modulation. Examples are setups where two LCDs are used to compensate amplitude-phase
cross modulation\cite{Neto1995, Kelly1998}, and double pixel setups where two pixels are combined to a
macropixel\cite{Birch2000, Arrizon2003, Bagnoud2004}. Each of this techniques has its own limitations. The
use of two LCDs introduces alignment and synchronization issues. Dual pixel schemes put specific demands
on the modulation capacities of the LCD, such as requiring phase-only modulation\cite{Bagnoud2004},
amplitude-only modulation\cite{Birch2000} or a phase modulation range of $2\pi$ \cite{Arrizon2003}.

\begin{figure}
\centering
  \includegraphics[width=.6\textwidth]{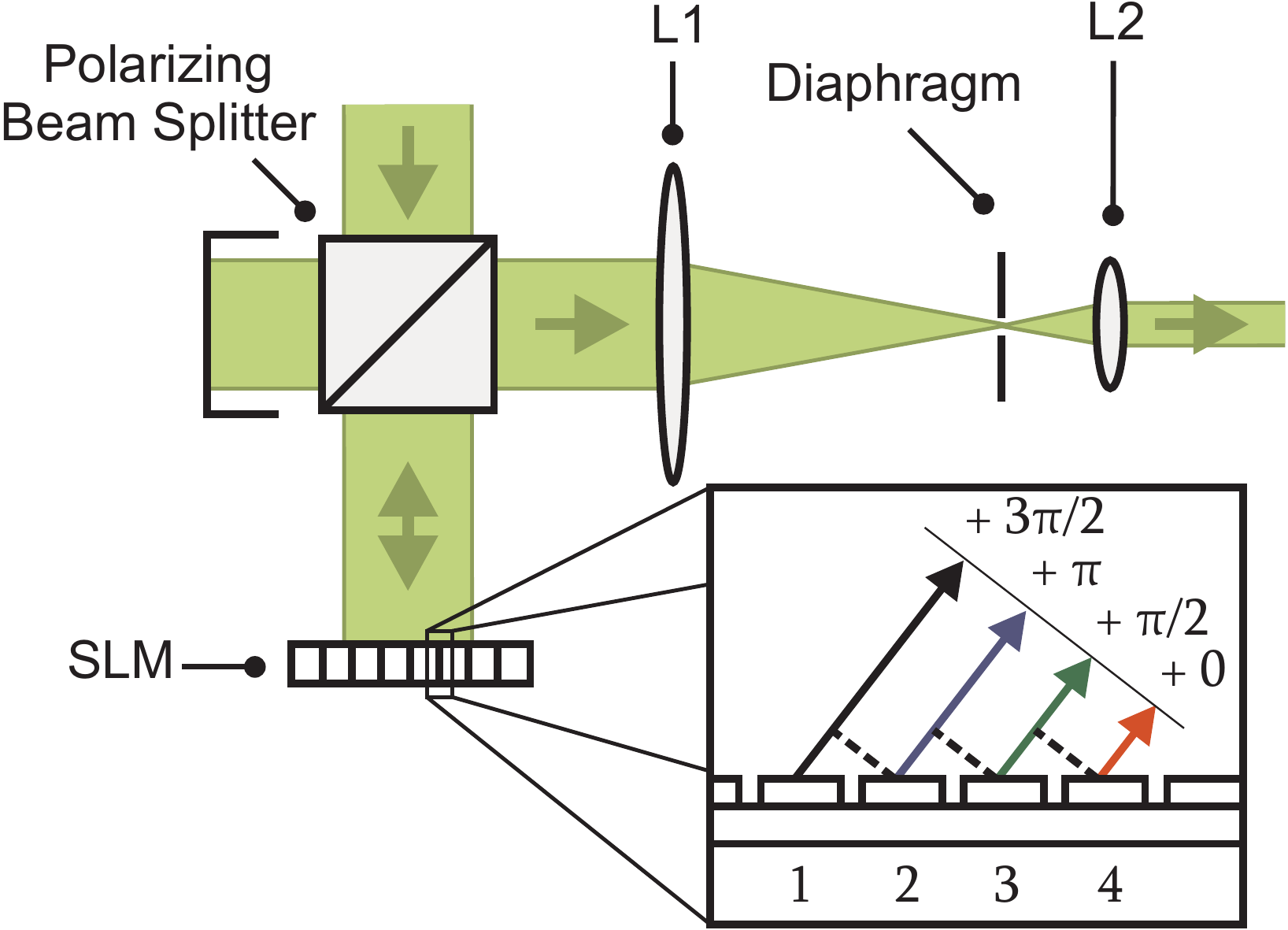}\\
  \caption{Experimental setup to decouple phase and amplitude modulation.
  Four neighboring pixels form one macropixel which can modulate amplitude
  and phase. In the plane of observation neighboring pixels are $\pi/2$ out
  of phase (inset). A spatial filter combines the
  pixels into one macropixel. $L_1$, $L_2$, lenses with a focal distance of $\mm{250}$
  and $\mm{200}$, respectively.}\label{fig:setup-phase-amplitude}
\end{figure}

We developed a method that combines four pixels into a macropixel. With this method, full spatial
amplitude and phase modulation can be achieved with any LCD. The setup used for this modulation scheme is
shown in Fig.~\ref{fig:setup-phase-amplitude}. A monochromatic beam of light is incident normal to the SLM
surface. The modulated light is reflected from the SLM. We choose an observation plane at which the
contribution of each neighboring pixel is $\pi/2$ out of phase, as is seen in the inset. A spatial filter
removes all higher harmonics from the generated field, so that four neighboring pixels are merged into one
macropixel.

By choosing the correct combination of pixel values, any complex value of the total field can be
synthesized. The electric field in a macropixel, $E_\text{sp}$, can be written as the sum of the fields,
$E_1$, $E_2$, $E_3$, and $E_4$, coming from the four different pixels. Behind the spatial filter,
$E_\text{sp}$ is given by
\begin{align}
E_\text{sp} &= E_1 \exp\left(\frac32 i\pi\right) + E_2 \exp(i\pi) + E_3 \exp\left(\frac12 i\pi\right)
+ E_4,\\
&= (E_{4r} - E_{2r}) + i (E_{3r} - E_{1r}) + (E_{1i} - E_{3i}) + i
(E_{4i} - E_{2i}),
\end{align}
where the indices $r$ and $i$ refer to the real and the imaginary
part of the field. The voltages on pixels 2 and 4 are chosen such
that $E_{4i} - E_{2i} = 0$, and in the same way the voltages on
pixels 1 and 3 are chosen such that $E_{1i} - E_{3i} = 0$. Equation
1 is now reduced to
\begin{equation}
E_\text{sp} = (E_{4r} - E_{2r}) + i (E_{3r} - E_{1r})
\end{equation}
The separate pixels are programmed so that the fields are given by
\begin{align}
E_1 &= -A\sin(i\phi) + i \Delta_1,\label{eq:lcd-macropixel-values-first} \\
E_2 &= -A\cos(i\phi) + i \Delta_2, \\
E_3 &= A\sin(i\phi) + i \Delta_1, \\
E_4 &= A\cos(i\phi) + i \Delta_2,\label{eq:lcd-macropixel-values-last}
\end{align}

\begin{figure}
\centering
  \includegraphics[width=\smallimage]{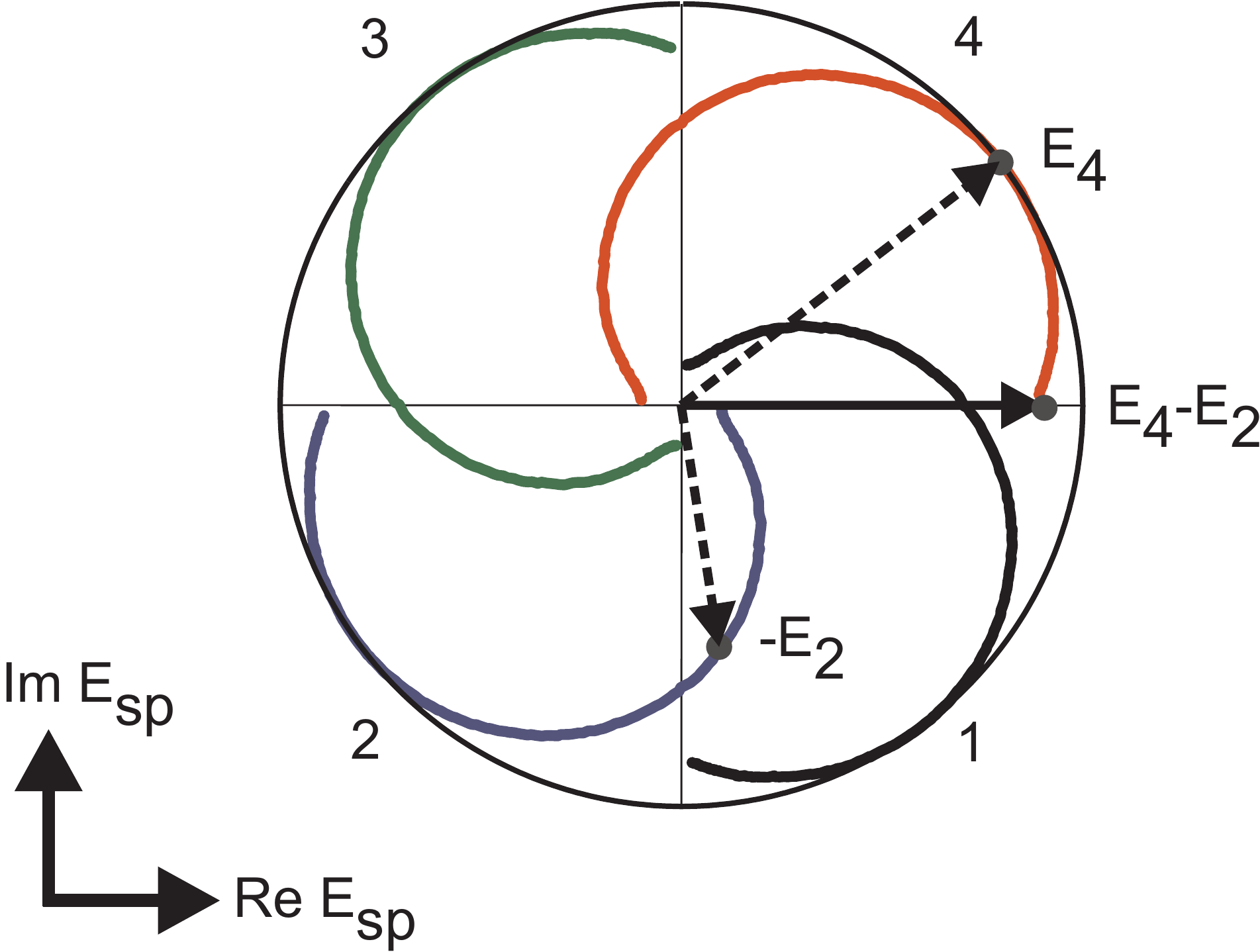}\\
  \caption{Modulation amplitude response of four pixels that form a macropixel.
  Pixels have a $\pi/2$ phase shift with respect to each other.
  The four pixels synthesize any complex value: $E_2$ and $E_4$ generate
  the real part of the field; $\Im (E_4$-$E_2) = 0$. $E_1$ and $E_3$ form
  the imaginary part (not shown).}\label{fig:macropixel-curves}
\end{figure}

\noindent with $A$ and $\phi$, respectively, the desired amplitude and phase. The imaginary parts $\Delta$
cancel. The desired complex value is now synthesized by the real values of the field at the four different
pixels. From the geometrical construction shown in Fig.~\ref{fig:macropixel-curves} it can be seen how we
modulate a value on the real axis, $E_{4r} - E_{2r}$, by choosing $E_{4i} = E_{2i}$. It is always possible
to find pixels values with exactly opposite imaginary parts and different real parts. The only requirement
posed on the SLM is that at least one of the field components has to vary when the pixel voltages are
changed.

The decoupled amplitude and phase modulation was used with a
$\lambda=\nm{532}$ diode pumped solid state laser\footnote{Coherent
Compass M315-100, max. 100~mW, intra cavity doubled, diode pumped
Nb:YAG laser.} for the experiments in
Chapter~\ref{cha:focusing-inside}. For the experiments in
Chapters~\ref{cha:diffraction-limit}, \ref{cha:communication} and
\ref{cha:dorokhov-experiment}, a $\lambda=\nm{632.8}$ helium neon
laser was used\footnote{JDS Uniphase 1125/P, 5mW polarized HeNe
laser.}. In Chapter~\ref{cha:dorokhov-experiment} the wavefront
synthesizer was extended to modulate two polarizations by simply
replacing the beam dump (see Fig.~\ref{fig:setup-phase-amplitude})
by a second LCD.

\subsection{Demonstration of amplitude and phase modulation\label{sec:amplitude-phase-experiment}}
We tested the new modulation technique with the same characterization method as was used to measure the
modulation curve of the LCD (see Section~\ref{sec:lcd-characterization}), with the difference that we now
use macropixels instead of actual pixels. We programmed the modulator so that the macropixels formed a
Ronchi grating. The notches were set at half of the maximum amplitude with a phase offset of zero. The
phase and amplitude of the rules was varied. A detector recorded the light intensity in the $0^\text{th}$
diffraction order\footnote{The $0^\text{th}$ diffraction order of the macropixel grating is the direction
of the modulated light when all \emph{macropixels} are set to the same amplitude and phase. This is not
the same as the $0^\text{th}$ diffraction order of the modulator, which is the direction of the light when
all \emph{actual pixels} are set to the same voltage.}. These experiments were performed at a wavelength
of $\nm{532}$.

\begin{figure}
\centering
  \includegraphics[width=\smallimage]{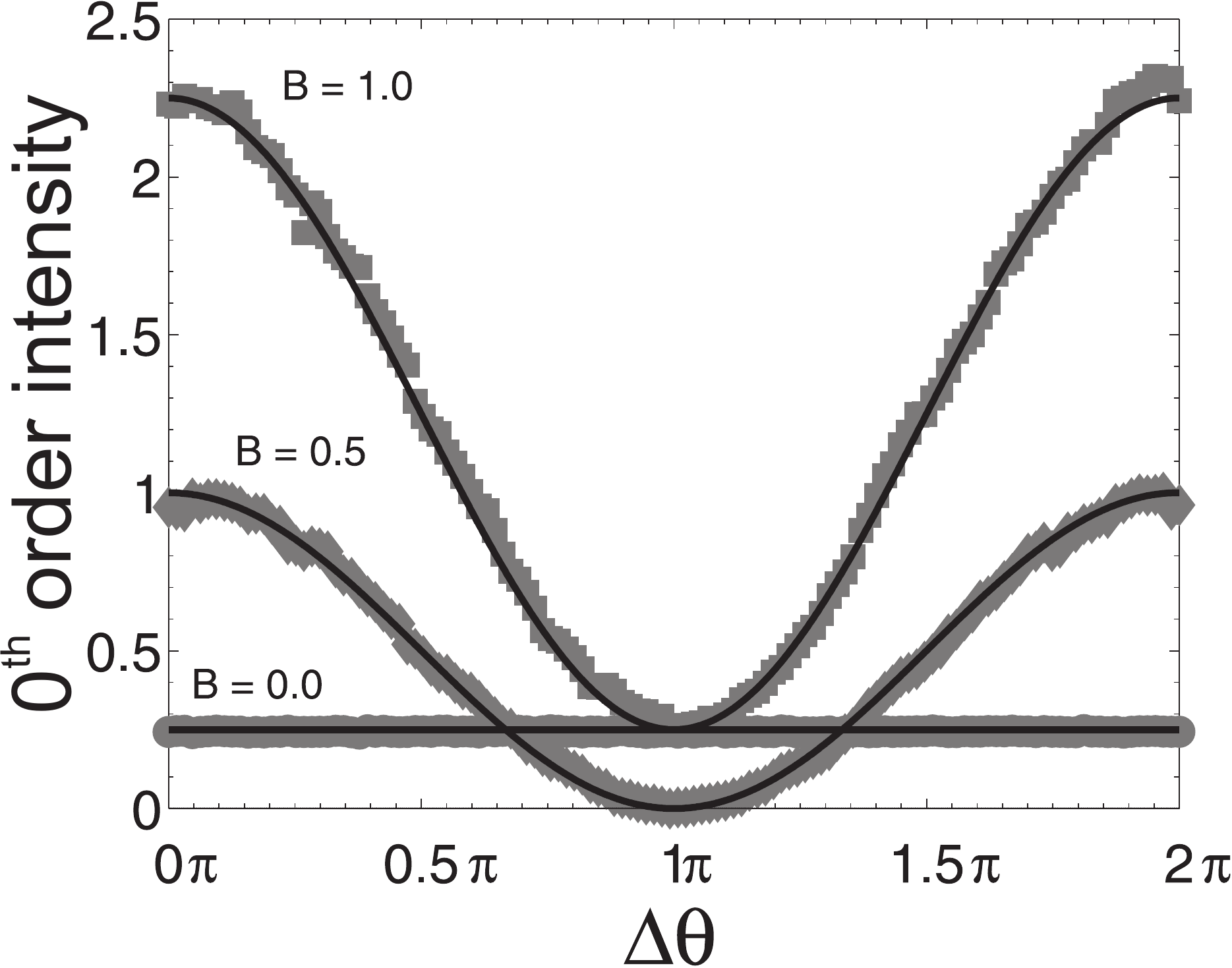}\\
  \caption{Intensity in 0$^\text{th}$ order grating mode as a function
  of the set phase difference $\Delta\theta \equiv \theta_B-\theta_A$.
  Notches of the grating are kept at amplitude $A=0.5$, phase $\theta_A=0$.
  Amplitude $B$ and phase $\theta_B$ of the rulers is varied. Solid
  curves, expected intensity for perfect modulation.
  Intensities are referenced to $I_0 = 4.56\cdot 10^3$ counts/second.}\label{fig:results-phi-amp-diffraction}
\end{figure}

We observed that the intensity in the $0^\text{th}$ diffraction
order varied as the cosine of the phase difference between the
notches and the rules of the grating, just as is expected from
Eq.~\eqref{eq:I-LCD-characterization}. We repeated the experiment
with different amplitudes in the rules of the grating. All
experimental result overlaps almost perfectly with the theoretical
curves (see Fig.~\ref{fig:results-phi-amp-diffraction}). From this
agreement, we conclude that a full $2\pi$ phase shift is achieved.
Moreover, at $\pi$ phase shift the intensity in the $0^\text{th}$
order vanishes, which indicates that the field of the notches has
the same magnitude and opposite sign as the field of the rules of
the grating.

\begin{figure}
\centering
  \includegraphics[width=\smallimage]{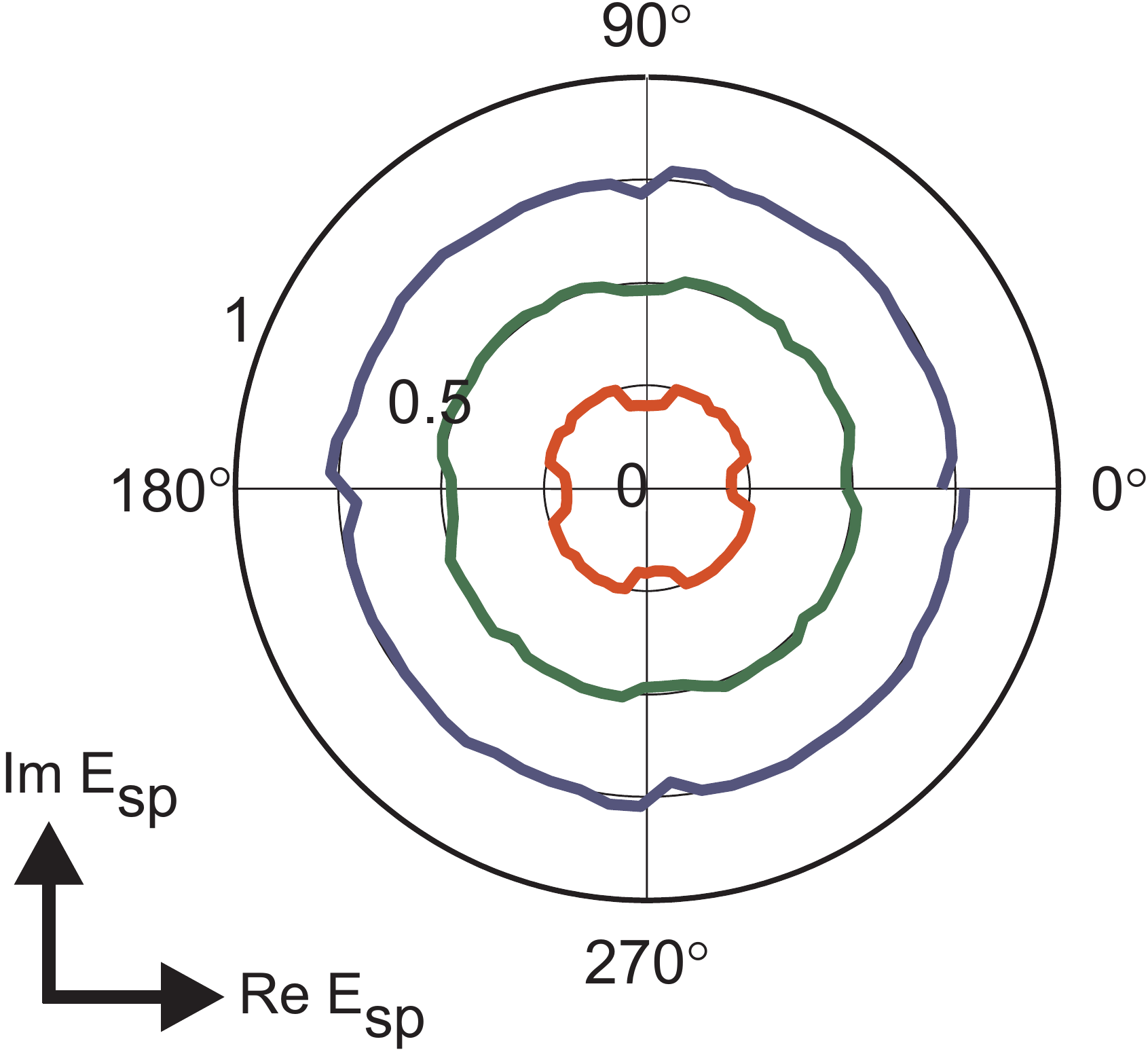}\\
  \caption{Independent phase and amplitude modulation. Curves show
  the measured relative amplitude $A/\sqrt{I_0}$ as a function of the
  programmed phase. Relative amplitudes are set to $0.25$, $0.5$, and
$0.75$. $I_0 = 19.7\cdot10^3$ counts/second.}\label{fig:results-2pi-constant-amplitude}
\end{figure}
To determine the amount of amplitude-phase cross-modulation quantitatively, we measured the intensity in
the $0^\text{th}$ diffraction order when all macropixels are set to the same field and amplitude. When the
phase is cycled from $0$ to $2\pi$, the observed intensity remains virtually equal. From the results shown
in Fig.~\ref{fig:results-2pi-constant-amplitude}, we find that the amplitude is constant to within
$2.5\%$, which is a significant improvement over the $21\%$ cross-modulation observed with the
phase-mostly modulation scheme. The small amplitude variations are periodic with a $\pi/2$ period. This
periodicity is understandable, since after $\pi/2$ rotation of the phase, the same pattern is programmed
on the LCD shifted by one pixel (see
Eqs.\eqref{eq:lcd-macropixel-values-first}-~\eqref{eq:lcd-macropixel-values-last}). Unlike the
phase-mostly modulation method, $0$ and $2\pi$ are equivalent, which means that there are no
discontinuities at phase wraps. The amplitude variations are probably the result of dynamic cross-talk
between neighboring pixels. This effect is discussed in the following section.

Although the amplitude of a macropixel is not affected by its phase, neighboring macropixels do effect
each other. A detailed analysis of this effect is found in Ref.~\citealt{Hsueh1978}. The effect is similar
to ordinary diffraction: the beams originating from the two macropixels expand through diffraction and
interfere with each other. It was found that, typically, a completely random wavefront carries
approximately $85\%$ of the intensity of a plane wave. The rest of the intensity is clipped by the iris
diaphragm. In Chapter~\ref{cha:focusing-inside} this effect was compensated for by measuring a reference
intensity with a random wavefront. In Chapter~\ref{cha:dorokhov-experiment} we used a more accurate method
where the sample is translated.

\subsection{Transient behavior\label{sec:lcd-transient}}

\noindent We now investigate how fast a pixel of the LCD can be switched. The image on the modulator is
updated with a refresh rate of 60 frames per second. To allow for real time operation, we reconfigured the
gamma lookup table of the LCD driver electronics. The table was configured so that a pixel value linearly
corresponds to an amplitude modulation on the real or imaginary axis. Furthermore, the conversion from a
polar representation (amplitude and phase) to the real and imaginary parts (see
Eqs.~\eqref{eq:lcd-macropixel-values-first}-\eqref{eq:lcd-macropixel-values-last}) is performed in real
time by the video acceleration hardware. In the process of displaying a new image the following sequence
of events can be identified:
\begin{enumerate}
  \item The control program loads new matrices for the amplitude and phase
  to the video hardware
  \item The video hardware scales and translates the matrices to
  screen coordinates. Then it performs the necessary calculations to
  convert amplitude and phase to pixel values and prepares the new image in a
  background buffer.
  \item The control program waits to just before the start of a new
  frame (the so called vertical retrace period)
  to swap the background buffer with the foreground buff\-er.
  \item After every vertical retrace the video hardware sends the
  image to the light modulator over a digital visual interface (DVI)
  link.
  \item The light modulator hardware receives the image and converts
  pixel values to voltages with the use of an internal gamma lookup table
  \item The light modulator hardware drives a matrix of transistors
  on the back of the LCoS display according to a pulse width modulation (PWM) scheme.
\end{enumerate}
To measure the transient behavior of the LCD, we repeatedly switched the whole display from the minimum to
the maximum voltage and back, each time waiting $\ms{500}$ between switches. A camera was configured to
measure the intensity in the center of the modulated beam with a shutter time of $\ms{1}$. The delay
between the vertical retrace and the camera trigger was varied from $\ms{0}$ to $\ms{120}$.

\begin{figure}
\centering
  \includegraphics[width=\smallimage]{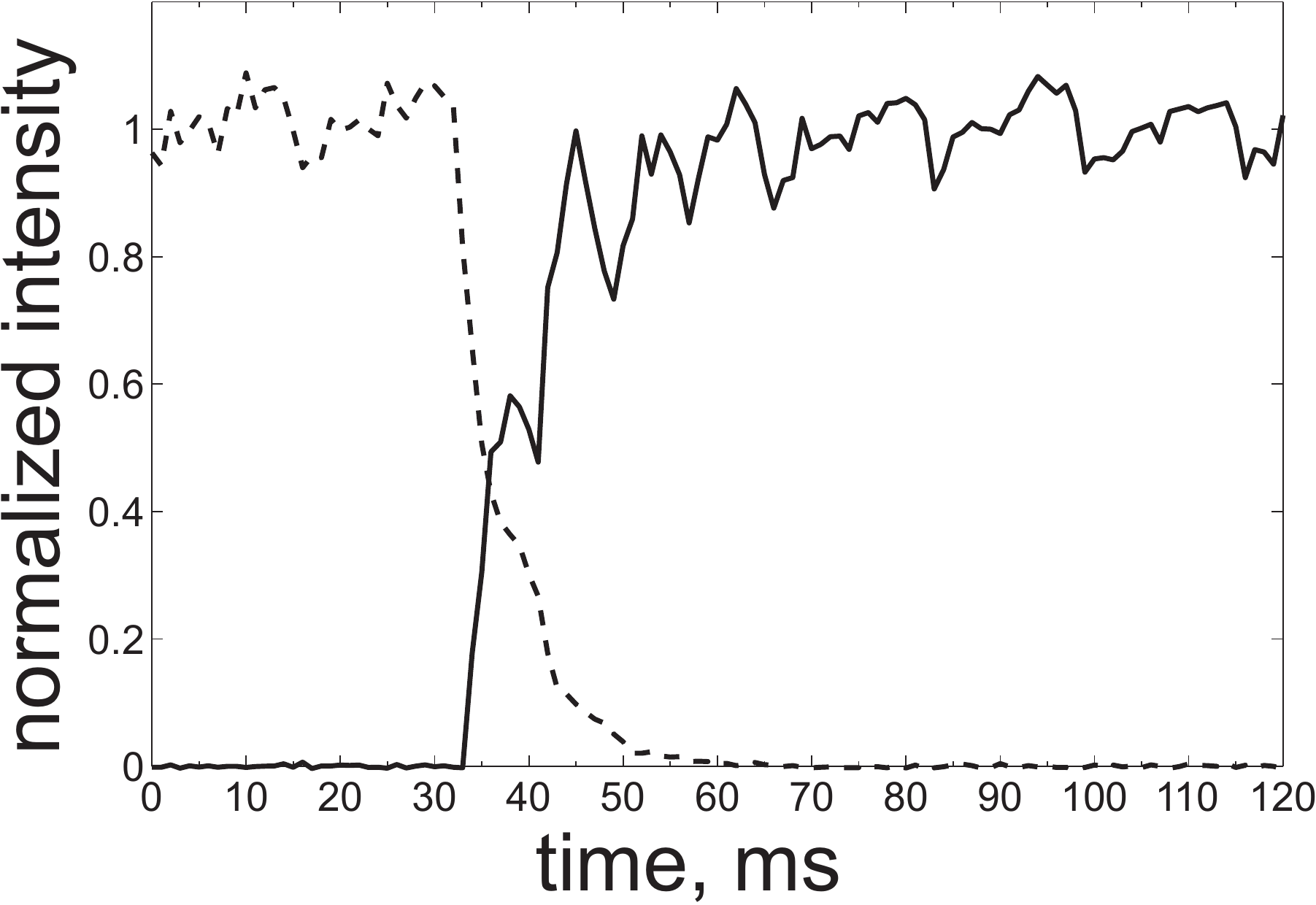}\\
  \caption{Transient intensity response of the LCD for switching the amplitude from $0$ to maximum (solid curve)
  and for switching it back to $0$ (dashed curve). During the first $\ms{33}$ (2 frames) the
  display does not respond at all. The intensity is normalized to the average intensity in the `on' state.}\label{fig:lcd-timing}
\end{figure}

Figure~\ref{fig:lcd-timing} shows the measured switching response of the LCD. We recognize three distinct
periods. During the first two frames (from $0$ to $\ms{33}$) the image on the LCD does not change. We call
this period the idle time $T_\text{idle}$. During the idle time the image is transferred from the computer
to the modulator. In the second period the voltage over the pixels changes and the liquid crystal
molecules reorient, which takes a certain time $T_\text{settle}\approx \ms{50}$. In the third period, the
liquid crystal molecules have reoriented completely. However, the signal still oscillates with a period of
exactly half a frame ($\ms{8.3}$). These oscillations are the result of how the LCD hardware drives the
pixels (see e.g. \cite{patent6067065}). Each pixel is switched on and off according to a PWM code. A
storage capacitor at each pixel integrates the total current to achieve the desired average voltage over
the liquid crystal. After half a frame, the controller reverses the voltage to avoid a DC current that
would damage the liquid crystal. This switching scheme results in rapid oscillations in the reflected
light.

When two neighboring pixels have a different voltage, there is a field gradient at their border. Due to
the pulse modulation scheme, the gradient will oscillate. The magnitude of this undesired effect depends
on the PWM code of each of the pixels. For example, consider the field response for a varying phase and a
constant amplitude of 0.25 (the smallest circle in Fig.~\ref{fig:results-2pi-constant-amplitude}). The
response curve shows small jumps in the amplitude at phase values of $-15^\circ$ and $+15^\circ$, these
jumps are repeated every $90^\circ$.

To understand these jumps, we examine the bit code of adjacent
pixels in a macro\-pix\-el. For a phase of $14^\circ$, pixel 1 and 2
have a value of $255$ and $230$. At a phase of $15^\circ$, the first
pixel values has changed to $256$ and the second pixel is still at
$230$. Although the change from $255$ to $256$ appears to be small,
these numbers have a completely different bit-pattern (011111111 and
100000000 respectively). Therefore, the field gradient between pixel
1 and pixel 2 will differ significantly between the two situations,
resulting in a jump in the field response.

The transient switching characteristics of the LCD and the temporal
oscillations put special demands on the timing of the detection.
This issue is addressed in Section~\ref{sec:detection}.

\subsection{Projecting the wavefront}
\begin{figure}
\centering
  \includegraphics[width=\textwidth]{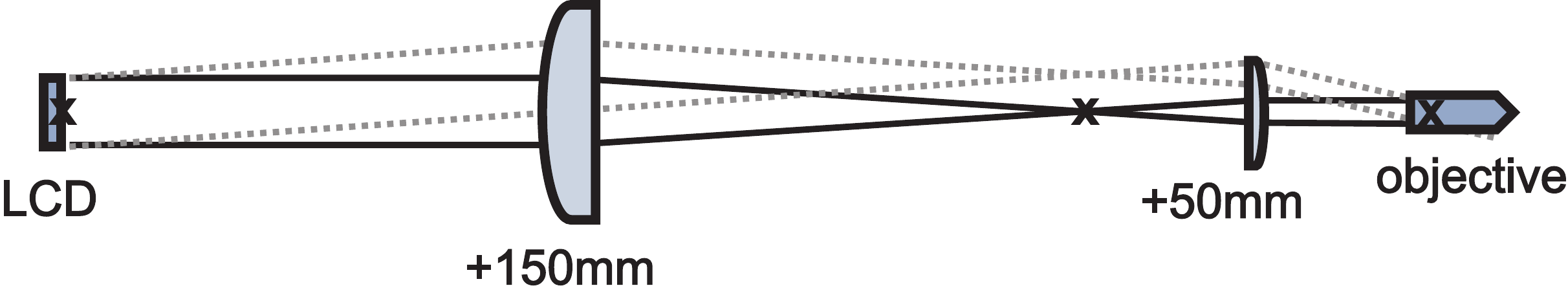}\\
  \caption{An imaging telescope is used to project the surface of the LCD
  to the entrance pupil of the microscope objective. Crosses indicate focal
  points of the lenses.}\label{fig:setup-telescope}
\end{figure}
\noindent In most of the experiments in this thesis, the shaped
wavefront is focused on the sample by means of a microscope
objective. The surface of the LCD is imaged onto the entrance
aperture of the microscope objective with an imaging telescope. When
combined phase and amplitude modulation is used (see
Section~\ref{sec:amplitude-phase}) there is a pinhole in the focus
of the telescope to spatially filter the generated wavefront. The
telescope also serves to demagnify the beam coming from the LCD to
the size of the microscope objective's aperture. It is essential
that a telescope with two positive lenses is used so that beams that
leave the LCD at an angle are also imaged onto the aperture of the
objective (see Fig.~\ref{fig:setup-telescope}). When the lenses are
aligned properly, the entrance pupil of the objective corresponds to
a circular area on the LCD. Pixels outside this area do not
contribute to the field on the sample and can, therefore, be skipped
in the optimization procedure.

In this configuration, a pixel on the LCD corresponds to an angle in
the focal plane of the objective. When we group pixels on the LCD
together in blocks, we reduce the angular resolution at the sample
surface and, thereby, reduce the size of the projected spot. The
aperture of the objective is always filled completely. Therefore,
when a high $\NA$ objective is used, the number of segments in the
incident wavefront is approximately equal to the number of
mesoscopic channels on the sample surface. If, for example, all
pixels are grouped into a single segment, light is focused to a
diffraction limited spot encompassing exactly one mesoscopic
channel.

If the sample is not positioned in the focal plane of the objective,
or when a low $\NA$ objective is used, the wavefront synthesizer
illuminates a spot that supports more scattering channels than there
are control segments in the incident wavefront. Thus the incident
field cannot completely be defined by the wavefront synthesizer. It
turns out that, in general, this limitation has little effect on how
well the propagation of light is controlled (see for instance
Fig.~\ref{fig:enhancement}).

\section{Detection\label{sec:detection}}
\noindent During the wavefront optimization procedure, a detector monitors the intensity in the target.
Different detectors were used for this purpose. In this section we discuss the detectors that were used,
as well as their relevant properties.

The first experiments were performed with a photodiode. A photodiode has an excellent dynamic range and a
fast response. The major drawback of using a photodiode is that it is very hard to select exactly a single
speckle. For this reason we started using cameras. During optimization the camera image is integrated over
a disk with a software defined radius and position. The diameter of the disk is chosen to be slightly
smaller than the speckles that are visible on the initial camera image. Using a camera also allowed us to
easily define multiple targets and to monitor the intensity in the background around the optimized
speckle.

We have used three different models of cameras. The first model is the Allied Vision Technologies Dolphin
F-145B. This camera is an all purpose charge coupled device (CCD) camera that is connected to the computer
with a IEEE 1394 link (firewire). To increase the dynamic range of the camera, the shutter time was varied
by the control program. In Chapter~\ref{cha:dorokhov-experiment} the required dynamic range was so high
that changing the shutter time was not sufficient. Instead, the computer controlled a motorized
translation stage that automatically placed a neutral density filter in front of the camera.

We used the Dolphin camera for all experiments, except for the
fluorescence measurements described in
Chapter~\ref{cha:focusing-inside}. For that experiment, we started
with a Hamamatsu ORCA electron multiplying CCD (EMCCD) camera. This
camera is cooled with a Peltier element to reduce the dark current.
Also, the signal is amplified on the CCD chip to overcome readout
noise when the signal is very weak. The Hamamatsu camera was
connected to a dedicated computer with a CamLink interface. However,
although the Hamamatsu camera has a high sensitivity, it was not
possible to measure small variations in the signal intensity. It
turned out that the overall background intensity of the camera image
varied on a frame by frame basis. This so called baseline drift
problem was solved by using a different EMCCD camera. The Andor Luca
DL658M is a cooled EMCCD camera that connects to the USB 2.0 bus. It
has a baseline clamping feature that eliminates almost all baseline
drift.

The linearity of all cameras was confirmed experimentally. We also
recorded a background image for every experiment. For most
experiments it was sufficient to subtract the average value of the
background image from the measured signal. However, the experiments
in Chapter~\ref{cha:dorokhov-experiment} required a measurement of
the exact intensity distribution over the whole camera. Therefore,
the full background image was subtracted pixel by pixel. Moreover,
in that experiment we corrected for the approximately 30\% lower
sensitivity of the camera close to the corners of the CCD chip. The
lower sensitivity is probably the result of a minute misalignment of
the microlenses on this chip.

\subsection{Timing\label{sec:measurement-timing}}
\noindent In all our experiments it turned out that the speed at which the optimal wavefront can be
constructed is the limiting factor. Therefore, we want to measure as quickly as possible. In
Section~\ref{sec:lcd-transient} we saw that the wavefront oscillates with a period of $\ms{8.3}$. To avoid
noise due to aliasing, the shutter time of the cameras ($T_\text{meas}$)was always set to a multiple of
this value.

In Section~\ref{sec:lcd-transient} we observed that it takes some
time for the image on the LCD to start changing ($T_\text{idle}$)
and then it takes some more time for the image to stabilize
($T_\text{settle}$). To maximize the number of measurements per
second and to further reduce aliasing effects, we synchronized the
camera with the wavefront synthesizer.

\begin{figure}
\centering
  \includegraphics[width=\mediumimage]{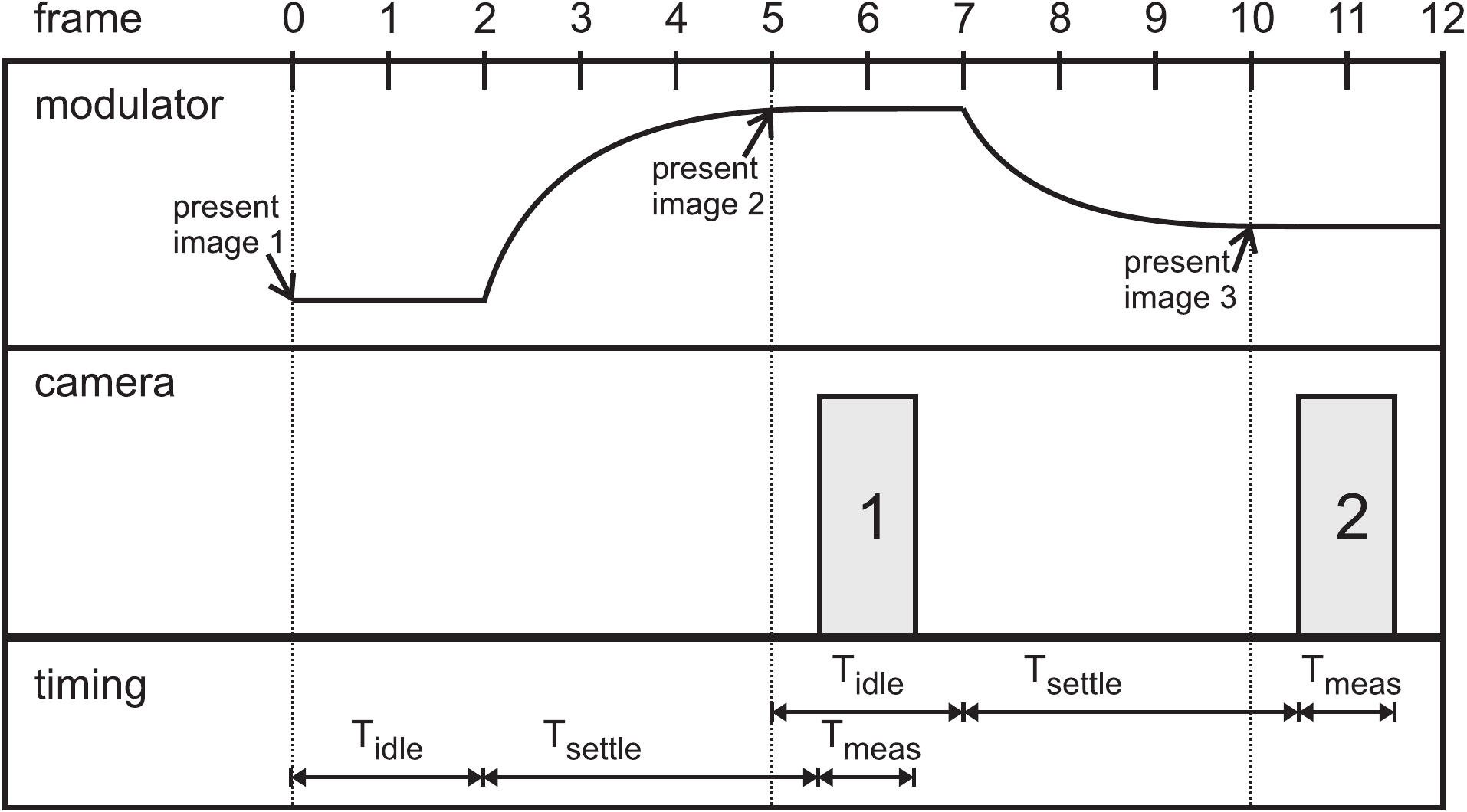}\\
  \caption{Timing diagram of a series of measurements. Time (in frames of $\ms{16.7})$ is indicated
  on the top axis. The curve in the topmost box symbolizes the switching behavior of the modulator. The
  numbered rectangles in the box below stand for the time that the camera is recording an image.
  In the lower box the relevant time intervals for synchronization are drawn.
  }\label{fig:measurement-synchronization}
\end{figure}

A timing diagram of a series of measurements is given in
Fig.~\ref{fig:measurement-synchronization}. In this diagram,
$T_\text{idle}=\ms{33}=2\;\text{frames}$ and
$T_\text{settle}=\ms{57}=3.4\;\text{frames}$. The first image is
presented during the vertical retrace period of the modulator (frame
0). At frame 2, the image starts to change. Then, at frame 5, the
second image is presented, although the measurement for the first
image has not even started. Since it takes two more frames for the
LCD to react, the image on the LCD is stable from frame 5 to frame
7. We trigger the camera after $T_\text{idle}+T_\text{settle} =
5.4\;\text{frames}$. Even if there is some jitter in the timing, the
camera has finished measuring before the image on the LCD starts
changing (frame 7).

With this tight timing scheme, we can perform a measurement each 5
frames (instead of each 7 frames). The idle time $T_\text{idle}$
only affects the first measurement of a sequence. In the experiments
we perform a sequence of 5 to 10 synchronized measurements for each
segment.

\section{Stability\label{sec:stability}}
\noindent The optimized wavefront is unique to a sample. When the sample is moved, the optimized wavefront
does not fit the sample anymore. Therefore, during the course of an optimization, the sample has to be
stable with sub-wavelength accuracy. The stability demands for the rest of the apparatus are high as well.
In this section we discuss the most important stability considerations for our experiments.

The whole apparatus was built with the requirements of interferometric stability in mind. We built it on
an actively levelled, damped optical table\footnote{TMC CleanTop II 780 series, 783-655-12R with 14-426-35
support.}. Moreover, only high quality opto-mechanical components were used\footnote{Most opto-mechanical
components were manufactured by Siskiyou and Thorlabs. For the beam expander, the laser, the LCDs, and the
sample, special stable mounts were designed.}. The sample was mounted on a flexure stage to avoid the
creep problems that are intrinsic to stages with ball bearings. The mechanical stability of the system was
tested by tapping the optical components while monitoring the position of the focused light with
sub-micron accuracy. If the focus did not return to exactly the same position after tapping a component,
the mount of the component was replaced by a more stable one.

The temperature and humidity of the setup were not controlled. Therefore, it is likely that thermal drift
and hygroscopic expansion negatively affect the stability. After switching on the system, it takes a few
hours to stabilize completely. During this time the laser, the cameras, the LCD, and the computer
controlled translation stages warm up. After this warmup time the total system is, typically, stable
enough to perform an optimization that takes an hour.

We found that air turbulence causes fluctuations in the feedback
signal. These fluctuations are in the order of a few percent in the
interesting frequency range of about 1~Hz. To reduce air turbulence,
a box was built around the experiment. This box reduces stray light
as well. As the laser is the most important source of hot, turbulent
air, it needed to be placed outside the box. Of course, all
components with fans (the cooled camera and the LCD control
electronics) were also placed outside the box.

Finally, there is a constraint of the stability of the wavelength of
the laser. To calculate the required stability, we estimate the
average path length through a sample. For example: a typical sample
has a mean free path of the path of $\ell=\mum{0.7}$, and a
thickness of $L=15\ell$. Then, the diffuse path length $s$ is in the
order of $s=15^2\ell=\mum{160}$. At a wavelength of $\nm{532}$, a
typical path is 300 wavelengths long. When a wavelength change
results in a $\pi$ phase shift over this path length, the wavefront
optimization fails. In this example, the wavelength needs to be
stable to within a nanometer (or expressed in inverse centimeters,
better than $31\;\text{cm}^{-1}$), which is absolutely no problem
for a temperature stabilized Nb:YAG laser or a HeNe laser.

\section{Samples\label{sec:samples}}
\noindent For our experiments we used a large diversity of strongly scattering objects. In many ways, our
method for controlling the propagation of light in such objects does not depend on the optical parameters
of the sample. There are, however, some practical limitations that restrict what samples we can use:

\begin{description}

\item[Stability] The optimal wavefront for focusing light through a strongly scattering object uniquely
depends on the exact configuration of the scatterers. Therefore, we are limited to solid samples. We
observed that after optimization the signal decreases with a typical timescale of approximately one hour,
a value that is probably limited by thermal drift in the setup rather than by the samples.

\item[Absorption] For our method to work, there has to be at least some light on the detector before the
optimization procedure is started. If the absorption in the sample
is too strong, there will be no feedback signal to optimize. For
most experiments it should be possible to use weakly absorbing
materials as long as the initial transmission is detectable.
However, for the experiment in Chapter~\ref{cha:dorokhov-theory}, we
expect that absorption decreases the effect that we are interested
in. Therefore, we only used non-absorbing materials in all our
experiments.

\item[Thickness] The optimization method detects the intensity in a single scattering channel. The number
of independent channels is roughly equal to $(2 L)^2/(\lambda/2)^2$,
where $L$ is the thickness of the sample and $\lambda$ is the
wavelength. Moreover, the total transmission scales as
$\ell_\text{tr}/L$, with $\ell_\text{tr}$ the transport mean free
path for light in the medium. Therefore, the total optical power in
a single speckle scales with $L^{-3}$. For this reason, most of our
samples were made relatively thin ($\sim \mum{10}$). However,
successful optimizations have been performed on samples that were up
to $\mm{1.5}$ thick (a baby tooth, see
Table~\ref{tab:enhancements}).

\item[Scattering strength] We want to study light propagation in non-absorbing opaque scattering media
where all transmitted light is diffuse. A scattering object is opaque when it is thicker than a few
transport mean free paths. We require that $L > 4\elltr$. In this regime, the fraction of ballistic
(non-diffuse) transmission is less than $\exp(-4)\approx 2\%$. To keep the number of channels as low as
possible, it is advantageous to use thin samples with a low $\elltr$. There are two ways to minimize
$\elltr$. First of all, the scale of the disorder should be comparable to the wavelength of light in the
medium. Secondly, the index contrast should be as high as possible. Good candidates for making strongly
scattering samples are pigment particles made of a high index material (TiO$_2$, ZnO). These particles
have a typical size of $\sim\nm{200}$.

\item[Flatness and homogeneity] Our method for controlling propagation of light does not depend on the
flatness or the homogeneity of the samples. However, for systematically analyzing the experimental results
it is highly desirable to have samples that have the same thickness over the whole sample area ($<20\%$
variations). Furthermore, the samples should be homogeneous in composition and certainly not have any
holes.

\item[Special requirement: doping] For the experiment that is described in
Chapter~\ref{cha:focusing-inside} it was required to place fluorescent nanospheres in the scattering
medium.

\item[Special requirement: substrate] The experiment in Chapter~\ref{cha:dorokhov-experiment} required
that the samples were on a thin glass substrate to allow two high $\NA$ microscope objectives to focus
both on the front and the back of the sample. Working without substrate all together was not possible
because the substrate provides structural stability to the sample.

\item[Special requirement: effective refractive index] The experiment in
Chapter~\ref{cha:dorokhov-experiment} also required the number of channels to be as low as possible.
Therefore, we used thin, strongly scattering samples. Moreover, we chose to use ZnO pigment in an air
matrix because of its relatively low effective refractive index. A low refractive index reduces reflection
at the sample boundaries and, thereby, reduces the size of the diffuse spot and the number of independent
scattering channels.

\end{description}

\noindent In conclusion, for most experiments the requirements on the samples are not very stringent and
allow for a wide range of materials to be used. We successfully applied our wavefront shaping method to
daisy petals, porous gallium phosphide, TiO$_2$ pigment, ZnO pigment, white airbrush paint, eggshell,
stacked layers of 3M Scotch tape, paper and even a baby tooth.

\subsection{Sample preparation method}
\noindent We developed a spray painting technique to fabricate
layers of strongly scattering material. The technique allows the
fabrication of flat, homogeneous layers with a thickness of around
$\mum{5}$ and more. Moreover, it is very easy to use different
materials or to dope the sample with fluorescent markers. We first
give the recipe for ZnO samples with embedded fluorescent
nanospheres. These samples were used for the experiment in
Chapter~\ref{cha:focusing-inside}. Then we explain how different
samples were made using the same technique.

\begin{description}
    \item[1. Substrate cleaning] Standard $\mm{40}\times\mm{24}$ microscope cover slips with a thickness
    of $\mum{160}$ were used as a substrate. The substrates were first rinsed with acetone to remove any
    residual organic material. Then they were rinsed with isopropanol, a solvent that is known to leave no
    drying stains, and left to dry.

    \item[2. Paint preparation] First, the nanosphere suspension\footnote{
        Duke Scientific red fluorescent nanospheres. Diameter $\mum{0.30\pm 5\%}$. Suspension with
        $1\%$ solids in water. $6.7\cdot10^{11}$ spheres / ml. Dyed with FireFly$^\text{TM}$
        excitation maximum $\nm{542}$, emission maximum $\nm{612}$.}
    was diluted by a factor of $10^5$ ($\ul{2.5}$ suspension in $\ml{250}$ water).
    Then, a suspension was made by mixing $\gram{2.5}$ of ZnO powder\footnote{
        Sigma-Aldrich Co. Zinc Oxide powder, $<\mum{1}, 99.9\%$ ZnO.
        With a scanning electron microscope (SEM) the average
        grain size was determined to be $\nm{200}$}
    with $\ml{7.3}$ water. The suspension was stirred on a roller bank for 30 minutes and then placed in an
    ultrasonic bath for 15 minutes. Finally, $\ml{0.73}$ of the diluted nanosphere solution was added.
    The suspension was again stirred for $30$ minutes and then placed in the ultrasonic bath for 15 seconds.

    \item[3. Spray painting] The paint was sprayed onto the substrate with an airbrush\footnote{Evolution
    Solo Airbrush from Harder \& Steenbeck, $\mm{0.4}$ needle diameter.}. The airbrush was operated at an
    air pressure of 2.3 bar. The paint was sprayed from approximately $\cm{20}$ distance to allow for a
    homogeneous coverage. The empty substrates were taped to an underground with an inclination of approximately
    $45^\circ$. Spraying covered these substrates with a thin wet film of paint. Directly after spraying, the
    samples were left to dry horizontally for two hours.
\end{description}

\begin{figure}
\centering
  \includegraphics[width=\wideimage]{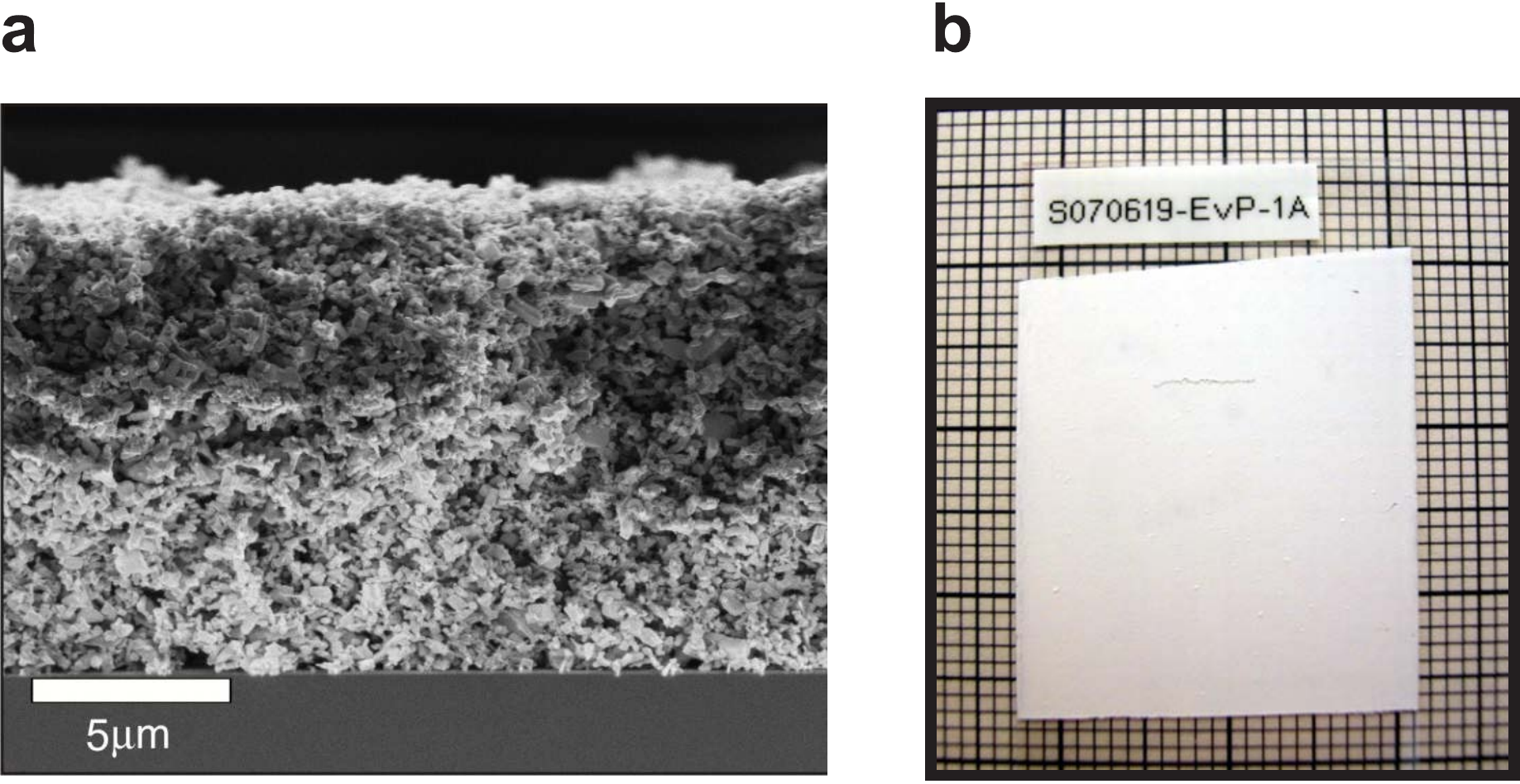}\\
  \caption{\subfig{a} Scanning electron microscopy image of a ZnO sample (side view).  Both
  the sample-air interface (top) and the sample-glass interface (bottom) are visible.
  \subfig{b} Photograph of a ZnO sample (on a background of millimeter paper).}\label{fig:ZnO-samples}
\end{figure}

\noindent The resulting samples are homogeneous, flat up to a variation of $\mum{1}$, and opaque (see
Fig~\ref{fig:ZnO-samples}). The adhesion between the ZnO and the glass is remarkably good, especially
since no binding agent was used at all. The thickness of the samples was determined around a scratch in
the sample surface using optical microscopy or Dektak profilometry. At $\lambda=\nm{532}$, the mean free
path is $\mum{0.7\pm0.2}$, which was determined from total transmission measurements (see
Section~\ref{sec:measurements-mfp}).

Various other suspensions were used to make samples. The most homogeneous samples were obtained with a
mixture of 2.5 parts airbrush paint\footnote{Hansa pro-color opaque white. Pigmented, water-based acrylic
polymer paint. Pigment: TiO$_2$ PW 6, CI\#77891 (specifications from manufacturer material safety data
sheet).} to 1 part acrylic medium\footnote{Schminke 50602 acrylic airbrush thinner}. We also created
samples using a suspension of 5~g of rutile TiO$_2$ pigment\footnote{Sachtleben Rutil R210 \#1060377/040}
in 10 ml airbrush thinner. Both the samples with airbrush paint and the samples with rutile pigment are
weakly fluorescent, which makes it hard to distinguish between the probe signal and the background
fluorescence. Therefore, for the experiments in Chapter~\ref{cha:focusing-inside} we used ZnO samples,
which do not fluoresce in the relevant wavelength area.

\section{Control program\label{sec:control-program}}
\noindent All elements of the experiment are controlled and synchronized by a control program. This
program manages frame grabbing, detector triggering, video hardware acceleration, optimization algorithms,
user input, realtime visual feedback, data storage and more. To keep a program of this scope maintainable
we used a component based development strategy. Instead of thinking of the program as a whole, we
separated it into self-contained components that offer a predefined service. An excellent introduction
into the ideas behind of component based programming and agile software development can be found in
Ref.~\citealt{Martin2003}.

Each component implements one or more interfaces. An interface is a specification of a set of properties
(publicly accessible data) and methods (executable functions) with a certain functionality. For example,
the interface of a camera component defines a method for triggering the camera, and it defines properties
such as the shutter time and the region of interest. Interfaces are defined in the most general terms
possible and hide all details of the implementation.

The only way to interact with a component is through its external interface. This gives the possibility to
create a different component that implements the same interface with different underlying details, for
example using a different low-level camera driver. From the outside, these two components are
indistinguishable. This means that we can exchange these two components transparently.\cite{Liskov1994}

The interchangeability of components also makes it possible to
extend the functionality of code without changing it. This
tantalizing idea is known as the Open-Closed
principle\cite{Martin1996}. For example, the first algorithm we
wrote for shaping a wavefront was developed with a photodiode
detector and a phase-mostly modulation configuration. Because of the
abstraction that was used, we were able to extend this algorithm to
work with an EMCCD camera and use the phase-amplitude modulation
scheme simply by connecting two different components to the
algorithm. Not a single line in the code of the algorithm was
changed.

\subsection{LabView and C++ mixed programming}
\noindent We chose a mixed programming environment for developing the control program. We used C++ to
implement reusable components for the detectors, algorithms and modulation schemes. Then, we used LabView
to construct the measurement procedure for a specific experiment. The rationale behind this separation to
make writing a measurement procedure as simple as possible.

LabView is a visual scripting environment that allows rapid development of simple measurement procedures.
Moreover, it is very easy to create a user interface for setting the parameters of the experiment and for
giving dynamic feedback in the form of graphs and false-color images. However, LabView is not suitable for
creating complex programs or for performing time critical operations. Also, LabView code usually has a
limited reusability.

C++, on the other hand, is a programming language that provides a low level control over the hardware and
allows for writing time critical code. Moreover, C++ is a structured programming language that offers many
mechanisms for code reuse and abstraction. Additionally, the Microsoft Visual Studio programming
environment allows easy navigation through a complex project and contains various useful tools such as an
excellent debugger.

\subsection{Component Object Model}
\noindent The Component Object Model (COM) is a standard for writing reusable components. A very
accessible introduction to COM is given in Ref.~\citealt{Brockschmidt1995}. COM was introduced by
Microsoft to enable interprocess communication and dynamic object creation in a way that is independent of
the programming language that is used. Nowadays, COM and ActiveX (one of the extensions to COM), are
supported on various operating systems.

COM/ActiveX components are listed in the Windows registry. The registry contains information about the
location of the components executable code, as well as a description of the interface(s) of the component.
In almost any modern programming language, including LabView, it is possible to select a component from
the registry and use it as if it was written in that programming language. It is even possible create a
COM object on a remote computer and access it as if it is a local object. The operating system makes sure
that all function calls are transparently relayed over the network connection. We successfully used this
feature to access the Hamamatsu EMCCD camera that was operated by a dedicated computer.

\subsection{Global structure of the control program}
\begin{figure}
\centering
  \includegraphics[width=\wideimage]{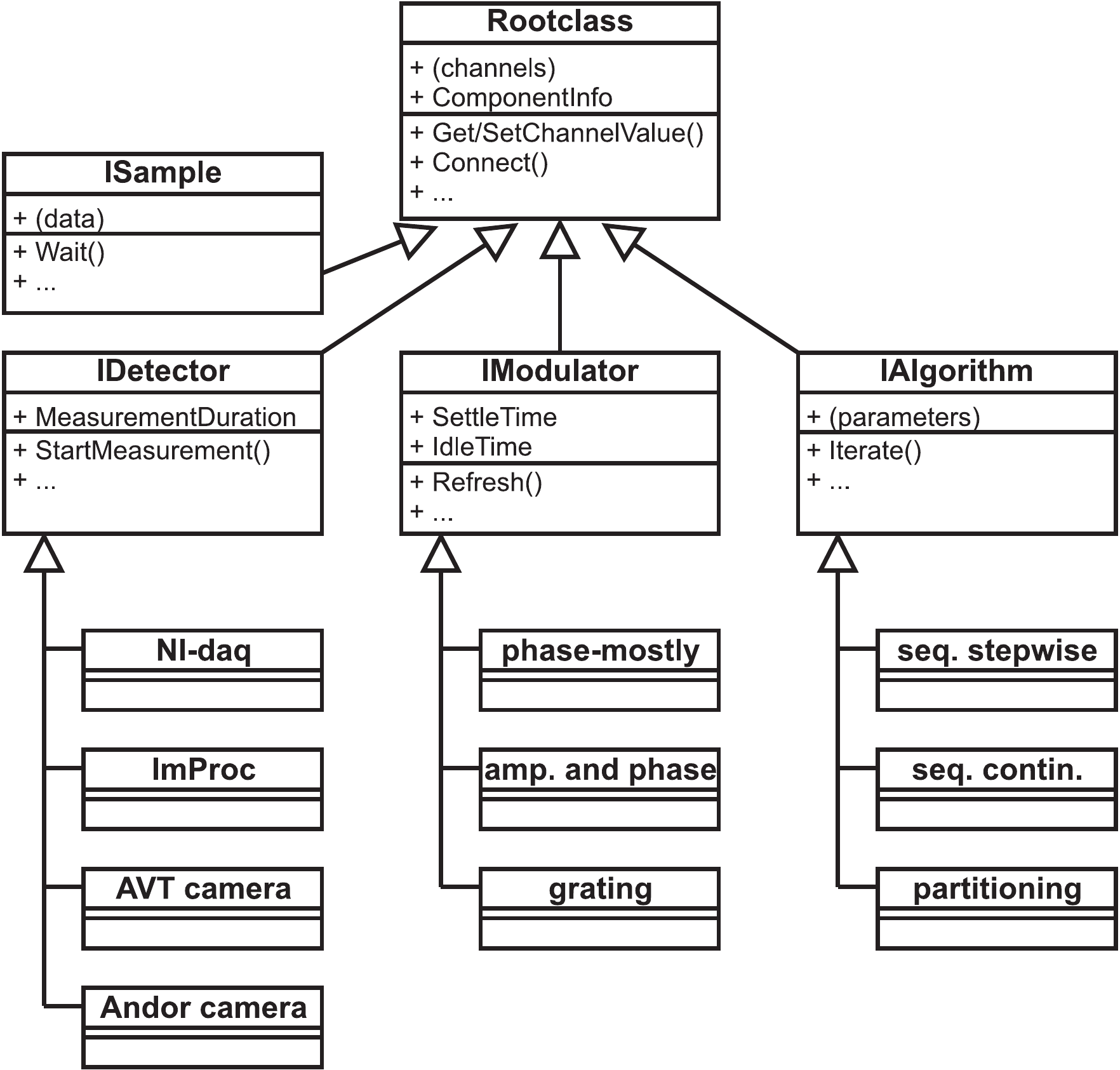}\\
  \caption{Simplified class diagram for the most important components
  of the control program. Blocks indicate classes. The top part of a box
  contains the class name, the middle part contains its properties
  and the lower part its methods. Arrows denote inheritance.
  E.g. AVT camera is an implementation of IDetector,
  which, in turn, derives from Rootclass.
  }\label{fig:uml-class-diagram}
\end{figure}
\noindent We defined four interfaces for describing different functions in the control program. These
interfaces are `IDetector' (for accessing detectors and cameras), `IModulator' (for programming the light
modulator), `IAlgorithm' (that implements a wavefront shaping algorithm) and `ISample' (for storing and
transferring measurement data). All interfaces have some common functionality that is embedded in a common
root class. A simplified class diagram of the interface hierarchy is shown in
Fig.~\ref{fig:uml-class-diagram}. We first discuss these interfaces and then explain how these components
are connected to form a measurement program.

\begin{description}
    \item[Rootclass (IDataAccess)] We constructed a common base interface for all components. It defines functions for
    accessing the component's data in a uniform way. All publicly accessible data of a component is
    organized in named `channels'. Access to the data is, in many ways, independent of the data type.
    For example
    \verb'SetChannelValue("Shutter", 1000)' sets the shutter time of a camera to $\us{1000}$
    and \verb'SetChannelValue("Phase", 0)' sets the phase of all segments on a modulator to 0.
    Some data channels represent connections to a different component (more on this below).\\
    The property \verb"ComponentInfo" contains a text string with the description of the component,
    including the names and values of all data channels as well as the information of all connected components.
    All measurement programs save this string to the disk so that the values of all parameters in the experiment
    are stored.
    \item[IDetector and ISample] The detector interface encompasses everything that returns a signal. Its channels correspond
    to settings of the detector, such as the shutter time for a camera. When \verb"StartMeasurement()" is
    called, the detector returns a sample object containing all measured data. Data acquisition happens
    asynchronously: the sample object is returned before the measurement and the data
    processing is complete. Only when \verb"GetChannelValue()" is called on one of the data channels of
    the sample the computer waits until that data is available. It is also possible to explicitly wait for
    the measurement to finish by calling \verb"Wait()" on the sample.\\
    The detector has a property \verb"MeasurementDuration", which is the maximum time that the
    measurement will take. This property is used for synchronizing the modulator and the measurements.\\
    There are many implementations of the IDetector interface. We wrote a component for each of the types
    of cameras and for the National Instruments data acquisition hardware (NI-daq). We also wrote image processing
    components (ImProc) that analyze a camera image and return a scalar value. Such `virtual
    detector' components are completely interchangeable with real detectors since they have the same
    interface.
    \item[IModulator] The modulator interface standardizes a light modulator. A modulator
    component has channels that contain matrices for the phase and, depending on the modulation scheme,
    amplitude. The properties \verb"SettleTime" and \verb"IdleTime" affect the synchronization of the
    measurements (see Section~\ref{sec:measurement-timing}).
    The function \verb"Refresh()" is called to display the new image after the phase and/or amplitude channels have been changed.\\
    For testing and characterization purposes, different types of modulator components have been created,
    such as the one that displays a Ronchi grating.
    \item[IAlgorithm] The algorithm component contains code for the feedback procedure. The
    three different algorithms that are listed in Fig.~\ref{fig:uml-class-diagram} are discussed in
    Chapter~\ref{cha:algorithms}. The algorithm is operated by connecting a detector and a modulator
    object and then repeatedly calling \verb"Iterate()".
\end{description}

\begin{figure}
\centering
  \includegraphics[width=\wideimage]{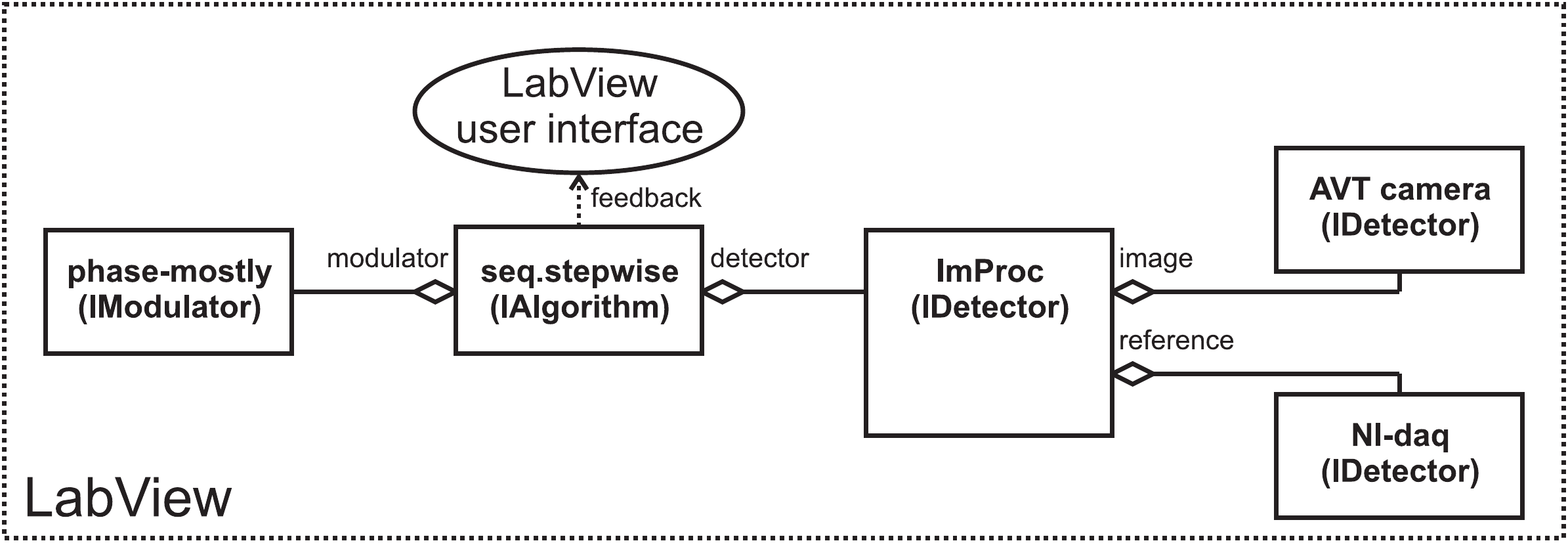}\\
  \caption{Structure of a typical wavefront optimization program.
  Rectangles, ActiveX components written in C++ are instantiated in a
  LabView environment and connected to each other;
  lines, connection between a component and a channel (diamond) on a different
  component;
  ellipse, a user interface constructed in LabView program
  provides feedback during optimization.
  }\label{fig:detector-tree}
\end{figure}

\noindent The LabView program instantiates the required objects and sets parameters such as the shutter
time, the number of segments, the position of the target, etc. Then, the program connects the components
together. The structure of a typical wavefront optimization program is shown in
Fig.~\ref{fig:detector-tree}. A modulator and an image processing node are connected to the algorithm. The
image processing node takes two inputs, one for the camera and one for a reference detector. The node
averages the camera image over the target area and divides this value by the reference signal to
compensate for variations in the incident intensity. The image processing node also takes care that the
camera and the reference detector are triggered simultaneously. A LabView user interface displays the
current status of the algorithm (percentage done, optimal wavefront settings measured so far, etc.).

It proved to be relatively easy to quickly build a program with these predefined components. Moreover,
most of the LabView code could also be reused in different experiments by simply exchanging components.
The design is open for extension; different cameras, algorithms, target functions and modulation schemes
were added without affecting other parts of the program. All in all, the flexible design helped in quickly
setting up an experiment.

\section{Conclusions and outlook\label{sec:outlook}}
\noindent In this chapter, we described the setup that was used for our wavefront shaping experiments. The
complete apparatus was designed from scratch. We discussed the considerations that played a role in
designing the experiment. Especially the wavefront synthesizer turned out to be more involved than the
cartoon in Fig.~\ref{fig:intro-principle} suggests. We first used a state-of-the-art method to achieve
phase-mostly modulation with a commercial liquid crystal display. Then, we developed a new method that
allows full control over both phase and amplitude of the light. We also discussed a versatile fabrication
method that we developed for making samples. This method allowed us to make samples with different
pigments and to embed fluorescent nanospheres.

We wrote a control program that executes the optimization algorithm and handles all data flow in the
experiment. The program was designed using a component based development strategy. All detailed
programming code was packaged in reusable and interchangeable components. Therefore, it was easy to
quickly try alternative measurement schemes, to test new algorithms, and to replace detectors and cameras
in a late stage of the project.

The setup that is discussed in this chapter is capable to create an
optimal wavefront in a matter of minutes. Although this is an
impressive achievement by itself, our method cannot yet be applied
to media with rapidly moving scatterers such as perfused skin or
turbid fluids. We believe that it is possible to speed up the
measurements considerably. Currently the speed is limited by the
response time of the LCD ($\sim \ms{50}$). However, recently
adaptive optics systems for turbulence correction have already been
demonstrated to reach a correction speed of 800~Hz.\cite{Baker2004}
Modulators that use ferroelectric liquid crystals (FLC) or
transparent lanthanum-modified lead zirconate titanate (PLZT) as a
switching medium can have a response time that is as short as
$\us{3}$. Even faster switching is achieved with micromachined
mirrors; such light modulators were shown to have a response
frequency of up to 10~MHz.\cite{Griffith2007} Typical frequencies
for a centimeter of living human tissue are in the order of
$1-100$~kHz\cite{Vakoc2005,Li2005} and we believe that these rates
can be achieved with our method in the near future.

\enlargethispage*{0.5\baselineskip}
\bibliography{../../bibliography}
\bibliographystyle{Ivo_sty}

\setcounter{chapter}{2}

\chapter{Focusing coherent light through opaque strongly
scattering media\label{cha:focusing-through}}
%
\begin{abstract}
We report focusing of coherent light through opaque scattering materials, by control of the incident
wavefront. The multiply scattered light forms a focus that is up to a factor 1000 brighter than the normal
diffuse transmission.

\noindent[This chapter has been published as: I.~M. Vellekoop and A.~P. Mosk, Opt. Lett. \textbf{32},
2309--2311 (2007)]
\end{abstract}

\enlargethispage{-\baselineskip}

\noindent Random scattering of light is what makes materials such as white paint, milk, or human tissue,
opaque. In these materials, repeated scattering and interference distort the incident wavefront so
strongly that all spatial coherence is lost.\cite{Sebbah2001} Incident coherent light diffuses through the
medium and forms a volume speckle field which has no correlations on a distance larger than the wavelength
of light. The complete scrambling of the field makes it impossible to control light propagation using the
well established wavefront correction methods of adaptive optics (see e.g. \cite{Tyson1998}).

We demonstrate focusing of coherent light through disordered scattering media by the construction of
wavefronts that invert diffusion of light. Our method relies on interference and is universally applicable
to scattering objects regardless of their constitution and scattering strength. We envision that, with
such active control, random scattering will become beneficial, rather than detrimental, to
imaging\cite{Sebbah2001} and communication\cite{Derode2003,Simon2001,Lerosey2007}.

\begin{figure}
  \centering
  \includegraphics[width=\wideimage]{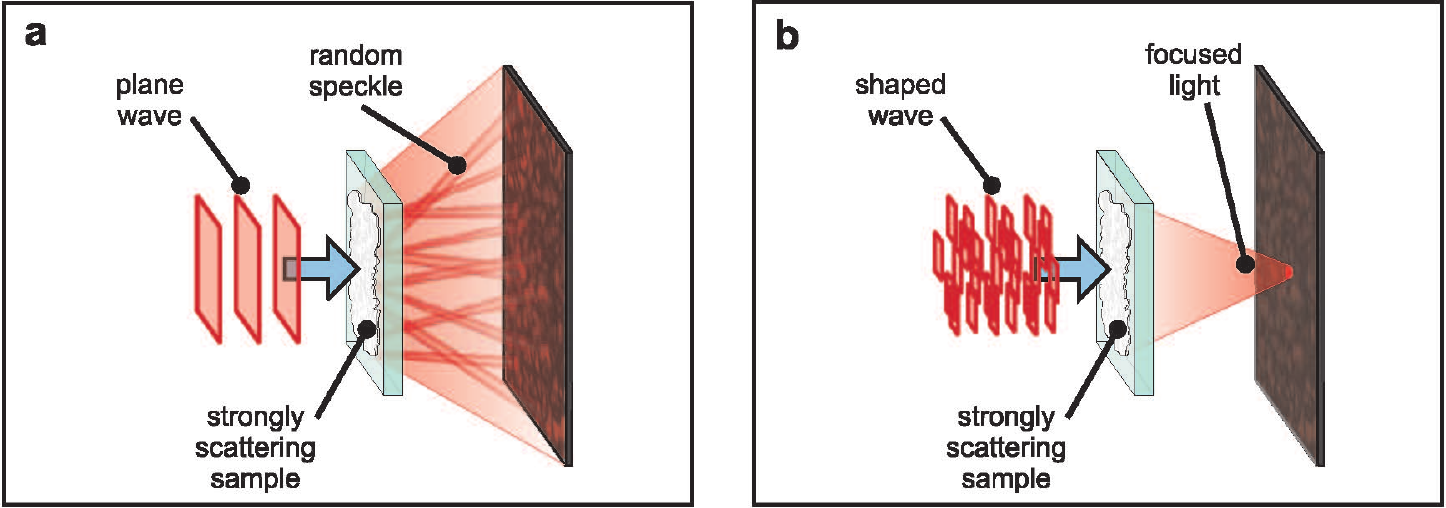}\\
  \caption{Design of the experiment. \subfig{a} A plane wave is focused on a disordered medium, a speckle pattern
  is transmitted. \subfig{b} The wavefront of the incident light is shaped so that scattering makes the light
  focus at a predefined target.} \label{fig:overview}
\end{figure}

Figure~\ref{fig:overview} shows the principle of the experiment.
Normally, incident light is scattered by the sample and forms a
random speckle pattern (Fig.~\ref{fig:overview}a). The goal is to
match the incident wavefront to the sample, so that the scattered
light is focused in a specified target area
(Fig.~\ref{fig:overview}b).

\section{Experiment}
\begin{figure}
  \centering
  \includegraphics[width=\mediumimage]{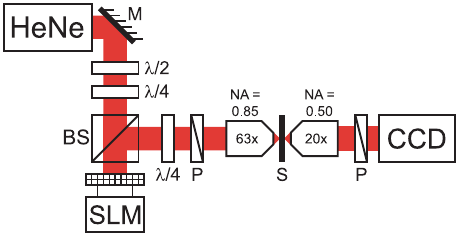}\\
  \caption{Schematic of the apparatus. A 632.8 nm HeNe laser beam is expanded and reflected off a
  Holoeye LC-R 2500 liquid crystal spatial light modulator (SLM). Polarization optics select a phase-mostly
  modulation mode. The SLM is imaged onto the entrance pupil of the objective with a 1:3 demagnifying lens
  system (not shown). The objective is overfilled; we only use segments that fall inside the pupil. The
  shaped wavefront is focused on the strongly scattering sample (S) and a CCD-camera images the transmitted
  intensity pattern. $\lambda/4$, quarter wave plate. $\lambda/2$, half
  wave plate. M, mirror. BS, 50\% non-polarizing beam splitter. P, polarizer.}
  \label{fig:ch3_setup}
\end{figure}

\noindent The experimental setup for constructing such wavefronts is
shown in Figure~\ref{fig:ch3_setup}. Light from a 632.8 nm HeNe
laser is spatially modulated by a liquid crystal phase modulator and
focused on an opaque, strongly scattering sample. The number of
degrees of freedom of the modulator is reduced by grouping pixels
into a variable number ($N$) of square segments. A charge coupled
device (CCD) camera monitors the intensity in the target focus and
provides feedback for an algorithm that programs the phase
modulator.

\sloppy

We performed first tests of inverse wave diffusion using rutile (TiO$_2$) pigment, which is one of the
most strongly scattering materials known. The sample consists of an opaque, $\mum{10.1}$ thick layer of
rutile\cite{Kop1997} with a transport mean free path of $\mum{0.55 \pm 0.10}$ measured at $\lambda =
632.8$ nm. Since in this sample the transmitted light is scattered hundreds of times, there is no direct
relation between the incident wavefront and the transmitted image.\cite{Pappu2002,Goodman2000}

\myfussy

\begin{figure}
  \centering
  \includegraphics[width=\textwidth]{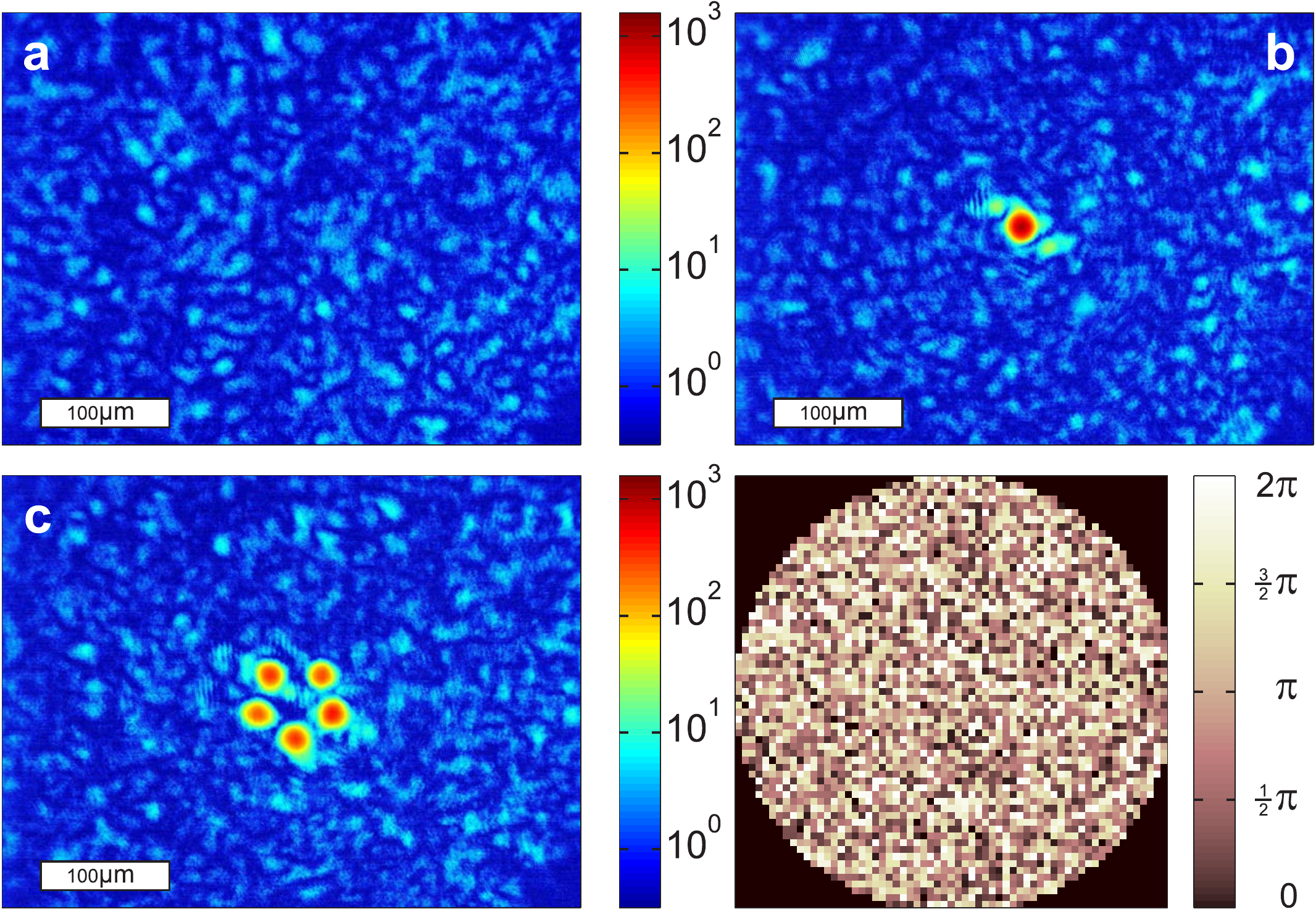}\\
  \caption{Transmission through a strongly scattering sample
  consisting of TiO$_2$ pigment. \subfig{a} Transmission micrograph with an unshaped
  incident beam. \subfig{b} Transmission after optimization for focusing at a single target.
  The scattered light is focused to a spot that is
  1000 times brighter than the original speckle pattern. \subfig{c}
  Multi-beam optimization. The disordered medium generates five sharp foci at
  the defined positions. Figures~\ref{fig:speckle}a to \ref{fig:speckle}c are presented on the same logarithmic color
  scale that is normalized to the average transmission before optimization.
  \subfig{d} Phase of the incident wavefront used to form Fig.~\ref{fig:speckle}c.} \label{fig:speckle}
\end{figure}

In Fig.~\ref{fig:speckle} we show the intensity pattern of the
transmitted light. In Fig.~\ref{fig:speckle}a we see the pattern
that was transmitted when a plane wave was focused onto the sample.
The light formed a typical random speckle pattern with a low
intensity. We then optimized the wavefront so that the transmitted
light focused to a target area with the size of a single speckle.
The result for a wavefront composed of 3228 individually controlled
segments is seen in Fig.~\ref{fig:speckle}b, where a single bright
spot stands out clearly against the diffuse background. The focus
was over a factor 1000 more intense than the non-optimized speckle
pattern. By adjusting the target function used as feedback it is
also possible to optimize multiple foci simultaneously, as is shown
in Fig.~\ref{fig:speckle}c where a pattern of five spots was
optimized. Each of the spots has an intensity of approximately 200
times the original diffuse intensity. In Fig.~\ref{fig:speckle}d we
show the phase of the incident wavefront corresponding to
Fig.~\ref{fig:speckle}c. Neighboring segments are uncorrelated,
which indicates that the sample fully scrambles the incident
wavefront.

\section{Algorithm}
The algorithm that constructs the inverse diffusion wavefront uses the linearity of the scattering
process. The transmitted field in the target, $E_m$, is a linear combination of the fields coming from the
$N$ different segments of the modulator,

\begin{equation}
E_m = \sum^N_{n=1}t_{mn}A_n e^{i\phi_n},\label{eq:scattering}
\end{equation}

\noindent where $A_n$ and $\phi_n$ are, respectively, the amplitude
and phase of the light reflected from segment $n$. Scattering in the
sample and propagation through the optical system is described by
the elements $t_{mn}$ of the unknown transmission matrix. Clearly,
the magnitude of $E_m$ will be the highest when all terms in
Eq.~\eqref{eq:scattering} are in phase. We determine the optimal
phase for a single segment at a time by cycling its phase from 0 to
2$\pi$. For each segment we store the phase at which the target
intensity is the highest. At that point the contribution of the
segment is in phase with the already present diffuse background.
After the measurements have been performed for all segments, the
phase of the segments is set to their stored values. Now the
contributions from all segments interfere constructively and the
target intensity is at the global maximum. A pre-optimization with a
small number of segments significantly improves the signal to noise
ratio. This method is generally applicable to linear systems and
does not rely on time reversal symmetry or absence of absorption.
Although mathematically this algorithm is the most efficient, in
noisy experimental conditions adaptive learning
algorithms\cite{Goldberg1989} might be more effective and an
investigation in such algorithms is on its way (also see
Chapter~\ref{cha:algorithms}).

\begin{figure}
  \centering
  \includegraphics[width=\mediumimage]{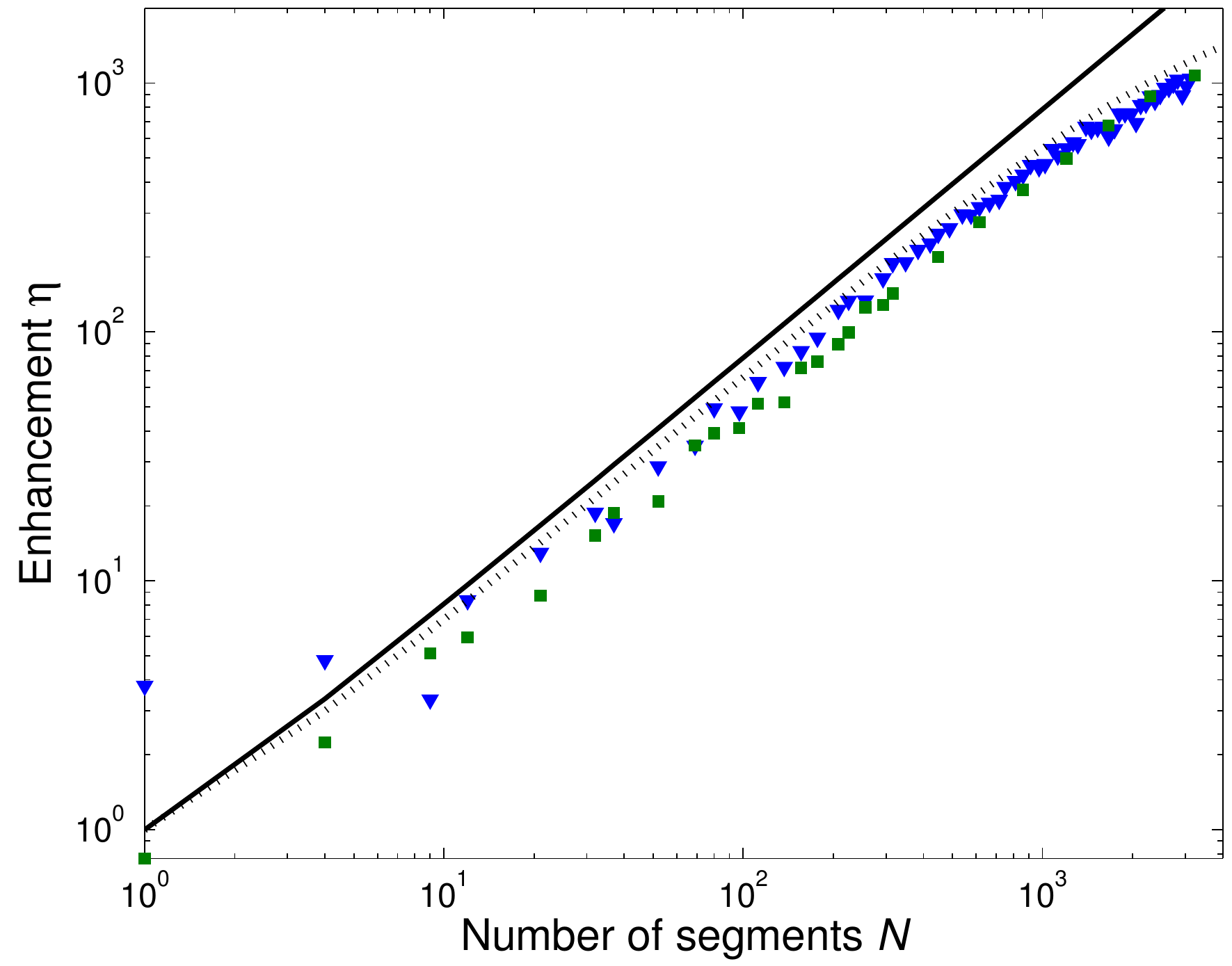}\\
  \caption{Measured intensity enhancement as a function of the number of
  segments. Squares, sample in focus, triangles, sample $\mum{100}$ behind focus, solid line, ideal enhancement (Eq.~\eqref{eq:scaling}), dotted line, corrected for residual
  amplitude modulation and finite persistence time of $T_p = 5400$s. The experimental uncertainty is of the order of the symbol size.}
  \label{fig:enhancement}
\end{figure}

\section{Scaling of the enhancement}
The maximum intensity enhancement that can be reached is related to
the number of segments that are used to describe the incident
wavefront. For a disordered medium the constants $t_{mn}$ are
statistically independent and obey a circular Gaussian
distribution\cite{Goodman2000,Garcia1989,Beenakker1997,Pendry1992}
and the expected enhancement $\eta$ -- defined as the ratio between
the optimized intensity and the average intensity before
optimization -- can be calculated,

\begin{equation}
\eta = \frac{\pi}{4}(N-1)+1.\label{eq:scaling}
\end{equation}

\noindent It was assumed that all segments of the phase modulator contribute equally to the total incident
intensity. We expect the linear scaling behavior to be universal as Eq.~\eqref{eq:scaling} contains no
parameters. Also, since we are free to choose the basis for Eq.~\eqref{eq:scattering}, we expect to find
the same enhancement regardless of whether the target is a focus or a far-field beam and regardless of how
the shaped wavefront is projected onto the sample. Interesting correlations between the transmission
matrix elements, which will cause corrections on Eq.~\eqref{eq:scaling}, are predicted when N approaches
the total number of mesoscopic channels.\cite{Beenakker1997, Pendry1992} With our current apparatus we are
far from this regime and no deviation from Eq.~\eqref{eq:scaling} is expected.

We tested the universal scaling behavior implied by Eq.~\eqref{eq:scaling} by changing $N$. Using the same
TiO$_2$ sample as before, the algorithm was targeted to construct a collimated beam. In
Fig.~\ref{fig:enhancement} the enhancement is plotted as a function of the number of segments for
different focusing conditions. The linear relation between the enhancement and the number of segments is
evident until the enhancement saturates at $\eta=1000$. All measured enhancements were slightly below the
theoretical maximum. This is understandable since all perturbations move the system away from the global
maximum. The main reason for deviations from the optimal wavefront is residual amplitude modulation in the
phase modulator, which introduced an uncontrolled bias in the field amounting to 14\% of the total
intensity.

The saturation of the enhancement is the result of slow changes in the speckle pattern. This instability
effectively limited the number of segments for which the optimal phase could be measured. We estimate that
the effective enhancement decreases to $\eta_\mathrm{eff} = \eta / (1+NT/T_p)$, where $T=1.2$s is the time
needed for one measurement and the persistence time $T_p=5400$s is the timescale at which the speckle
pattern of the TiO$_2$ sample remains stable. Depending on the environmental conditions, $T_p$ can be
considerably higher and enhancements of over two thousand have been measured overnight.

\begin{table}
  \centering
  \begin{tabular}{cccccc}
  \hline
    Sample                  & L (\textmu m)    & $\eta\;(\pm15\%)$   & $N$\\
  \hline
    Rutile TiO$_2$          & 10.1$\pm$0.3  & 1080  & 3228\\
    Daisy petal, fresh      & 43$\pm$5      &   64  & 208\\
    Daisy petal, dried      & 37$\pm$5      &  630  & 1664\\
    Chicken eggshell        & 430$\pm$30    &  250  & 3228\\
    Human tooth             & 1500$\pm$100  &   70  & 208\\
  \hline
    \end{tabular}
    \caption{Measured intensity enhancement for different materials. $L$, sample thickness ($\pm$ surface roughness); $\eta$, maximum enhancement reached; $N$, number of segments used by the algorithm to describe the wavefront.}
    \label{tab:enhancements}
\end{table}

To verify the universal applicability of inversion of wave diffusion, we used a variety of materials of
natural origin. Table~\ref{tab:enhancements} lists the intensity enhancement for different materials we
used. Although the samples vary in thickness, composition and scattering strength, they were all able to
focus a properly prepared wavefront to a sharp spot. The intensity enhancement varies between 60 and 1000.
The main reason for this variation is that the persistence time is not the same for all materials.

\section{Conclusion}
\noindent In summary, our results show that precise control of diffuse light is possible using an optimal,
non-iterative algorithm; light can be directed through opaque objects to form one or multiple foci. The
brightness of the focal spot is explained by a model based on statistical optics. We expect inverse wave
diffusion to have applications in imaging and light delivery in scattering media, possibly including metal
nano\-struc\-tures\cite{Stockman2002}. Dynamic measurements in biological tissue are possible when the
time required for achieving a focus can be reduced to below 1 ms per segment\cite{Vakoc2005,Li2005}; we
estimate that this timescale is technologically possible with the use of fast phase
modulators\cite{Hacker2003}. Furthermore, the high degree of control over the scattered light should
permit experimental verification of random matrix theories for the transport of
light\cite{Beenakker1997,Pendry1992}.
\bibliography{../../bibliography}
\bibliographystyle{Ivo_sty}

\setcounter{chapter}{3}

\chapter{The focusing resolution of opaque
lenses\label{cha:diffraction-limit}}\enlargethispage{-2\baselineskip}

\begin{figure}
\centering
    \includegraphics[width=\textwidth]{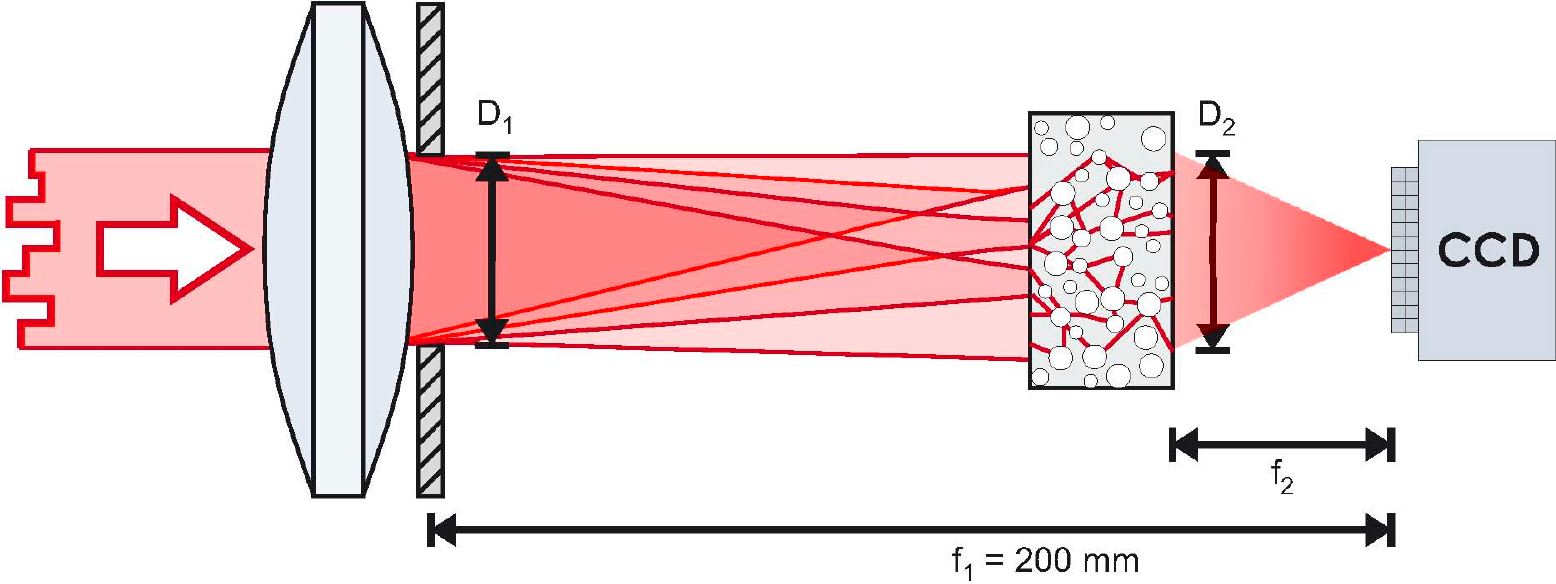}\\
    \caption{Schematic of the experimental geometry. Light from a phase modulator is imaged on the center
    plane of a lens (imaging telescope not shown). The numerical aperture of the lens is controlled by a pinhole. A CCD
    camera is positioned in the focal plane of the lens. Between the camera and the lens a strongly
    scattering sample is placed. The sample can be moved to change the distance to the camera.
    }\label{fig:setup-diffraction-limit}
\end{figure}

\noindent In this chapter, we investigate both experimentally and theoretically the focusing resolution of
the opaque lenses that were introduced in Section~\ref{sec:opaque-lenses-definition}. For every optical
system there is a limit on how sharp it can focus light. The smallest focus that can theoretically be
achieved with an aberration free lens is called the diffraction limit. The diffraction limit is determined
by the aperture and focal distance of the lens. Here we examine how these concepts translate to the
situation of an opaque lens. We use the experimental geometry that is shown in
Fig.~\ref{fig:setup-diffraction-limit} to measure the focal width of an opaque lens and then compare this
width to the theoretical diffraction limit of an ordinary lens.

In Section ~\ref{sec:opaque-lens-wavefront-shaping} we explain how
scattering inside an opaque lens helps focusing. The focusing
resolution of an opaque lens is determined experimentally in
Section~\ref{sec:exp-resolution}. In Section~\ref{sec:exp-corr} we
compare the shape of the focus with the measured speckle correlation
function. To theoretically understand the focusing capacity of
opaque lenses, a matrix model does not suffice. Therefore, we
developed a continuous field description of opaque lenses in
Section~\ref{sec:continuous}. This new theoretical framework
explains the experimental results. Moreover, the theory predicts the
existence of the optical equivalent of the acoustical time-reversal
lenses that focus sound beyond the diffraction
limit\cite{Derode2002}.

\section{Wavefront shaping with an opaque lens\label{sec:opaque-lens-wavefront-shaping}}
In practice, it is extremely difficult to design (systems of) lenses that have almost no
aberrations.\cite{Longhurst1973, Hecht1998} A good microscope objective, for instance, contains up to
twenty lens elements for compensating various aberrations.\cite{NikonComa} An alternative approach to
achieve a diffraction limited focus is to use adaptive optics to compensate for the aberrations in the
lens system.\cite{Tyson1998} However, the adaptive optics system that generates the incident wavefront may
have many limitations, such as a maximum number of control segments, residual phase-amplitude
cross-modulation, a limited maximal wavefront curvature, etcetera. All such limitations cause systematical
deviations from the diffraction limited focus.

We found that, surprisingly, the shape of the focus of an opaque lens is not affected by experimental
limitations of the wavefront modulator. Paradoxically, the scattering medium makes that we have apparently
more control over the scattered field than we have over the incident field. A similar effect has been
observed in time-reversal experiments using ultrasound and microwaves: a limited number of broadband
transmitters suffices to generate a diffraction limited focus through a disordered medium (see
\cite{Fink1999} for an excellent review on time-reversal focusing). Although our experiments are performed
at a single (optical) frequency, many of the concepts developed for time-reversal experiments are still
valid.

\begin{figure}
\centering
  \includegraphics[width=\tinyimage]{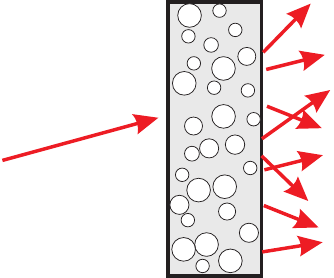}\\
  \caption{Scattering helps wavefront shaping and focusing. A single incident
  beam contributes to all transmitted wave vectors. As a result, a sharp focus
  can be generated even if the wavefront synthesizer only controls light in
  a small solid angle.}\label{fig:scattering-helps}
\end{figure}

In Fig.~\ref{fig:scattering-helps}, we illustrate how scattering
improves focusing. Disordered scattering in the sample randomly
redistributes the incident light over all scattering channels (see
e.g. \cite{Goodman2000, Sebbah2001, Pappu2002}). In a microwave
experiment, it was demonstrated that even evanescent channels are
excited.\cite{Lerosey2007} Scattering makes that light from a single
segment of the incident beam contributes to all wave vectors of the
desired field. By tuning the relative phases of all segments, their
contributions to the desired field add up constructively. The more
segments are used, the better the desired field can be approximated.
The constructed field will never be perfect as the number of degrees
of freedom in the transmitted field cannot exceed the number of
control segments. However, thanks to the disordered scattering any
deviation in the generated field is, on average, uniformly
distributed over all outgoing channels, resulting in a low overall
intensity contribution. Imperfections in the generated wavefront
cause a loss of contrast with the speckle background.

\section{Measured focusing resolution of an opaque
lens}\label{sec:exp-resolution}
\noindent The experimental geometry used for analyzing the focus of an opaque lens is shown in
Fig.~\ref{fig:setup-diffraction-limit}. The sample is a $\mum{6}$ layer of airbrush paint on a standard
microscope cover slip. The wavefront illuminating the sample is shaped to increase the intensity on a
single pixel of the camera. The algorithm that shapes the wavefront is the same as was used in
Chapter~\ref{cha:focusing-through}; it does not take into account the shape of the focus. The distance
between the sample and the camera can be varied to analyze the focusing resolution of the opaque lens at
different focal lengths.

We compare the measured focal width to the calculated diffraction
limit. The optimal focus of a homogeneously illuminated lens with a
circular aperture is an Airy disk.\cite{Longhurst1973} The half
width at half maximum of the Airy disk is $w=0.51\lambda/\NA$, where
$\NA$ is the numerical aperture of the lens. This radius is almost
equal to the Abbe diffraction limit\cite{Abbe1873} of
$w=\lambda/(2\NA)$. To calculate the diffraction limit of the opaque
lens, we need to determine its numerical aperture
$\NA_2=D_2/(2f_2)$, where $D_2$ is the diameter of the illuminated
area on sample and $f_2$ is the focal length of the opaque lens.

The diameter of the spot on the sample depends on what wavefront is
generated. When the glass lens in front of the aperture is
illuminated with a plane wave, the diameter of the spot on the
sample equals $D_2 = 2 f_2 \NA_1$, where the numerical aperture of
the glass lens is $\NA_1=D_1/ (2 f_1)=5\cdot 10^{-3}$ with
$D_1=\mm{2.1}$ the diameter of the aperture and $f_1=\mm{200}$ the
focal length of the lens.

When the modulator is programmed to generate a shaped wavefront, diffraction at the lens aperture will
cause the beam to diverge. Segments of the phase modulator are imaged onto the aperture to squares with
sides of length $D_s= \mm{0.03}$. The angle of diffraction for a square source equals $\alpha \equiv
(4\lambda)/(\pi^2 D_s) = 0.008$ rad (see e.g. \cite{Siegman1986}). Combining the effects of the focusing
and diffraction, we find the spot diameter
\begin{equation}
D_2 = 2 f_2 \NA_1 + 2 (f_1-f_2) \alpha,\label{eq:dspot}
\end{equation}
where $f=\mm{200}$ is the focal length of the lens. The radius (half
width at half maximum intensity) of the diffraction limited focus at
the detector plane is given by
\begin{equation}
w = \frac{\lambda f_2}{2 f_2 \NA_1 + (f_1-f_2)
\alpha}.\label{eq:theory-FWHM}
\end{equation}
\begin{figure}
\centering
    \includegraphics[width=\smallimage]{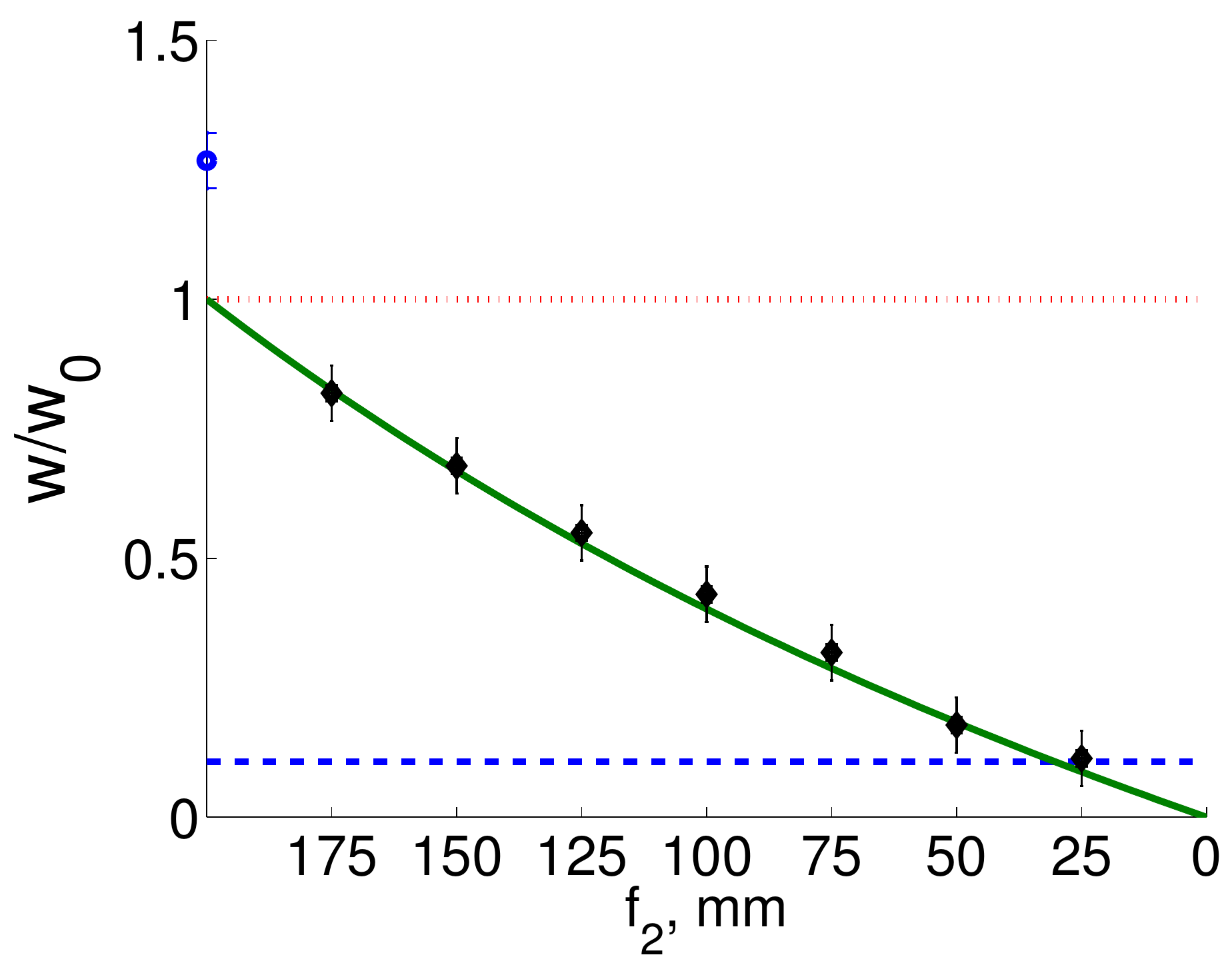}\\
    \caption{Relative focal width $w$ as function of
sample-camera distance $f_2$. Diamonds, measured values with a
sample present; solid curve, theoretical diffraction limit; circle
(at $f_2=\mm{200}$), width of the focus after optimizing without
sample; dashed line, size of a single pixel of the camera, detection
limit. All widths are relative to the diffraction limit of the glass
lens (dotted line, $w_0=\mum{31}$).}\label{fig:z-scan}
\end{figure}

The experimental results are displayed in Fig.~\ref{fig:z-scan}. All
widths are relative to the diffraction limit of the glass lens
($w_0=\mum{31}$). Without sample, the sharpest focus that was
obtained by optimizing the wavefront has a width of $1.2$ times this
diffraction limit. The fact that the diffraction limit was not
reached clearly demonstrates the limitations of the wavefront
synthesizer. With a sample, however, these limitations do not
degrade the sharpness of the focus. The width of the focus is
exactly the diffraction limit that is predicted by
Eq.~\eqref{eq:theory-FWHM}.

\section{Measured relation between the focus and the speckle correlation function}\label{sec:exp-corr}
\begin{figure}
\centering
  \includegraphics[width=\smallimage]{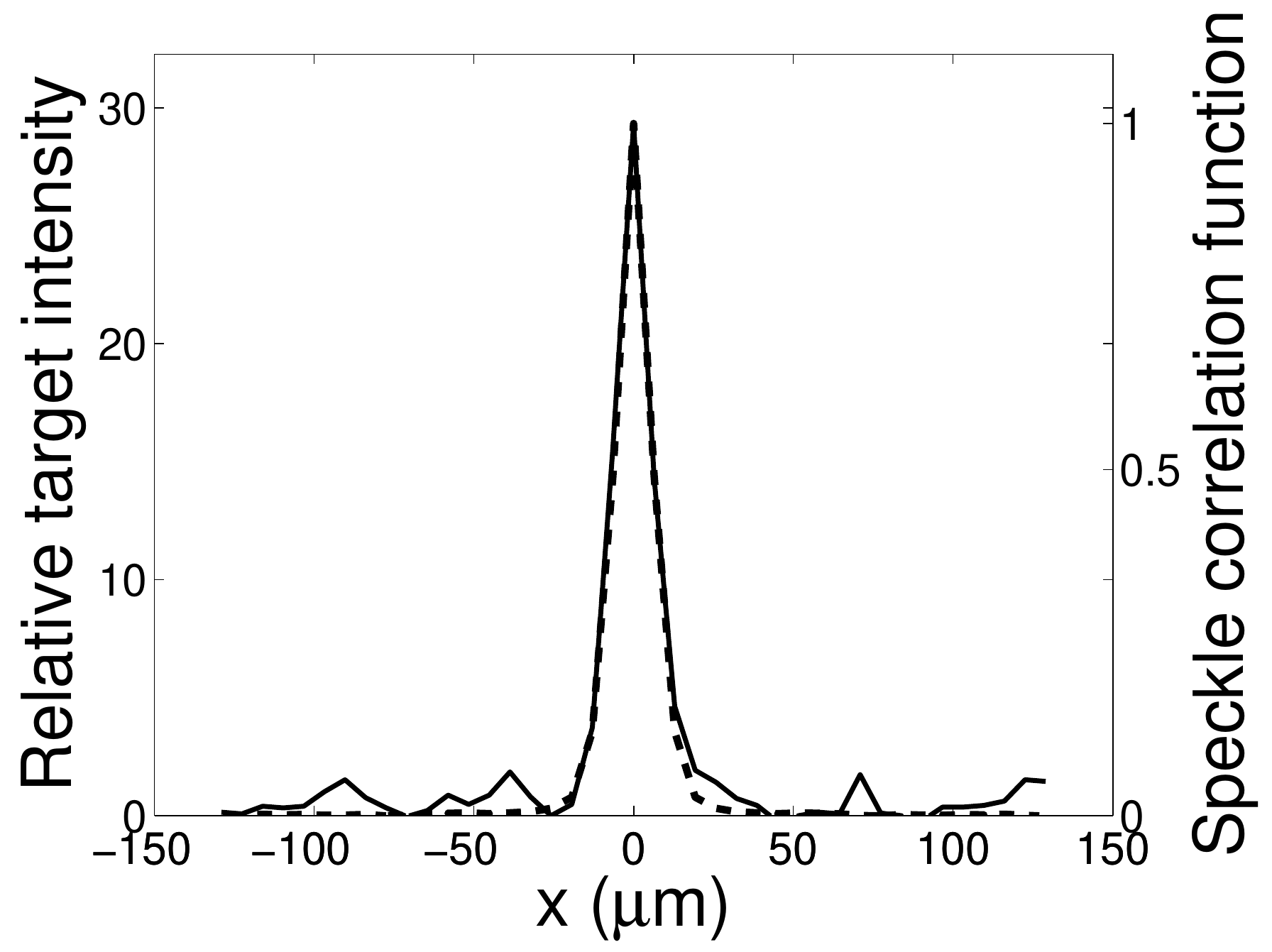}\\
  \caption{Intensity profile of the focus at $y=0$ (solid curve) and speckle correlation function (dashed curve)
  for an $\mum{8}$ thick layer of airbrush paint creating a focus at $\mm{26}$ distance. The speckle correlation function
  was measured with a randomly generated incident wavefront.}\label{fig:tio2z}
\end{figure}

\noindent It appears that the focus of an opaque lens always has the same size as a typical speckle. In
this section, we analyze this effect by comparing the measured profile of the focus to the measured
speckle correlation function. The speckle correlation function $C$ is a measure for the shape of a typical
speckle.\cite{Goodman2000} It is defined as
\begin{equation}
C(x_1, y_1; x_2, y_2) \equiv \frac{\avg{I(x_1,y_1) I(x_2,
y_2)}}{\avg{I(x_1,y_1)}\avg{I(x_2,y_2)}}-1\label{eq:CII-def},
\end{equation}
where the brackets denote spatial averaging over all speckles. We
experimentally investigated the relation between the speckle
correlation function and the generated focus. First, we used the
geometry as shown in Fig.~\ref{fig:setup-diffraction-limit} with the
sample as close to the camera as possible ($\mm{26}$). The resulting
spot has a radius of $\mum{6.6\pm3}$, which is hardly larger than
the size of a single pixel on the camera ($\mum{6.45}$). The speckle
correlation function and the intensity profile of the spot overlap
exactly, as can be seen in Fig.~\ref{fig:tio2z}.

\begin{figure}
\centering
  \includegraphics[width=\smallimage]{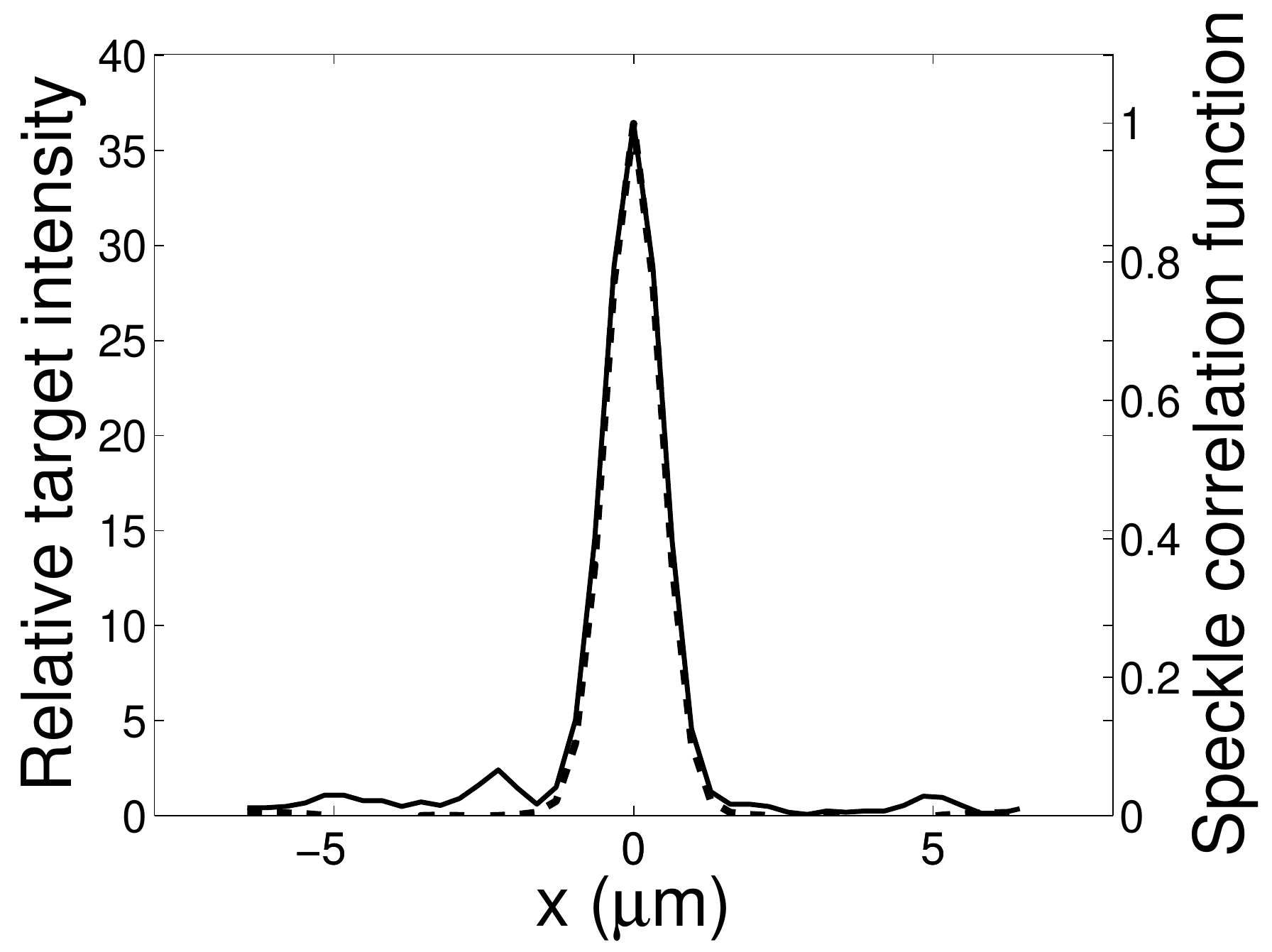}\\
  \caption{Intensity profile of the focus at $y=0$ (solid curve) and speckle correlation function (dashed curve)
  for an egg shell of $\mum{430\pm30}$ thick creating a focus at $\mm{3.5\pm0.5}$ distance.}\label{fig:egg}
\end{figure}

\begin{figure}
\centering
  \includegraphics[width=\smallimage]{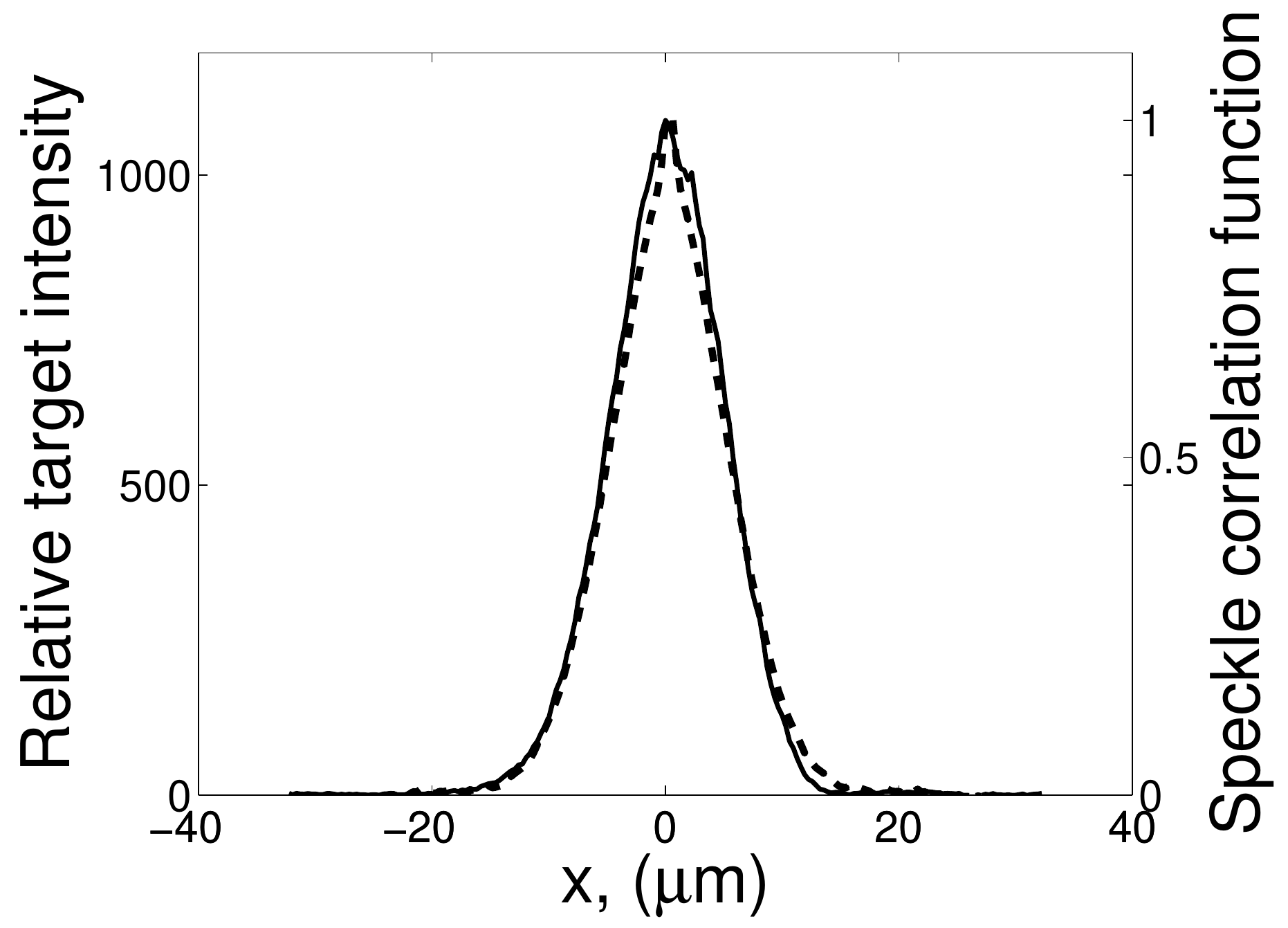}\\
  \caption{Intensity profile of the focus at $y=0$ (solid curve) and speckle correlation function (dashed curve)
  for a $\mum{10.1\pm0.3}$ thick sample of white paint creating a focus at $\mm{3.5\pm0.5}$ distance. The central point of the correlation function
  was removed since it was affected by measurement noise in the camera image.}\label{fig:tio2}
\end{figure}

The same experiment was performed in a different geometry and with different samples. This time the
incident light was focused using a microscope objective and the transmitted light was imaged using a
second objective (see Chapter~\ref{cha:focusing-through}). For a relatively thick sample (egg shell,
$\mum{430\pm30}$ thick) we placed the detecting objective at a distance of $\mm{3.5\pm0.5}$ and found a
very narrow focus of $\mum{1.1\pm0.2}$ (see Fig.~\ref{fig:egg}). In Fig.~\ref{fig:tio2}, the results for a
thinner sample ($\mum{10.1\pm0.3}$ of rutile TiO$_2$ pigment) in the same configuration is shown. For this
sample, we find a wider focus of $\mum{11\pm0.2}$. In both cases, the profile of the focus overlaps with
the speckle correlation function.

\section{Continuous field theory for opaque
lenses}\label{sec:continuous}
\noindent After optimizing to create a focus, a fraction of the field at the back of the sample forms a
converging beam, while the rest of the field forms a random speckle. In
Chapter~\ref{cha:focusing-through}, we introduced a matrix model to explain the intensity of the focus of
an opaque lens. Although that model proved suitable for describing the intensity in the focus of an opaque
lens, it cannot predict the shape of the focus. We now develop a continuous field formalism that explains
both the intensity and the shape of the focus. This formalism also explains the similarity of the focus
profile and the speckle correlation function that was observed in the experiment.

\subsection{Continuous field formalism}
Light propagates from the modulator to the sample, then it is
transmitted by the sample, and finally it reaches the focal plane
where the target detector is placed. We separate light propagation
into two parts. The first part starts at the modulator and ends at
the back of the sample. The second part is the propagation from the
back of the sample to the focal plane.

We define coordinates in the focal plane as $\rb$ ($z_b=f_2$ as defined in
Fig.~\ref{fig:setup-diffraction-limit}), on the back surface of the sample as $\rk$ ($z_k=0$), and in the
plane of the phase modulator as $\ra$. The target for optimization is at point $\trb$. The field at the
back of the sample is given by
\begin{equation}
E(\rk) = \sum_a^N t_a(\rk) E_a\label{eq:E-rk},
\end{equation}
where the summation is over all segments of the phase modulator. We
assume the field to be constant within a segment, so that we can
write $E(\ra)=E_a$. Just behind the modulator, light propagation is
nearly perpendicular to the surface of the modulator. We can,
therefore, use the Fresnel-Kirchhoff diffraction formula (see e.g.
\cite{Born1993, Fowles1989}) to find coefficient $t_a(\rk)$
\begin{equation}
t_a(\rk) = \frac{i}{\lambda}\iint_{S_a} \ud^2\ra G(\rk, \ra).
\end{equation}
The integration is over the surface of a single segment of the light
modulator ($S_a$) and $G(\rk, \ra)$ is the unknown Green function
for propagating from the modulator to the back of the sample. Light
propagation from the back of the sample towards the focal plane is
described by the free space Green function $g(\rb-\rk)$
\begin{equation}
E(\rb) = \iint \ud^2\rk g(\rb-\rk)
\frac{i}{\lambda}E(\rk)\label{eq:Erb}.
\end{equation}
Field propagation from a single segment $a$ to the focal plane is
described by a transmission coefficient $t_a(\rb)$
\begin{equation}
t_a(\rb) \equiv \iint \ud^2\rk g(\rb-\rk)
\frac{i}{\lambda}t_a(\rk)\label{eq:ta-rb-definition}.
\end{equation}
The role of $t_a(\rb)$ is similar to that of $t_{ba}$ in the matrix
formalism. However, $t_a(\rb)$ is a continuous function of the
spatial coordinate and also gives the shape of the focus of the
opaque lens.

\subsection{Optimized field}\label{sec:optimized-field}
We consider the case where the incident field is shaped with a modulator that perfectly modulates phase
and amplitude. In Chapter~\ref{cha:dorokhov-theory}, we will see that the results for imperfect modulation
or phase only modulation are equal up to a constant prefactor. After optimizing for focusing at point
$\trb$, the field generated by the modulator is (see also Chapter~\ref{cha:dorokhov-theory})
\begin{equation}
\optim{E}_a = E_0 t^*_a(\trb),\label{eq:optimal-field}
\end{equation}
where the tilde sign indicates the value of $E_a$ after
optimization. $E_0$ is a constant that fixes the average intensity
at the modulator to $\Iin$, it is defined as
\begin{align}
E_0 &\equiv \frac{\Iin}{\sqrt{\frac{1}{N}\Iavg(\trb)}},\\
\Iavg(\trb)&\equiv \Iin \sum_a^N |t_a(\trb)|^2\label{eq:Iavg-def}.
\end{align}
A physical interpretation of $\Iavg(\trb)$ is that it is the
intensity in the focal plane averaged over all possible
configurations of the light modulator that have an average intensity
of $\Iin$.

The field at the back of the sample after optimization is calculated
by substituting Eq.~\eqref{eq:ta-rb-definition} into
Eq.~\eqref{eq:optimal-field} and then substituting the result in
Eq.~\eqref{eq:E-rk}
\begin{equation}
\optim{E}(\rk) = \sum_a^N t_a(\rk) \left(E_0 \iint \ud^2\rk'
g(\trb-\rk')\frac{i}{\lambda}t_a(\rk')\right)^*\label{eq:E_k-opt}.
\end{equation}
The term $t_a(\rk) t^*_a(\rk')$ fluctuates rapidly with the distance
$\Dr\equiv \rk'-\rk$. The ensemble average of this product is the
field-field correlation function\cite{Pnini1989}
\begin{equation}
\avg{t_a(\rk) t^*_a(\rk')}=\avg{|t_a(\rk)|^2}\frac{\sin(k_0 |\Dr|)}{k_0
|\Dr|}\exp{\left(-|\Dr|/(2\ell)\right)}\label{eq:tt-corr},
\end{equation}
where $k_0\equiv 2\pi/\lambda$ and $|t_a(\rk)|^2$ is the intensity
transmission coefficient from segment $a$ to point $\rb$. Its
average value $\avg{|t_a(\rk)|^2}$ depends on the geometry of the
experiment and the thickness of the sample. Because the correlation
function Eq.~\eqref{eq:tt-corr} is very sharp, we can take
$g^*(\trb-\rk')$ out of the integral in Eq.~\eqref{eq:E_k-opt}.
Integrating Eq.~\eqref{eq:tt-corr} over $\rk'$ gives.
\begin{equation}
\iint \ud^2\rk' \avg{t_a(\rk)
t^*_a(\rk')}=\avg{|t_a(\rk)|^2}2\pi\frac{1}{k_0^2+1/(2\ell)^2}
\approx \avg{|t_a(\rk)|^2} \frac{\lambda^2}{2\pi},
\end{equation}
where we used that $4(k_0\ell)^2\gg1$. This assumption is valid even
for very strongly scattering samples. The area given by
$\lambda^2/(2\pi)$ corresponds to the size of a single scattering
channel (see e.g. \cite{Boer1995}). The average value of the
optimized field at the back of the sample now reduces to
\begin{equation}
\avg{\optim{E}(\rk)} = g^*(\trb-\rk) C_0 \Iavg(\rk),
\label{eq:E_k-opt-avg-smooth}
\end{equation}
where $C_0\equiv i\lambda E_0 / (2\pi\Iin)$ and $\Iavg(\rk)\equiv\Iin \sum_a^N |t_a(\rk)|^2$, just as in
Eq.~\eqref{eq:Iavg-def}.

In the geometry as shown in Fig.~\ref{fig:setup-diffraction-limit}, $\Iavg(\rk)$ is constant in a disk
with a diameter $D$ given by Eq.~\eqref{eq:dspot} and zero outside of the disk. In this case,
Eq.~\eqref{eq:E_k-opt-avg-smooth} becomes
\begin{equation}
\avg{\optim{E}(\rk)} = g^*(\trb-\rk) C_0 \Iavg \text{circ}_D(\rk),
\label{eq:E_k-opt-avg}
\end{equation}
with
\begin{equation}
\text{circ}_D(\rk) \equiv
\begin{cases}
0 & x_k^2+y_k^2 > D^2/4\\
1 & \mathrm{otherwise}.
\end{cases}
\end{equation}
Equation~\eqref{eq:E_k-opt-avg} tells us that, up to a constant
proportionality factor, the average field at the back of the sample
is the complex conjugate of the free space propagator. In other
words, the transmitted field is the same \emph{as if} the sample
time-reversed a wave coming from the target focus. Hence, the
transmitted light focuses on the target as well as is physically
possible for a system with an aperture given by
$\text{circ}_D(\rk)$.

In a different geometry, for example when the incident light is
focused sharply on a thick sample, the intensity distribution
$\Iavg(\rk)$ will be a smooth function of the position. In such a
geometry, an opaque lens differs from an ordinary lens. With an
ordinary lens the amplitude profile at the back of the lens is equal
to the \emph{amplitude} profile of the \emph{incident} illumination.
For an opaque lens, however, the amplitude of the controlled field
at the back of the sample is proportional to the \emph{intensity}
distribution at the \emph{back} of the sample. In the following
section we will see how to make use of this difference.

\subsection{Better than diffraction limited
focusing}\label{sec:superlenses}
\begin{figure}
\centering
  \includegraphics[width=\wideimage]{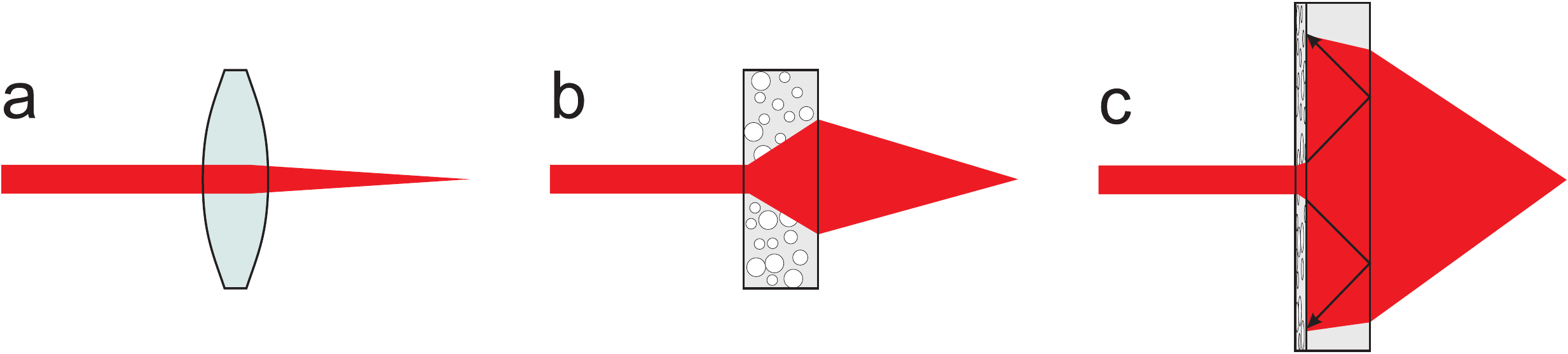}\\
  \caption{Better than diffraction limited focusing with an opaque lens.
  \subfig{a} An ordinary lens at best focuses light to the diffraction limit
  given by the diameter of the incident beam. \subfig{b} An opaque lens focuses light
  more sharply than the diffraction limit because diffusion in the
  lens increases the diameter of the transmitted converging
  beam. \subfig{c} Opaque wafer lens made out of gallium phosphide.
  Reflection at the GaP-air interface fold back the optical paths and improve the focusing
  resolution even further.}\label{fig:superlenses}
\end{figure}
In time-reversal microwave experiments it was observed that scattering improves the focusing resolution
beyond the diffraction limit by `folding' the propagation paths in a waveguide.\cite{Derode2002} In this
section, we show that a similar dramatic improvement of the focusing resolution can also be obtained in an
optical system with a slab geometry.

With an ordinary lens, the intensity profile of the transmitted is
the same as that of the incident light. For a given diameter of the
incident beam, the lens at best focuses to the diffraction limit
$\lambda f/D$ (see Fig.~\ref{fig:superlenses}a).

With an opaque lens, however, diffusion in the lens makes that the
transmitted spot is larger than the incident beam
(Fig.~\ref{fig:superlenses}b). Therefore, an opaque lens focuses
light more sharply than the best possible ordinary lens. The
intensity distribution $\Iavg(\rk)$ at the back of the sample
depends on the thickness and mean free path of the sample, as well
as on the extrapolation ratios that describe the boundary
conditions. The thicker the scattering medium, the sharper the
focus.

The benefit of a thick scattering medium was observed in the experiment. Under the same experimental
conditions, the focus of a $\mum{430}$ thick opaque lens (Fig.~\ref{fig:egg}) was ten times narrower than
the focus of a $\mum{10.1}$ thick opaque lens (Fig.~\ref{fig:tio2}).

A proposed special example of an opaque lens is shown in
Fig.~\ref{fig:superlenses}c. A thin layer on one side of a gallium
phosphide (GaP, n=3.3) wafer is made porous. The porous gallium
phosphide scatters the light into a $2\pi$ solid angle. However, due
to the high index contrast at the back of the wafer, most of the
transmitted light is internally reflected back to the scattering
layer, where it is scattered by the porous GaP for a second time. By
folding the optical paths this way, the spot at the back of the
sample increases dramatically, and the focusing resolution improves
accordingly.

We now estimate the factor by which a GaP wafer lens improves the
diffraction limit. The critical angle for total internal reflection
at the GaP-air interface is $\theta_c=17.6^\circ$. For a wafer with
a thickness of $L$, the incident spot needs to have a diameter of at
least $2\tan \theta_c L$ for the reflected internally reflected
light to overlap the original spot. We neglect the contribution of
light that is reflected at an angle of over $45^\circ$ and we take
into account only the first reflection at the GaP-air interface.
Even with this pessimistic estimate, a $\mm{0.5}$ thick opaque a GaP
wafer lens already expands a $\mm{0.32}$ beam to $\mm{4}$ before
focusing it. These realistic numbers indicate that a $\mm{0.5}$
thick wafer can replace a $6\times$ beam expander and a perfectly
corrected microscope objective. In a way, this wafer is a miniature
Cassegrain objective.

\subsection{Intensity profile of the focus}
We use the general result in Eq.~\eqref{eq:E_k-opt-avg-smooth} to calculate the intensity distribution in
the focal plane. Using Eqs.~\eqref{eq:Erb} and \eqref{eq:E_k-opt-avg-smooth} we find
\begin{equation}
\optim{E}(\rb) =\frac{i}{\lambda} \iint \ud^2\rk g(\rb-\rk)
g^*(\trb-\rk) C_0 \Iavg(\rk).
\end{equation}
In the paraxial limit (see e.g. \cite{Fowles1989}), we can approximate
\begin{equation}
g(\rb-\rk) g^*(\trb-\rk) \approx \frac{e^{i\Phi}}{f_2^2}\exp\left(\frac{i k_0}{f_2}\left[x_k (x_\bopt-x_b)
+y_k (y_\bopt-y_b)\right]\right),
\end{equation}
with $f_2$ the focal distance of the opaque lens. We isolated phase
factor $\Phi\equiv x_\bopt^2-x_b^2+y_\bopt^2-y_b^2$. We now have
\begin{equation}
\avg{\optim{E}(\rb)} =  \frac{ie^{i\Phi}}{\lambda f_2^2} C_0 \iint
\ud^2\rk \Iavg(\rk) \exp\left(\frac{i k_0}{f_2}\left[x_k
(x_\bopt-x_b) +y_k
(y_\bopt-y_b)\right]\right)\label{eq:Fresnel-focusing}.
\end{equation}
In Eq.~\eqref{eq:Fresnel-focusing} we recognize a 2-dimensional
Fourier transform. We define the Fourier transformed intensity $U_0$
as
\begin{equation}
U_0(\rb) \equiv \iint \ud^2\rk \Iavg(\rk) \exp\left(\frac{i
k_0}{f_2}\left[x_k (x_\bopt-x_b) +y_k (y_\bopt-y_b)\right]\right)
\label{eq:U0-def}.
\end{equation}
To conform to literature (see e.g. \cite{Goodman2000}), we adopt the notation
\begin{equation}
\Fourier{\Iavg(\rk)} \equiv U_0(\rb).\label{eq:Fourier-def}
\end{equation}
Now, the intensity in the focal plane is given by expanding $C_0$
\begin{equation}
\optim{I}(\rb) = \frac{1}{(2\pi f_2^2)^2}\frac{N}{\Iavg(\trb)}
|\Fourier{\Iavg(\rk)}|^2\label{eq:focus-intensity},
\end{equation}
which confirms the observation made in
Section~\ref{sec:optimized-field} that the opaque lens perfectly
focuses a field that is proportional to the intensity distribution
at the back of the sample. The intensity exactly at the center of
the focus equals
\begin{equation}
\optim{I}(\trb) = \left(\frac{\Ptot}{2\pi f_2^2}\right)^2
\frac{N}{\Iavg(\trb)},
\end{equation}
where $\Ptot\equiv \iint \ud^2\rk \Iavg(\rk)$ is the average total
power that is transmitted through the sample. Before optimization,
the power is distributed over $2\pi$ solid angle. Therefore,
$\Iavg(\trb)=\Ptot/(2\pi f_2^2)$ and
\begin{equation}
\optim{I}(\trb) = N \Iavg(\trb),
\end{equation}
which means that the intensity in the target has increased by exactly a factor $N$. For phase only
modulation, the enhancement reduces to $1+(N-1)\pi/4$ (see Chapter~\ref{cha:dorokhov-theory}). The matrix
model used in Chapter~\ref{cha:focusing-through} predicts the same intensity enhancement. However, the new
formalism clearly shows that the enhancement needs to be measured exactly at the maximum of the focus and
not, for instance, averaged over the whole speckle. This experimentally important detail was not clear
from the matrix model. Moreover, the new formalism predicts the exact shape of the focus.

\subsection{Connection with speckle correlation function}
When the sample is thicker than a transport mean free path, the
field at the back surface of the sample is uncorrelated on length
scales larger than the wavelength of light (also see
Eq.~\eqref{eq:tt-corr}) and, equivalently, the transmitted light has
no preferential direction. For such a sample, in the paraxial limit
the correlation function (Eq.~\eqref{eq:CII-def}) only depends on
the coordinate difference $\Delta \rb \equiv \rb - \vr_{b'}$. The
van Cittert-Zernike theorem tells us that, under these conditions,
the correlation function is given by\cite{Cittert1934, Zernike1938,
Goodman2000}
\begin{equation}
C(\Delta x_b, \Delta y_b) = \frac{1}{\Ptot^2}\left|\Fourier{
\avg{I(\rk)}}\right|^2\label{eq:Cittert-Zernike},
\end{equation}
where $I(\rk)$ is the diffuse intensity distribution at the back
surface of the sample and $\Fourier{\avg{I(\rk)}}$ is defined as in
Eq.~\eqref{eq:Fourier-def}. When the phase modulator is configured
to generate a random field, $\avg{I(\rk)}=\avg{\Iavg(\rk)}$. In this
case, it follows from Eq.~\eqref{eq:focus-intensity} that,
\begin{equation}
\optim{I}(x, y, f_2) = C(x, y) N \frac{\Ptot}{2\pi
f_2^2}{\Iavg(\trb)}. \label{eq:spot-theorem}
\end{equation}
Equation~\ref{eq:spot-theorem} predicts that the intensity profile of the focus is exactly equal to the
speckle correlation function (up to a constant prefactor). This identity is striking because the intensity
in the focus is the square of an amplitude, whereas the correlation function is an product of intensities.
Our continuous field model explains the results of Section~\ref{sec:exp-corr}, where this identity was
observed experimentally in different geometries.

\section{Conclusion}
\noindent We observed experimentally that opaque lenses focus light
to a diffraction limited spot. This experimental observation is
explained by a continuous field model that was developed in
Section~\ref{sec:continuous}. The model provides a simple method to
predict the focusing resolution of an opaque lens for different
geometries. For instance, a thin opaque lens that is illuminated
with a flat-top profile focuses light to a diffraction limited spot.
Theoretically, the focusing resolution can even be improved beyond
the diffraction limit by making use of the diffusion in a thick
opaque object. Extremely sharp focusing can be achieved with a GaP
wafer lens where the optical paths are folded back by total internal
reflection.

The shape of the focus is exactly the same as the correlation function of the speckle that is generated by
the opaque lens. This experimentally important observation was explained with the continuous field model
for opaque lenses.

An opaque lens focuses light to a diffraction limited spot or even sharper. Additionally, opaque lenses
are extremely thin and extremely cheap compared to a microscope objective. Therefore, disposable opaque
lenses might replace conventional microscope objectives, for example in applications that require tight
focusing in a polluting or corroding environment, or when space is tight.

\bibliography{../../bibliography}
\bibliographystyle{Ivo_sty}

\setcounter{chapter}{4}
\chapter{Demixing light paths inside disordered metamaterials\label{cha:focusing-inside}}

\begin{abstract}
We experimentally demonstrate the first method to focus light inside
disordered photonic metamaterials. In such materials, scattering
prevents light from forming a geometric focus. Instead of geometric
optics, we used multi-path interference to make the scattering
process itself concentrate light on a fluorescent nanoscale probe at
the target position. Our method uses the fact that the disorder in a
solid material is fixed in time. Therefore, even disordered light
scattering is deterministic. Measurements of the probe's
fluorescence provided the information needed to construct a specific
linear combination of hundreds of incident waves, which interfere
constructively at the probe.

\noindent[This chapter has been published as: I.~M. Vellekoop, E.~G. van Putten, A. Lagendijk, and A. P.
Mosk, Opt. Express \textbf{16}, 67--80 (2008)]
\end{abstract}

\noindent Photonic metamaterials are materials that gain their optical properties from their
nanostructure, rather than from their chemical composition. Their sub-wavelength elements interact very
strongly with the electromagnetic field and provide unprecedented control over the propagation of
light\cite{Leonhardt2006, Pendry2006}. Exciting perspectives of metamaterials include field enhancement in
photonic crystal cavities\cite{Yoshie2004}, negative refractive index optics\cite{Lezec2007, Dolling2007,
Shalaev2007}, and sub-wavelength imaging\cite{Veselago1968, Liu2007}. Perhaps even more amazing is the use
of metamaterials to cloak objects, making them completely invisible for a certain frequency of
light.\cite{Leonhardt2006, Pendry2006} While these perspectives require near-perfectly manufactured
metamaterials, intentionally disordered metamaterials also find applications such as sub-wavelength
focusing\cite{Lerosey2007} and enhanced non-linear conversion\cite{Baudrier-Raybaut2004, Stockman2004}.
Furthermore, disordered metamaterials are of fundamental interest because of their analogy to electrical
conductors\cite{Beenakker1997, Pendry1992}, their quantum optical properties\cite{Lodahl2005}, and
mesoscopic transport phenomena\cite{Storzer2006, Zhang2007, Sebbah2002}. Recently it was shown
experimentally that localization induced by disorder can confine and guide light.\cite{Schwartz2007}

\begin{figure}\centering
\includegraphics[width=\wideimage]{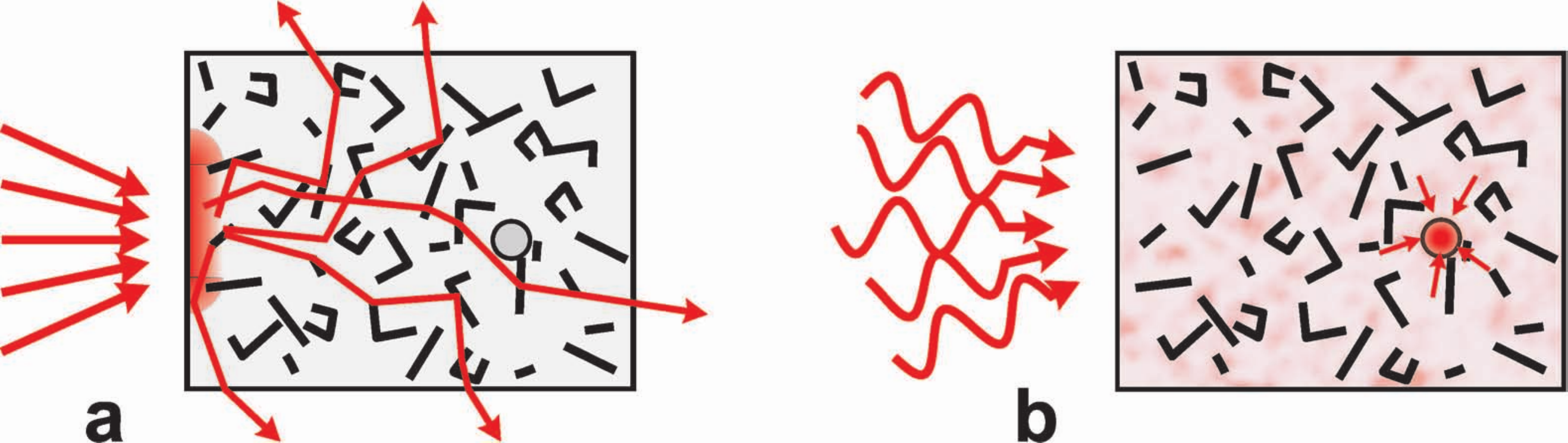}\\
\caption{Principle of channel demixing. \subfig{a} Conventional way of illuminating a metamaterial sample
or device. In order to get light to the area inside the metamaterial indicated by the circle, one might
try to focus light using an ordinary lens. However, the randomly oriented nanoparticles in the material
mix the incident scattering channels in a complicated way so that light does not travel directly to the
target. Incident rays will not converge to a geometrical focus. \subfig{b} Our new way of illuminating a
specific point inside a metamaterial. The wave nature of light is used as an advantage. When the phase
delay for each of the incident scattering channels is set correctly, the channels demix inside the sample;
multi-channel interference makes the light focus at the desired point.} \label{fig:principle}
\end{figure}

To direct light to a desired position in a metamaterial requires
thorough knowledge of the propagation of light in these structures.
Since metamaterials are strongly photonic, even weak manufacturing
imperfections strongly influence
propagation\cite{Vlasov2000,Hughes2005,Koenderink2005}, which makes
it even harder to get light to the right place. For disordered
metamaterials it is, even in theory, impossible to have {\it a
priori} knowledge of light propagation. To treat the metamaterial as
a homogeneous medium and pretend to focus somewhere inside
(Fig.~\ref{fig:principle}a) is an erroneous approach; in reality,
the structure of the material prevents light from forming any focus
at all.

We demonstrate experimentally that light can be delivered to a target inside a disordered photonic
metamaterial, or in fact any strongly scattering solid, without having any {\it a priori} knowledge of the
optical properties of the system. Our approach is based on the fact that the structure of the metamaterial
mixes light entering through different scattering channels (a scattering channel is an angular or spatial
mode of the optical field\cite{Beenakker1997}). The mixing process is complex and stochastic due to
disorder. Even so, reciprocity dictates that there always exists a linear combination of incident channels
that demix inside the metamaterial to focus at the target. In our experiment we find this unique
configuration and focus light on a target that is deeply embedded in a disordered photonic structure
(Fig.~\ref{fig:principle}b).

It is well known that the inverse wave problem is ill posed. In other words, even in principle optical
measurements cannot be used to reconstruct the three-dimensional structure of a photonic metamaterial.
Likewise, optical reflection and transmission measurements cannot reveal the combination of channels that
demix inside the material.

There exist different methods for focusing waves in the presence of scattering. In transparent media where
geometrical optics applies, delivery of light can be achieved very effectively using a variety of adaptive
optics wavefront correction techniques (see e.g. Ref.~\citealt{Tyson1998}). For example, adaptive optics
is commonly used to reduce astronomical seeing\cite{Roddier1997}, to improve the resolution of ophthalmic
imaging\cite{JOSA-retinal-imaging}, or for femtosecond pulse shaping\cite{Weiner2000}. Unfortunately,
these adaptive optics techniques break down in nanostructured media where geometrical optics is not
applicable.

For ultrasound and microwaves time-reversal has proven highly effective\cite{Fink1999,Lerosey2007} in
refocusing waves. Light waves can be refocused back to their source by phase
con\-ju\-ga\-tion\cite{Fisher1983}, or they can be focused behind a scattering region by using feedback
from an external detector\cite{Vellekoop2007thesis}. All these methods require physical access to the
target position, to place a transceiver, an extensive non-linear medium, or a detector. Inside a photonic
metamaterial, with structures smaller than the wavelength of light, these methods cannot be used.

\begin{figure}\centering
\includegraphics[width=\textwidth]{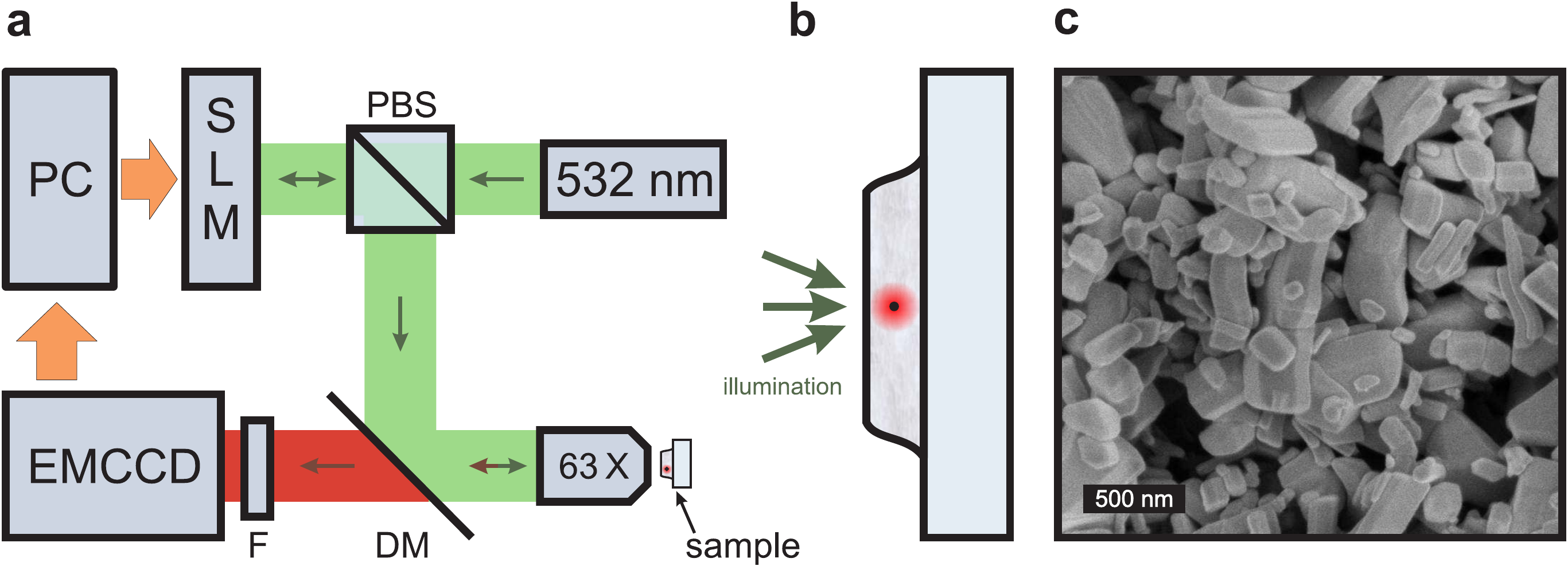}
\caption{Experimental setup and samples. \subfig{a} Simplified
schematic of the experiment. A 532 nm laser is expanded and
illuminates a spatial light modulator (SLM) that spatially modulates
the phase of the reflected light. The SLM is imaged onto the back
aperture of a 63x microscope objective that focuses the modulated
light on a sample. Fluorescence light is collected by the same
objective and imaged with an EMCCD camera. A computer drives the SLM
and analyzes the EMCCD images. PBS, polarizing beam splitter; DM,
dichroic mirror; F, fluorescence filter. Lenses were omitted from
the schematic. \subfig{b} Geometry of the samples. A low
concentration of 300-nm diameter fluorescent spheres is dispersed in
a zinc oxide pigment, deposited on a glass substrate. \subfig{c} SEM
image of a sample.} \label{fig:setup-focusing-inside}
\end{figure}

\begin{figure}\centering
\includegraphics[width=\textwidth]{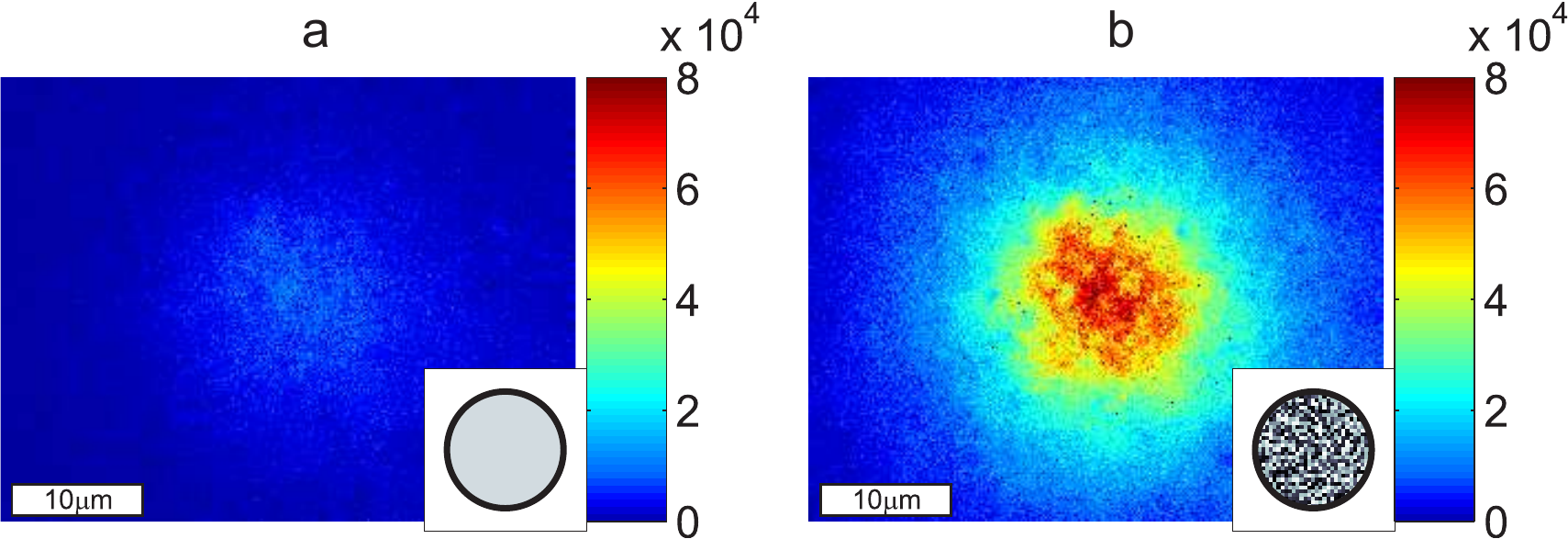}
\caption{Experimental demonstration of channel demixing. \subfig{a} Fluorescence image of a 300 nm sphere,
embedded in ZnO pigment at a depth of $\mum{9.7}$. Conventional illumination after a scan to find the
position where the fluorescence is maximal; approximately 3 times the speckle average intensity. Inset,
phase of the incident wavefront (plane wave). \subfig{b} Our channel demixing method. Same sphere and
sample position as in \subfigref{a}, the phases of the incident channels were set to the measured optimal
values. The resulting fluorescence intensity has increased by a factor of 22 with respect to the average
intensity. Inset, phase of the incident wavefront. Inside the sample this seemingly random wavefront
transforms into a focus. Intensities are in counts per second.} \label{fig:results}
\end{figure}

We have developed an approach that does allow focusing of light
inside a metamaterial. Our method combines an advanced feedback and
control system with a nanoscale fluorescent probe that is situated
at the desired location. The power of the emitted light is a measure
for the intensity at the position of the probe. In our experiment
(Fig.~\ref{fig:setup-focusing-inside}a), we used a liquid crystal
spatial light modulator (SLM) that spatially modulates the phase of
light coming from a green (532 nm) continuous wave laser. The pixels
of the SLM are grouped into 640 square segments that are imaged onto
the back aperture of a microscope objective. In this configuration,
at the focal plane of the objective each segment corresponds to a
solid angle encompassing approximately one mesoscopic scattering
channel. For each of these incoming channels the phase can be set
individually using the SLM. The microscope objective illuminates a
sample that is mounted on a xyz-translation stage. The fluorescence
light emitted by the probe inside the sample is collected by the
same microscope objective and imaged with an electron multiplying
charge coupled device (EMCCD) camera placed in a confocal reflection
configuration. Our samples consist of a white layer of zinc oxide
(ZnO) pigment on a glass substrate
(Fig.~\ref{fig:setup-focusing-inside}b). The pigment grains have an
average diameter of 200 nm, resulting in a sub-wavelength disordered
structure (Fig.~\ref{fig:setup-focusing-inside}c). The pigment layer
is up to $\mum{32}$ thick, and all samples are fully opaque. The
transport mean free path was measured by fully analyzing transport
of light through a series of samples (see
Section~\ref{sec:Experimental-details}). The samples were found to
be strongly photonic with a mean free path of $\mum{0.7\pm0.2}$,
comparable to the wavelength of light. Fluorescent spheres with a
diameter of $\nm{300\pm 15}$ are sparsely dispersed through the
sample.

Conventional focusing of light does not work in disordered photonic media. The absence of a ballistic
focus was confirmed by performing an extensive 3-dimensional scan of the sample with respect to the
microscope objective (see Section~\ref{sec:Experimental-details}). The only change in the fluorescence
emission was the result of the volume speckle field in the medium that is formed by scattered
monochromatic light. As the sample moves, the weak fluorescence emission fluctuates randomly with the
sample position. We consequently find that the maximal emission found by scanning the sample is only
approximately 3 times the typical emission. In numerical simulations we find a similarly weak enhancement
(Section~\ref{sec:Theoretical-details}). The fluorescence emission with the sample positioned at the
brightest speckle is shown in Fig.~\ref{fig:results}a.

We now cycle the relative phase of a single incident channel between 0 and $2\pi$ while keeping the phase
of all other channels constant. The intensity at the probe is the result of interference between the
modulated channel and the other channels, and therefore has a sinusoidal dependence on the phase. From the
phase and the amplitude of the sinusoid we obtain the complex propagation amplitude from the modulated
channel to the point inside the metamaterial. The complete set of propagation amplitudes contains all
information that is needed to calculate the propagation of any incident wave towards the target. The
optimal configuration of incident channels is found by taking the complex conjugate of the propagation
amplitudes, so as to fully compensate for the mixing of the channels that will occur.

Figure~\ref{fig:results}b shows the strong increase in fluorescence emission that was achieved with
channel demixing. By constructing the optimal configuration of incident channels, the intensity
distribution inside the sample changed from random speckle to a pattern specially tailored for focusing
light at the target. We find that the amplitudes for neighboring channels were uncorrelated, which is an
indication that the channels were originally fully mixed. Light from different incident channels mixed
while propagating in the medium and demixed again at the target position to form a focus. This focus was a
factor of 22 more intense than the typical intensity and a factor of 6 more intense than the maximum that
was achieved by searching for the brightest speckle.

\begin{figure}\centering
\includegraphics[width=\smallimage]{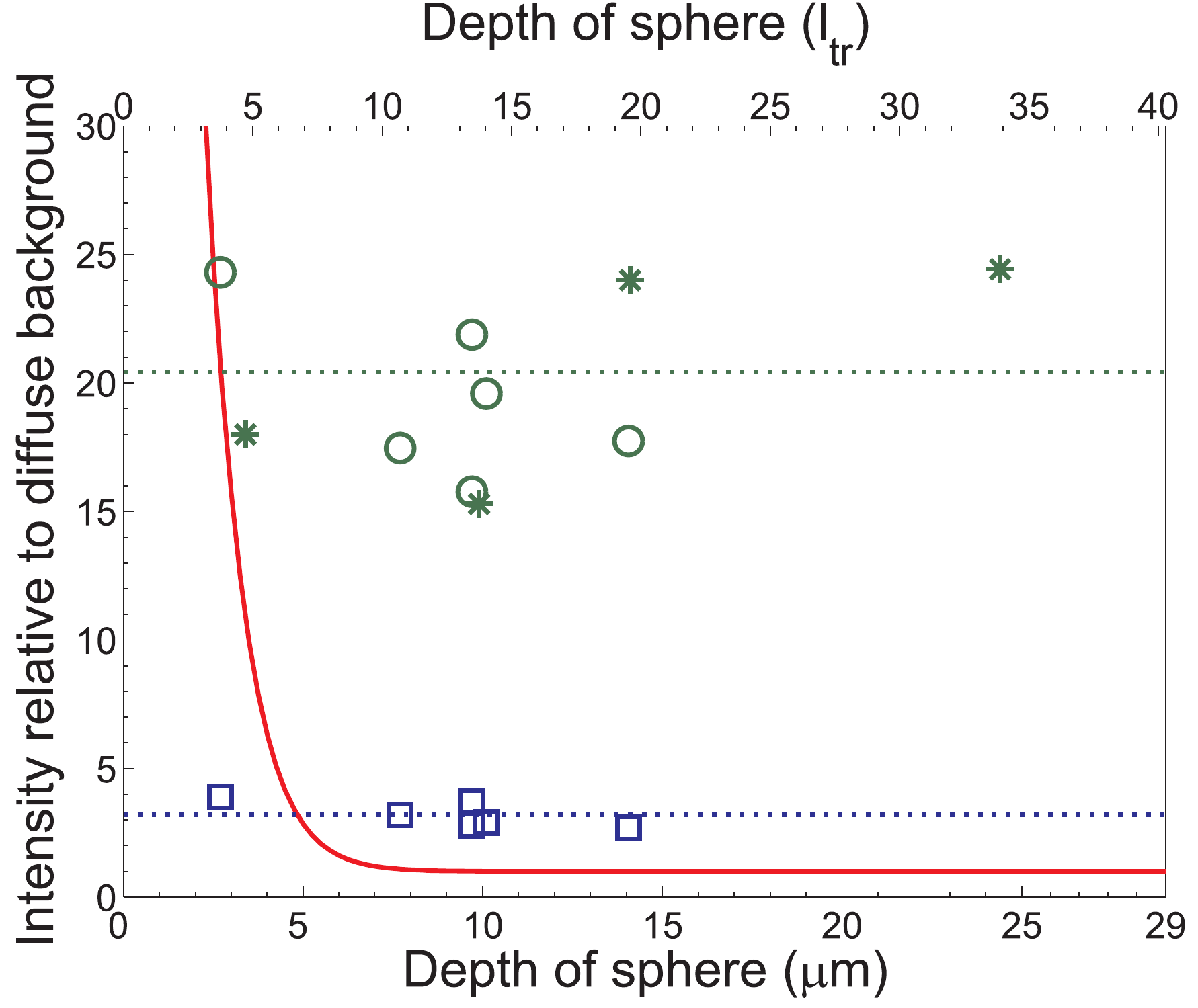}
\caption{Measured intensity for targets at different depths inside a
$\mum{29\pm3}$ thick sample. Stars, results of our channel demixing
technique, starting from a random sample position. An intensity was
reached that is on average 20.4 times higher than the diffuse
background (dotted line). Circles, results of our channel demixing
technique starting on a bright speckle, demonstrating that starting
from a bright spot is not advantageous. Squares, maximum intensity
that was achieved by searching for the brightest speckle. An average
intensity increase of only 3.2 was obtained (dotted line). Solid
line, theoretical best case speckle average using adaptive optics,
assuming that all geometrical aberrations are corrected perfectly.
All measurements are normalized to the diffuse background.}
\label{fig:depthdata}
\end{figure}

The experiment was repeated for targets at different depths inside a $\mum{29\pm3}$ thick sample. In
Fig.~\ref{fig:depthdata} the target intensity that was reached with channel demixing is compared to the
maximum intensity reached by translating the sample. Channel demixing increases the intensity at the
target by a factor of on average 20.4, regardless of the depth of the target. It makes no difference
whether the sample was meticulously placed on the brightest speckle first or whether the channel demixing
procedure was started without a preceding scan of the sample position. For completeness, we also plot the
calculated residual ballistic focusing component that can be expected when the sphere is placed exactly in
the geometrical focus. For this calculation, we optimistically assumed that all geometric aberrations are
perfectly corrected by an adaptive optics system\cite{Tyson1998}. As can be seen in
Fig.~\ref{fig:depthdata}, adaptive optics could be effective for targets that are only a few mean free
paths deep. When the target lies deeper, scattering on the sub-wavelength structure of the material
destroys even a perfectly corrected geometrical focus. Using channel demixing, however, we consequently
find that the increase does not depend on the position of the target. Therefore, we conclude that
scattering in no way limits the depth at which light can be delivered inside the metamaterial.

Theoretically, in the current experimental configuration, channel demixing has the possibility to achieve
an intensity ratio that is approximately four times as high as the measured value of 20.4 (see
Section~\ref{sec:Theoretical-details}). In practice, however, we are limited by photobleaching of the
probe. We expect to achieve even higher intensity ratios as more stable probes, such as quantum dot
assemblies\cite{Han2001}, become available.

The measured set of complex propagation amplitudes characterizes propagation of light towards a
sub-wavelength target inside the metamaterial. For this measurement no physical access to the target is
required; it is not even required to know where the target probe exactly is. In both ordered and
disordered metamaterials these amplitudes contain information about the structure and the optical quality
of the metamaterial.\cite{John1988, Hughes2005, Shapiro2005, Weaver2005} This information, which was not
experimentally accessible before, makes channel demixing a unique method for characterizing any
metamaterial. Our experimental results call for the development of theoretical approaches that treat light
propagation inside disordered photonic metamaterials.\cite{note-BeenakkerPendry}

We have shown that light can be focused on a target inside a photonic metamaterial that completely
scrambles the incident waves. Since the illumination is selective, channel demixing can be used to
increase the contrast with a fluorescent background. A different application would be to selectively
activate photochemical processes by concentrating light at a point inside a scattering material. Finally,
the use of a fluorescent probe permits selectively focusing light on fluorescent areas, which is highly
desirable in biological imaging, and may be a key technology for selectively illuminating individual cells
in disordered biomaterials such as human skin tissue.

\begin{sectionappendix}
\section{Experimental details}\label{sec:Experimental-details}

\subsection{Apparatus}
\begin{figure}\centering
\includegraphics[width=\mediumimage]{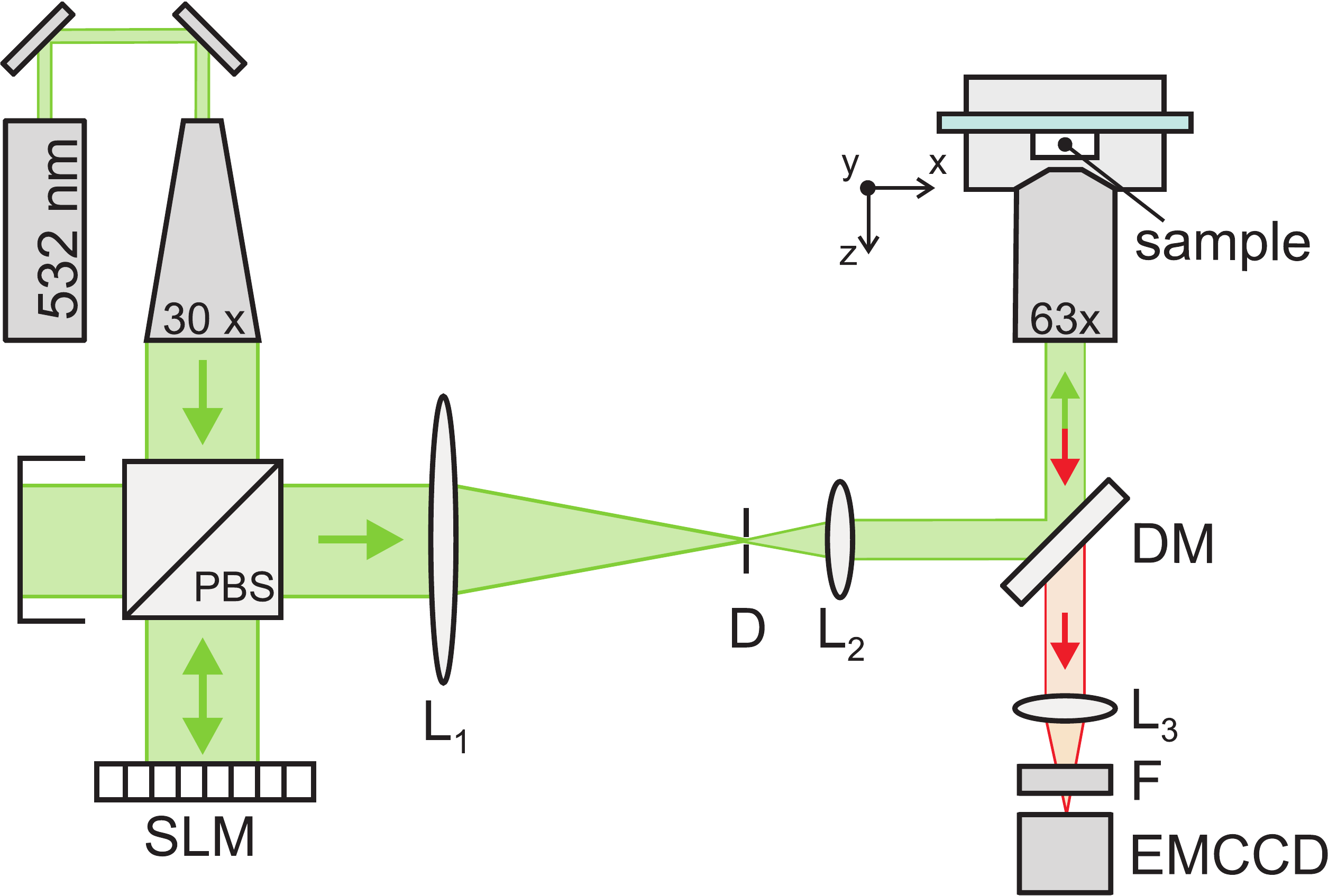}
\caption{Experimental setup. The left part consists of a laser (532 nm) that is expanded by a 30x beam
expander and modulated with a spatial light modulator (SLM). A 1:2 demagnifying telescope images the SLM
on the back aperture of a 63x microscope objective. The objective focuses the light on a sample that is
mounted on a XYZ piezo positioning stage. A dichroic mirror (DM) and a bandpass filter (F) block the
excitation light. The fluorescence emission is imaged on an electron multiplying CCD camera (EMCCD). D,
iris diaphragm; PBS, polarizing beam splitter cube; L$_1$, L$_2$, L$_3$, lenses with a focal distance of
200 mm, 100 mm and 150 mm, respectively. Some folding mirrors and beam attenuation optics were
omitted.}\label{fig:setup-focusing-inside-detailed}
\end{figure}

\noindent The goal of the experiment is to illuminate a disordered medium in such a way that light focuses
at a predefined target deep inside. The experimental setup consists of a spatial light modulator that is
used to adjust the illumination, and a confocal fluorescence microscope that indirectly measures the
intensity at the target. A detailed schematic of the setup is given in
Fig.~\ref{fig:setup-focusing-inside-detailed}. Light from a continuous wave 532 nm laser (Coherent Compass
M315-100) is expanded using a 30x beam expander. The expanded beam illuminates a Holoeye LC-R 2500 spatial
light modulator (SLM) with a resolution of 1024 x 768 pixels. The pixels are grouped into square segments
of 20 x 20 pixels. We have independent control over the phase and amplitude of each of the
segments.\cite{Putten2008thesis} During this experiment, the amplitude of all segments is kept constant.
The SLM is imaged onto the back aperture of a microscope objective using a 1:2 demagnifying telescope. In
total, 640 segments fall within the aperture of the objective, which means that we control 640 degrees of
freedom. The microscope objective (Zeiss infinity corrected 63x, NA=0.95 Achroplan air) focuses the light
on the sample, thereby mapping the segments to channels in k-space. The sample is mounted on a
XYZ-nanopositioning stage (Physik Instrumente P-611.3S NanoCube) to allow for a precise 3D-scan of the
geometrical focus.

\sloppy

The returned fluorescence light is separated from the excitation
light using a combination of a dichroic mirror (Semrock
FF562-Di02-25x36) and a bandpass filter (Semrock FF01-593/40-25). An
electron multiplying CCD camera (Andor Luca DL-658M) images the
fluorescence emission. A computer sums the intensity in the camera
image to find the total power in the diffuse spot of fluorescence
light. For each of the segments, measurements are performed at phase
delays of $0, 2\pi/5, 4\pi/5, 6\pi/5$, and $8\pi/5$ and a sinusoid
is fitted to these measurements. Before advancing to the next
segment, the phase delay is reset to 0.

\myfussy

\subsection{Measurement sequence}\label{sec:timetrace} \noindent
\noindent Channel demixing is achieved by measuring the propagation coefficients from the incident
channels towards the embedded target. After all propagation coefficients are measured, it is known which
linear combination of scattering channels focuses at the target. To quantify the effectiveness of this
method, we measure the intensity at the probe position before and after the demixing procedure.

Since the field inside the medium is a completely random speckle
pattern, the initial intensity is a random variable. A single
measurement will not provide a reliable reference. Therefore, we
measure the speckle averaged initial intensity $I_0$ by generating
500 different random combinations of incident channels and
determining the average fluorescence intensity. This way,
fluctuations caused by volume speckle are averaged out and a
reliable measure for the initial intensity is found.

Next, the propagation constants for the different scattering channels are measured. This measurement takes
approximately 18 minutes. During these measurements, the dye in the sphere photobleaches, which causes the
response of the probe to decrease. Since we are interested in how much excitation light is focused on the
target position, we have to take into account the reduced response of the probe particle
\begin{equation}
\eta_\text{measured} = \frac{I_\text{opt}}{I_0} \frac{I_1}{I_2},
\end{equation}
where $I_\text{opt}$ is the fluorescence intensity directly after
constructing the optimal combination of scattering channels. $I_1$
is the fluorescence intensity at the start of this measurement step
(directly after the reference measurement), and $I_2$ is the
intensity at the end of the measurements (directly before the SLM is
programmed with the optimal configuration). The ratio $I_1/I_2$ -
which is between 1.0 and 1.8 when the sample is positioned at a
typical position and between 1.5 and 3.9 when the sample is
positioned at the brightest speckle - corrects for the fact that
photobleaching decreases the sensitivity of the probe.

\begin{figure}\centering
\includegraphics[width=\smallimage]{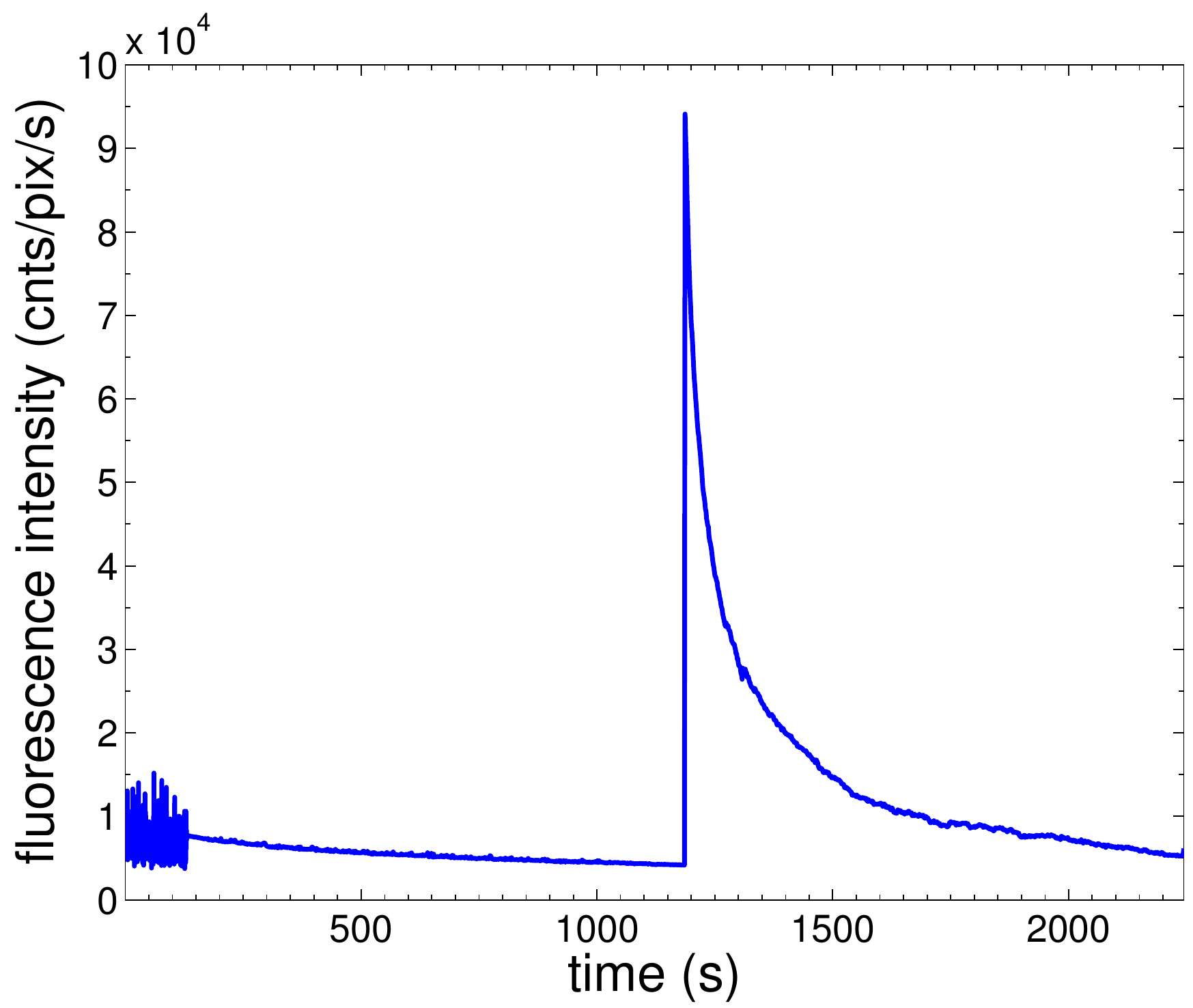}
\caption{Evolution of the fluorescence intensity during the measurements. The measurement procedure has
three phases. 0-130 s, reference measurement; 130-1187 s, measurement of propagation coefficients; at 1187
s, construction of optimal wavefront; 1187-2244 s, fluorescence measurement with optimal
wavefront.}\label{fig:timetrace}
\end{figure}

Figure~\ref{fig:timetrace} shows how the fluorescence intensity evolves during the measurement sequence.
During the reference measurement, the intensity fluctuates rapidly. In the next step, the propagation
constants are measured. Since we only cycle the phase of a single channel at a time, the illuminating
wavefront is practically constant. During this step the fluorescence intensity decreases slowly due to
photobleaching. After the measurements have been performed for all segments, the phase modulator is
programmed with the optimal phases and the fluorescence intensity jumps to a higher value as the incident
light now interferes constructively at the target. After this jump, the wavefront is kept constant. The
reason that the fluorescence intensity decreases is that the higher incident intensity on the sphere
causes the dye to bleach more rapidly.

\subsection{Sample preparation and characterization}
\noindent We demonstrated how interference of scattered light was
used to focus light deep inside disordered metamaterials. Our samples are strongly scattering, that is,
the mean free path for light is of the order of the wavelength. Inside the sample, fluorescent probes mark
the target positions. Typically, the probes are over 10 mean free paths deep, which makes it impossible to
focus light on them using conventional means (Also see Section~\ref{sec:intensity-inside}). In this
section we address the fabrication and characterization of these samples.

\subsubsection{Sample preparation} \noindent Our samples consist of a layer of Zinc Oxide pigments (Aldrich
Zinc Oxide  $< \mum{1}$, 99.9\%). The average particle diameter is $\nm{200}$ (measured using scanning
electron microscopy). The fluorescent probes are made from polystyrene and have a specified diameter of
$\nm{300\pm 15}$ (Duke Scientific red fluorescent nanospheres dyed with FireFly$^\text{TM}$). An aqueous
suspension of pigment and spheres was sprayed on a standard microscope cover slide using a commercial
airbrush (Harder \& Steenbeck Evolution). The thickness of the dried samples was measured with an optical
microscope by making a scratch in the surface. The background fluorescence of the ZnO was shown to be
negligible compared to the fluorescence of the dyed nanospheres.

\subsubsection{Measurement of the mean free
path}\label{sec:measurements-mfp} \noindent Transport of light in scattering media is quantified by the
transport mean free path $\ell$. A method for obtaining $\ell$ is to measure the total (diffuse)
transmission of light using an integrating sphere. $\ell$ then follows from the relation\cite{Rivas1999}
\begin{equation}
\frac{1}{T} = \frac{L + z_{e1} + z_{e2}}{\ell +
z_{e1}},\label{eq:TT}
\end{equation}

\begin{figure}\centering
\includegraphics[width=\smallimage]{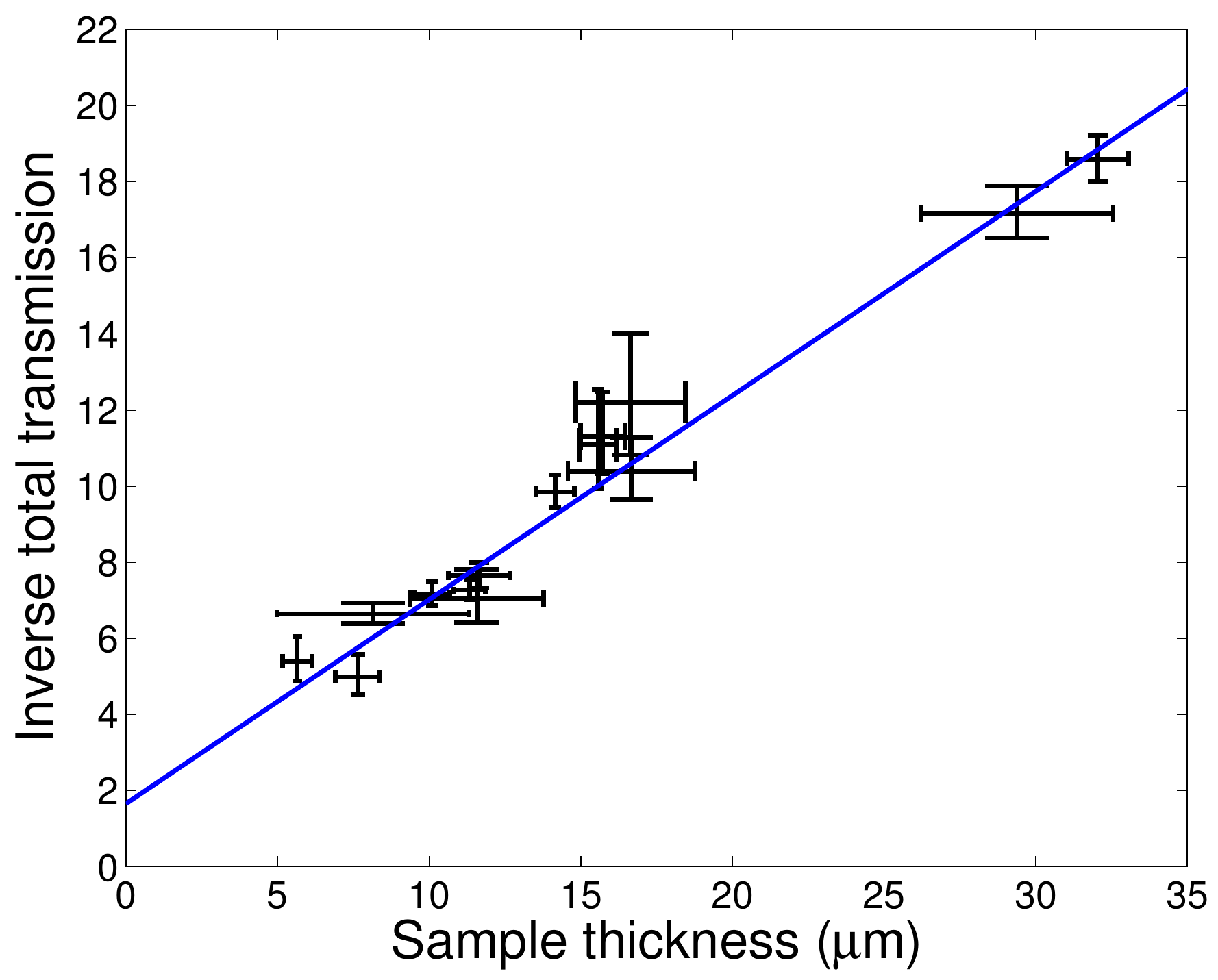}
\caption{Inverse total transmission through layers of ZnO pigment. The solid line is a fit of
Eq.~\eqref{eq:TT} with $\ell = \mum{0.72}$, $z_{e1} = \mum{1.15}$, and $z_{e2}= \mum{1.24}$. Error bars
indicate the standard deviation of the measured thickness and total transmission at different positions on
the sample.} \label{fig:TT}
\end{figure}

\noindent where $T$ is the total intensity transmission, $L$ is the sample thickness and $z_{e1}$,
$z_{e2}$ are the extrapolation lengths that describe boundary effects on the front and back surfaces of
the sample respectively. The extrapolation lengths follow from the effective refractive index of the
medium, $n_\text{eff}$.\cite{Vera1996} We measured the total transmission through a set of similar samples
with varying thickness. The inverse total transmission for an ensemble of 14 samples is plotted in
Fig.~\ref{fig:TT}. The samples range in thickness from $\mum{6}$ to $\mum{32}$. As expected, the inverse
total transmission increases linearly with sample thickness. We obtain a good linear fit for $\ell =
\mum{0.7\pm 0.2}$ and $n_\text{eff} = 1.4 \pm 0.1$. The extrapolation lengths can then be calculated from
$n_\text{eff}$ and we find $z_{e1}=\mum{1.15 \pm 0.2}$ and $z_{e2}=\mum{1.24 \pm 0.2}$.

\subsubsection{Determination of the depth of the fluorescent probe}
\noindent For each experiment we select a different probe sphere at
a different depth inside a sample. To determine the depth of the
probe, we image the diffuse spot of emitted light and fit this data
with diffusion theory. The accuracy of methods that rely on
inverting the diffusion equation is limited to about one mean free
path due to noise.\cite{McLean1995}

The intensity inside a disordered medium with a source at depth
$z_0$ follows from diffusion theory. We solved the diffusion
equation for a slab of thickness $L$ using the Dirichlet boundary
conditions $I(-z_{e1}) = 0$, and $I(L+z_{e2})=0$ \cite{Carslaw1959,
Chandrasekhar1960}. The extrapolation lengths $z_{e1}$ and $z_{e2}$
account for the reflections at the front and the back of the sample,
respectively. A closed form solution is found using Fourier
transformed traversal coordinates $\mathbf{q_\perp} \equiv (q_{\perp
x}, q_{\perp y})$.
\begin{equation}
I_d(\mathbf{q_\perp}, z) =
    \begin{cases}
        J_\text{in}\frac{\displaystyle\sinh{(q_\perp [L_e-z-z_{e1}])}\sinh{(q_\perp [z_0+z_{e1}])}}{\displaystyle D q_\perp \sinh{(q_\perp L_e)}} & z > z_0\\
        J_\text{in}\frac{\displaystyle\sinh{(q_\perp [L_e-z_0-z_{e1}])}\sinh{(q_\perp [z+z_{e1}])}}{\displaystyle D q_\perp \sinh{(q_\perp L_e)}} & z \leq z_0\\
    \end{cases}
    \label{eq:diffusion-equation-solution}
\end{equation}
where $I_d$ is the diffuse energy density at depth $z$, $q_\perp \equiv \sqrt{q_{\perp x}^2+q_{\perp
y}^2}$, and $J_\text{in}$ is the power of the source in Watts. $D$ is the diffusion constant and $L_e
\equiv L + z_{e1} + z_{e2}$.

To obtain the depth of the source, we Fourier transform the microscope image of the fluorescence light and
fit Eq.~\eqref{eq:diffusion-equation-solution} with z=0 using $z_0$ and an intensity prefactor as fit
parameters.

\subsection{3-dimensional scan results}
\begin{figure}\centering
\includegraphics[width=\smallimage]{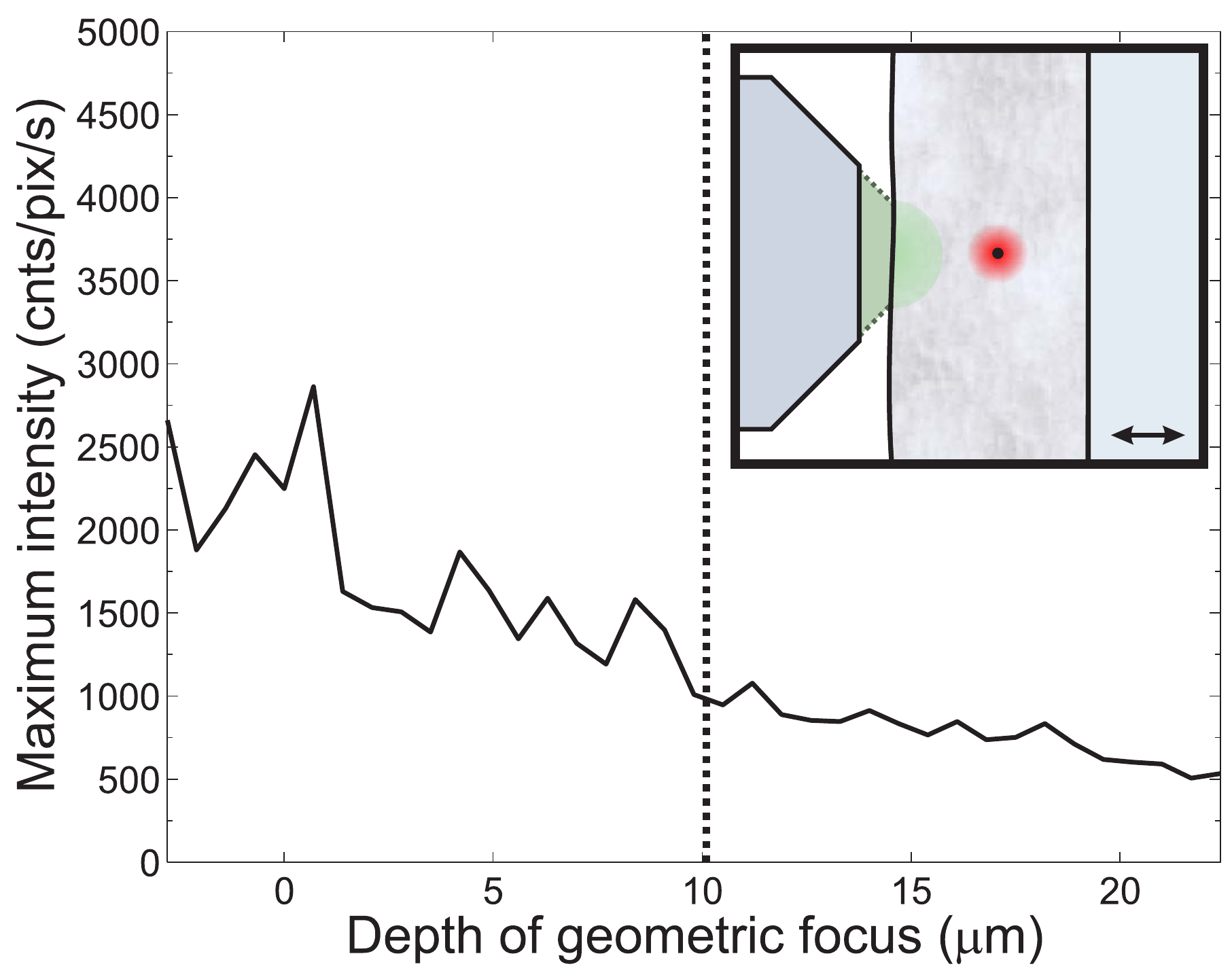}
\caption{Determination of the absence of a geometrical focus. The sample is moved with respect to the
objective (inset). For each depth, the sample is scanned in a plane parallel to the surface to find the
maximum fluorescence intensity (solid line). In homogeneous media, there would be a sharp peak at the
point where the fluorescent sphere is in the geometrical focus of the microscope objective (dotted line).
In disordered photonic media, a geometric focus cannot be formed. The highest intensity is found when
`focusing' just below the sample surface.} \label{fig:zscan}
\end{figure}

\noindent Conventional focusing does not work in strongly scattering materials such as our samples. The
absence of a ballistic focus was confirmed by performing an extensive 3-dimensional scan of the sample
with respect to the microscope objective. The fluorescence emission strongly fluctuates with the sample
position. These fluctuations are the result of a volume speckle field in the medium that is formed by
scattered monochromatic light. As the sample moves, the volume speckle fluctuates randomly. For each depth
of the geometrical focus, we record the brightest speckle intensity. Figure~\ref{fig:zscan} shows the
results of a typical 3-D scan. The depth of the geometrical focus was estimated to be $n_\text{eff} z$,
with $n_\text{eff}=1.4$ as measured, and $z$ the position of the translation stage. No sign of a ballistic
focus was observed. In other words, it is not possible to focus light on the target using conventional
means.

\stepcounter{sectionapp}
\section{Analysis of the channel demixing method}\label{sec:Theoretical-details}
\noindent The optical field $E$ at a point $\rb$ inside the medium
is given by
\begin{equation}
E(\rb) = \iiint \ud^3\ra G(\rb, \ra) S(\ra),\label{eq:green}
\end{equation}
where $G$ is Green's function for propagating from the sources
$S(\ra)$ to point $\rb$. In a disordered medium, $G$ is stochastic
and, therefore, initially unknown. We treat each segment of the
phase modulator as a sheet source with amplitude $A$ and phase
$\phi$. Since the phase modulator is in good approximation
illuminated homogeneously, all amplitudes $A$ are equal. By
integrating over the surface area $S$ of each of the $N$ segments,
Eq.~\eqref{eq:green} is discretized,
\begin{align}
E(\rb) &= \sum_a^N \iint_{S_a} \ud^2\ra G(\rb, \ra) A e^{i \phi_a},\\
&\equiv A \sum_a^N t_{ba} e^{i \phi_a}.\label{eq:field-before}
\end{align}

\noindent When the phase of a single segment $a$ of the phase
modulator is changed, the intensity at point $\rb$ responds as
\begin{equation}
I(\rb)\equiv |E(\rb)|^2 = I_{0b} + 2 A \mathrm{Re} (E_{b\bar{a}}^*
t_{ba} e^{i \phi_{a}}),\label{eq:intensity-during}
\end{equation}
with
\begin{align}
I_{0b} &\equiv \left|E_{b\bar{a}}\right|^2 + A^2 |t_{ba}|^2,\\
E_{b\bar{a}} &\equiv A \sum_{a'\neq a}^N t_{ba'} e^{i\phi_{a'}}
\approx E(\rb).
\end{align}
Since the number of segments $N$ is large, $E_{b\bar{a}}\approx
E(\rb)$ and, therefore, equal for each of the segments. By repeating
this procedure for all segments $a$, we measure the coefficients
$t_{ba}$ up to an unknown common prefactor $E(\rb)$.

Once all coefficients $t_{ba}$ are known, our channel demixing
method maximizes $E(\rb)$ by setting $\phi_a = -\text{arg} (t_{ba})$
for all segments. This configuration of phases gives the global
maximum of Eq.~\eqref{eq:field-before},
\begin{equation}
E_\text{max}(\rb) = A \sum_a^N |t_{ba}|.\label{eq:field-after}
\end{equation}
A comparison with Eq.~\eqref{eq:field-before} shows that now all amplitudes are summed with the same phase
and that all incident channels will interfere constructively at the target.

\subsection{Maximum enhancement - scalar waves,
simplified case}
\noindent The theoretical increase in intensity that can be achieved with channel demixing depends on the
number of modulator segments $N$ and the statistical properties of coefficients $t_{ba}$. In disordered
systems it is often allowed to assume that the coefficients are independently drawn from a single circular
Gaussian distribution.\cite{Goodman2000,Garcia1989,Beenakker1997,Pendry1992,Mirlin1998} Under this
assumption, the expected ratio between the original diffuse intensity and the maximally achievable
intensity equals
\begin{equation}
\eta = \frac\pi4(N-1)+1.\label{eq:enhancement-max}
\end{equation}
where $N$ can never exceed the number of mesoscopic scattering channels. The number of contributing
scattering channels depends on the depth of the target. The deeper the target is embedded in the
scattering material, the more channels will contribute to the mixing process. We estimate the number of
channels that contribute to the target field by examining the size of the diffuse spot of the fluorescence
light. The number of scattering channels in a spot with surface area $A$ is given by\cite{Tiggelen1993,
Boer1995} $N_\text{max} = 2 \pi A / \lambda^2$. In Fig.~\ref{fig:results}b, the full width at half maximum
of the diffuse spot is $\mum{15}$, which leads to $A=\mum{177}^2$ and $N_\text{max} = 3.9\cdot 10^3$. In
our experiment $N = 640 \ll 3.9\cdot 10^3$ and neighboring segments were uncorrelated.

\subsection{Maximum enhancement - finite size probe}
\noindent Ideally, the sphere that is used as a probe is only sensitive to the field of a single
polarization at a single point. In reality, however, the fluorescent sphere probes the summed intensity of
three polarizations in a finite volume. Therefore, it is as if we have multiple targets for which we try
to increase the intensity simultaneously. As a result, the extra intensity is divided over all targets.
The ratio between the original diffuse intensity and the maximally achievable intensity now equals
\begin{equation}
\eta_\text{M} = \frac\pi4 \frac{N-1}{M}+1,\label{eq:enhancement-max-mm}
\end{equation}
where $M$ is the number of targets. We determine $M$ by calculating how many orthogonal fields can exist
in the volume of the probe sphere. Using Mie theory we find that for spheres of $300\pm15$ nm the three
electric dipole modes and the three magnetic dipole modes all give an equally strong contribution to the
fluorescence power. The sphere is too small to support higher order (quadrupole) modes, so we estimate
that the probe contains 6 optical modes. Using Eq.~\eqref{eq:enhancement-max-mm} with $M=6$ and $N=640$,
we find an expected theoretical enhancement of 85. The experimental enhancement was approximately 25\% of
this value. We believe that the most important reason for not reaching the theoretical maximum enhancement
is photobleaching. The channel demixing method favors the mode that initially is the brightest.
Unfortunately, this target mode also photobleaches the fastest. Since we correct only for the average
photo bleaching rate (see Section~\ref{sec:timetrace}), the measured enhancement will systematically be
lower than the actual enhancement.

\subsection{Maximum enhancement of speckle scan}
\noindent When the sample is moved, the speckle pattern in the
medium fluctuates. As a result, the probe fluorescence will
fluctuate too. Therefore, when a 3D-scan of the sample position is
performed, at some points the intensity will be somewhat higher than
the diffuse background. We used numerical simulations to calculate
the expected maximum enhancement that this scanning technique can
achieve. We assumed Rayleigh statistics for the speckle field and
calculated the summed intensity in $6$ independent optical modes at
$N_r$ different positions. The simulation was repeated 1000 times
and for each run the maximum speckle intensity was stored. The
number of independent speckles, $N_r$, is limited by the diffuse
intensity envelope of the light inside the material. We estimate
that $N_r$ is between 1000 and 10000, depending on the depth of the
probe. With these numbers, the simulations predict that the
brightest speckle is 2.9 to 3.4 times as intense as the diffuse
background.

\subsection{Diffuse and ballistic intensities inside the medium\label{sec:intensity-inside}}
\subsubsection{Diffuse intensity}
\noindent When a slab of diffusive material is illuminated, the incident light can be described by a
diffuse source at depth $z_0 = \ell$ in the medium.\cite{Akkermans1986} The maximum diffuse intensity at
depth $z$ is found by inverse Fourier transforming Eq.~\eqref{eq:diffusion-equation-solution} with
$x=y=0$.
\begin{equation}
I_d(x=0, y=0, z) = \frac{1}{2\pi}\int_0^\infty \ud q_\perp I_d(q_\perp, z) S(q_\perp)
q_\perp.\label{eq:Id-integral}
\end{equation}
Where $S(q_\perp)$ accounts for the spatial extent of the source. We modelled the source as a Gaussian
spot that encompasses 640 scattering channels; $S(q_\perp)=\exp(-q_\perp^2/0.43)$.
Eq.~\eqref{eq:Id-integral} was evaluated numerically. The diffuse intensity was used to normalize the
theoretical curve for the perfect ballistic intensity in Fig.~\ref{fig:depthdata}.

\subsubsection{Perfect ballistic intensity} \noindent We assume that
all geometrical abberations are perfectly corrected and that all the
incident light is directed towards a sphere in the geometrical
focus. While travelling through the disordered medium, scattering
and diffraction on sub-wavelength structures attenuate the incident
beam according to the Lambert-Beer law,
\begin{equation}
I_b = \frac{e^{-z/\ell_{sc}}}{A v}, \label{eq:ballistic-intensity}
\end{equation}
where $I_b$ is the ballistic energy density at the sphere, $A$ is the cross-section area of the sphere,
$z$ is the depth of the sphere, $v$ is the velocity of light inside the medium and $\ell_{sc}$ is the
scattering mean free path for light. The parameters $\ell_{sc}$ and $v$ have not been measured directly
for our samples. We can, however, give an upper limit for the ballistic intensity by overestimating
$\ell_{sc} = \ell$ and underestimating $v = 3 D / \ell$. The ballistic intensity was normalized by the
diffuse intensity and plotted in Fig.~\ref{fig:depthdata}.

\end{sectionappendix}
\bibliography{../../bibliography}
\bibliographystyle{Ivo_sty}

\setcounter{chapter}{5}
\chapter{Exploiting the potential of disorder in optical communication\label{cha:communication}}


\noindent Scattering of light is a long recognized problem that
causes losses in optical telecommunication, and decreases the
resolution of imaging systems\cite{Ishimaru1978}. However,
impressive experiments utilizing ultrasound and radio waves have
shown that, for these waves, scattering and interference can be used
as an advantage.\cite{Simon2001} Disordered scattering is used for
ultrafast radiographic communication\cite{Foschini1996} and for
sub-wavelength focusing of microwaves\cite{Lerosey2007}.

Unfortunately, these revolutionary methods have so far not been
available in the optical regime. The reason is that recording and
playing back electrical fields at optical frequencies is close to
impossible. Therefore, time-reversal - which is highly effective for
sound and radio waves - is not a realistic option. In this chapter,
we show that the potential of scattering can be exploited without
time-reversal. A self-learning optical system uses scattering to
improve its resolution beyond the limit imposed by diffraction at
its aperture. Thanks to the increased resolution, multiple
communication channels can coexist in the space that is normally
needed to support one unscattered communication channel.

We first explain the principle of the experiment and present our results. Then, in
Section~\ref{sec:comm-experiment}, we give a more detailed description of the apparatus and the
experimental procedure.

\section{Increasing the information density}
\begin{figure}
\centering
\includegraphics[width=\wideimage]{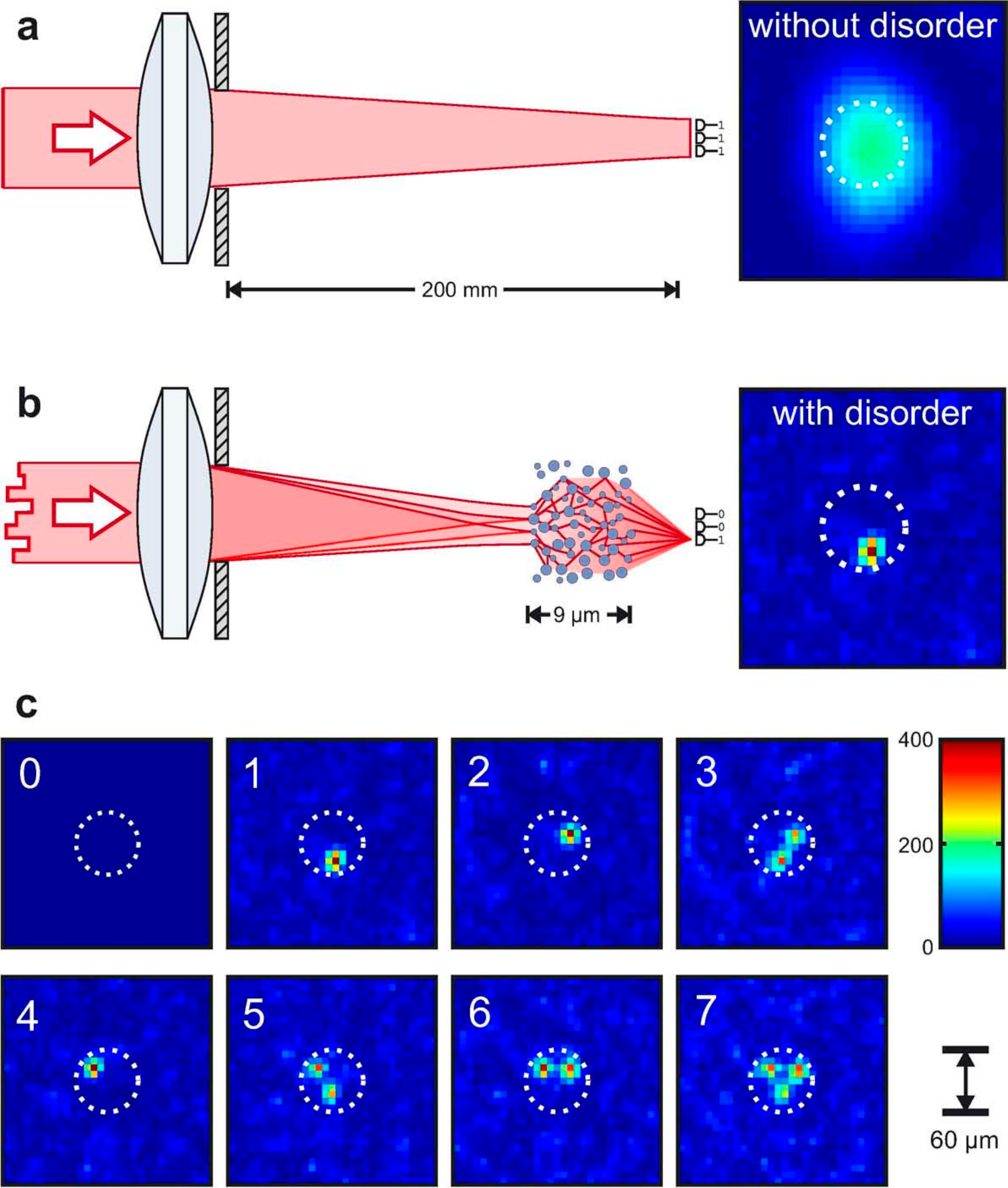}
\caption{\subfig{a} An optical system with a limited numerical aperture cannot focus light sharply, all
three detectors receive the same signal. \subfig{b} Scattering improves the diffraction limit. The
incident light is spatially shaped to match a scattering fingerprint. Scattered light focuses sharply to a
point. Now, each detector can be addressed individually. \subfig{c} Measured transmitted intensity for eight different
linear combinations of incident wavefronts. Any three-bit pattern can be generated
in the area previously needed for one bit.\label{fig:communication}}
\end{figure}

Figure~\ref{fig:communication}a shows an optical system where a lens
focuses monochromatic light on a camera. Due to the limited
numerical aperture (NA=1/200) of the system, the spot will at best
have a diffraction limited width of $\mum{60}$ (full width at half
maximum). While aberrations in the optics can efficiently be removed
using adaptive optics\cite{Tyson1998}, the resolution of an adaptive
optics system is still limited by its numerical aperture.

We choose three individual pixels of the camera to define three
independent receivers. The spacing between the pixels is less than
the diffraction limit of the lens. When the beam path is blocked by
a disordered scattering medium (a $\mum{9}$-thick layer of
airbrushed white paint), the image on the camera changes
dramatically. Instead of a focal spot, we now record a disordered
speckle pattern that is the result of random interference. Although
the scattered light does not focus, its features are much sharper
than the size of the original diffraction limited spot because the
scattering process increases the effective numerical aperture of the
optical system.

To control the propagation of light through the disordered environment, we developed a learning algorithm
that acquires an optical fingerprint (a spatial amplitude and phase pattern) for each of the three
receivers. The optical fingerprint uniquely describes propagation through the scattering medium towards
the receiver. When the wavefront of the incident light matches this fingerprint, the scattered light
converges to a focus at the receiver (Fig.~\ref{fig:communication}b) and simultaneously the intensity on
the other two receivers decreases.

We used a recently developed technique to spatially shape both phase and amplitude of a laser beam with a
high accuracy.\cite{Putten2008thesis} This technique allowed us to construct any linear superpositions of
fingerprints. Propagation through the scattering medium separates the superposition into separated foci on
the receivers. All patterns of three bits can be generated, as is shown in Fig.~\ref{fig:communication}c.
A thin layer of white paint has increased the information density by a factor of three.

In conclusion, our method for matching optical fingerprints takes highly successful ideas from microwave
technology and introduces them to the optical regime. Apart from improving optical communication,
applications include sub-wavelength focusing with ordered\cite{Luo2003} or disordered\cite{Lerosey2007}
metamaterials.

\section{Experimental details\label{sec:comm-experiment}}

\begin{figure}
\centering
\includegraphics[width=\smallimage]{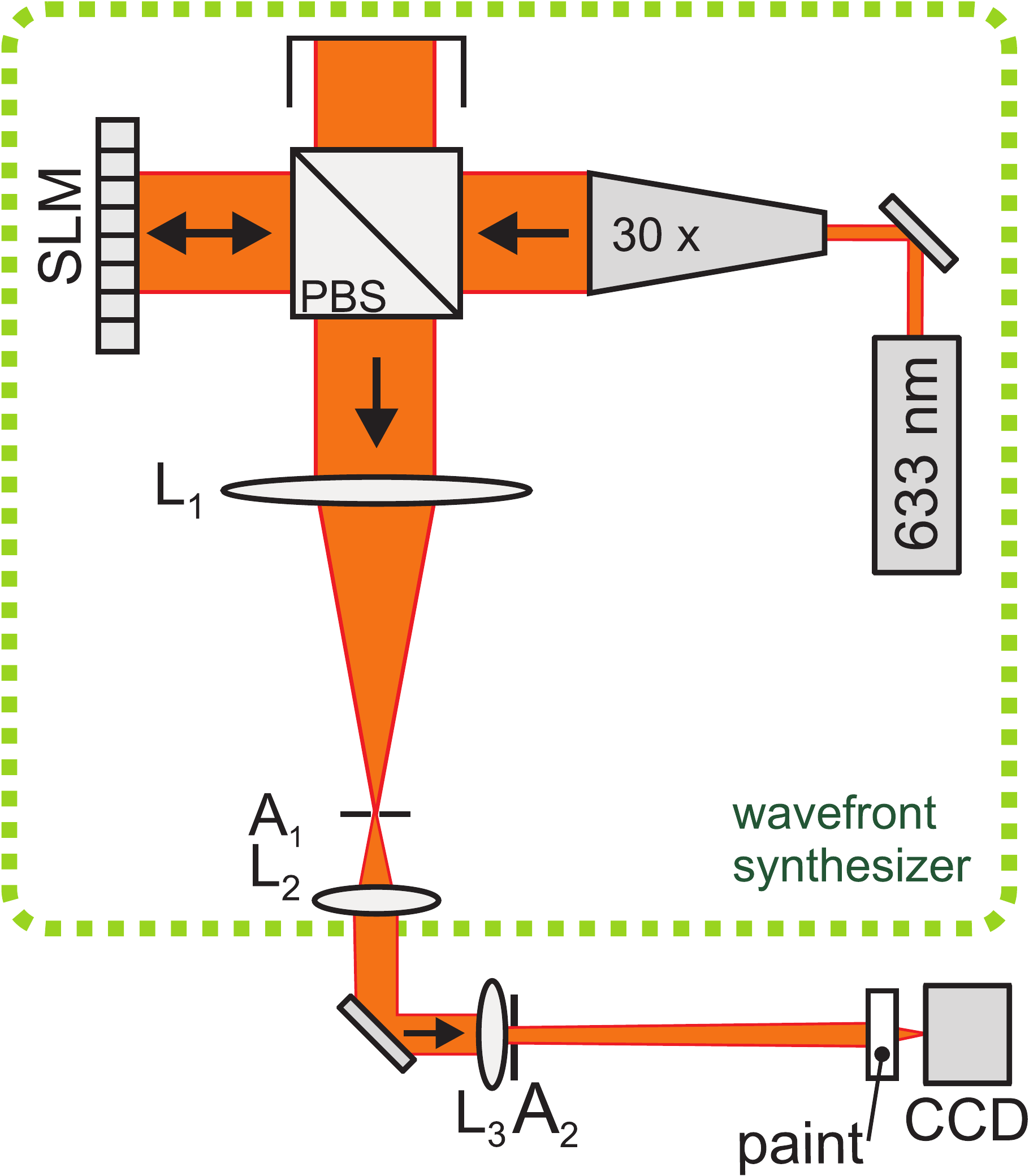}
\caption{Schematic of the experimental setup. The first part of the
setup consists of a wavefront synthesizer that modulates amplitude
and phase of a 632.8~nm Helium-Neon laser. The laser beam is
expanded by a 30x beam expander and modulated with a spatial light
modulator (SLM). A 1:5 demagnifying telescope images the SLM on the
surface of lens L$_3$. A pinhole in the focal point of the telescope
(A$_1$) acts as a spatial filter, which allows the SLM to modulate
both amplitude and phase (see text). The second part is the actual
experiment, consisting of a camera (CCD) that is placed in the focal
plane of lens L$_3$. The numerical aperture of the lens is lowered
using a 2.1-mm diameter aperture (A$_2$). At 26~mm from the camera a
$\mum{9}$-thick strongly scattering layer of white paint is placed.
A$_1$, A$_2$, apertures; PBS, polarizing beam splitter cube; L$_1$,
L$_2$, L$_3$, lenses with a focal distance of 250~mm, 50~mm and
200~mm, respectively. Some folding mirrors and beam attenuation
optics were omitted.}\label{fig:setup-communcitation}
\end{figure}

\noindent The setup consists of a low numerical aperture optical system and a wavefront synthesizer that
spatially modulates light from a 632.8~nm Helium-Neon laser (see Figure~\ref{fig:setup-communcitation}).
Wavefront modulation is achieved using a Holoeye LC-R~2500 spatial light modulator (SLM). The SLM is of
the twisted nematic type and by itself does not allow independent control over the phase and amplitude of
the reflected light. We developed a technique that combines four pixels into a macropixel of which both
phase and amplitude can be controlled fully.\cite{Putten2008thesis}

Without scattering, the optical system is not able to create a sharp
focus, regardless of the wavefront that the wavefront synthesizer
produces. When we place a layer of strongly scattering paint in the
optical path, light is scattered. The scattered light forms a random
speckle pattern with sharp features. To focus light to a single
point, we generate a wavefront that precisely matches the scattering
matrix of the layer of paint. The scattered light now interferes
constructively in a single point.

To construct the matching wavefront, or `fingerprint', we use a
method that is similar to the technique described in
Chapter~\ref{cha:focusing-through}. One by one the phase of a
segment of the phase modulator is cycled from 0 to $2\pi$.
Meanwhile, the intensity at the desired focal point is monitored
with a camera. From the resulting interference signal we deduce the
phase and amplitude of that segment's contribution to the target
focus. This measurement is repeated for all segments. From these
measurements, the unique\cite{Pappu2002} optical fingerprint that
causes constructive interference in the target focus is constructed.
In contrast to the method presented in
Chapter~\ref{cha:focusing-through}, we also control the amplitude of
the incident light. We set the amplitude of each of the segments to
be proportional to the measured amplitude of that segment's
contribution to the target focus. In this way the overlap of the
incident field with the fingerprint of the medium is optimal.
Segments that contribute little to the focus are effectively
switched off so that the background intensity outside of the focus
is reduced and contrast is improved. Theoretically, amplitude and
phase modulation increases the contrast with the diffuse background
by a factor of $4/\pi$ with respect to phase only modulation.
Moreover, thanks to the combined amplitude and phase control,
fingerprints for different foci can be superposed to generate any
desired bit pattern.

\bibliography{../../bibliography}
\bibliographystyle{Ivo_sty}

\setcounter{chapter}{6}
\chapter{Phase control algorithms for focusing light through turbid media\label{cha:algorithms}}
\begin{abstract}
Light propagation in materials with microscopic inhomogeneities is affected by scattering. In scattering
materials, such as powders, disordered metamaterials or biological tissue, multiple scattering on
sub-wavelength particles makes light diffuse. Recently, we showed that it is possible to construct a
wavefront that focuses through a solid, strongly scattering object. The focusing wavefront uniquely
matches a certain configuration of the particles in the medium. To focus light through a turbid liquid or
living tissue, it is necessary to dynamically adjust the wavefront as the particles in the medium move.
Here we present three algorithms for constructing a wavefront that focuses through a scattering medium. We
analyze the dynamic behavior of these algorithms and compare their sensitivity to measurement noise. The
algorithms are compared both experimentally and using numerical simulations. The results are in good
agreement with an intuitive model, which may be used to develop dynamic diffusion compensators with
applications in, for example, light delivery in human tissue.

\noindent[This chapter has been accepted for publication in Optics Communications.]
\end{abstract}


\begin{figure}
\centering
  \includegraphics[width=\wideimage]{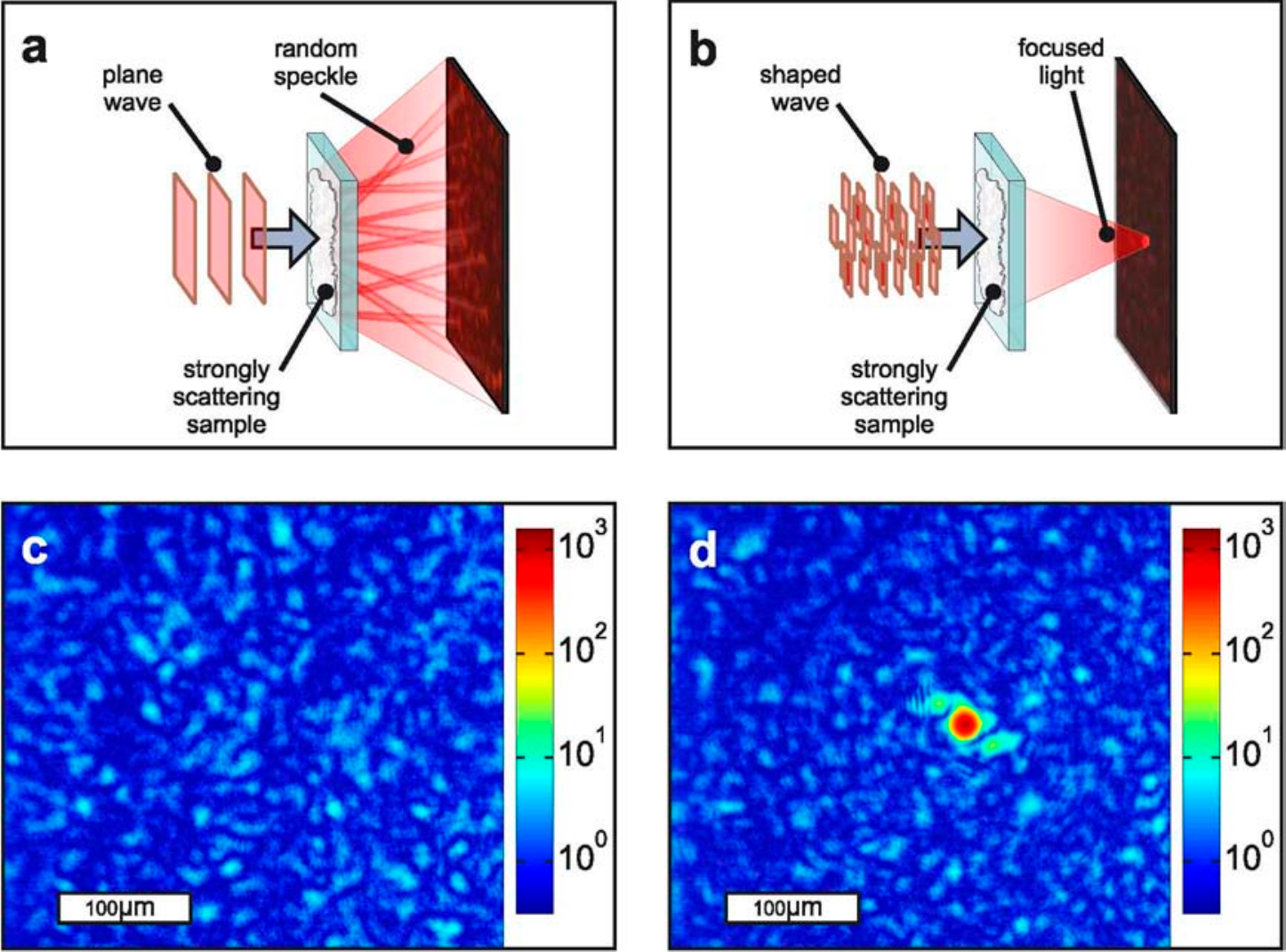}\\
  \caption{Principle and experimental results of inverse wave diffusion. \subfig{a} A multiply scattering object
  destroys the spatial coherence of incident light. \subfig{b} When the same object is illuminated
  with a specially constructed matching wavefront, the transmitted light focuses to a tight spot. \subfig{c} Recorded intensity
  transmission of an unshaped wave through a $\mum{10}$ thick layer of TiO$_2$ pigment. \subfig{d} Intensity
  transmission through the same sample with a shaped wavefront.}\label{fig:inverse-diffusion}
\end{figure}

\noindent Materials such as paper, white paint or human tissue are
non-transparent because of multiple scattering of
light.\cite{Milne1921,Chandrasekhar1960,Ishimaru1978} Light
propagating in such materials is diffuse. Recently, we have shown
that coherent light can be focused through diffusive media yielding
a sharp, intense focus.\cite{Vellekoop2007thesis} Starting with the
situation where a scattering object (a layer of TiO$_2$ pigment with
a thickness of approximately 20 transport mean free paths)
completely destroys the spatial coherence of the incident light
(Fig.~\ref{fig:inverse-diffusion}a, \ref{fig:inverse-diffusion}c),
we controlled the incident wavefront to exactly match scattering in
the sample. Afterwards, the transmitted light converged to a tight,
high contrast focus (Fig.~\ref{fig:inverse-diffusion}b,
\ref{fig:inverse-diffusion}d). These matched wavefronts experience
inverse diffusion, that is, they gain spatial coherence by
travelling through a disordered medium.

For a given sample of scattering material, there is a unique
incident wavefront that makes the object optimally focus light to a
given point. Like a speckle pattern, this wavefront is disordered on
the scale of the wavelength of light. This wavefront cannot be
constructed from a small number of smooth base functions, which
unfortunately renders the efficient algorithms used in adaptive
optics (see e.g. \cite{Tyson1998}) ineffective. In
Ref.~\citealt{Vellekoop2007thesis}, we presented an algorithm that
finds the optimal wavefront when the sample is perfectly stationary
and the noise level is negligible. To find applications in, for
example, fluorescence excitation or photodynamic therapy, the
wavefront has to be adjusted dynamically as the scatterers in the
sample move. In this paper, we present two additional algorithms,
that dynamically adjust the wavefront to follow changes in the
sample. The performance of the algorithms is in good agreement with
numerical simulations and with an analytical model. We show that the
new algorithms are superior to the original algorithm when the
scatterers in the sample move or when the initial signal to noise
ratio is poor.

Wave diffusion is a widely encountered physical phenomenon. The use
of multiple scattered waves is the subject of intensive study in the
fields of, for instance, ultrasound
imaging\cite{Fink1999,Lobkis2001,Borcea2005}, radio and microwave
antennas\cite{Lerosey2007,Foschini1996},
seismography\cite{Shapiro2005}, submarine
communication\cite{Kuperman1998}, and surface
plasmons\cite{Stockman2002}. While the algorithms discussed in this
paper were developed for spatial phase shaping of light, they can be
used for any type of wave and apply to spatial phase shaping as well
as to frequency domain phase shaping (also known as coherent
control, see e.g. \cite{Herek2002}) as the concepts are the same.

This article is organized as follows. First the key concepts of inverse diffusion are introduced and the
three different algorithms are presented. Then the experimental apparatus is explained and the measured
typical performance of the algorithms is compared. In the subsequent section, we will compare the
experimental results with numerical simulations and analyze the data in terms of noise and stability of
the scatterers. Finally, we will analytically explain the characteristic features of the different
algorithms and discuss their sensitivity to noise.

\section{Algorithms for inverse diffusion}\label{sec:algorithms}
\begin{figure}
\centering
  \includegraphics[width=\wideimage]{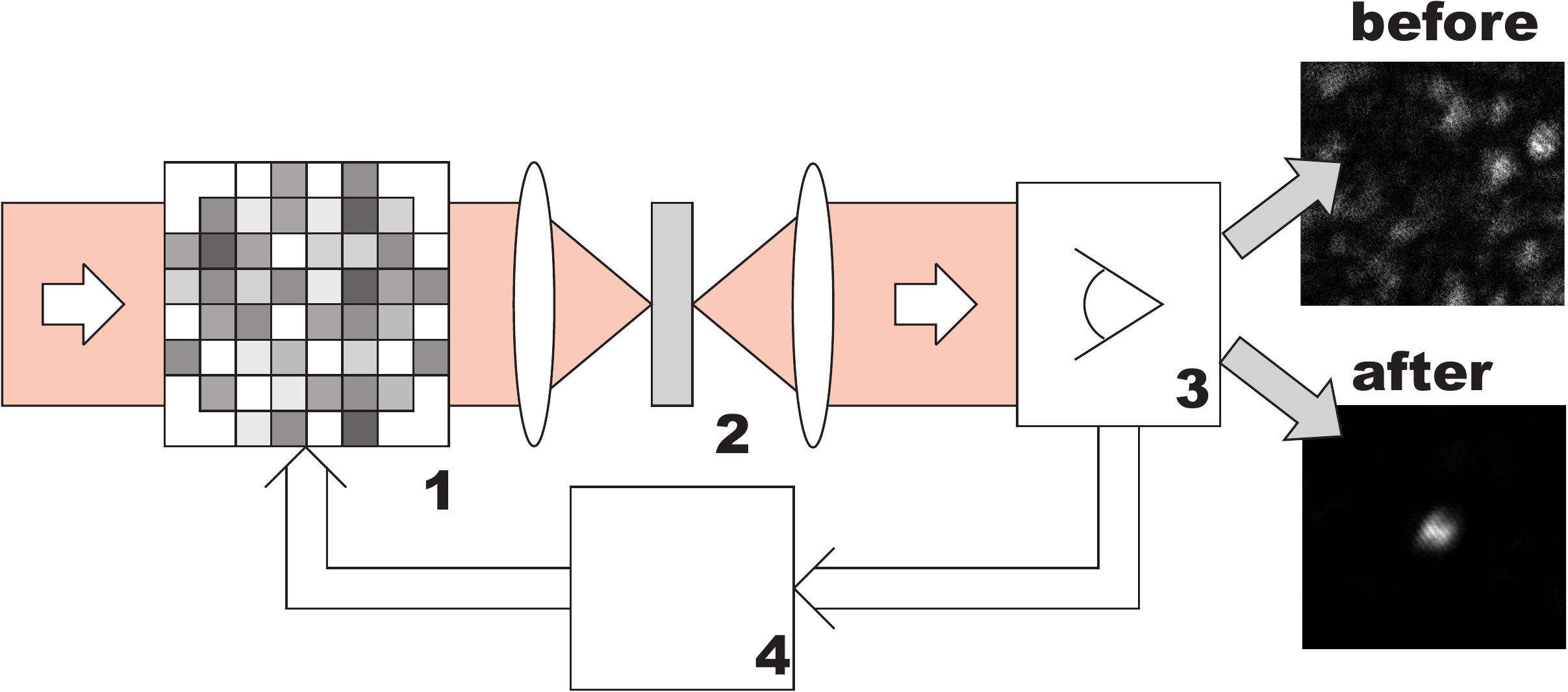}\\
  \caption{Feedback loop for achieving inverse diffusion. An incident monochromatic beam is shaped using a spatial light modulator (1)
  and projected on a non-transparent multiply scattering object (2). A detector (3) detects the amount of transmitted light that reaches
  the target area. A feedback algorithm (4) uses the signal from the detector to program the phase modulator. Before the algorithm is
  started, the transmitted light forms a random speckle pattern. The algorithm changes the incident wave
  to increase the intensity in the target area. After a few iterations, the transmitted light focuses on the target.}\label{fig:block-diagram}
\end{figure}

\noindent The key elements of an inverse diffusion setup are a multiply scattering sample, a spatial light
modulator, an optimization algorithm and a detector, as shown in Fig.~\ref{fig:block-diagram}. The sample
can be anything that scatters light without absorbing it. We will consider only samples that are thicker
than approximately 6 transport mean free paths for light. Light transmitted through these samples is
completely diffuse and the transmitted wavefront is completely scrambled, i.e., it has no correlation with
the incident wavefront.\cite{Pappu2002} We assumed that the intensity is low so that non-linear effects
are negligible.

\enlargethispage*{\baselineskip}

The incident wavefront is constructed using a spatial phase modulator. The modulator consists of a
2D-array of pixels that are grouped into $N$ equally sized square segments. A computer sets the phase
retardation for each of the segments individually to a value between $0$ and $2\pi$. The optimization
algorithms program the phase modulator based on the detector output. Initially, each of the segments
randomly contributes to the field at the detector. The algorithms find the unique configuration of
retardations for which all contributions are in phase at the target. As follows from the triangle
inequality, the magnitude of the field at the detector is at a global maximum for this optimal
configuration. Since the sample completely scrambles the incident wavefront, all segments of the wavefront
are scattered independently and the optimal wavefront will not be smooth.

Behind the sample is a detector that provides feedback for the
algorithm. The detector defines the target area where the intensity
is maximized. The field at the detector is the result of
interference from scattered light originating from the different
segments of incident wavefront. When the phase of one or more
segments is changed, the target intensity responds sinusoidally. We
sample the sine wave by taking 10 measurements. The process of
capturing a single sine wave and possibly adjusting the phase
modulator accordingly is called an iteration.

The amount of control we have over the propagation of light in the disordered system is quantified by the
signal enhancement. The enhancement $\eta$ is defined as
\begin{equation}
\eta \equiv \frac{I_N}{\avg{I_0}},
\end{equation}
where $I_N$ is the intensity in the target after optimization and
$\avg{I_0}$ is the ensemble averaged transmitted intensity before
optimization. In a perfectly stable system, the enhancement is
proportional to $N$ \cite{Vellekoop2007thesis}, meaning that the
more individual segments are used to shape the incident wavefront,
the more light is directed to the target. In practice, however, the
enhancement is limited by the number of iterations that can be
performed before the sample changes too much. We define the
persistence time $T_p$ as the decay time of the field autocorrelate
of the transmitted speckle, which is a measure for the temporal
stability of the sample. The persistence time depends on the type of
sample and on environmental conditions. Typical values of $T_p$
range from a few milliseconds in living tissue\cite{Briers1995} to
hours for solid samples in laboratory conditions. The other relevant
timescale is the time required for performing a single iteration,
$T_i$. In our experiments, we operate the phase modulator at just
below 10 Hz and take ten measurements for each iteration; we have
$T_i\approx 1.2$s.

We will now present three algorithms we used to invert wave
diffusion. The advantages and disadvantages of the algorithms are
discussed briefly and will be analyzed in detail later.

\begin{figure}
\centering
  \includegraphics[width=\mediumimage]{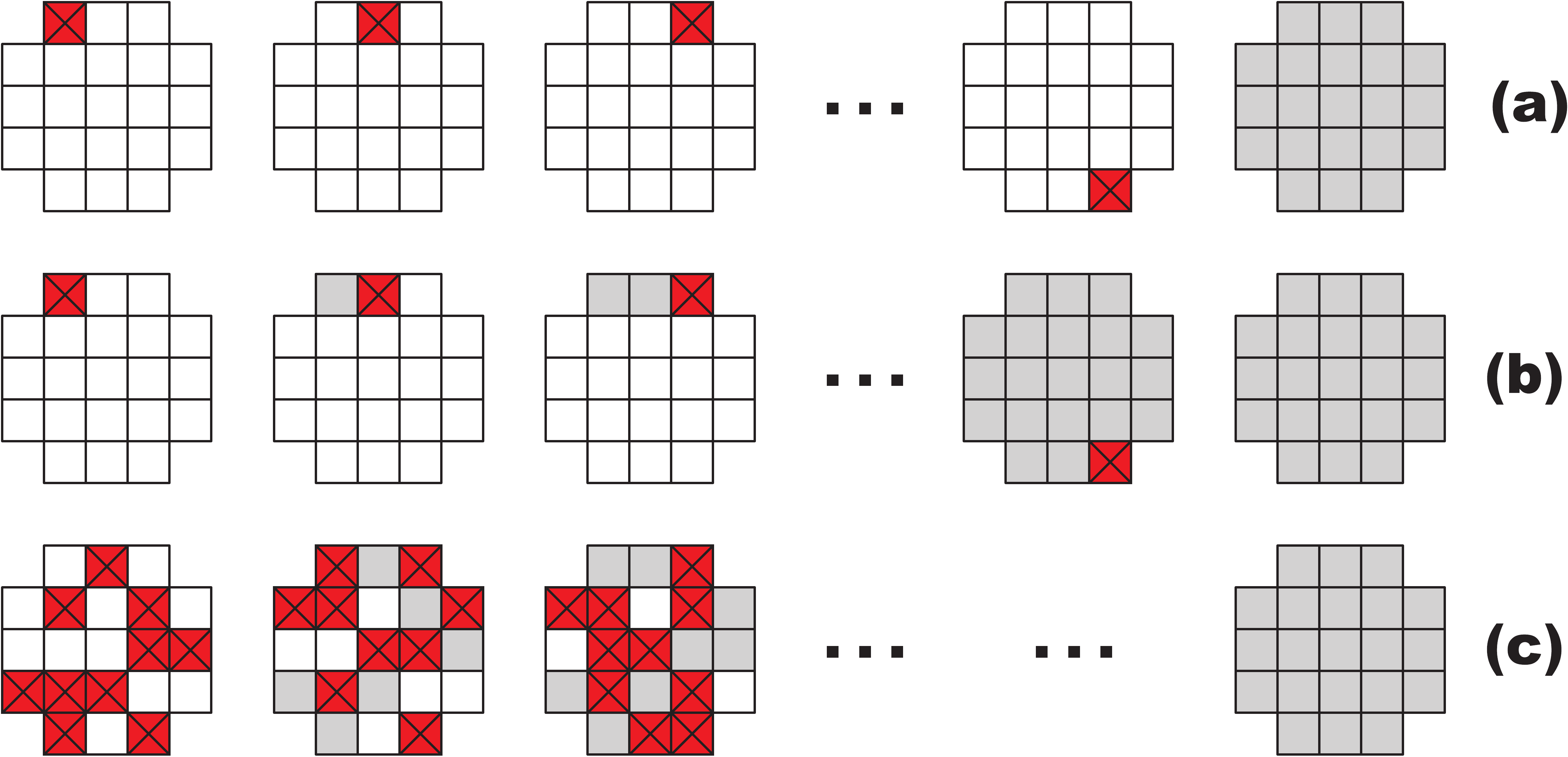}\\
  \caption{Principle used in the three different optimization algorithms. \subfig{a} For the stepwise sequential algorithm, all segments are addressed sequentially
  (marked squares). After the optimal phase is measured for all segments, the modulator is updated to construct the optimal
  wavefront (light gray squares). \subfig{b} The continuous sequential algorithm is equal to the first algorithm, except that the modulator is updated after each iteration.
  \subfig{c} The partitioning algorithm randomly selects half of the segments and adjusts their overall phase. The modulator is updated after each measurement. }\label{fig:algorithms-explained}
\end{figure}

\subsection{The stepwise sequential algorithm\label{sec:alg-stepwise-sequential}}
\noindent The stepwise sequential algorithm that was used in Ref. \cite{Vellekoop2007thesis} is very
straightforward. It is based on the fact that the field at the detector is a linear superposition of the
contributions from all segments. This means that we can construct the optimal wavefront by optimizing each
of the segments individually. The computer consecutively cycles the phase of each of the $N$ segments from
0 to $2\pi$. The feedback signal is monitored and the phase for which the target intensity is maximal is
stored. At this optimal phase, the contribution of this single segment is in phase with the background
field (the mean contribution coming from all other segments). The phase retardation of the segment is
reset to $0$ before continuing with the next segment. This way, the background field remains unchanged.
Only after all iterations are performed, the phase of each segment is set to this optimal value (see
Fig.~\ref{fig:algorithms-explained}a). Now, all contributions have the same phase as the original
background field.

In absence of measurement noise or temporal instability, algorithm 1
is guaranteed to find the global maximum in the lowest number of
iterations possible. However, when $N T_i\gg T_p$, the speckle
pattern decorrelates before all measurements are performed and the
algorithm will not work. Therefore, it is important to adjust the
number of segments to the persistence time.

\subsection{The continuous sequential algorithm}
\noindent The continuous sequential algorithm is very similar to the
stepwise sequential algorithm except for the fact that the phase of
each segment is set to its maximum value directly after each
measurement (see Fig.~\ref{fig:algorithms-explained}b). This
approach has two advantages. First of all, the algorithm runs
continuously and dynamically follows changes in the sample's
scattering behavior. Furthermore, the target signal starts to
increase directly, which increases the signal to noise ratio of
successive measurements. It is still necessary to adjust $N$ to the
persistence time $T_p$.

\subsection{The partitioning algorithm}
\noindent The two sequential algorithms change the phase of one
segment at a time. As one segment contributes only a small fraction
of the wavefront, the signal to noise ratio is low and the initial
convergence is slow. As an alternative, we propose a partitioning
algorithm that simultaneously changes the phase of a whole group of
segments (see Fig.~\ref{fig:algorithms-explained}c). The amplitude
of the interference signal is maximal when the group contains half
of the segments. At each iteration the segments are divided randomly
over two equally sized partitions. Then, the target intensity is
maximized by changing the phase of one partition with respect to the
other. This process is repeated indefinitely, and the modulator is
randomly repartitioned every time.

Since the phase of many segments is changed simultaneously, the
initial increase in intensity will be fast and the feedback signal
will be high. Therefore, this algorithm is expected to be less
sensitive to noise and to recover from disturbances more rapidly.

\section{Experiment}

\begin{figure}
\centering
  \includegraphics[width=\mediumimage]{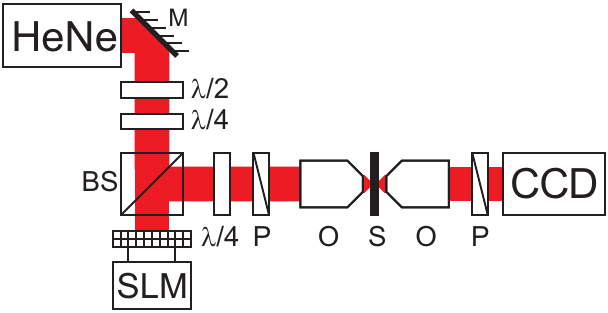}\\
  \caption{Experimental apparatus used for inverting diffusion. Light from a HeNe laser is spatially modulated by
  a liquid crystal spatial light modulator (SLM). Wave plates and a polarizer (P) are used to generate and select the polarization
  state for which the modulator works in phase-mostly mode. The shaped beam is demagnified by a factor of 3 (telescope not shown) and
  focused on the sample (S). A reference detector monitors the total intensity falling on the sample. A microscope objective, a polarizer and a CCD-camera
  are used to detect the intensity in the target focus, a few millimeters behind the
  sample. M, mirror; BS, non-polarizing 50\% beam splitter.
  }\label{fig:setup-algorithms}
\end{figure}

\noindent The different algorithms were tested experimentally using
the setup shown in Fig.~\ref{fig:setup-algorithms}. In our case the
scattering medium is a $\mum{10}$ thick layer of rutile TiO$_2$
pigment \cite{Kop1997} with a mean free path of $\mum{0.55\pm0.1}$,
determined by measuring the total transmission at a wavelength of
$\nm{632.8}$. This sample is illuminated by a $\nm{632.8}$ HeNe
laser. The laser beam is expanded and spatially modulated by a
Holoeye LC-R 2500 liquid crystal light modulator (LCD). The
modulator is a reflective twisted nematic liquid crystal on silicon
(TN-LCoS) device. Such devices are designed for intensity
modulation. However, by choosing an appropriate combination of the
incident polarization and the analyzed polarization, the LCD can be
used as a phase modulator with a minimal residual intensity
modulation\cite{Davis2002}. We characterized the LCD using the
method described in \cite{Remenyi2003} and select an optimal
combination of elliptical polarizations to achieve a phase-mostly
modulation mode with $2\pi$ phase modulation and a maximum of $21\%$
residual intensity modulation. The shaped beam is focused on the
sample using a 63x objective with a numerical aperture (NA) of 0.85.
A 20x objective (NA=0.5) images a point that is approximately $3.5$
mm behind the sample onto a 12-bit CCD camera (Allied Vision
Technologies Dolphin F-145B). This point is the target area where we
want the light to focus. A computer integrates the intensity in a
circular area with a radius of 20 pixels (corresponding to
$\mum{129}$ in the focal plane of the objective). This target area
is smaller than a typical speckle spot. Using this signal as
feedback, the computer programs the phase modulator using one of the
algorithms described above.
\begin{figure}
\centering
\includetwographics{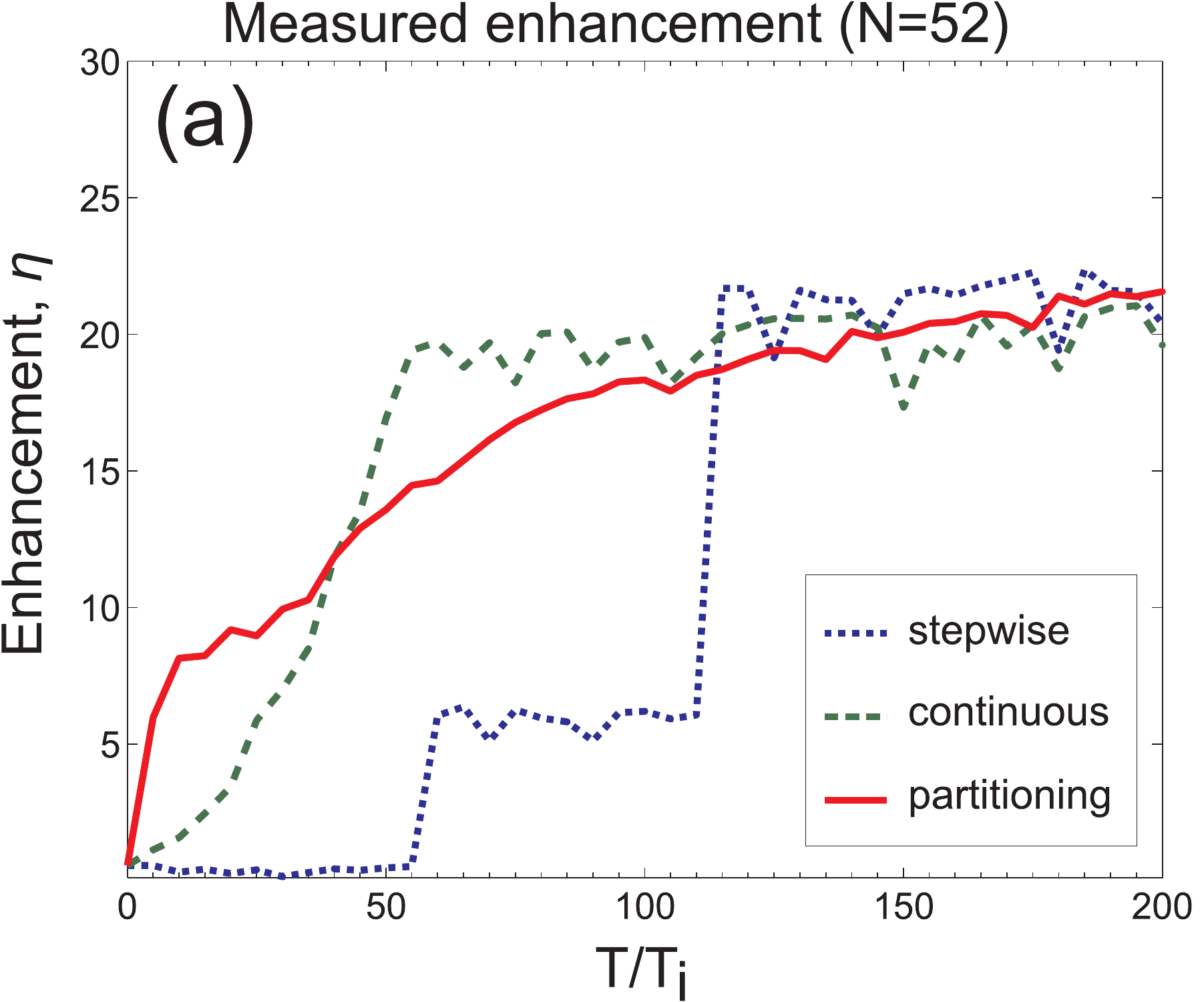}{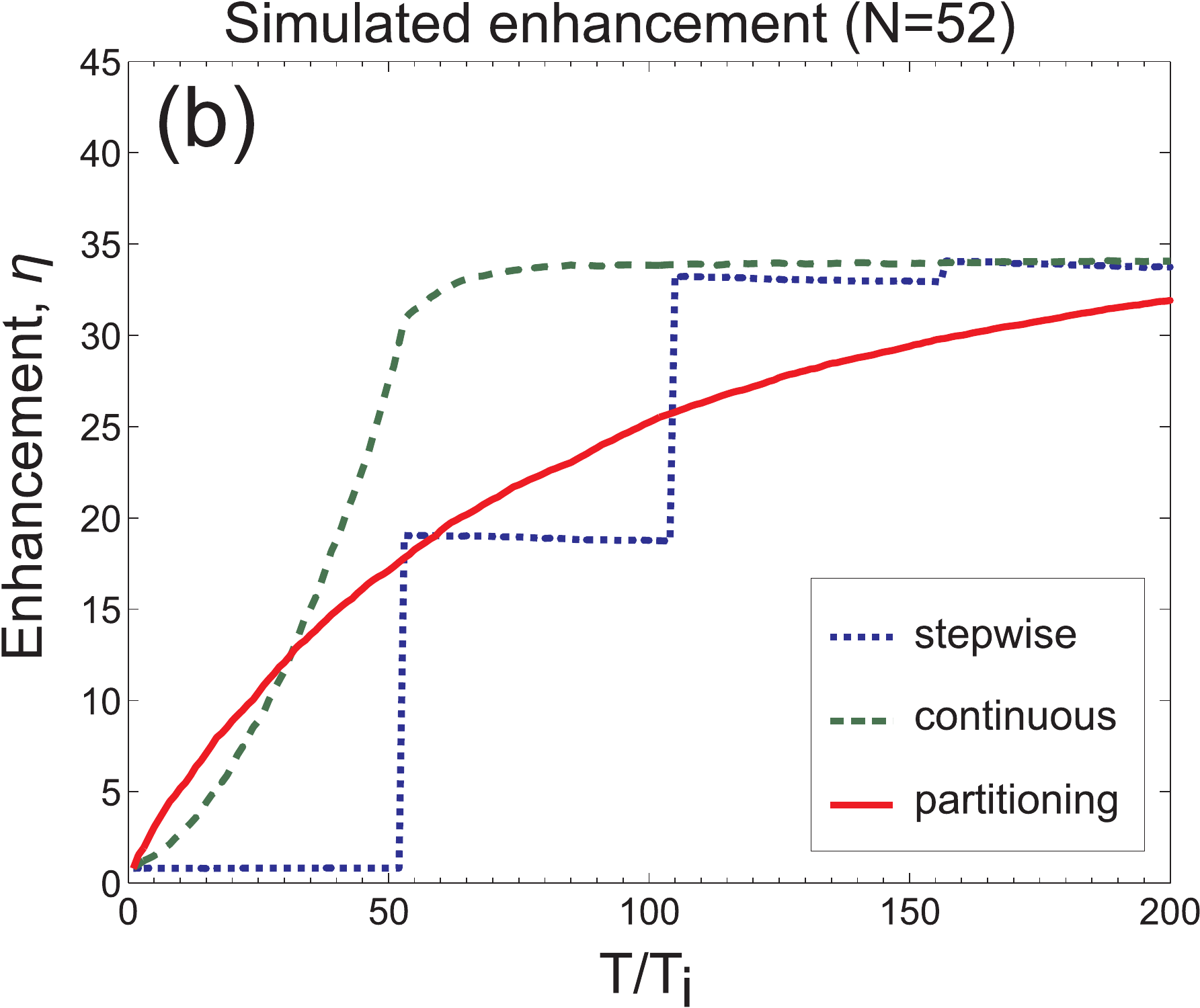}\\
  \caption{\subfig{a} Typical runs of the stepwise sequential algorithm (dotted curve), the continuous sequential algorithm (dashed curve)
  and the partitioning algorithm (solid curve). All algorithms were run with $N=52$.
  The sequential algorithms were repeated four times. \subfig{b} Simulation results for $N=52$ averaged over 64 runs. The
  simulation captures the main features of the three algorithms, but predicts a higher maximum enhancement. } \label{fig:low-N}
\end{figure}

We first run the three different algorithms with $N=52$. Since $T_p/T_i \gg 52$, we do not expect to see
decoherence effects. In total, 208 iterations were performed, which means that the sequential algorithms
ran four times consecutively. The results of the optimization procedures is shown in
Fig.~\ref{fig:low-N}a. Although the three algorithms reach the same final enhancement of intensity, there
are significant differences between the algorithms. The enhancement for the stepwise sequential algorithm
increases in discrete steps because the phase modulator is only reprogrammed every $N$ iterations. During
the first $N$ iterations, the target signal is low and the algorithm suffers from noise. The saturation
enhancement is reached after the second update (after $2N$ iterations). The continuous sequential
algorithm and the partitioning algorithm both start updating the wavefront immediately and, therefore,
have a higher initial increase of the signal. The continuous sequential algorithm is the first algorithm
to reach the saturation enhancement (after $N$ iterations). The partitioning algorithm has the fastest
initial increase in the target signal. It is, however, the last algorithm to reach the saturation
enhancement since the final convergence is very slow.

\begin{figure}
\centering
\includetwographics{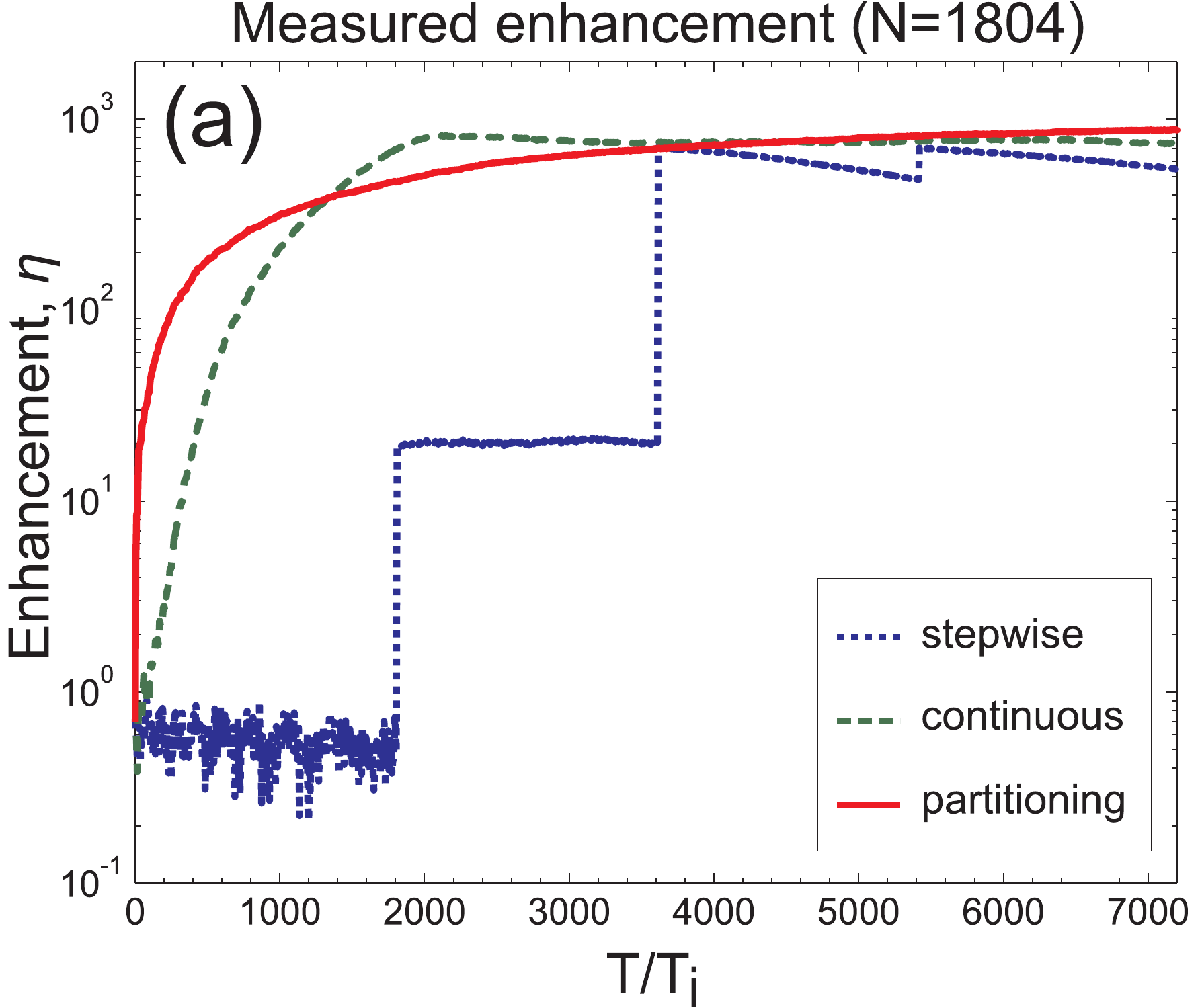}{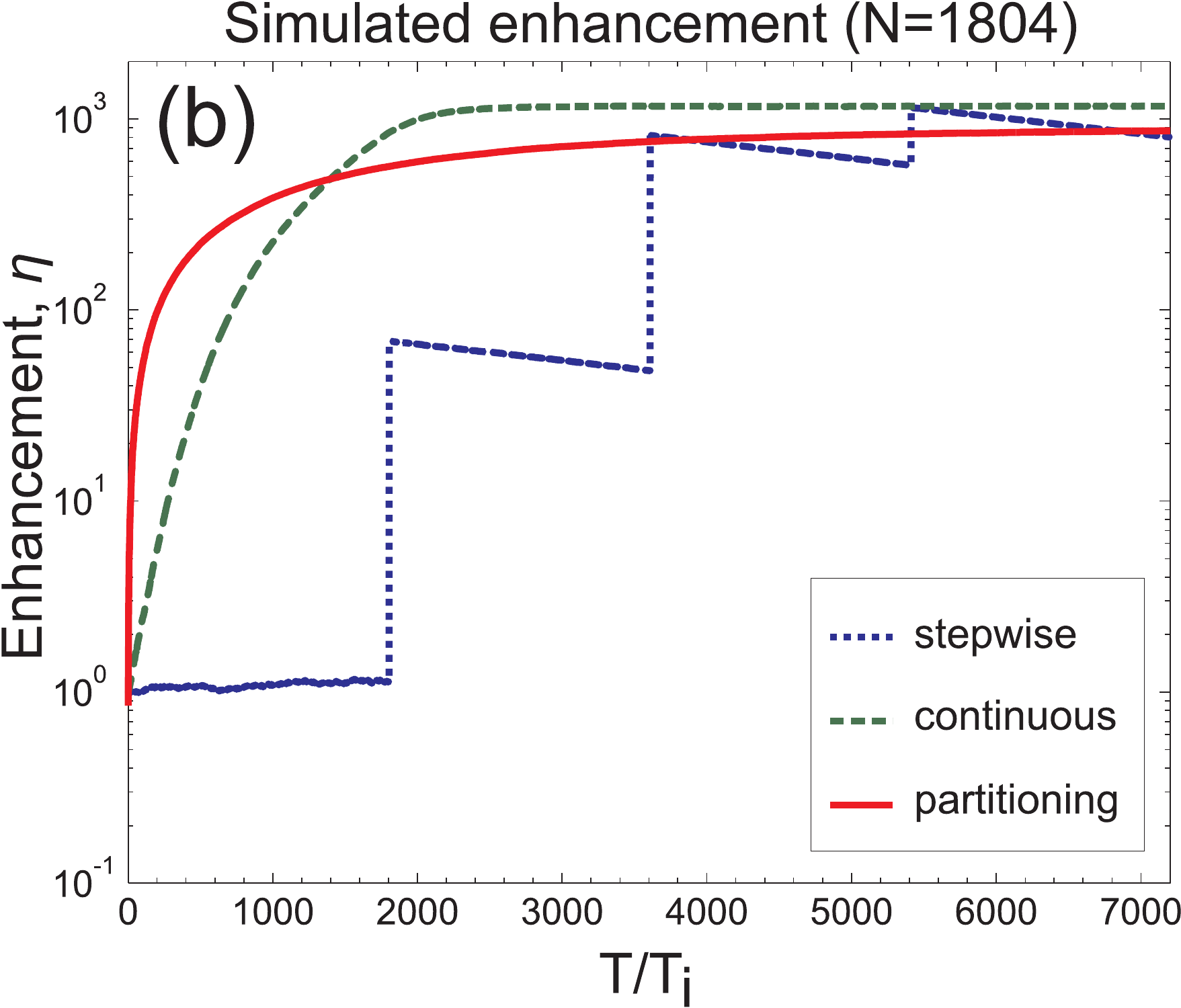}
  \caption{\subfig{a} Typical runs of the stepwise sequential algorithm (dotted curve), the continuous sequential algorithm (dashed curve)
  and the partitioning algorithm (solid curve). All algorithms were run with $N=1804$. The enhancement are plotted on a logarithmic scale.
  \subfig{b} Simulation results for $N=1804$ averaged over 64 runs.
  }\label{fig:high-N}
\end{figure}

When the number of segments in the wavefront is increased, we expect to find a higher target intensity.
Figure~\ref{fig:high-N}a shows the experimental results for $N=1804$ on a logarithmic scale. The final
intensity enhancement is approximately 40 times higher than in Fig.~\ref{fig:low-N}a. A further difference
is that in this situation the effects of decoherence are no longer negligible. This effect is most clearly
visible with the stepwise sequential algorithm. The phase modulator is updated after each $N$ iterations
and between the updates the intensity decays exponentially with a $1/e$ decay of about $T_p/T_i=5000$
iterations.

The convergence behavior of the three algorithms is similar to the
experiment that is shown in Fig.~\ref{fig:low-N}. The partitioning
algorithm clearly causes a higher signal enhancement during the
first 1000 iterations. The initial increase in the enhancement is
linear with a slope of 0.37. Initially, this linear increase is far
superior to the quadratic increase obtained with the continuous
sequential algorithm.

We conclude that both new algorithms are valuable improvements over the original stepwise sequential
algorithm. These algorithms are far less sensitive to noise and the target signal is kept at a constant
value even in the presence of decoherence. The partitioning algorithm has the fastest initial increase and
therefore will recover from disturbances most rapidly.

\section{Simulations}
\noindent In order to obtain a better understanding of the effects
of noise and fluctuations on the performance of the different
algorithms we perform numerical simulations. The disordered medium
is represented by a transmission matrix with elements drawn from a
circular Gaussian distribution (more details on the matrix
representation can be found below). Decoherence is modelled by
adding a small perturbation to the transmission matrix after every
measurement. Finally, measurement noise is in\-clud\-ed by adding a
random value to the simulated detector signal.

Figures~\ref{fig:low-N}b and \ref{fig:high-N}b show the simulated enhancement for a system with $T_p/T_i
=5000$. Every iteration, ten measurements are performed for phase delays between $0$ and $2 \pi$. To each
of these measurements, Gaussian noise with a standard deviation of $0.3 I_0$ was added. The magnitude of
the noise is comparable to experimental observations. The three different algorithms were run with $N=52$
and $N=1804$ to simulate the experiments shown in Fig.~\ref{fig:low-N}a and Fig.~\ref{fig:high-N}a.

The simulations are in good qualitative agreement with the
experimental data. The result for the stepwise sequential algorithm
shows that the effects of noise and decoherence are simulated
realistically. Furthermore, the initial signal increase and the long
time convergence behavior correspond to the measured results. The
only significant difference is the 20\% to 50\% higher enhancement
reached in the simulations. A possible explanation for this
difference is the residual amplitude cross-modulation in our phase
modulator. Due to this cross-modulation, the optimal wavefront
cannot be generated exactly. Furthermore, the amplitude modulation
decreases the accuracy of the measurement of the optimal phase. The
partitioning algorithm is less sensitive to this last effect since
the cross-modulation is averaged over many segments with different
phases. Since the simulations capture the overall behavior of the
algorithms very well, we can use them to extrapolate to situations
with a lot of noise and strong decoherence or, on the other hand, to
perfectly stable systems.

\section{Analytical expressions for the enhancement}
\noindent In this section, we analyze the performance of the
algorithms with analytical theory and compare these results to the
simulations. We describe scattering in the sample with the
transmission matrix elements, $t_{mn}$. This matrix couples the
fields of the incident light and the transmitted light.
\begin{equation}
E_m = \sum_n^N t_{mn} A_n e^{i\phi_n},
\end{equation}
where the $\phi_n$ is the phase of the $n$th segment of the phase modulator. Assuming that the modulator
is illuminated homogeneously, all incoming channels carry the same intensity. We write $A_n = 1/\sqrt{N}$
to normalize the total incident intensity. Elements $E_1, E_2, \ldots$ correspond to single scattering
channels of the transmitted light. Since we are interested in focusing light to a single spot, we need to
consider only a single transmission channel, $E_m$. The intensity transmitted into channel $m$ is given by
\begin{equation}
|E_m|^2 = \frac1N\left|\sum_n^N t_{mn}
e^{i\phi_n}\right|^2\label{eq:intensity-base}.
\end{equation}
Regardless of the values of the elements $t_{mn}$ of the transmission matrix, the intensity $|E_m|^2$ has
its global maximum when the phase modulator exactly compensates the phase retardation in the sample for
each segment, i.e. $\phi_n=-\arg(t_{mn})$. The target intensities before optimization ($I_0$) and after an
ideal optimization ($I_\text{max}$) are given by
\begin{equation}
I_0 = \frac1N\left|\sum_n^N t_{mn}\right|^2 \label{eq:intensity-before},
\end{equation}
and
\begin{equation}
I_\text{max} = \frac1N \left(\sum_n^N \left| t_{mn} \right|\right)^2\label{eq:intensity-after}.
\end{equation}

For a disordered medium the elements of $t_{mn}$ are independent and have a Gaussian
distribution.\cite{Goodman2000,Garcia1989,Webster2004,Beenakker1997} Rewriting
Eq.~\eqref{eq:intensity-after} gives
\begin{align}
\avg{I_\text{max}} &= \avg{\frac1N\sum^N_{n, k\neq n}
|t_{mn}||t_{mk}| + \frac1N \sum_n^N |t_{mn}|^2}, \label{eq:intensity-after-intermediate}\\
& = \avg{I_0}\left[(N-1)\frac{\pi}{4}+1\right], \label{eq:intensity-after-expanded}
\end{align}
where the angled brackets denote ensemble averaging over disorder. Eq.~\eqref{eq:intensity-after-expanded}
predicts that the expected maximum enhancement for an ideally stable, noise free system linearly depends
on the number of segments $N$. For $N \gg 1$, we have $\eta \approx \pi N /4$.

\subsection{Performance in fluctuating environments}
\noindent In reality, the sample will not be completely stable. Whether this instability is due to a drift
of the sample position, movement of the scatterers, changing humidity or any other cause, the transmission
matrix will fluctuate over time. In the simulations, we modelled decoherence by repeatedly adding a small
perturbation to each of the matrix elements.
\begin{equation}
t_{mn}\rightarrow \frac{1}{\sqrt{1+\delta^2}}(t_{mn} +
\xi)\label{eq:decoherence-principle},
\end{equation}
where $\xi$ is drawn from a complex Gaussian distribution with mean 0 and standard deviation $\delta$. The
prefactor normalizes the transformation so that $\avg{|t|^2}$ remains constant. By substituting the
continuous limit of Eq.~\eqref{eq:decoherence-principle} in Eq.~\eqref{eq:intensity-after-intermediate},
we find an analytic expression for the effect of decoherence,

\begin{equation}
\avg{I_N} = \avg{I_0} \left[\frac{\pi}{4N}\left(\sum^N_n
e^{-T_n\delta^2/(2T_i)}\right)^2+O(1)\right],\label{eq:intensity-decoherence}
\end{equation}
where $T_n$ is the time that has past since the phase of segment $n$ was measured. This simple model
explains the exponential decay of the intensity that was observed in the measurements (see
Fig.~\ref{fig:high-N}a) and predicts a decay time of $T_p= T_i/\delta^2$.

We now calculate the maximum enhancement that can be reached with
the continuous sequential algorithm in the presence of decoherence.
Because the phases of the segments are measured sequentially, at any
given time the values for $T_n/T_i$ are equally spaced between $1$
and $N$. From Eq.~\eqref{eq:intensity-decoherence} we find a maximum
intensity enhancement of
\begin{equation}
\eta_N\equiv\frac{\avg{I_N}}{\avg{I_0}}=
\frac{\pi}{4N}\left(\frac{1-e^{-N
T_i/(2T_p)}}{e^{T_i/(2T_p)}-1}\right)^2+O(1).\label{eq:intensity-alg12}
\end{equation}

The maximum enhancement for both sequential algorithms is the same.
However, since the stepwise algorithm only updates the projected
wavefront after $N$ iterations, the enhancement decreases
exponentially between updates.

\begin{figure}
\centering
  \includegraphics[width=\smallimage]{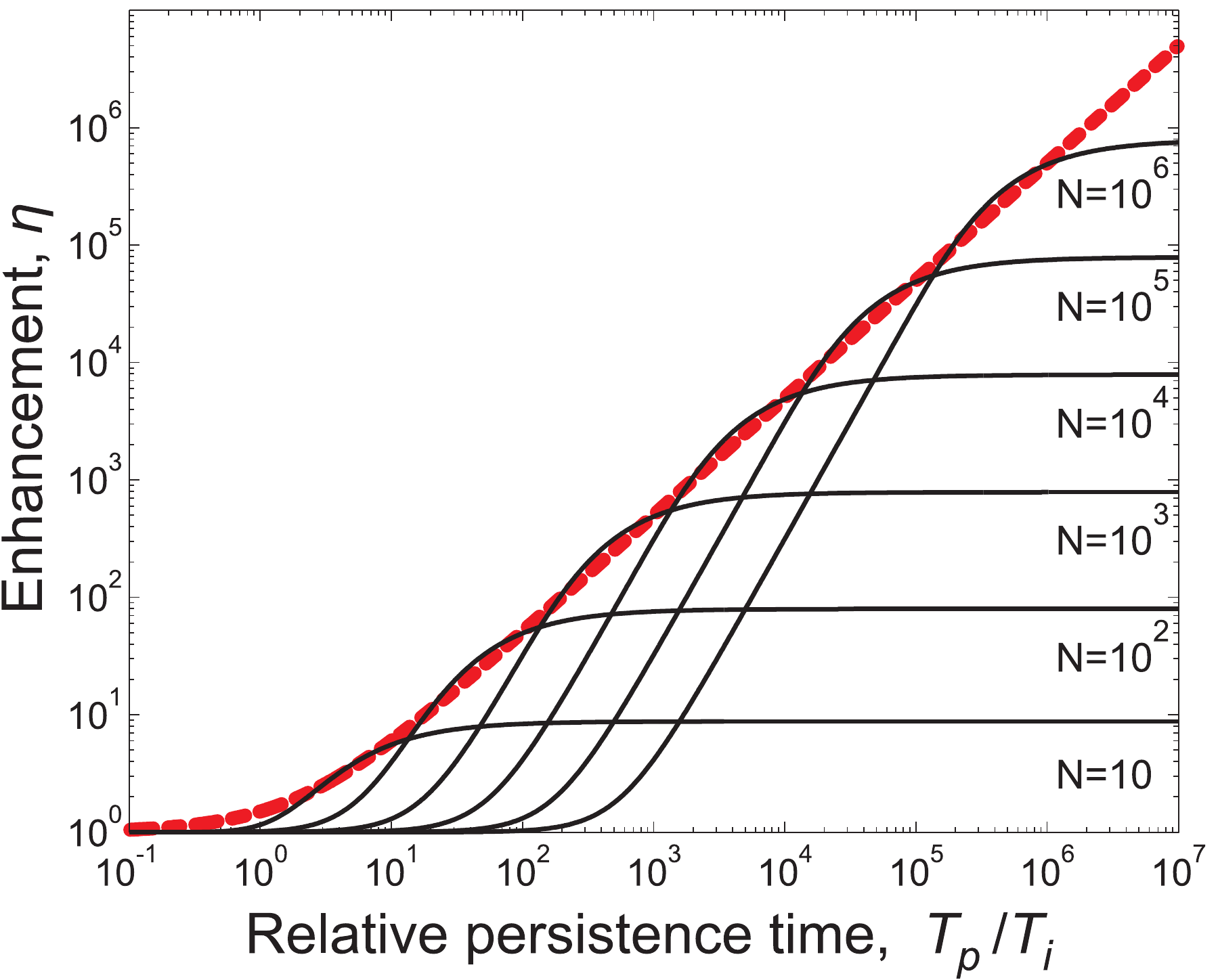}\\
  \caption{Theoretical maximum enhancement as a function of coherence time for different algorithms. The solid lines represent the maximum enhancement
  that can be obtained using the sequential algorithms. The enhancement depends on the number of segments used in the
  algorithm. The dashed line shows the enhancement for the partitioning algorithm where $N\gg T_p/T_i$.}\label{fig:algorithm-comparison-decoherence}
\end{figure}

In Fig.~\ref{fig:algorithm-comparison-decoherence} the enhancement for different values of $N$ is plotted
versus $T_p/T_i$. When the persistence time is large ($T_p/T_i \gg N$), decoherence effects do not play a
role and the enhancement linearly depends on $N$ as was seen in Eq.~\eqref{eq:intensity-after-expanded}.
For $T_p/T_i < N$, however, the enhancement decreases because the speckle pattern decorrelates before all
iterations are performed and the enhancement drops to zero. As a consequence, the sequential algorithms
only perform optimal when $N$ is adjusted to $T_p$. When $T_p$ is known a-priori, this optimum for $N$ can
be found by maximizing Eq.~\eqref{eq:intensity-alg12}. We find
\begin{equation}
N_\text{opt}=W T_p/T_i\label{eq:N-optimal},
\end{equation}
where $W\approx 2.51$ is the solution of $\exp(W/2)=1+W$. The maximal enhancement achievable with
sequential algorithms follows by substituting Eq.~\eqref{eq:N-optimal} into Eq.~\eqref{eq:intensity-alg12}
and equals $\eta_\text{opt} = 0.640 T_p/T_i$.

With the partitioning algorithm $\eta$ increases by $1/2$ each iteration of the algorithm (see
Section~\ref{sec:alg3}). As long as $N \gg T_p/T_i$, the enhancement saturates at $\eta = T_p/(2 T_i) +
1$, when the increase is exactly cancelled by the effect of decoherence. The most important difference
with the sequential algorithms, is that the enhancement reached with the partitioning algorithm does not
depend on $N$. In Fig.~\ref{fig:algorithm-comparison-decoherence} it can be seen that the partitioning
algorithm outperforms the sequential algorithms for almost all combinations of $T_p$ and $N$. The
sequential algorithm only give a slightly higher enhancement when they are fine-tuned for a known
persistence time ($N = 2.51 T_p/T_i$). In most situations, $T_p$ is not known a-priory or varies over time
and the partitioning algorithm will be preferable.

\begin{figure}
\centering
  \includegraphics[width=\smallimage]{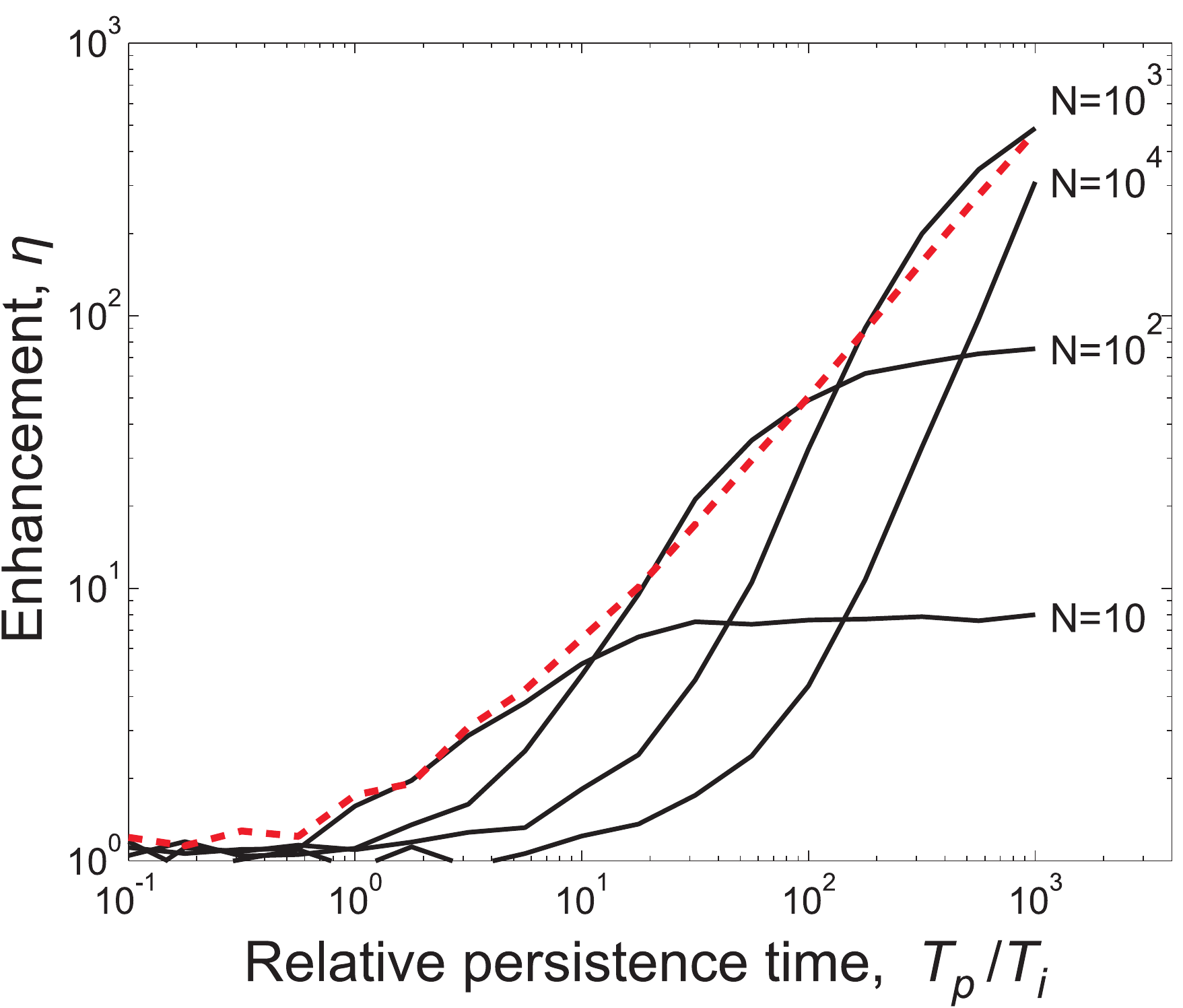}\\
  \caption{Simulated effect of decoherence on the sequential algorithms (solid curve) and on the partitioning algorithm (dashed
  curve). The simulations are averaged over 25 runs. Only when $N \approx 2.51 T_p/T_i$, the sequential algorithms perform slightly better than the partitioning
  algorithm.}\label{fig:algorithm-decoherence-simulation}
\end{figure}

Our analytical results for all three algorithms are supported by
numerical simulations (see
Fig.~\ref{fig:algorithm-decoherence-simulation}). The simulations
exactly reproduce the theoretical curves shown in
Fig.~\ref{fig:algorithm-comparison-decoherence}. For the simulations
of the partitioning algorithm we used $N=4096$, a number that can
easily be reached with a LCD phase modulator. Again, the
partitioning algorithm can be seen to have good overall performance,
whereas the sequential algorithms only work well for certain
combinations of $N$ and $T_p$.

In conclusion, the maximum enhancement that can be reached linearly depends on the sample's persistence
time. For the sequential algorithms $\eta = 0.64 T_p/T_i$, but only when $N$ is precisely adjusted to
$T_p$. The partitioning algorithm has $\eta = 0.5 T_p/T_i$, as long as $N$ is large enough. Using these
analytical relations, the performance of each of the three algorithms in different experimental situations
can easily be estimated.

\section{Effect of Noise}
\begin{table}
\centering
\begin{tabular}{|l|c|c|c|}
\hline
          & stepwise   & continuous & partitioning\\
          & sequential & sequential &\\
\hline signal & $2 I_0 \sqrt{1/N}$ & $2 I_0 \sqrt{\eta/N}$ & $\eta I_0$\\
bias & $I_0$ & $\eta I_0$ & $\eta I_0$\\
relative shot noise SNR & $2\sqrt{I_0/N}$ & 2$\sqrt{I_0/N}$ & $\sqrt{\eta I_0}$\\
rms phase correction & $\sqrt{3}\pi$ & $\sqrt{3}\pi$ & $\sqrt{2/\eta}$\\
\hline
\end{tabular}
\caption{Signal and noise characteristics of the three algorithms. The rms phase correction is a measure
for the required sensitivity.}\label{tab:noise}
\end{table}

\noindent Measurement noise affects the measured phases.
Noise-induced errors in the phases lead to a reduction of the
enhancement, $\eta$. We will now compare the signal-to-noise ratio
(SNR) of the three different algorithms. In a single iteration of an
algorithm, the phase of one or more segments is varied, while the
phase of the other segments is kept constant. The intensity at the
detector equals
\begin{equation}
I(\Phi) = I_A + I_B + 2\sqrt{I_A I_B} \cos{(\Phi - \Phi_0)},\label{eq:signal}
\end{equation}
where $I_B$ is the intensity at the target originating from the modulated segments, $I_A$ is the target
intensity caused by light coming from the other segments, $\Phi$ is the phase that is varied and $\Phi_0$
is the unknown optimal value for the phase. The last term in Eq.~\eqref{eq:signal} is the signal that is
relevant for measuring $\Phi_0$. There first two terms constitute a constant bias.

Table~\ref{tab:noise} lists the magnitudes of the signal and the bias for each of the three algorithms. If
the detection system is photon shot noise limited, the noise is proportional to the square root of the
bias. When, on the other hand, constant noise sources such as readout noise or thermal noise are dominant,
the SNR is directly proportional to the signal magnitude. Table~\ref{tab:noise} also shows the root mean
square (rms) phase correction that is applied during each iteration of the corresponding algorithm. The
rms phase correction is a measure for the required accuracy of the measurements.

With the stepwise sequential algorithm, $I_A \approx I_0$ and on average $I_B = I_0/N$. Since the initial
diffuse transmission $I_0$ is low and $N$ can be very high, the SNR is low. The continuous sequential
algorithm has a higher SNR since the overall intensity at the detector increases while the algorithm
progresses and $I_A \approx \eta I_0$. Assuming the dominant noise source is constant, the SNR will
increase as the enhancement becomes higher. Therefore, the algorithm can be accelerated by decreasing the
integration time of the camera as the algorithm advances. In case the detection system is photon shot
noise limited, the SNR remains constant during the optimization since both the signal and the shot noise
scale as $\sqrt{\eta I_0}$.

\sloppy

The highest SNR is achieved with the partitioning algorithm. Since
we always change the phase of half of the segments, $I_A \approx I_B
\approx \eta I_0/2$, resulting in a maximal signal. Unlike the
sequential algorithms, the SNR does not depend on $N$. Therefore the
number of segments can be increased without suffering from noise.
Like with the continuous sequential algorithm, the integration time
can be adjusted dynamically to optimize the speed/SNR tradeoff.
Although the partitioning algorithm has the highest SNR, the
magnitude of the phase corrections decreases as the algorithm
progresses. The required accuracy in measuring $\Phi_0$ increases at
the same pace as the SNR increases.

\myfussy

The partitioning algorithm is very sensitive to measurement errors since, when the measured $\Phi_0$ has
an error, half of the segments will be programmed with the wrong phase. In the extreme case where the
error equals $\pi$ the enhancement completely disappears in a single iteration. A simple and effective
solution to this problem is to keep the previous configuration of the phase modulator in memory. When an
optimization step causes the signal to decrease, the algorithm can revert to the saved configuration.

\section{Simultaneously optimizing multiple targets}
\noindent For some applications one wishes to focus on multiple targets at once. To achieve multi-target
focusing, one could use the sum of the intensities in all targets as feedback. In this case the feedback
signal would equal (compare Eq.~\ref{eq:signal})
\begin{align}
I_\text{M}(\phi) &\equiv \sum_m^M I_m(\phi)\\
&= \sum_m^M I_{mA}+I_{mB}+2\sqrt{I_{mA}I_{mB}}\cos(\Phi -
\Phi_{m0}), \label{eq:signal-mm}
\end{align}
where the summation is over all $M$ targets. Each of the cosine terms in Eq.~\eqref{eq:signal-mm} is
weighted by the square root of the intensity contribution of the modulated segments $I_{mB}$ and the
contribution of the unmodulated segments $I_{mA}$. This weighting makes that targets that initially are
more intense are optimized more effectively. While the intensity increases in all targets, the enhancement
will not be distributed equally over the targets and the distribution depends on the starting conditions.

With the sequential algorithms, it is possible to distribute the
enhancement equally over all targets by using a different feedback
signal. If we use the summed square root of the intensities in the
targets as feedback, we have a feedback `amplitude'
\begin{align}
A_M(\phi) &\equiv \sum_m^M \sqrt{I_m(\phi)}\\
&\approx \sum_m^M\sqrt{I_{mA}}+\sqrt{I_{mB}}\cos(\Phi - \Phi_{m0}),
\label{eq:signal-mm-sqrt}
\end{align}
where we used that $I_{mA}\gg I_{mB}$ for the sequential algorithms. For these algorithms, $I_{mA}$ is
approximately equal to the initial intensity in target $m$. The feedback signal in
Eq.~\eqref{eq:signal-mm-sqrt} is only proportional to the propagation amplitude $\sqrt{I_{mB}}$ of the
modulated segment to the target. Therefore, the signal does not depend on the initial conditions and the
intensity enhancement will be distributed evenly over all targets.

\section{Conclusion}
\noindent Three different algorithms for inverting wave diffusion were presented. The algorithms were
compared experimentally, with numerical simulations and using analytical theory. We found good agreement
between experimental data, simulations and theory. Moreover, the simulations and theory can be used to
predict the performance in different experimental situations.

The effectiveness of the algorithms was quantified by the enhancement. It was seen that the enhancement
depends on the number of segments $N$ and the relative persistence time $T_p/T_i$. For the sequential
algorithms to have optimal performance, it is required to adjust $N$ to match $T_p$. This means that these
algorithms need a-priori knowledge of the system. The partitioning algorithm does not need this knowledge
and always performs close to optimal. Moreover, the algorithm causes the enhancement to increase the most
rapidly of the three investigated methods. All in all, this algorithm is a good candidate for applying
inverse diffusion in instable scattering media such as living tissue. In the future, learning algorithms
(see e.g. \cite{Goldberg1989,Judson1992}) might be developed to further improve the performance of inverse
diffusion, for instance by dynamically balancing the trade-off between signal to noise ratio and
measurement speed.

The maximum enhancement linearly depends on the number of measurements that can be performed before the
speckle pattern decorrelates ($T_p/T_i$). The faster the measurements, the higher the enhancement. In our
current system, the speed is limited by the response time of the LCD. Fast micro mechanical phase
modulators have a mechanical response time of about $\us{10}$ (see e.g. \cite{Hacker2003}), which allows a
$10^4$ times faster operation than with our current system. At such high speeds, the response time of the
operating system of the control computer will become a bottleneck. Since the computation time required for
the algorithm itself is negligible, this bottleneck can be overcome by using a real-time operating system,
or by replacing the computer with a programmable logic device. In perfused tissue, a typical decorrelation
timescale is 10 ms \cite{Briers1995}, which means that an enhancement of about 50 should be possible with
currently available technology.

\begin{sectionappendix}
\section{Calculation of the performance of the partitioning algorithm}\label{sec:alg3}
\noindent In this section we calculate the development of the enhancement of the partitioning algorithm
under ideal conditions. During one iteration of the partitioning algorithm, the phase modulator is
randomly split into two groups ($A$ and $B$), each containing half of the segments. The relative phase
($\Phi$) of group $B$ is cycled from 0 to $2\pi$. During this cycle, the target intensity is given by
\begin{equation}
I(\Phi) = \left| E_{mA} + E_{mB}e^{i\Phi}\right|^2,\label{eq:random-intensity-before}
\end{equation}
where $E_{mA}$ is the contribution of the segments in group $A$ to the target field
\begin{equation}
E_{mA} = \sum_{n\in A} \sqrt{\frac{\avg{I_0}}{N}}\xi_{mn},\label{eq:partitioning-A-xi}
\end{equation}
with
\begin{equation}
\xi_{mn} \equiv \sqrt{\frac{1}{\avg{I_0}}} t_{mn} e^{i\phi_n},
\end{equation}
and similar for $E_{mB}$. The coefficients $\xi_{mn}$ are initially random and distributed according to a
normalized circular Gaussian distribution, meaning that $\avg{\xi}=0$ and
$\avg{(\Re{\xi})^2}=\avg{(\Im{\xi})^2}=1/2$. As the algorithm proceeds, the phases $\phi_n$ are adjusted
and the distribution gradually changes to a Rayleigh distribution when a high enhancement is reached. The
average value $\avg{\xi}$ increases from $0$ to $\sqrt{\pi/4}$ as all contributions are aligned to be in
phase. At any moment during the optimization, $\avg{\xi} = \sqrt{(\eta-1)/(N-1)}$ and $\avg{|\xi|^2}=1$.

\begin{figure}
\centering
\includegraphics[width=\wideimage]{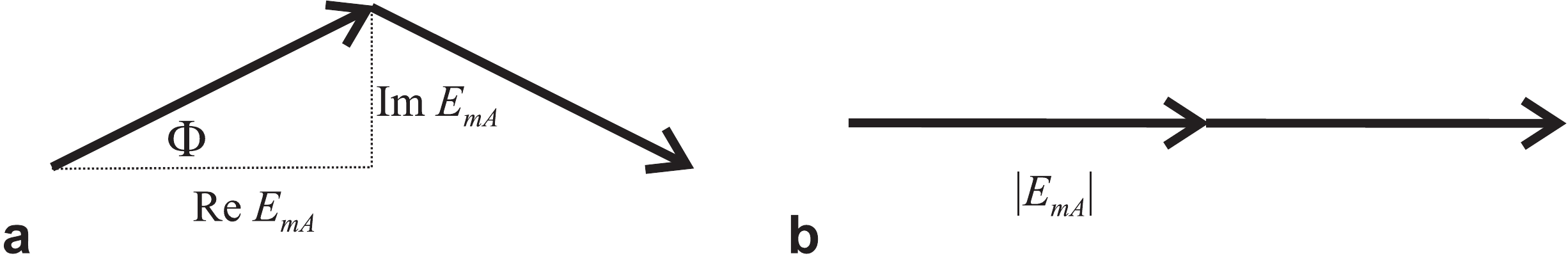}
\caption{Complex plane representation of the partitioning algorithm. \subfig{a} Before the iteration the
contributions from $A$ and $B$ are not exactly in phase. \subfig{b} After the iteration, the contributions
are aligned and the resulting intensity is higher.}\label{fig:vectors}
\end{figure}

Figure~\ref{fig:vectors} gives a graphical representation of a single iteration. Before the iteration,
$E_{mA}$ and $E_{mB}$ have a different phase. Without loss of generality, we choose the phase of
$(E_{mA}+E_{mB})$ to be $0$. The intensity before the iteration is given by
\begin{equation}
I_\text{before} = \left( \Re E_{mA} + \Re E_{mB} \right)^2.
\end{equation}
After the iteration, $\Phi$ is set to the value that caused the highest target intensity, which means that
$E_{mA}$ and $E_{mB}$ are now in phase. The target intensity then equals
\begin{equation}
I_\text{after} = \left( |E_{mA}| + |E_{mB}| \right)^2,\label{eq:random-intensity-after}
\end{equation}
which is higher than or equal to before the iteration.

We now calculate the average intensity gained in a single iteration. We consider the regime where already
a few iterations have been done ($\eta \gg 1$). In this regime, we can approximate
\begin{align}
|E_{mA}| &= \sqrt{(\Re E_{mA}) ^2 + (\Im E_{mA})^2}\\
&\approx \Re E_{mA} + \frac{(\Im E_{mA})^2}{2\Re E_{mA}}.
\end{align}
Using this result in Eq.~\eqref{eq:random-intensity-after} gives
\begin{multline}
I_\text{after} = \left( \Re E_{mA} + \Re E_{mB} \right)^2 + (\Im E_{mA})^2 + (\Im E_{mB})^2 +\\
 + \frac{\Re{E_{mA}}}{\Re{E_{mB}}}(\Im E_{mB})^2 + \frac{\Re{E_{mB}}}{\Re{E_{mA}}}(\Im E_{mA})^2 +\\
+ \frac{(\Im E_{mA})^4}{4(\Re E_{mA})^2}+ \frac{(\Im E_{mB})^4}{4(\Re E_{mB})^2} + \frac{(\Im E_{mA} \Im
E_{mB})^2}{2\Re E_{mA}\Re E_{mB}}
\end{multline}
where the terms on the last line can be neglected. If $N\gg1$, then
$\Re{E_{mB}}/\Re{E_{mA}}\approx 1$. The intensity gain for the
iteration $\Delta I \equiv I_\text{after}-I_\text{before}$ is now
found to be
\begin{equation}
\Delta I = 2(\Im E_{mA})^2 + 2(\Im E_{mB})^2
\end{equation}\label{eq:random-intensity-increase}
We are primarily interested in the regime $N\gg\eta\gg1$), where the algorithm picks up the main part of
the final enhancement. In this regime, $\avg{\xi} \ll 1$ and the probability distribution of $\xi$ is
still close to the original Gaussian distribution. Therefore, $\avg{(\Im{\xi})^2}\approx 1/2$ and it
follows from Eq.~\eqref{eq:partitioning-A-xi} that
\begin{equation}
\Delta I = \frac12\avg{I_0}
\end{equation}
Therefore, we expect the intensity enhancement $\eta$ to increase with $1/2$ after each iteration of the
algorithm. With this information, we also calculate the typical phase adjustment that is performed in each
iteration. From Fig.~\ref{fig:vectors} it follows that the root mean square phase adjustment equals
\begin{equation}
\Phi_{\text{rms}} \equiv \sqrt{\avg{\Phi^2}} = \sqrt{\avg{\frac{(\Im
E_{mA})^2}{(\Re E_{mA})^2}}} = \sqrt{\frac{2}{\eta}}
\end{equation}
When $\eta$ approaches its maximum, all contributions are almost
completely in phase and $\avg{(\Im{\xi})^2}$ vanishes. In this
regime, the algorithm becomes less and less effective, as was seen
in simulations and experiments (see Fig.~\ref{fig:high-N}). Adaptive
switching to one of the other algorithms will ensure optimal
convergence.

\end{sectionappendix}
\bibliography{../../bibliography}
\bibliographystyle{Ivo_sty}

\setcounter{chapter}{7}
\chapter{Transport of light with an optimized wavefront}\label{cha:dorokhov-theory}

\noindent In our experiments we construct a wavefront that maximizes
the intensity at a given target position. In this chapter, we
investigate the transport properties of light with such an optimized
wavefront. We introduce the concept of `active' measurements to
describe a measurement where the wavefront is tailored to the
specific realization of disorder in the sample. We show that this
new type of experiments is very promising as it could be used to
find signatures of absorption and possibly Anderson localization of
light, or to test mesoscopic transport theories and selectively
study extended modes in a sample.

So far, we have used a simple \emph{ad hoc} matrix model (Eq.~\eqref{eq:scattering}) to describe
scattering and focusing through a disordered medium. This model is based on the assumption of uncorrelated
transmission coefficients (UTC). In reality, the constraint on energy conservation makes that the matrix
elements are correlated. Such correlations can be described by random matrix theory (RMT). RMT predicts
that it is possible to generate an incident wavefront that is fully transmitted through an opaque object,
regardless of the thickness of the object.

In Section~\ref{sec:RMT} we give a brief introduction to RMT and discuss how it can be applied to describe
scattering in a disordered optical waveguide. In Section~\ref{sec:new-observables} we use RMT to describe
the transport of light in an active measurement. We calculate experimentally observable quantities and
compare the situations with the matched and unmatched incident wavefronts. Furthermore, we discuss the
physical relevance of these `actively measured' quantities and compare the results with predictions made
with the UTC model. The total transmission of an optimized wavefront is found to be an observable that
characterizes the correlations in the scattering matrix. Finally, in
Section~\ref{sec:realistic-conditions}, we extend the RMT model so that it applies to a slab geometry and
accounts for other experimental parameters, such as the resolution of the light modulator. We use the
extended model to calculate how experimental limitations affect the measurements and we show how they can
be compensated for.

\section{Random matrix theory}\label{sec:RMT}
\noindent Random matrix theory is a very general mathematical theory for statistically describing
extremely complex systems. Some excellent reviews on RMT can be found in Refs.~\citealt{Mehta1967,
Beenakker1997} and \citealt{Guhr1998}. The theory has been developed by Wigner\cite{Wigner1951} to
describe statistics of eigenvalues of many-body systems. It was originally employed to describe the
properties of excited states of atomic nuclei and also proved very successful in calculating the
statistics of electron scattering in mesoscopic conductors and quantum dots\cite{Dorokhov1984, Mello1988,
Akkermans2007, Datta2007}. Weaver\cite{Weaver1989} was the first to study the application of RMT to
acoustical waves instead of quantum mechanical wave functions. In pioneering experiments in
acoustics\cite{Schroeder1954, Weaver1989} and microwaves\cite{Stoeckmann1990, Doron1990} the distribution
of eigenfrequencies in irregular cavities was measured. RMT proved very successful in predicting the
statistical properties of eigenfrequency spacing in these systems.\cite{Guhr1998} Although most results of
RMT are for monochromatic waves and, therefore, do not describe the dynamics of scattering, methods for
extending the matrix treatment to describe time dependence are being developed\cite{Brouwer1997,
Tiggelen1999, Schomerus2001, Skipetrov2004}.

More recently, RMT has been used to calculate correlations of light
scattering in disordered optical waveguides (see e.g.
\cite{Beenakker1997}) and random lasers\cite{Molen2007}. Because
only very general concepts are used, it is very easy to translate
results for one type of waves to a different system. This also means
that experimental observations in one field of physics are directly
relevant for many other disciplines.

The basic assumption of RMT is that the system is so complex that
you might as well use a completely random Hamiltonian matrix to
describe it. In Wigner's original approach\cite{Wigner1951}, the
`randomness' of the matrix is invariant under unitary
transformations of Hilbert space. In terms of light scattering, this
invariance means that the scattered light has no spatial or angular
correlation with the incident light whatsoever. The only other
constraint that Wigner imposed on the random matrices is that there
is a conserved quantum number that imposes a symmetry condition on
the matrix. In the case of scattering this condition is that the
energy is conserved (no absorption). Usually, an optical system also
has time-reversal symmetry, which puts an extra constraint on the
random matrix. Due to these constraints the elements of the matrix
have to be statistically correlated. Typically, the number of
scattering channels is large and the correlation between any pair of
channels is negligible. Only when many scattering channels are
involved, the effects of the correlations become visible.

\subsection{Random matrix theory in a waveguide geometry}
\begin{figure}
\centering
  \includegraphics[width=\mediumimage]{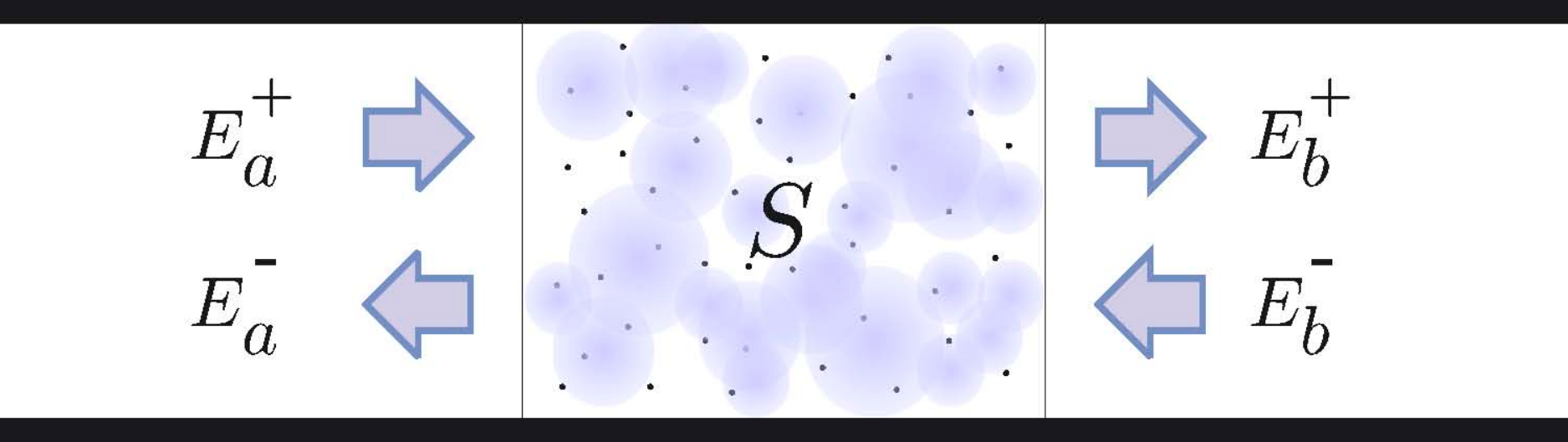}\\
  \caption{Scattering in a waveguide geometry.
  The scattering matrix $S$ couples incident waveguide modes $E^+_a$ and $E^-_b$
  to outgoing waveguide modes $E^-_a$, $E^+_b$. In a transmission experiment, $E^-_b=0.$}\label{fig:S-matrix}
\end{figure}

\noindent We first explain how RMT is applied to a simple waveguide geometry as is shown in
Fig.~\ref{fig:S-matrix}. To both sides of a strongly scattering disordered sample a waveguide supporting
exactly $N$ propagating waveguide modes (channels) is connected. Evanescent fields in the waveguide are
not taken into consideration. The sample scatters incident light into reflected and transmitted waveguide
channels.

Scattering by the sample is represented by the scattering matrix $S$ that connects the incident field
$E^+_a$, $E^-_b$ to the channels leaving the sample ($E^-_a$ and $E^+_b$). Channels labelled $a$ are
channels on the left side of the sample and Channels labelled $b$ are channels on the right side of the
sample. The scattering matrix $S$ is defined as
\begin{equation}
\begin{bmatrix}
E^-_a\\
E^+_b
\end{bmatrix}
=S
\begin{bmatrix}
E^+_a\\
E^-_b
\end{bmatrix}.
\end{equation}
We assume that the sample does not absorb any light and, therefore,
$S$ is unitary. When, in addition, scattering in the sample is
reciprocal, time-reversal symmetry is obeyed and $S$ is both unitary
and symmetric. Although all our experimental systems have
time-reversal symmetry, unitarity alone is sufficient for the
derivations made in this chapter to be valid. The scattering matrix
can be split into four matrices describing reflection and
transmission
\begin{equation}
\begin{bmatrix}
E^-_a\\
E^+_b
\end{bmatrix}
=
\begin{bmatrix}
r^{-+}_{aa} & t^{--}_{ab}\\
t^{++}_{ba} & r^{+-}_{bb}
\end{bmatrix}
\begin{bmatrix}
E^+_a\\
E^-_b
\end{bmatrix}.
\end{equation}
From now on, we only consider transmission from left (channels $a$)
to right (channels $b$) as described by the matrix $t^{++}_{ba}$, in
short $t$. RMT tells us that the restriction of energy conservation
gives rise to a statistical structure in the transmission matrix.
This structure is visible in the singular value decomposition (SVD)
of the transmission matrix. Like any matrix, the transmission matrix
$t$ can be decomposed as
\begin{equation}
t = U \tau V\label{eq:svd},
\end{equation}

\begin{figure}
\centering
  \includegraphics[width=\mediumimage]{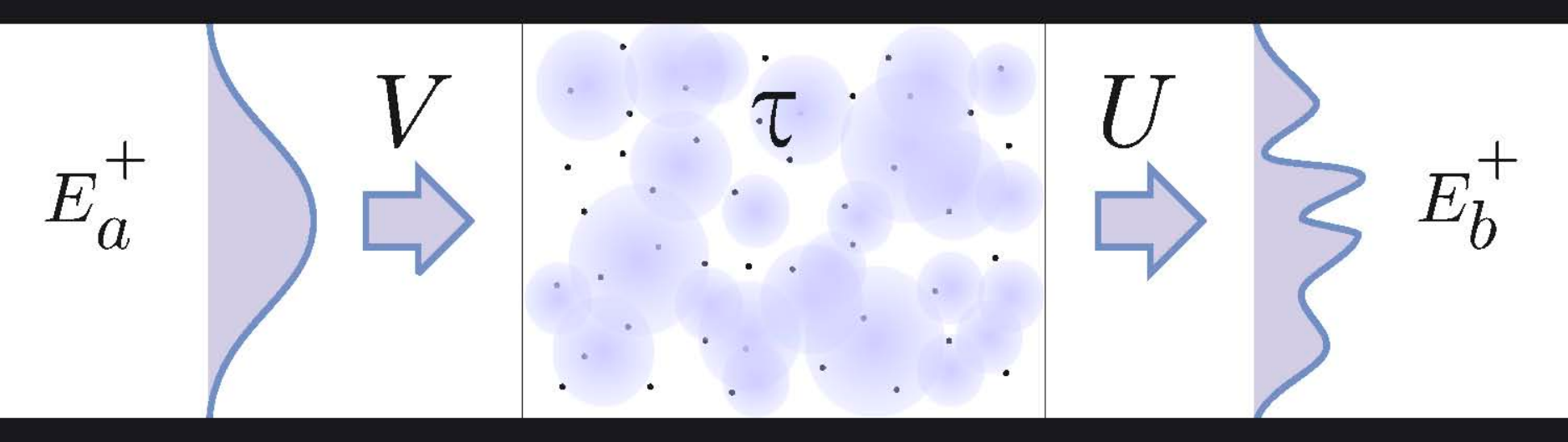}\\
  \caption{Transmission experiment in a waveguide geometry.
  The transmission matrix of the system is decomposed in three
  matrices:
  $V$ maps the incident light to eigenmodes of the sample,
  $\tau$ gives amplitude transmission coefficients for the eigenmodes.
  $U$ maps the eigenmodes back to real space on the other side of the sample.
}\label{fig:waveguide-N}
\end{figure}

\noindent where $U$ and $V$ are unitary matrices and $\tau$ is a diagonal matrix with real, non-negative
elements. The physical interpretation of Eq.~\eqref{eq:svd} is illustrated in Fig.~\ref{fig:waveguide-N}.
Matrix $V$ maps the incident light from real space coordinates onto transmission eigenmodes of the sample.
The intensity transmission coefficients for each of these eigenmodes are given by the diagonal elements of
$\tau^2$. Matrix $U$ maps the eigenmodes back to spatial coordinates at the back of the sample.

In RMT $U$ and $V$ are random unitary matrices with elements that
have a circular Gaussian distribution. The statistical properties of
these matrices are invariant to any unitary transform of the Hilbert
space.\cite{Wigner1951} In other words, when we replace $U$ (or $V$)
by $U'=BU$ ($V'=BV$), with $B$ any unitary matrix, the ensemble
properties do not change. It, therefore, does not matter whether a
channel is defined as a waveguide mode, an incident plane wave or a
position on the sample surface. Moreover, these constraints on $U$
and $V$ make that these matrixes are statistically independent of
$\tau$.

\subsection{Distribution of transmission eigenvalues}
\begin{figure}
\centering
  \includegraphics[width=\smallimage]{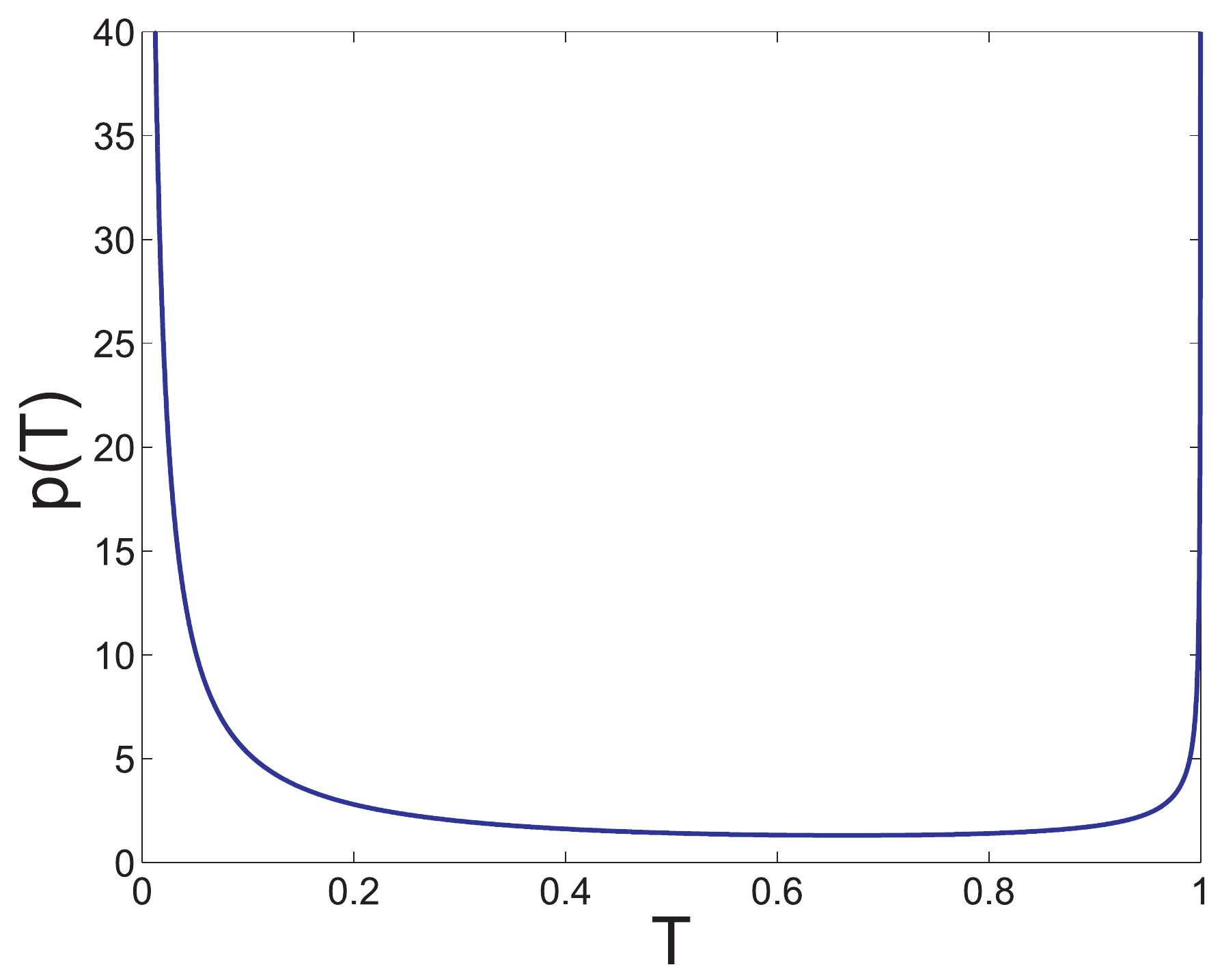}\\
  \caption{Probability density of the intensity transmission eigenvalues $T\equiv\tau^2$ as derived in Ref.~\citealt{Dorokhov1984}.}\label{fig:dorokhov}
\end{figure}
\noindent Intuitively, one might expect that all modes in a random
system are roughly equivalent and the distribution of
$T\equiv\tau^2$ is peaked around the total transmission of the
sample. However, as was shown first by Dorokhov\cite{Dorokhov1984},
the eigenvalues have a probability density that is given by
\begin{equation}
p(T) = \avg{T}\frac{1}{2T\sqrt{1-T}}\label{eq:dorokhov}.
\end{equation}
This probability density is plotted in Fig.~\ref{fig:dorokhov}. The
eigenvalue distribution is bimodal: almost all eigenvalues are
either $1$ or $0$. A common terminology is to speak of `open' and
`closed' channels. Light in an open channel is almost fully
transmitted ($T\approx 1$), whereas light in closed channels is
almost completely reflected ($T\approx 0$). The prediction that all
channels are either open or closed is also known as the maximal
fluctuation theorem\cite{Pendry1990, Pendry1992}.

The distribution in Eq.~\eqref{eq:dorokhov} is the result of the restriction of energy conservation that
is imposed on the scattering matrix. Due to this restriction, a mechanism called eigenvalue repulsion
spreads out the values of $1/\xi$ to a uniform distribution.\cite{Pendry1990, Stone1991, Beenakker1997}
Here, $\xi$ is the localization length that is associated with a mode in the sample. The localization
length relates to the transmission of a sample by $T=\cosh^{-2}(L/\xi)$. The uniform distribution of
$1/\xi$ transforms into the distribution of $T$ as is given by Eq.~\eqref{eq:dorokhov}. Originally, RMT
was used to derive Eq.~\eqref{eq:dorokhov} for a waveguide. However, later it was shown that this
probability density function is more generally valid and also applies to a slab
geometry.\cite{Nazarov1994}

Equation~\eqref{eq:dorokhov} has divergencies at $0$ and at $1$. The divergency at $0$ is not integrable.
In fact, there is a minimum value for the transmission coefficient of a mode\cite{Beenakker1997,
Rossum1999}, which is given by $T=\cosh^{-2}(1/\avg{T})$. This cutoff normalizes the probability
distribution of $T$. The physical interpretation of the cutoff is that with $\avg{T}\approx{\ell/L}$, the
localization length of a mode cannot be shorter than the mean free path $\ell$. In practice, the cutoff
for $T$ is often ignored since most of the time one is only interested in the moments of $T$. These
moments depend very little on the cutoff.\cite{Rossum1999}

\subsection{The effect of refractive indices}
Random matrix theory is most conveniently formulated in a waveguide
geometry with an equal number of waveguide channels on both side of
the sample. In an experiment, the refractive index in front of the
sample is often different from the refractive index at the back.
Moreover, the sample itself has an effective refractive index that
may differ from that of its environment. The number of channels
scales with the refractive index squared. Therefore, the number of
channels in the sample is not the same as the number of channels in
the connected waveguides.

To account for the different number of scattering channels on both
sides of the medium we assume that all channels are statistically
equivalent. The number of channels is determined by the refractive
index in each of the media. At an interface, scattering randomly
redistributes the light over all channels. When the light enters a
medium with a lower refractive index, some linear combinations of
channels are completely reflected. This reflection is the random
equivalent of total internal reflection.

In general the transmission coefficient at the surface of a
disordered medium is angle dependent\cite{Lagendijk1989, Zhu1991,
Vera1996}. If we define channels as incident angles, different
channels are not statistically equivalent. Therefore, we chose to
define channels in real space coordinates. Since angular effects are
equal for all channels, the channels are still statistically
equivalent. Furthermore, we assume that the refractive index
contrasts are low and that scattering is strong, so that neighboring
channels can be considered statistically independent.

We now consider three possible combinations of refractive indices
and analyze their effect on the RMT formalism.

\subsubsection{Matrix description of a sample in air}
\begin{figure}
\centering
  \includegraphics[width=\mediumimage]{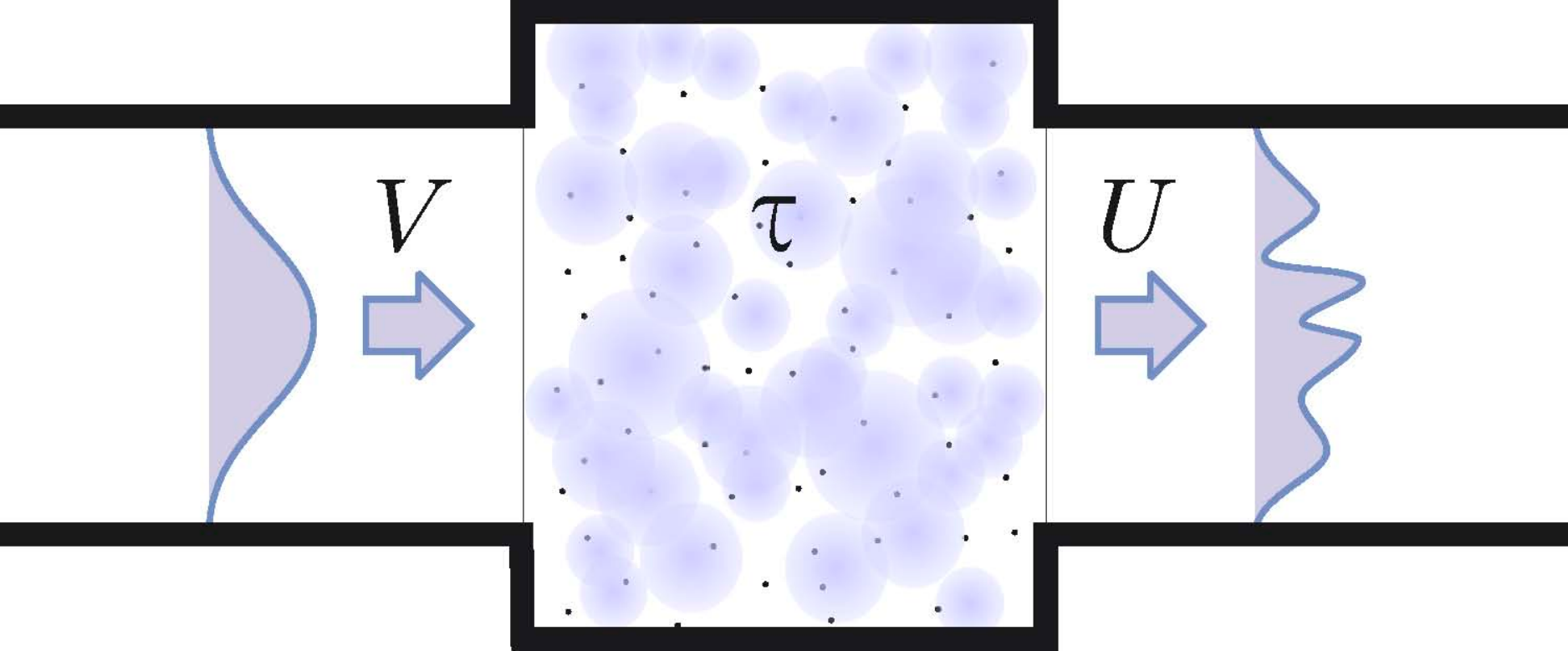}\\
  \caption{Diagrammatic representation of a sample in a medium with a
  lower refractive index. The sample supports more channels than the
  connecting waveguides and can, be thought of in a sense being `wider'.
}\label{fig:modesB}
\end{figure}

\noindent When a sample is surrounded by air, its effective refractive index is higher than that of the
surrounding medium. This means that the sample supports more channels than can possibly couple to the
outside world. The difference in supported channels is shown schematically in Fig.~\ref{fig:modesB}. Since
no light can couple to these bound modes, we ignore them and just use the formalism in Eq.~\eqref{eq:svd},
with $\tau$ a random subset of the actual full $\tau$ matrix of the sample.

\subsubsection{Matrix description of a sample in a dense medium}
\begin{figure}
\centering
  \includegraphics[width=\mediumimage]{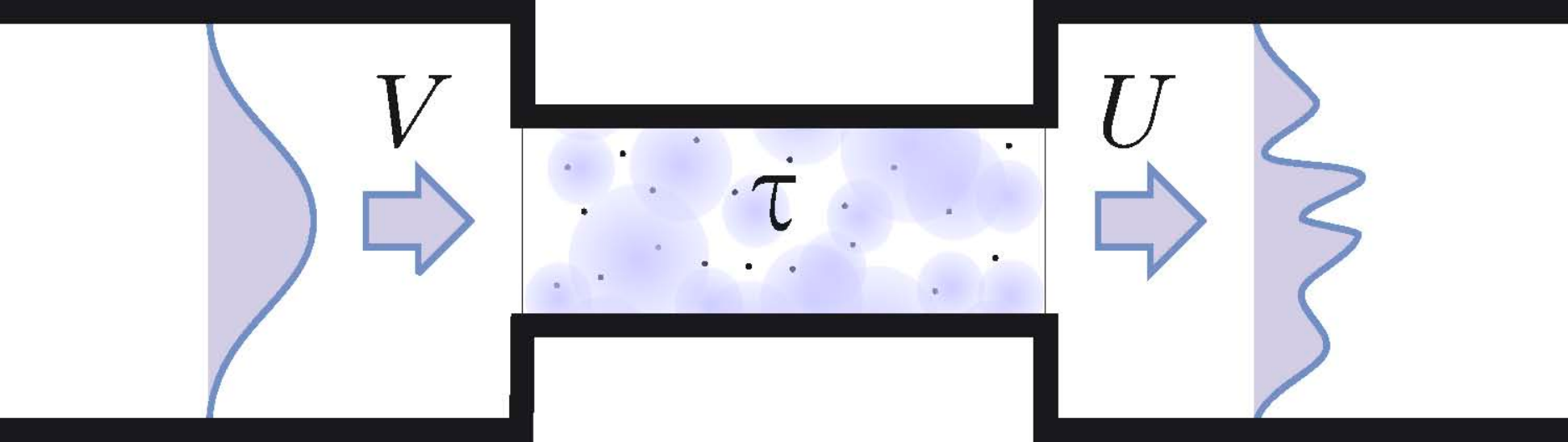}\\
  \caption{Diagrammatic representation of a sample in a medium with a
  higher refractive index. The sample supports fewer channels
  than the connecting waveguides and can,
  be thought of in a sense being `narrower'. Not all incident channels can couple into the sample.
  }\label{fig:modesC}
\end{figure}

\noindent In the situation where the refractive index of the sample is lower than that of its environment
not all incident channels can couple into the sample. Or, more correctly, some linear combinations of
incident channels have no support in the sample. As is depicted in the diagram in Fig.~\ref{fig:modesC},
the sample acts as a bottleneck. This situation could be modelled by using the formalism in
Eq.~\eqref{eq:svd} with a matrix $\tau$ that is extended with extra zeros for channels that cannot exist
inside the sample.

\subsubsection{Matrix description of a sample on a substrate}
\begin{figure}
\centering
  \includegraphics[width=\mediumimage]{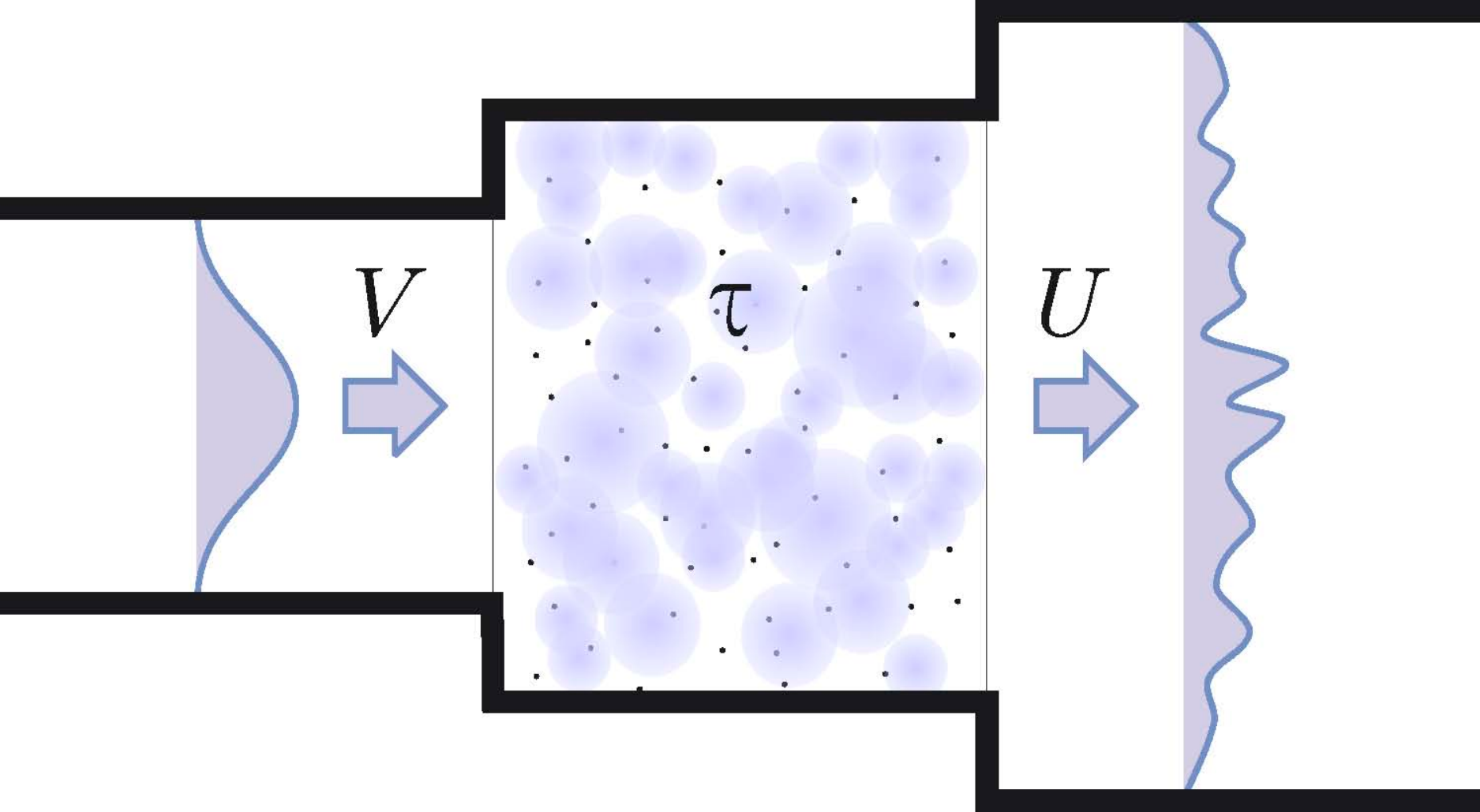}\\
  \caption{Diagrammatic representation of a sample on a substrate
  with a higher refractive index than that of the sample.
  All channels on the low index side of the sample can couple
  to the sample and, in turn, all channels in the sample
  can couple to the substrate.}\label{fig:modesD}
\end{figure}

\noindent When the sample is on a substrate with a high refractive
index, it is sandwiched between a low refractive index medium and a
high refractive index medium. In this experimentally relevant
situation, the number of incident channels is not equal to the
number of transmitted channels and $t$ is a rectangular $M\times N$
matrix. Here $N$ is the number of incident channels and $M$ is the
number of transmitted channels. In the singular value decomposition
of $t$, $U$ is a unitary $M\times M$ matrix, $V$ is a unitary
$N\times N$ matrix, and $\tau$ is a rectangular diagonal $M\times N$
matrix. The number of diagonal elements in $\tau$ equals $K\equiv
\min{(M,N)}$. When the sample is illuminated through the low index
waveguide, as is shown in Fig.~\ref{fig:modesD}, there will be no
total internal reflection as the light travels from a low refractive
index towards a high refractive index. Since the case with a
rectangular $t$ matrix is the most general, we will use this case in
our derivations.

\section{A new class of experimental observables}\label{sec:new-observables}
\noindent A common way to experimentally study an optical system is
to measure its response to a fixed incident wavefront. For a
disordered system, the response will be stochastic. Useful results
are obtained by statistical methods such as ensemble averaging and
calculating correlation functions.

In our experiment, however, we actively adapt the incident wavefront
to each specific realization of disorder in the sample and
selectively excite an extended optical mode. This concept is
analogous to the approach taken in the field of coherent control,
where a femtosecond pulse is shaped in the frequency domain to
optimize the response of a chemical system\cite{Shapiro1986,
Bowman1990, Judson1992, Zeidler2001, Weinacht2002, Herek2002}. This
pulse selectively excites a chemical pathway that can then be
studied. For most systems described by RMT such control is not
possible and at best ensemble averaged quantities can be observed.

One of the potentials of our method becomes apparent from the distribution $p(T)$ in
Eq.~\eqref{eq:dorokhov}. This equation predicts that there always exist channels that have $T=1$. Using a
wavefront synthesizer we can, in principle, generate an incident wavefront that only couples to these open
channels. Such a wavefront should be fully transmitted through an otherwise opaque scattering medium.

Open channels correspond to optical modes that extend over the whole
thickness of the sample.\cite{Beenakker1997} With our active
measurement method, it is possible to selectively study the
properties (correlations, speckle statistics, delay time statistics,
behavior in random gain media, second harmonic generation, etcetera)
of extended modes without having a contribution from the localized
modes that correspond to closed channels. The relation between
localized modes and extended modes is of paramount importance in the
study of Anderson localization (where the number of extended modes
vanishes, see e.g. \cite{Beenakker1997}) and in the field of random
lasers\cite{Mujumdar2004}.

\begin{figure}
\centering
  \includegraphics[width=\mediumimage]{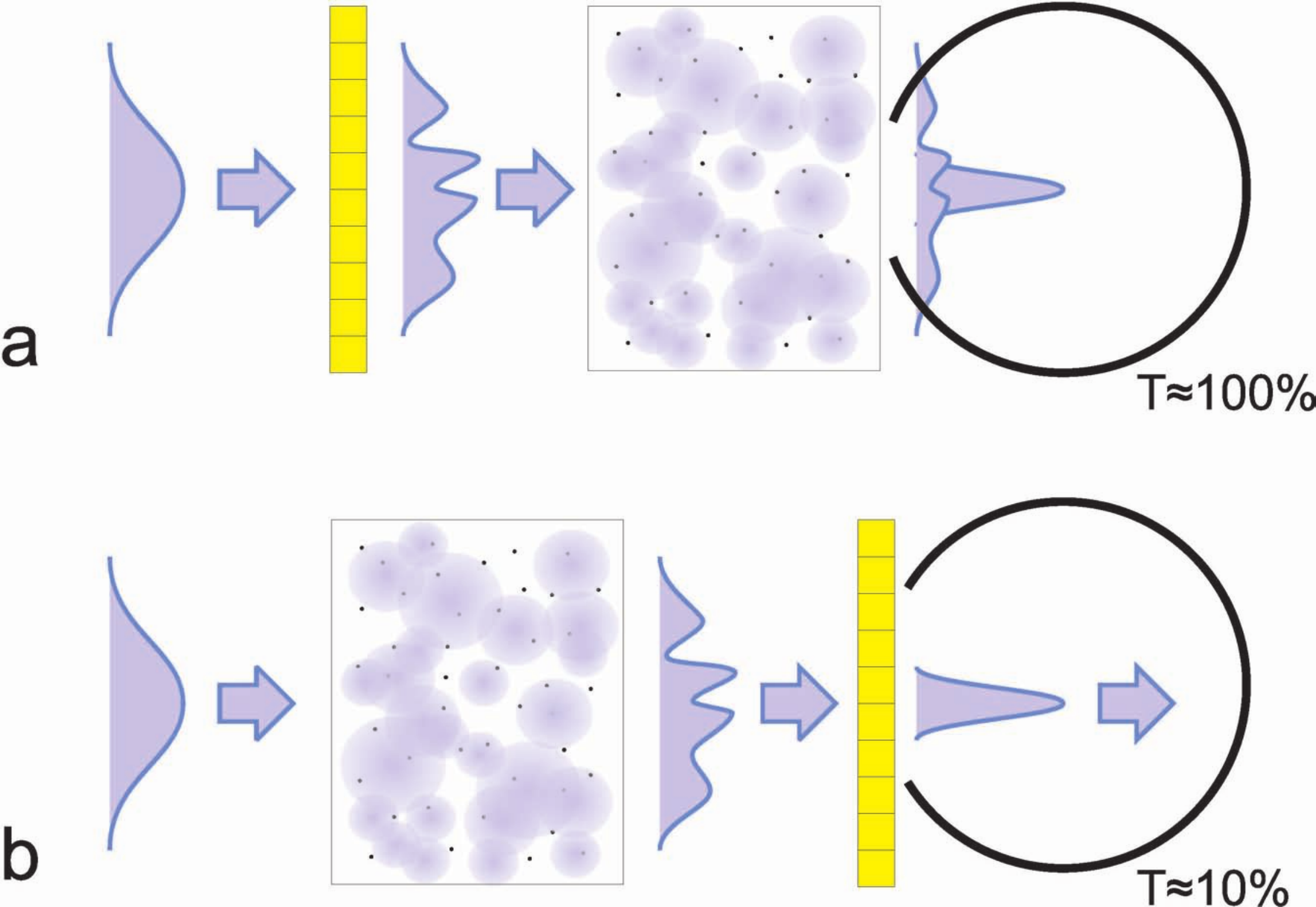}\\
  \caption{Difference between placing a light modulator in front of
  the sample or behind.
  \subfig{a} Light modulator in front of the sample.
  A sharp focus surrounded by a speckle pattern is formed and the
  total transmission in the order of $100\%$. \subfig{b} Light modulator
  behind the sample. All transmitted light is formed into a focus
  without speckle. The total transmission does not change and typically is in
  the order of $10\%$.}\label{fig:front-back}
\end{figure}

The possibility to perform active measurements once again
illustrates the fundamental difference between having a light
modulator in front of the sample or behind the sample. The two
situations are shown in Fig.~\ref{fig:front-back}. In
Fig.~\ref{fig:front-back}a the modulator is placed in front of the
sample and the incident wavefront is optimized for creating a focus
behind the sample. The optimal wavefront only couples to extended
modes in the sample and the total transmission increases. Since only
a fraction of all channels is open, there is no complete set of
basis vectors for forming any desired field behind the sample. As a
result, it is not possible to concentrate all light in the focus;
the rest of the transmitted light will form a random speckle
pattern. When a modulator is placed behind the sample, as is shown
in Fig.~\ref{fig:front-back}b, it could in principle focus all
transmitted light to a single point. However, light transport in the
sample is, of course, not affected.

To find a wavefront that excites only open channels, one could theoretically perform a singular value
decomposition of the full transmission matrix (see Eq.~\eqref{eq:svd}). However, measuring the complete
scattering matrix is experimentally very challenging since it contains tens of millions of elements.
Moreover, measurement noise would accumulate and disturb the singular value decomposition. A different,
much easier, approach is to create a focus at any position behind the sample. Since closed channels are
fully reflected, only open channels contribute to this focus. When we optimize the incident light to
create a maximally intense focus, we generate a wavefront that only overlaps with open channels.

In this section, we use the RMT model to calculate some
experimentally observable quantities and compare the result for an
unshaped wavefront (passive measurement) to the result of a
wavefront that is optimized for creating a focus (active
measurement). The results of the calculations that follow below are
summarized in Table~\ref{tab:I-results}. We first derive these
quantities for a waveguide geometry with a wavefront synthesizer
that perfectly controls the phase and amplitude of all incident
channels. This assumption greatly simplifies the calculations.
Later, in Section~\ref{sec:realistic-conditions}, we will see how
the general results for an open system or for imperfect modulation
can easily be found once the results for a perfect waveguide system
are known.

\begin{table}
  \centering
\begin{tabular}{|l|r@{}l|r@{}l|}
  \hline
               & \multicolumn{2}{c|}{passive}
               & \multicolumn{2}{c|}{active (optimized)}   \\
  \hline
single channel &$ \avg{I_\bopt} $&$=\avg{\tau^2}\frac{K}{MN}$
               &$ \avg{\optim{I}_\bopt}    $&$=\avg{\tau^2}\frac{K}{M}$\\
total          &$ \avg{\Ttot}   $&$=\avg{\tau^2}\frac{K}{N}$
           &$\avg{\tTtot}$&$=\frac{2}{3}$ \\
  \hline
\end{tabular}
\caption{Average transmission into a single channel and average
total transmission with and without actively shaped
wavefront.}\label{tab:I-results}
\end{table}

\subsection{Observables in passive measurements}
We now calculate average values for the total transmission through
the sample and for the intensity in a single transmission channel.
The incident wavefront is not adapted to the sample, which is the
normal, passive, way of measuring transmission coefficients. The
general equation for calculating the field in outgoing channel $b$
is
\begin{equation}
E_b = \sum_a^N t_{ba} E_a\label{eq:E_b-t}.
\end{equation}
Using the singular value decomposition in Eq.~\eqref{eq:svd}, we
obtain
\begin{equation}
E_b = \sum_a^N \sum_k^K U_{bk} \tau_{kk} V_{ka}
E_a.\label{eq:E_b-svd}
\end{equation}
Because RMT is symmetric to unitary transformations of the incident
field, we can use any arbitrary normalized field and achieve the
same outcome. For simplicity, we illuminate only incident channel
$\ain$ and have $E_a = \delta_{\ain a}$. The total transmission is
calculated using
\begin{equation}
\Ttot \equiv \sum_b^M I_b\label{eq:Ttot-def},
\end{equation}
with $I_b \equiv |E_b|^2$.  The total transmission follows from substituting Eq.~\eqref{eq:E_b-svd} into
Eq.~\eqref{eq:Ttot-def}
\begin{equation}
\Ttot = \sum_b^M \left|\sum_k^K U_{bk} \tau_{kk} V_{k\ain}\right|^2.
\end{equation}
After expanding the square and reordering terms, we have
\begin{equation}
\Ttot = \sum_b^M \sum_{k,k'}^K U_{bk}^*U_{bk'}^{\phantom{*}}
\tau_{kk}^{\phantom{*}} \tau_{k'k'}^{\phantom{*}}
V_{k\ain}^{\phantom{*}} V^*_{k'\ain}.\label{eq:Ttot-long}
\end{equation}
Unitarity of $U$ implies that $\sum_b^M
U_{bk}^*U_{bk'}^{\phantom{*}} = \delta_{kk'}$. Therefore,
Eq.~\eqref{eq:Ttot-long} reduces to
\begin{equation}
\Ttot = \sum_{k}^K \tau_{kk}^2
\left|V_{k\ain}^{\phantom{2}}\right|^2.\label{eq:Ttot}
\end{equation}
To find the ensemble average total transmission, we use the assumption from RMT that $V$ and $\tau$ are
statistically independent. We can, therefore, average Eq.~\eqref{eq:Ttot} over all realizations of
disorder to find
\begin{equation}
\avg{\Ttot} = \avg{\sum_{k}^K \tau_{kk}^2
\left|V_{k\ain}^{\phantom{2}}\right|^2} = \avg{\tau^2}
\avg{\sum_{k}^K |V_{k\ain}|^2} =
\avg{\tau^2}\frac{K}{N}.\label{eq:Ttot-avg}
\end{equation}
where the unitarity of $V$ was used in the final step. When the
refractive index behind the sample is higher than the refractive
index in front of the sample, the factor $K/N$ equals one. On
average, an incident field excites all eigenmodes with equal
probability. Therefore, the average total transmission equals
$\avg{\tau^2}$.

If, on the other hand the refractive index behind the sample is
lower, the factor $K/N$ is less than one. To understand this factor,
we examine the situation where there is no sample, just a direct
transition between the two media. When $N>M$, there are more
channels in the incident field than there are in the transmitted
field. The excess channels cannot exist in the medium behind the
sample and undergo total internal reflection. The fraction of
channels that does not experience total internal reflection is
$M/N$. When a sample is present, energy is distributed randomly over
all sample channels. Only a fraction $M/N$ of the light can leave
the sample.

Since in the RMT formalism all transmitted channels are equivalent, the average intensity in a single
transmitted channel is simply found by dividing Eq.~\eqref{eq:Ttot-avg} by $M$
\begin{equation}
\avg{I_b} = \avg{\tau^2}\frac{K}{MN}.
\end{equation}

\subsection{Observables in active measurements}
We now calculate the same quantities as in the previous section for
an active measurement where the incident wavefront is optimized to
maximize transmission into a single channel $\bopt$. We first
determine this optimal incident field $\optim{E}_a$ and then use it
to calculate the transmission into channel $\bopt$ as well as the
total transmission.

\subsubsection{Intensity in the optimized channel}
We now calculate the intensity in the target channel $\bopt$ after
an optimization. The incident field that maximizes the intensity in
this channel follows from the Cauchy-Schwartz inequality
\begin{equation}
I_\bopt = \left|\sum_a^N t_{\bopt a} E_a\right|^2 \leq \sum_a^N |t_{\bopt a}|^2 \sum_{a'}^N
|E_{a'}|^2,\label{eq:cauchy-schwartz}
\end{equation}
where the two sides are equal if and only if
\begin{equation}
\optim{E}_a = A_0 t_{\bopt a}^*,\label{eq:E_a-opt}
\end{equation}
where we denoted the optimal incident field after optimization with
a tilde sign. We are free to choose prefactor $A_0 \in \mathbb{C}$;
we choose $A_0$ so that $\optim{E}_a$ is normalized. Now, the
optimal value of $I_\bopt$ follows directly from
Eq.~\eqref{eq:cauchy-schwartz}
\begin{equation}
\optim{I}_\bopt = \sum_a^N |t_{\bopt a}|^2.\label{eq:Ib-opt-t}
\end{equation}
Using the singular value decomposition in Eq.~\eqref{eq:svd}, we
find
\begin{align}
\optim{I}_\bopt &= \sum_a^N \left|\sum_k^K U_{\bopt k} \tau_{kk}
V_{ka}\right|^2,\\
&= \sum_a^N \sum_{k,k'}^K U_{\bopt k}^* U_{\bopt k'}^{\phantom{*}}
\tau_{kk}^{\phantom{*}} \tau_{k'k'}^{\phantom{*}} V_{ka}^*
V_{k'a}^{\phantom{*}}\;.
\end{align}
We use the unitarity of $V$ to get rid of the sum over $a$
\begin{equation}
\optim{I}_\bopt = \sum_k^K \left|U_{\bopt k}^{\phantom{2}}\right|^2
\tau_{kk}^2.\label{eq:Ib-opt}
\end{equation}
Our model allows us to average over $U$ and $\tau$ separately to
find
\begin{equation}
\avg{\optim{I}_\bopt} =
\avg{\tau^2}\frac{K}{M}.\label{eq:Ib-opt-avg}
\end{equation}
In the situation where $M=K=N$, the fraction of the light that is transmitted into channel $\bopt$ is
equal to the total transmission before optimization (Eq.~\eqref{eq:Ttot-avg}).

\subsubsection{Total transmission}
We now calculate the transmitted field for the optimized incident
wavefront given by Eq.~\eqref{eq:E_a-opt}. Using
Eq.~\eqref{eq:E_b-svd} we find
\begin{equation}
\optim{E}_b = \sum_a^N \left(\sum_k^K U_{bk} \tau_{kk} V_{ka}\right)
A_0 \left(\sum_{k'}^K U_{\bopt k} \tau_{k'k'} V_{k'a}\right)^*.
\end{equation}
Unitarity of $V$ implies that $\sum^N_a
V_{ka}^{\phantom{*}}V^*_{k'a}=\delta_{kk'}$. We now have
\begin{equation}
\optim{E}_b = A_0 \sum_k^K U_{bk}^{\phantom{*}} U_{\bopt k}^*
\tau_{kk}^2.
\end{equation}
The total transmission after optimization now follows from
Eq.~\eqref{eq:Ttot-def},
\begin{equation}
\tTtot = A_0^2 \sum_b^M \left(\sum_k^K U_{bk}^{\phantom{*}}
U^*_{\bopt k} \tau_{kk}^2\right)\left(\sum_{k'}^K U^*_{bk'} U_{\bopt
k'}^{\phantom{*}} \tau_{k'k'}^2\right).
\end{equation}
Using unitarity of $U$ in the summation over $b$ we find
\begin{equation}
\tTtot = A_0^2 \sum_k^K \left|U_{\bopt k}^{\phantom{2}}\right|^2
\tau_{kk}^4.
\end{equation}
The prefactor $A_0$ was defined to normalize the incident wavefront.
Using Eq.~\eqref{eq:E_a-opt} we find
\begin{equation}
A_0^{-2} = \sum_a^N {|t_{\bopt a}|^2} = \sum_k^K \left|U_{\bopt k}^{\phantom{2}}\right|^2
\tau_{kk}^2\label{eq:A_0-def}\;,
\end{equation}
where the equality of Eqs.~\eqref{eq:Ib-opt-t} and \eqref{eq:Ib-opt} was used in the last step. Finally,
we have
\begin{equation}
\tTtot = \frac{\sum_k^K \left|U_{\bopt k}^{\phantom{4}}\right|^2
\tau_{kk}^4}{\sum_{k'}^K \left|U_{\bopt
k'}^{\phantom{2}}\right|^2\tau_{k'k'}^2}.
\end{equation}
Here we see that the numerator is a weighted average over $\tau^4$ and the denominator is a weighted
average over $\tau^2$ with equal weights. Averaging over $U$ gives
\begin{equation}
\avg{\tTtot} = \frac{\avg{\tau^4}}{\avg{\tau^2}} \equiv \bucket\label{eq:Ttot-opt}.
\end{equation}
We see that the total transmission after optimizing the incident
wavefront is equal to $\bucket$, which is defined as the ratio
between the fourth and the second moment of $\tau$. $\bucket$ does
not depend on the refractive indices outside the sample. Using the
distribution $p(T)$ that is predicted by RMT
(Eq.~\eqref{eq:dorokhov}), $\bucket$ can be calculated
\begin{equation}
\bucket = \frac{2}{3}.
\end{equation}
Surprisingly, $\bucket$ turns out to be a universal constant. A
fraction of $2/3$ of the power of an ideally optimized wavefront is
transmitted, regardless of the thickness of the sample.

\subsection{Comparison with the uncorrelated model}\label{sec:compare-UTC}
\noindent The RMT model is very similar to the uncorrelated
transmission coefficients (UTC) model in the sense that both models
use a matrix with statistically equivalent random elements. The
singular value decomposition that formed the start of our RMT
calculations can also be performed on the UTC matrix and all
observables can be calculated in exactly the same way. In effect,
the only difference between the two models is the distribution of
transmission eigenvalues. When we choose $\avg{\tau^2}$ to be equal
in both models, the only observable difference is the value of
$\bucket$.

RMT takes into account the correlations arising from the constraint
of energy conservation and gives the universal result of
$\bucket=2/3$. In the UTC model the matrix elements are
uncorrelated. For such a matrix the eigenvalues are expected to have
an exponential distribution.\cite{Kogan2000} In that case, $\bucket
= 2\avg{\tau^2}$. The power in the optimized channel increases to
$\avg{\tau^2}$, whereas the background transmission remains constant
at $\avg{\tau^2}$. The optimized transmission is twice the original
transmission and, therefore, it is not a universal constant in this
model.

Apparently, $\bucket$ is an accessible observable that depends on
the symmetry of the disordered system, which makes it very
interesting to determine it experimentally.

\subsection{Random matrix theory for thin samples}
We calculated $\bucket=2/3$ using the RMT eigenvalue distribution of
Eq.~\eqref{eq:dorokhov}. This equation is only valid in the limit of
a `thick' sample. To investigate when a sample can be considered
thick, we use the equation for the exact probability distribution of
transmission eigenvalues as calculated by
Beenakker\cite{Beenakker1997}. The exact equation is more involved
than its limit in the diffusion regime. Beenakker defines
$x\equiv{\textrm{arccosh}(1/\tau)}$ and then derives
\begin{equation}
P(x, s) = \frac{2}{\pi} \Im U(x-i0^+, s),
\end{equation}
with
\begin{equation}
s = \frac{4L}{3\ell_{tr}},
\end{equation}
and $U$ is the solution to the implicit equation
\begin{equation}
U(\zeta, s) = U_0(\zeta - s U(\zeta, s)),
\end{equation}

\begin{figure}
\centering
  \includegraphics[width=\smallimage]{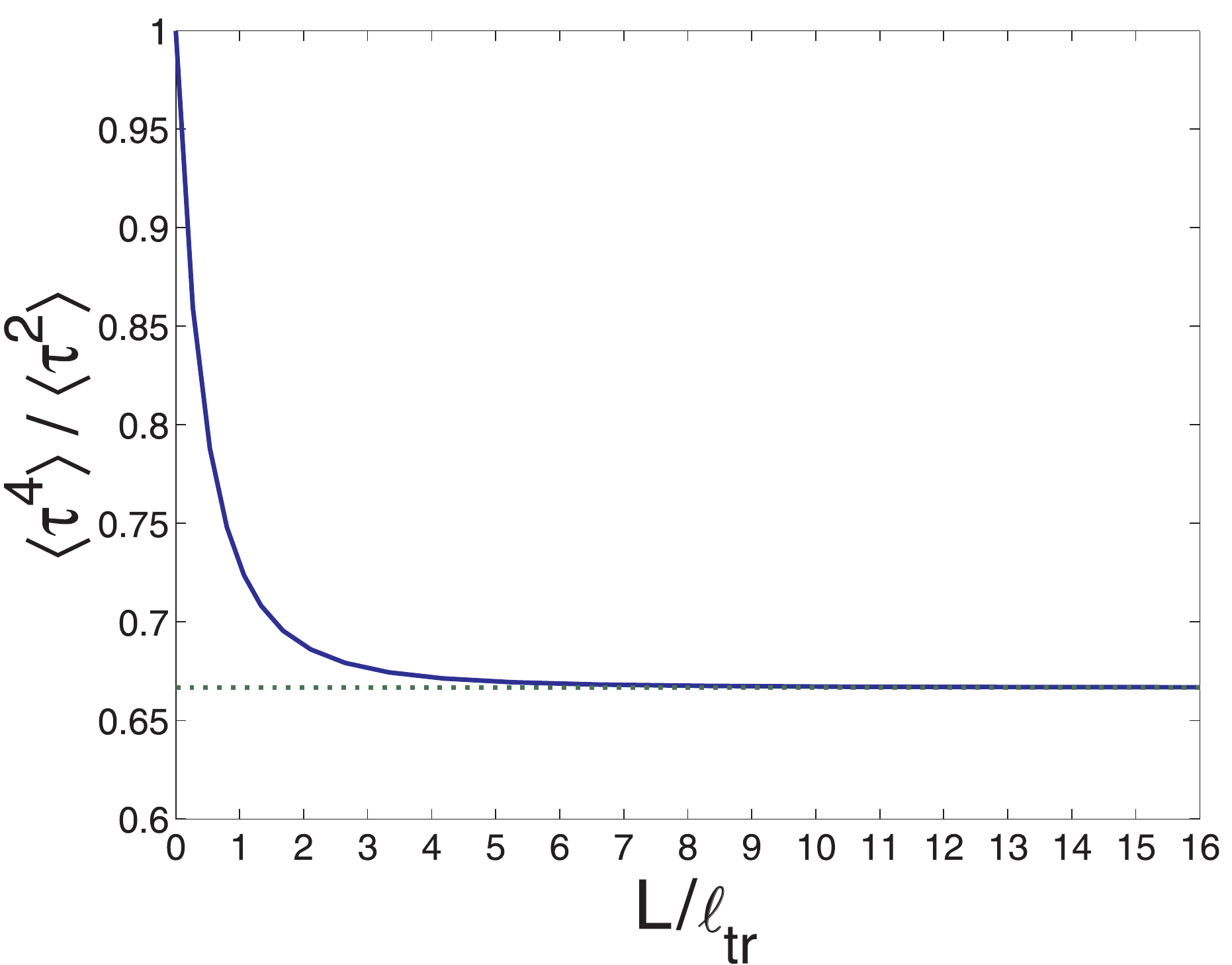}\\
  \caption{Ratio of the fourth moment to the second moment of the transmission eigenvalues ($\bucket$)
  as a function of sample thickness. With increasing thickness, $\bucket$
  rapidly converges to $2/3$.\cite{Beenakker1997}}\label{fig:tau-moment-ratio}
\end{figure}

\noindent with starting condition $U_0(\zeta) = \textrm{coth} (\zeta)$. We used these equations to
numerically evaluate $\bucket$ and we found that $\bucket$ converges to $2/3$ with increasing sample
thickness (Fig.~\ref{fig:tau-moment-ratio}). The convergence is very rapid: starting at $\bucket=1$ for an
infinitely thin sample, $\bucket$ is already at $0.7$ for samples that are only one mean free path thick.
Therefore, we conclude that samples in our experimental range can be considered thick and
Eq.~\eqref{eq:dorokhov} can be used.

\subsection{Non-diffusive behavior}
The distribution of eigenvalues depends on the transport physics in
the disordered medium. In the previous section, we saw that, for
very thin samples, the distribution depends on the thickness of the
sample.

In microwave experiments, it was observed that absorption in the
medium changes the eigenvalue distribution to an exponential
distribution.\cite{Mendez-Sanchez2003} This result can be understood
in the sense that there is no constraint of energy conservation in
such sys\-tems.\cite{Kogan2000} Therefore, the elements of the
transmission matrix are uncorrelated, just as in the UTC model.

In the limit of extremely strong scattering ($k\ell \approx 1$) interference effects dramatically change
the transport properties of waves.\cite{Anderson1953, Ioffe1960} This effect is known as Anderson
localization. The observation of Anderson localization of light\cite{Wiersma1997, Schuurmans1999,
Storzer2006} has proven very difficult, especially because absorption can, in many situations, have a
similar signature\cite{Scheffold1999}.

Both absorption and localization make that the extended modes
vanish, thereby decreasing $\bucket$. However, in quasi-1D systems
the effect of localization on the distribution of eigenvalues is
rather spectacular. At the localization threshold, eigenvalues start
 to cluster. This effect is called eigenvalue crystallization and
leads to an oscillating eigenvalue distribution (see e.g.
\cite{Beenakker1997}). The study of localization in open
3-dimensional systems is an ongoing field of research. Recent
results\cite{Skipetrov2006} predict the distribution of decay rates
of optical modes. It would be very interesting to know the
distribution of transmission eigenmodes in such a 3-dimensional
system and investigate if, by measuring $\bucket$, the difference
between localization and absorption can be established
experimentally.

\section{Random matrix theory in an optical experimental
situation}\label{sec:realistic-conditions} In this section, we extend random matrix theory of wave
scattering to describe a realistic optical experimental situation. First, the geometrical correlations
that are present in a slab geometry are analyzed and then the effect of limitations of the modulator is
taken into account. We find that a realistic experimental situation can be described with the degree of
amplitude control $\gamma$ that accounts for all deviations from the `ideal' waveguide geometry. The value
of the observables in a realistic situation are summarized in Table~\ref{tab:I-results-gamma}.

\begin{table}
  \centering
\begin{tabular}{|l|r@{}l|r@{}l|}
  \hline
               & \multicolumn{2}{c|}{passive}
               & \multicolumn{2}{c|}{active}   \\
  \hline
single channel &$ \avg{I_\bopt} $&$=\avg{\tau^2}\frac{K}{\Meff N}$
               &$ \avg{\optim{I}_\bopt}    $&$=\gamavg\avg{\tau^2}\frac{K}{M}$\\
total          &$ \avg{\Ttot}   $&$=\avg{\tau^2}\frac{K}{N}$
           &$\avg{\tTtot}$&$=\gamavg\frac{2}{3}+\left(1-\gamavg\right)\avg{\tau^2}\frac{K}{N}$ \\
  \hline
\end{tabular}
\caption{Value of some important observables in a realistic
experimental situation (also compare
Table.~\eqref{tab:I-results}).}\label{tab:I-results-gamma}
\end{table}

\subsection{Slab geometry}\label{sec:position}
\begin{figure}
\centering
  \includegraphics[width=\tinyimage]{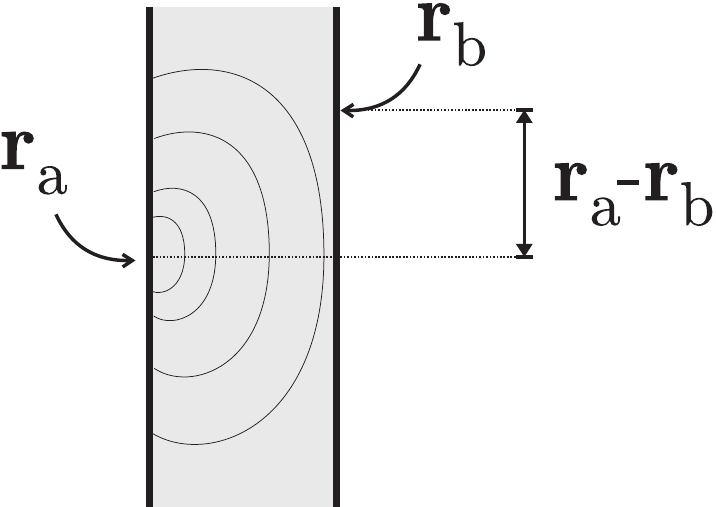}\\
  \caption{Slab geometry. Light diffuses from a point with surface coordinates $\ra$ on the front of the
  sample to a point $\rb$ at the back of the sample.}\label{fig:slab}
\end{figure}
\noindent In our experiments we have a slab geometry as is depicted in Fig~\ref{fig:slab}. In this
geometry, not all channels are equivalent. When light is focused on incident channel $a$, transmitted
channels directly opposite $a$ will receive more intensity than channels that are further away. To include
this geometric effect, we recognize that the average transmission coefficient is a function of the
relative position of the channels.
\begin{equation}
\avg{|t_{ba}|^2} =
f(|\rb-\ra|)\avg{\tau^2}\frac{K}{MN}\label{eq:position},
\end{equation}
where all channels are defined in $r$ space. $\ra$ and $\rb$ are the 2-D coordinates of, respectively, the
incident and the transmitted channels at the front and back sample surfaces. In a waveguide geometry
$f(|\trb-\ra|)=1$. In a slab geometry $f$ can be calculated using diffusion theory. It is a function that
starts at $1$ and decreases with distance $|\trb-\ra|$ with a typical decay length that is comparable to
the thickness of the sample. Furthermore, $f$ depends on the mean free path and the extrapolation lengths.
Since $f$ decreases with distance, channels that are further away from the target contribute less. The
effective total number of channels that contribute to the intensity in $\bopt$ is given by
\begin{equation}
\Neff = \sum^N_a f(|\trb-\ra|).
\end{equation}
$\Neff$ converges to a finite value in the limit
$K,M,N\rightarrow\infty$. In this limit, the waveguide model
transforms into a model for an infinite slab and all physical
quantities become independent of $N$. Similarly, the effective
number of transmitted channels at the back of the sample originating
from a point source at $\ralpha$ equals
\begin{equation}
\Meff = \sum^M_b f(|\rb-\ralpha|) = \frac{M}{N}\Neff.
\end{equation}
To introduce spatial dependency in the RMT formalism, we use the
singular value decomposition (Eq.~\eqref{eq:svd}) to find
\begin{equation}
\avg{|t_{ba}|^2} = \avg{\left|\sum^K_k U_{bk}\tau_{kk}
V_{ka}\right|^2}.
\end{equation}
Assuming that $\tau$ is still independent of $U$ and $V$, all
geometrical effects are accounted for by imposing the following
correlation between on $U$ and $V$
\begin{equation}
\avg{|U_{bk} V_{ka}|^2} =
f(|\rb-\ra|)\frac{1}{MN}\label{eq:position-svd}.
\end{equation}
In the previous sections, we (implicitly) used the assumption that
the matrices $U$ and $V$ are uncorrelated only in the calculation of
$\avg{I_b}$; all other observables in Table~\ref{tab:I-results} are
not affected by this correlation. In the active measurements, the
incident field is matched to the sample to maximize $I_b$. In a
passive measurement, however, the incident illumination can have any
arbitrary profile that only depends on the experimental
configuration. The average value for $I_b$ depends on this intensity
profile $I_a$
\begin{equation}
\avg{I_b} = \sum_a^N f(|\rb-\ra|)\avg{\tau^2}\frac{K}{MN} I_a\label{eq:Ib-position}.
\end{equation}
When the detector is further away from the sample than the diameter of the diffuse spot, it is convenient
to define far-field angular channels (far-field speckles). The number of such channels is $\Meff$ (also
see Ref.~\citealt{Boer1995} for a discussion on the number of independent channels). In that case,
Eq.~\eqref{eq:Ib-position} reduces to
\begin{equation}
\avg{I_b} = \avg{\tau^2}\frac{K}{\Meff N},
\end{equation}
where the total incident power was assumed to be normalized. All
other observables that were derived above are not affected by the
slab geometry. It is now possible to supplement these results with
spatial information. We can, for instance, calculate the intensity
profile of the optimal incident field (Eq.~\eqref{eq:E_a-opt})
\begin{equation}
\avg{|\optim{E}_a|^2} = \avg{A_0^2} \avg{|t_{\bopt a}|^2} =
\avg{A_0^2}
f(|\trb-\ra|)\avg{\tau^2}\frac{K}{MN}\label{eq:optimal-field-profile}.
\end{equation}
Below, we will use this intensity profile to determine how well a
generated field match\-es the optimal field.

\subsection{Wavefront modulation imperfections}\label{sec:imperfect}
\noindent So far, we have assumed that the incident wavefront is constructed with an infinitely large
perfect wavefront generator. All practical imperfections, such as a finite size modulator or phase only
modulation, are accounted for by a single parameter $\gamma$. We start by decomposing the incident field
into a contribution to the perfect incident field and an orthogonal part $\Delta E_a$
\begin{equation}
E_a = \gamma \optim{E}_a + \sqrt{1-|\gamma|^2} \Delta
E_a\label{eq:gamma-decomp},
\end{equation}
where $\gamma$ is the inner product of the optimal field and the
generated field
\begin{equation}
\gamma \equiv \sum_a^N \optim{E}_a^* E_a\label{eq:gamma-definition}.
\end{equation}
With an ideal modulator, the optimal field can be generated perfectly and $\gamma=1$. All experimental
limitations decrease $\gamma$. In the next section, we give some examples how to calculate $\gamma$ in
realistic situations. Now, we first calculate how $\gamma$ effects the observables $\optim{I}_\bopt$ and
$\tTtot$. Since $\Delta E_a$ is orthogonal to the optimal field, it does not contribute to the intensity
in the target focus. After optimization, the intensity in the focus is given by (compare
Eq.~\eqref{eq:Ib-opt-avg})
\begin{equation}
\avg{\optim{I}_\bopt} =
\gamavg\avg{\tau^2}\frac{K}{M}\label{eq:Ib-opt-realistic}.
\end{equation}
To calculate the total transmission, we use a similar approach
\begin{equation}
\avg{\tTtot} = \gamavg \bucket + \left(1-\gamavg\right)
\left(\avg{\tau^2}\frac{K}{N}-\epsilon\right)\label{eq:Ttot-realistic},
\end{equation}
where $\avg{\tau^2}K/N$ is the transmission coefficient of the
unoptimized field and $\epsilon$ is a small correction term takes
into account the fact that the total transmission of the orthogonal
field is slightly lower than the total transmission of a random
field since it does not include the contribution of transmitted
channel $\bopt$. Also, at the back of the sample the transmitted
fields corresponding to $\Delta E$ and $\optim{E}_a$ are not
completely orthogonal. Both effects are of order $1/\Neff$ and
partially cancel. Therefore it is safe to neglect $\epsilon$.

When the incident field is not shaped to fit the scattering matrix
of the sample, there will still be some accidental overlap with the
optimal field. For such an incident field, $\gamavg$ is of order
$1/\Neff$ and will be neglected.

\subsection{Examples of realistic experimental situations}
We now calculate $\gamavg$ for various experimental conditions. Eqs.~\eqref{eq:Ib-opt-realistic} and
\eqref{eq:Ttot-realistic} can then be used to obtain the intensity in the focus, as well as the total
transmission after optimization.

\subsubsection{Example 1: Finite size modulator}
Suppose the light modulator illuminates a finite area that contains
$N_s$ independent scattering channels. The modulator perfectly
modulates amplitude and phase of all these scattering channels. Then
\begin{equation}
\gamma = \sum_{a}^{N_s} \optim{E}_a^* E_a = A_0 A_f \sum_{a}^{N_s}
|t_{\bopt a}|^2\label{eq:gamma-finite-support},
\end{equation}
where $A_0$ normalizes the optimal field (see
Eq.~\eqref{eq:A_0-def}) and $A_f$ normalizes the actually generated
field. Using Eq.~\eqref{eq:position} to introduce position
dependence, we find
\begin{equation}
\gamavg = \frac{\sum_{a}^{N_s} f(|\trb-\ra|) }{\sum_{a}^{N}
f(|\trb-\ra|)} = \frac{\sum_{a}^{N_s} f(|\trb-\ra|)
}{\Neff}\label{eq:gamma2-size}.
\end{equation}
In the limit where the light modulator is very large $\gamma\rightarrow 1$. In the limit where the
modulator is much smaller than the diffuse spot described by $f$, the degree of control is
$\gamavg=N_s/\Neff$.

\subsubsection{Example 2: Limited resolution} We first calculate
$\gamavg$ for the situation where the light modulator has a limited
spatial resolution. We assume that each of the $N_s$ segments of the
phase modulator illuminates a spatial cluster of $N/N_s$ scattering
channels. We group the channels by segment and use
Eq.~\eqref{eq:gamma-definition} to find
\begin{equation}
\gamma = \sum_s^{N_s} \sum_{a\in \{a_s\}} \optim{E}_a^*
E_s^{\phantom{*}}\label{eq:gamma-segments},
\end{equation}
where $\{a_s\}$ is the collection of channels that is illuminated by
segment $s$. $E_s$ is the field that is generated by segment $s$ of
the modulator. The incident field is optimized for a whole cluster
at a time. After an optimization we have
\begin{align}
E_s &= A_s \sum_{a\in \{a_s\}} \optim{E}_{a}\label{eq:E-segments} \\
A_s &\equiv \left(\frac{N}{N_s}\sum_s^{N_s} \left|\sum_{a\in
\{a_s\}} \optim{E}_a\right|^2\right)^{-1/2}
\end{align}
where the prefactor $A_s$ normalizes the total power of the incident
field. Substituting Eq.~\eqref{eq:E-segments} in
Eq.~\eqref{eq:gamma-segments} we find
\begin{equation}
\gamma = A_s \sum_s^{N_s} \left|\sum_{a\in \{a_s\}}
\optim{E}_a\right|^2,
\end{equation}
\begin{equation}
|\gamma|^2 = \frac{N_s}{N}\sum_s^{N_s} \left|\sum_{a\in \{a_s\}}
\optim{E}_a\right|^2\\.
\end{equation}
Because elements of $\optim{E}_a$ are independent and $\optim{E}_a$
is normalized, we have
\begin{equation}
\avg{\optim{E}_{a}\optim{E}_{a'}}=\delta_{a'a}\frac{1}{N},
\end{equation}
which leads to
\begin{equation}
\gamavg = \frac{N_s}{N}.
\end{equation}

\subsubsection{Example 3: Phase only modulation}
We consider the case of a phase only modulator that perfectly controls the phase of all $N$ incident
waveguide channels. The amplitude of the channels is fixed at $1/\sqrt{N}$. For this experimental
configuration, we find

\begin{equation}
\gamma = \sum_{a}^N \optim{E}_a^* E_a^{\phantom{*}} =
\frac{1}{\sqrt{N}}\sum_{a}^N |\optim{E}_a|,
\end{equation}
\begin{equation}
\gamma^2 = \frac{1}{N}\sum_{a, a'\neq a}^N
|\optim{E}_a||\optim{E}_{a'}| +
\frac{1}{N}\sum_{a''}|\optim{E}_{a''}|^2.
\end{equation}
To average, we use the complex Gaussian probability function of
$\optim{E}_a$ to arrive at
\begin{equation}
\avg{\gamma^2} = \frac{\pi}{4} + \frac{1}{N}\left(1-\frac{\pi}{4}\right).
\end{equation}

\subsubsection{Example 4: Realistic experimental conditions}
In this example, we calculate the degree of control under realistic
experimental conditions. We analyze the geometry that was used for
measuring the data in Fig.~\ref{fig:enhancement} where the
enhancement as a function of the number of modulator segments is
plotted. In this geometry, no lenses were used behind the sample;
the camera was approximately $\cm{30}$ away from the sample surface.
The incident light is focused on the sample surface and illuminates
a spot that is much smaller than a millimeter. Analysis of the
measured $t_{ba}$ coefficients showed that, on average, all segments
contribute approximately equally to the amplitude of the target
(variations are within 15\%). Furthermore, limitations of the phase
modulator made that the generated field had a bias. On average, 14\%
of the incident intensity had a constant phase and could not be
shaped.

Since all segments contribute equally, we expect the enhancement to
be given by a combination of Example 2 and Example 3. Also taking
into account that 14\% of the incident light could not be shaped, we
find
\begin{equation}
\gamavg = 0.86 \frac{N_s-1}{\Neff}\frac{\pi}{4} +
\frac{1}{\Neff}\label{eq:gamma-experimental-geometry},
\end{equation}
which is an upper limit for the case where the sample is absolutely
stable and there is no influence of measurement noise. We can now
calculate the expected maximum intensity enhancement $\eta$
\begin{equation}
\eta \equiv \frac{\avg{\optim{I}_\bopt}}{\avg{I_\bopt}}.
\end{equation}
Using Table~\ref{tab:I-results-gamma} with
Eq.~\eqref{eq:gamma-experimental-geometry} yields
\begin{align}
\eta &= \gamavg \Meff \frac{N}{M}\\
&= 0.68 (N_s-1) + 1.
\end{align}
This result is the same as was found with the UTC model in Chapter~\ref{cha:focusing-through}. When all
segments contribute equally, the UTC model gives the correct results even if the sample is not in a
waveguide geometry. The measured enhancement ($\eta \approx 0.57 N_s$, see Fig.~\ref{fig:enhancement}) was
slightly lower than the theoretical value because of measurement noise and due to the fact that segments
in the center contribute slightly more than segments close to the edge of the phase modulator. This minute
effect could be included in $f$ in Equation~\eqref{eq:gamma2-size}, to calculate a value for $\gamavg$
that is a few percent lower. In the experiments described in Chapter~\ref{cha:focusing-through} the
background intensity did not increase noticeably. This result is explained by the fact that $N_s\ll \Neff$
and, therefore, $\gamavg \ll 1$. In this regime, the total transmission is hardly affected by an
optimization (see Eq.~\eqref{eq:Ttot-realistic}).

\section{Conclusion}
We developed a model for describing the transport of a waves that
are actively shaped to focus through a disordered sample. The new
model is based on RMT and, therefore, correctly takes into account
correlations between scattering channels.

The concept of `active' measurements was introduced. These
measurements are based on wavefront shaping. One of the most
interesting observables in an active measurement is the total
transmission of an optimized wavefront. The fraction of transmitted
power for such a wavefront turns out to be $2/3$, which is
universally true regardless of the thickness of the sample. Physical
processes such as absorption and Anderson localization change this
value, which makes measuring it extremely interesting.

All results were first calculated in an ideal waveguide geometry and
then translated to a realistic experimental slab geometry. This
two-step calculation process proved to be a very powerful and
versatile technique.

Apart from the observables that were calculated in this chapter, there is a multitude of interesting
quantities that can be measured with our active measurement strategy. Since most of the theoretical work
has been done on ensemble averages rather than on the subset of extended modes, many questions are still
unanswered.

\bibliography{../../bibliography}
\bibliographystyle{Ivo_sty}

\setcounter{chapter}{8} 
\chapter{Observation of open transport channels in
disordered optical systems\label{cha:dorokhov-experiment}}

When a beam of light impinges on a strongly scattering object, such as a layer of white paint, tiny
particles randomly scatter light in all directions. After a few scattering events, light is completely
diffuse. The transport of multiply scattered light is, in general, described very well by a diffusion
equation. However, the diffusion equation does not take into account interference effects. The inclusion
of interference in the description of disordered wave propagation has led to the discovery of fundamental
effects like enhanced backscattering\cite{Albada1985, Wolf1985}, universal conductance
fluctuations\cite{Boer1994, Scheffold1998} and Anderson localization of light\cite{Wiersma1997,
Storzer2006, Schwartz2007}.

Here we demonstrate experimentally a new interference effect that was first predicted by
Dorokhov\cite{Dorokhov1984}. Like the interference effects mentioned above, this effect was first derived
for the propagation of electron wave functions in a disordered waveguide (a metal wire) and then
translated to optics.\cite{Rossum1995, Beenakker1997} A disordered waveguide can be described as a
collection of orthogonal transport channels that each carry a fraction of the incident wave. Using random
matrix theory (RMT), Dorokhov showed that due to interference these channels are either open (transmission
close to 1) or closed (transmission close to 0).

The existence of open transport channels means that it is, in
principle, possible to construct a wave that is fully transmitted
through a normally opaque mesoscopic object, regardless of the
thickness of that object. In a mesoscopic electronic conductor such
control over the incident wave is not possible. For light, however,
we can shape the wavefront of the incident wave with a high
accuracy. In this chapter we show that the shaped wave selectively
couples to open transport channels, causing the transmission through
an opaque, strongly scattering sample to increase dramatically.


In Section~\ref{sec:dorokhov-theory-summary} we give the theoretical
relation for the total transmission of a shaped wavefront. Then, in
Section~\ref{sec:dorokhov-experiment}, the wavefront optimization
setup is described. The experimental results are presented in
Section~\ref{sec:dorokhov-results}, followed by a more detailed
description of the data analysis procedure and an estimation of the
experimental error margins in Section~\ref{sec:dorokhov-details}.

\section{Expected total transmission\label{sec:dorokhov-theory-summary}}
\noindent The average transport of light is described accurately by
a diffusion equation. On average, the total transmission coefficient
is given by $(\ell+z_{e1})/(L+z_{e1}+z_{e2})$, with $\ell$ the mean
free path for light, $L$ the thickness of the scattering sample and
$z_{e1}$ and $z_{e2}$ extrapolation lengths that account for
reflection at the sample interface\cite{Lagendijk1989, Zhu1991}.

To increase the total transmission above the value predicted by diffusion theory, we use a feedback
algorithm that shapes the incident wavefront. The algorithm maximizes the intensity in a target focus
behind the sample. Since only open transport channels contribute to the intensity in the target, the
ideally shaped wavefront selectively couples to these open channels. In the extreme case where all open
channels have a transmission coefficient of exactly $1$ (the so called maximum fluctuations
theorem\cite{Pendry1990, Pendry1992}), light with an optimal wavefront is completely transmitted.

Random matrix theory\cite{Dorokhov1984} predicts a distribution that
is close to the extreme situation of maximal fluctuations (see
Eq.~\eqref{eq:dorokhov}). We express the amount of fluctuations in
the distribution as $\bucket$, which is defined in
Eq.~\eqref{eq:Ttot-opt} and denotes the transmission coefficient of
an ideally shaped wavefront. For Dorokhov's distribution $\bucket =
2/3$. This universal value is independent of the thickness or the
mean free path of the sample.

To couple to open channels, a high degree of control over the incident wavefront is essential. When only a
small subset of the scattering channels is controlled, the transmission matrix elements are effectively
uncorrelated and no open channels can be found. Therefore, the experiment is designed to approximate the
ideal wavefront as closely as possible. Still, in any experiment there will be a difference between the
optimal wavefront and the field that is actually generated. This is especially true in an open system
where the optimal wavefront has infinite support. The total transmission of an incompletely optimized
wavefront equals (see Eq.~\eqref{eq:Ttot-realistic})
\begin{equation}
\tTtot \equiv \frac{\tPtot}{\Pin} =
|\gamma|^2\bucket+(1-|\gamma|^2)\Ttot\label{eq:Ttot-opt-experiment},
\end{equation}
where $\tPtot$ is the totally transmitted power of the incompletely
optimized wavefront, $\Pin$ is the incident power and
$\Ttot\equiv\Ptot/\Pin$, with $\Ptot$ the totally transmitted power
of an non-optimized wavefront. $|\gamma|^2$ is the degree of control
as defined in Eq.~\eqref{eq:gamma-definition}. In the case that we
have full control over the incident wavefront $|\gamma|^2\rightarrow
1$ and $\tTtot\rightarrow \bucket$.

The degree of control can be determined experimentally by measuring
the power in the target focus $\optim{P}_\bopt$ after optimization.
From Eq.~\eqref{eq:Ib-opt-realistic} we can make $|\gamma|^2$
explicit
\begin{equation}
|\gamma|^2 = \frac{\optim{P}_\bopt}{\Ptot}
\frac{n_2^2}{n_1^2}\label{eq:gamma-experiment},
\end{equation}
where $n_1=1$ and $n_2=1.52$ are the refractive indices of the medium in front of the sample (air) and
behind the sample (glass substrate) respectively. Since $|\gamma|^2$ can be determined experimentally,
Eq.~\eqref{eq:Ttot-opt-experiment} can be solved to find $\bucket$.

\section{Experiment\label{sec:dorokhov-experiment}}
\begin{figure}
\centering
  \includegraphics[width=\wideimage]{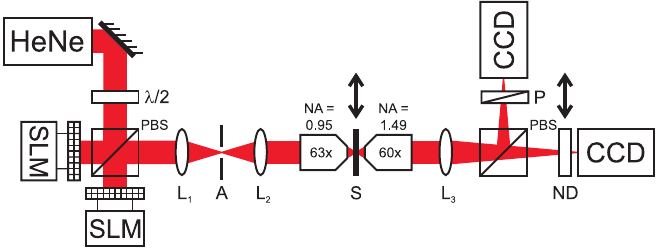}\\
  \caption{Experimental setup for finding open transport channels.
  HeNe, expanded $\nm{632.8}$ HeNe laser; $\lambda/2$, half waveplate; PBS,
  polarizing beam splitter cube; SLM, spatial light modulator. A, iris diaphragm; 63x, microscope
  objective; 60x, oil immersion microscope objective; S, sample; P, polarizer; ND, neutral density filter;
  L$_1$, L$_2$, L$_3$, lenses with focal length of respectively $\mm{250}$, $\mm{200}$ and $\mm{600}$.
  ND and S are translated by computer controlled stages.
}\label{fig:setup-dorokhov}
\end{figure}
\noindent The experimental apparatus is designed to have optimal control over the incident wavefront. The
setup is shown in Fig.~\ref{fig:setup-dorokhov}. An expanded beam from a $\nm{632.8}$ HeNe laser is
rotated to a $45^\circ$ linear polarization by a half waveplate and impinges on a polarizing beam splitter
cube. Both horizontal and vertical polarization are modulated using the macropixel modulation method that
was described in Section~\ref{sec:amplitude-phase}. The synthesized wavefront is focused onto the sample
with a high numerical aperture ($\NA$) objective\footnote{Zeiss 440068-0000-000 Achroplan 63x/$\NA=0.95$
without cover glass correction.} so that almost all angular channels are controlled. A high $\NA$
oil-immersion objective\footnote{Nikon MRD01691 CFI Plan Apochromat TIRF 60x/$\NA=1.49$} collects almost
all transmitted light. The immersion oil matches the refractive index of the substrate $n=1.52$ and,
therefore, eliminates reflections at the back surface of the substrate. The transmitted light is split
into horizontal and vertical polarizations by a beam splitter cube. A second polarizer improves the cube's
limited extinction ratio for reflected light. A camera measures the power of the horizontally polarized
light in an area with the size of a single speckle to provide feedback for the optimization algorithm. We
use the stepwise sequential algorithm (see Section~\ref{sec:alg-stepwise-sequential}) to optimize the
wavefront. A calibrated neutral density filter with a transmission of $1.4\cdot10^{-3}$ is automatically
placed in front of the camera to be able to measure the high intensity in the target focus after
optimization. The magnification of the detection system is $225\times$. At this magnification, individual
speckles are well separated on the camera. A second camera measures the transmission in the vertical
polarization channel. This camera is not used during the optimization procedure itself.

The sample is a layer of spray-painted ZnO particles (see
Section~\ref{sec:samples}) in an air matrix. The particles have an
average diameter of $\nm{200}$, which makes them strongly scattering
for visible light. The mean free path was determined by measuring
the total transmission and equals $\mum{0.85\pm0.15}$ at a
wavelength of $\nm{632.8}$. Two samples were used: one with a
thickness of $\mum{5.7\pm0.5}$, and one that is approximately twice
as thick ($\mum{11.3\pm0.5}$). The sample was mounted on a motorized
translation stage that translates the sample perpendicular to the
axis of the objectives between experiments. This way, a different
random configuration of scatterers could be used for each
experiment. The translation stage also provides a reliable way of
making reference measurements (see
Section~\ref{sec:dorokhov-results}).

\section{Results\label{sec:dorokhov-results}}
\begin{figure}
\centering
\includetwographics{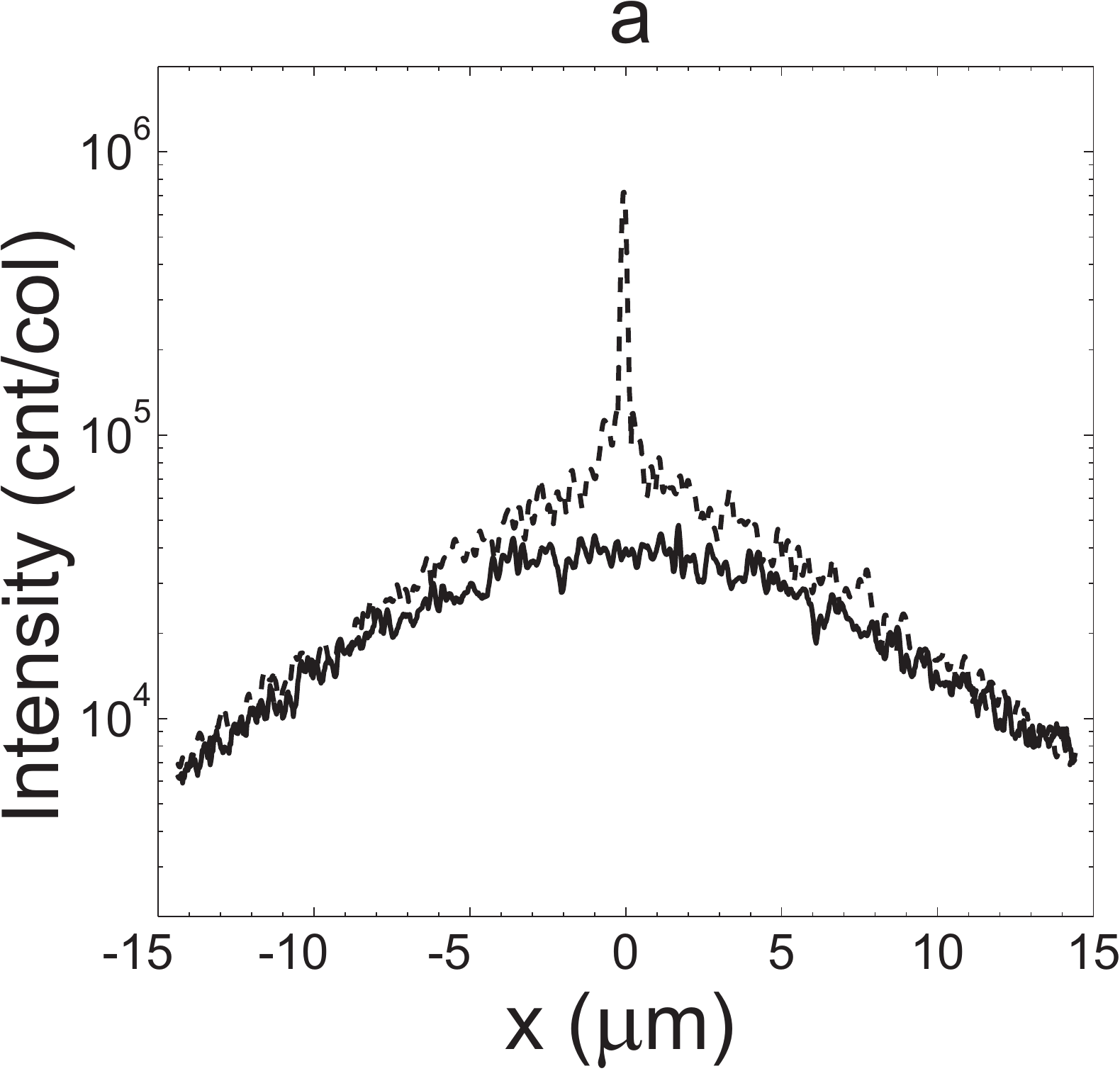}{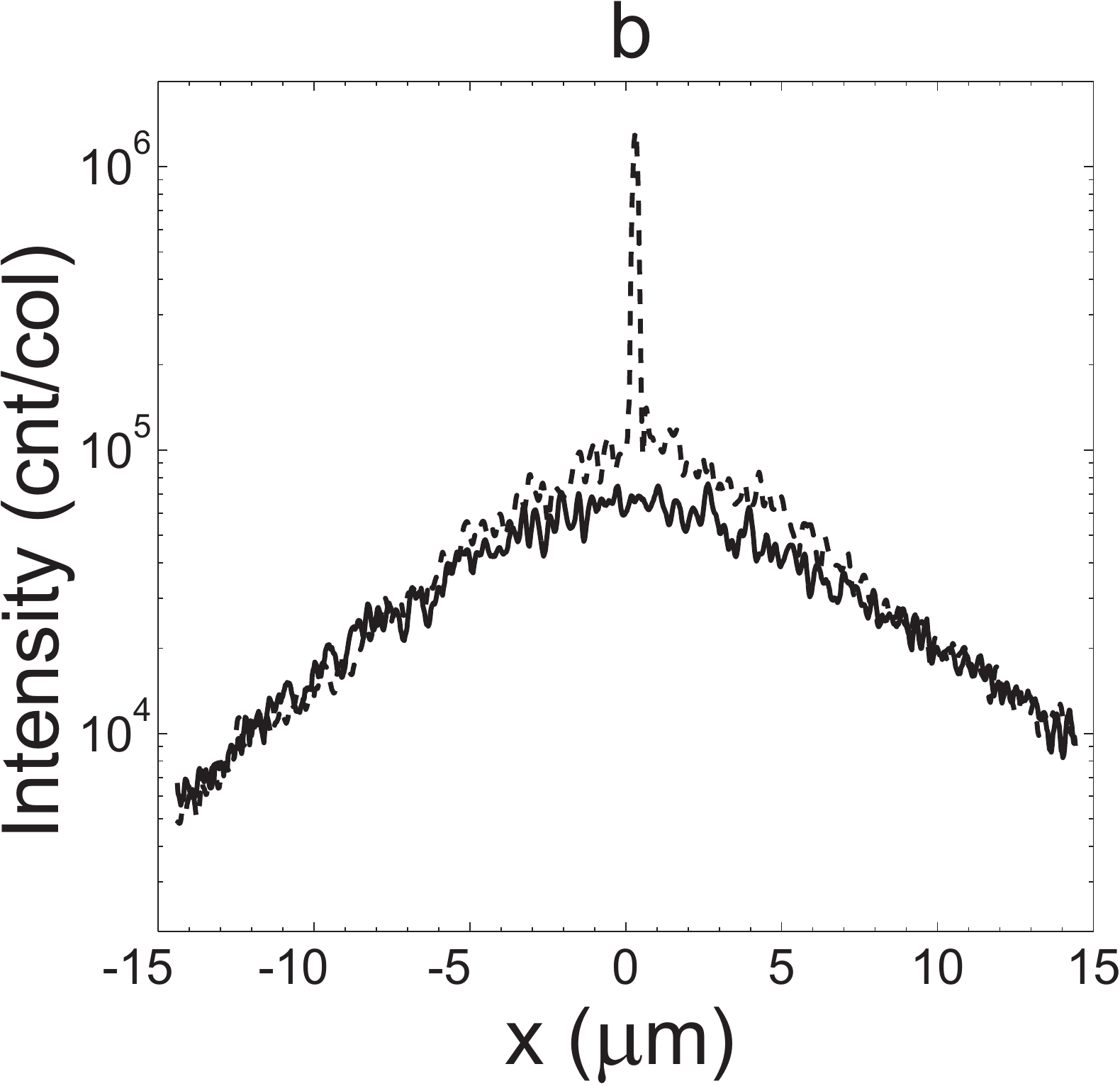}
  \caption{Logarithmic plot of the intensity transmission
  in the horizontal polarization channel.
  The intensity is integrated per column.
  \subfig{a} $\mum{11.3}$-thick layer of ZnO pigment.
  Solid curve, transmission of non-optimized wavefront;
  Dashed curve, transmission of optimized wavefront. Measurements with
  and without neutral density filter were combined to cover
  the full dynamic range.
  \subfig{b} Same for a $\mum{5.7}$-thick layer of ZnO pigment.
  }\label{fig:results-before-after-dorokhov}
\end{figure}

Optimizing the incident wavefront caused the intensity in the target
to increase dramatically. The increase in the target intensity was a
factor of $746\pm28$ for the $\mum{11.3}$-thick sample and a factor
of $671\pm14$ for the $\mum{5.7}$-thick sample in the particular
cases that are shown in
Fig.~\ref{fig:results-before-after-dorokhov}. In addition, the
intensity also increased outside the target area, even though the
algorithm only used the intensity in the target as feedback. This is
an important observation since it indicates that the transmission
matrix is correlated. By optimizing transmission to a single target,
the incident intensity is redistributed from closed channels to open
transport channels. If the elements of the transmission matrix had
been uncorrelated, the background intensity would not have changed
(also see Section~\ref{sec:compare-UTC}).

The increase in the background intensity is limited to an area of a
few micrometers around the target. The diameter of this area seems
to be roughly proportional to the thickness of the sample. This
makes sense as, in a slab geometry, propagation paths that are
further apart than the thickness of the sample are essentially
uncorrelated. The intensity increase of the background is maximal
just around the optimized spot. There, the intensity increased by a
factor of $4.1\pm0.7$ and $2.3\pm0.3$ for the thick and the thin
sample respectively\footnote{In
Fig.~\ref{fig:results-before-after-dorokhov} the intensity increase
appears to be lower since the figure shows the intensity integrated
over all columns.}. The total transmission in the background (not
including the optimized spot) increased a factor of $1.32\pm0.05$
and $1.26\pm0.08$, respectively.

\begin{figure}
\centering
  \includegraphics[width=\smallimage]{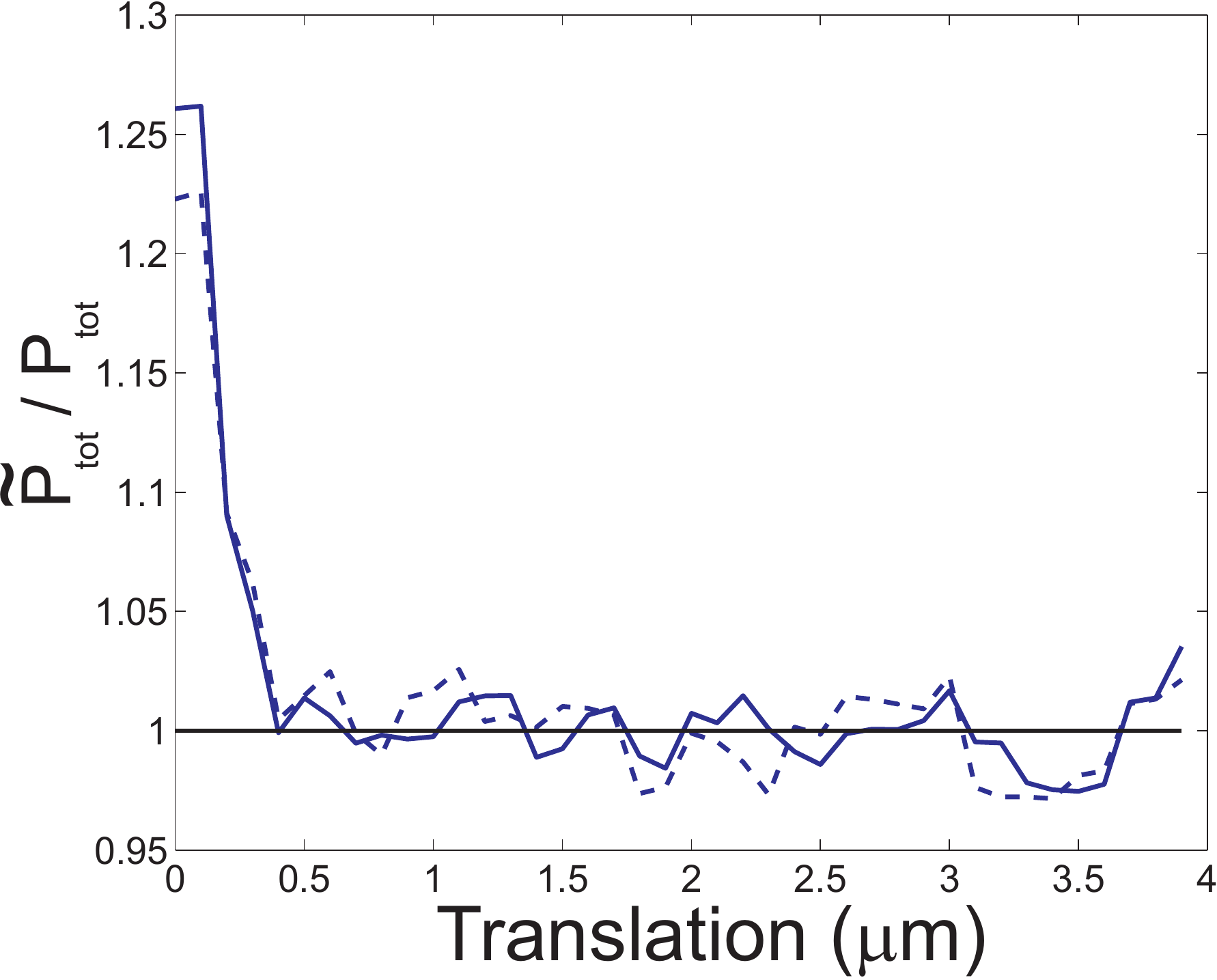}\\
  \caption{Typical total transmission after optimization
  as a function of sample displacement.
  Solid curve, horizontal polarization.
  Dashed curve, vertical polarization.
  Curves are normalized to the reference transmission (average intensity between
  $\mum{0.6}$ and $\mum{3.9}$ displacement).
  Fluctuations of the intensity are the result of speckle.
  }\label{fig:results-sample-translation}
\end{figure}

The algorithm optimizes the transmission through the whole optical
system that is between the light modulator and the camera. To make
sure that the extra intensity is indeed the result of an increased
transmissivity of the sample, we need to exclude the trivial effect
of an increased transmission trough e.g. the microscope objective.
Therefore, we used exactly the same wavefront for measuring the
non-optimized reference transmission as we used to measure the
optimized transmission. The only difference between the two
measurements is a slight translation of the sample. At the new
position, the scattering behavior of the sample is different and the
optimized wavefront has no special significance anymore.

The effect of translating the $\mum{11.3}$-thick sample is shown in
Fig.~\ref{fig:results-sample-translation}. Initially, the total transmission in both of the polarization
channels has increased significantly. The intensity on one of the cameras is slightly higher since it
includes the contribution of the optimized spot. Since the pixels in the target area saturate, in this
plot the difference is relatively small. After translating the sample over $\nm{600}$, the effect of
optimizing the wavefront is completely lost. The transmission is averaged over $34$ images taken at
displacements increasing from $\mum{0.6}$ to $\mum{3.9}$ to obtain the reference transmitted power for the
non-optimized situation ($\Ptot$). We observe a slight variation in the total transmission when the sample
is translated. The expected variance of fluctuations in the total transmission is $1/(2g)$ (see e.g.
Ref.~\citealt{Boer1995}), where $g\equiv N \Ttot$, with $N$ the number of participating scattering
channels. From the size of the diffuse spot we estimate $N \approx 4.5\cdot 10^3$. With a measured
transmission of $\Ttot=0.24$ (see Section~\ref{sec:dorokhov-details}), we expect fluctuations with a
standard deviation of $2\%$, which is consistent with the data.

\begin{figure}
\centering
  \includegraphics[width=\mediumimage]{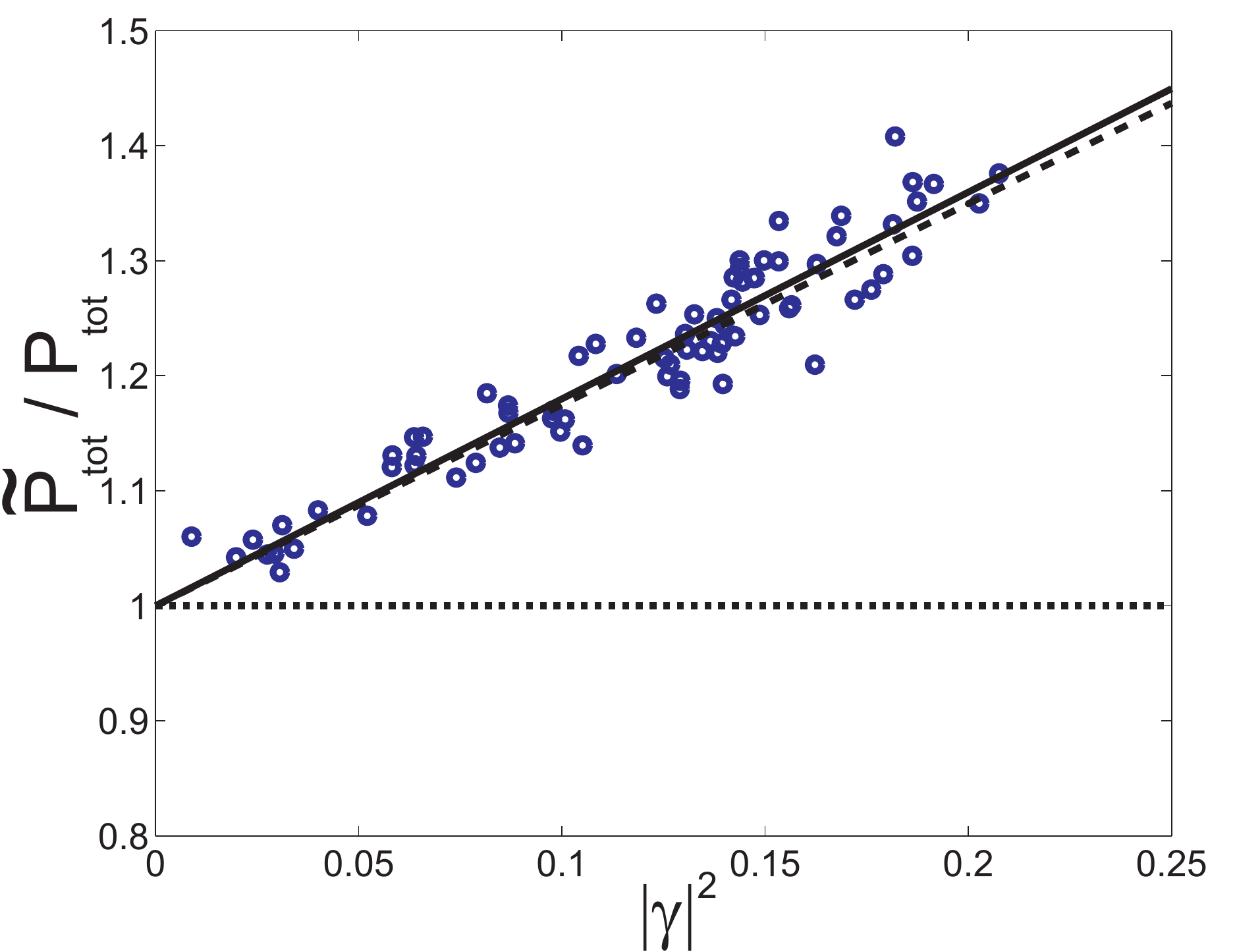}
  \caption{Normalized total transmission as a function of the degree of control
  $|\gamma|^2$ for a single sequence of measurements.
  Circles, measured transmission after constructing the incident wavefront;
  solid line, linear regression to the measurements;
  dotted line, diffusion theory;
  dashed line, random matrix theory, $\bucket=2/3$
  (there are no adjustable parameters).}\label{fig:results-dorokhov-relative-transmission}
\end{figure}

We performed automated sequences of optimizations. Between each measurement, the sample was translated to
find a new realization of disorder. For each of the measurements, we determined the degree of control
using Eq.~\eqref{eq:gamma-experiment}. In Fig.~\ref{fig:results-dorokhov-relative-transmission} we plotted
the increase in total transmission versus the degree of control for a $\mum{11.3}$-thick sample. The total
transmission increases linearly with $|\gamma|^2$. The maximum increase in transmission that was achieved
is 40\%. We conclude that, with increasing degree of control, a larger fraction of the incident intensity
couples to open channels, causing the total transmission to increase.

From a linear regression we find that the curve has a slope of 1.7.
With an original average transmission coefficient of $\Ttot=0.24$ we
expect a total transmission of $0.65$ when $|\gamma|^2=1$. This
result is very close to the expected value of $2/3$.

For this particular sequence, we reached a maximum control fraction
of $|\gamma|^2=0.21$. The algorithm used 3816 control segments and
phase only modulation. In this situation $|\gamma|^2$ can
theoretically not be higher than $(3816/N) (\pi/4) \approx 0.67$.
Even in a perfectly stable system, $|\gamma|^2$ will be lower due to
geometrical effects (see Section~\ref{sec:imperfect}), so
$|\gamma|^2=0.21$ is in line with expectations. The variations in
$|\gamma|^2$ are probably the result of drifting environmental
conditions that affect the stability of the setup. Although this
drift is undesired, it has the advantage of giving a wide range of
$|\gamma|^2$ to investigate.

We now combine the results from different measurements into a
universal plot. To do this, we subtracted the contribution of the
uncontrolled fraction of the incident wavefront to find the
controlled part of the transmission
\begin{equation}
\Tcontrol \equiv \tTtot-(1-|\gamma|^2)\Ttot
\label{eq:Ttot-compensated-experiment},
\end{equation}

\begin{figure}
\centering
  \includegraphics[width=\mediumimage]{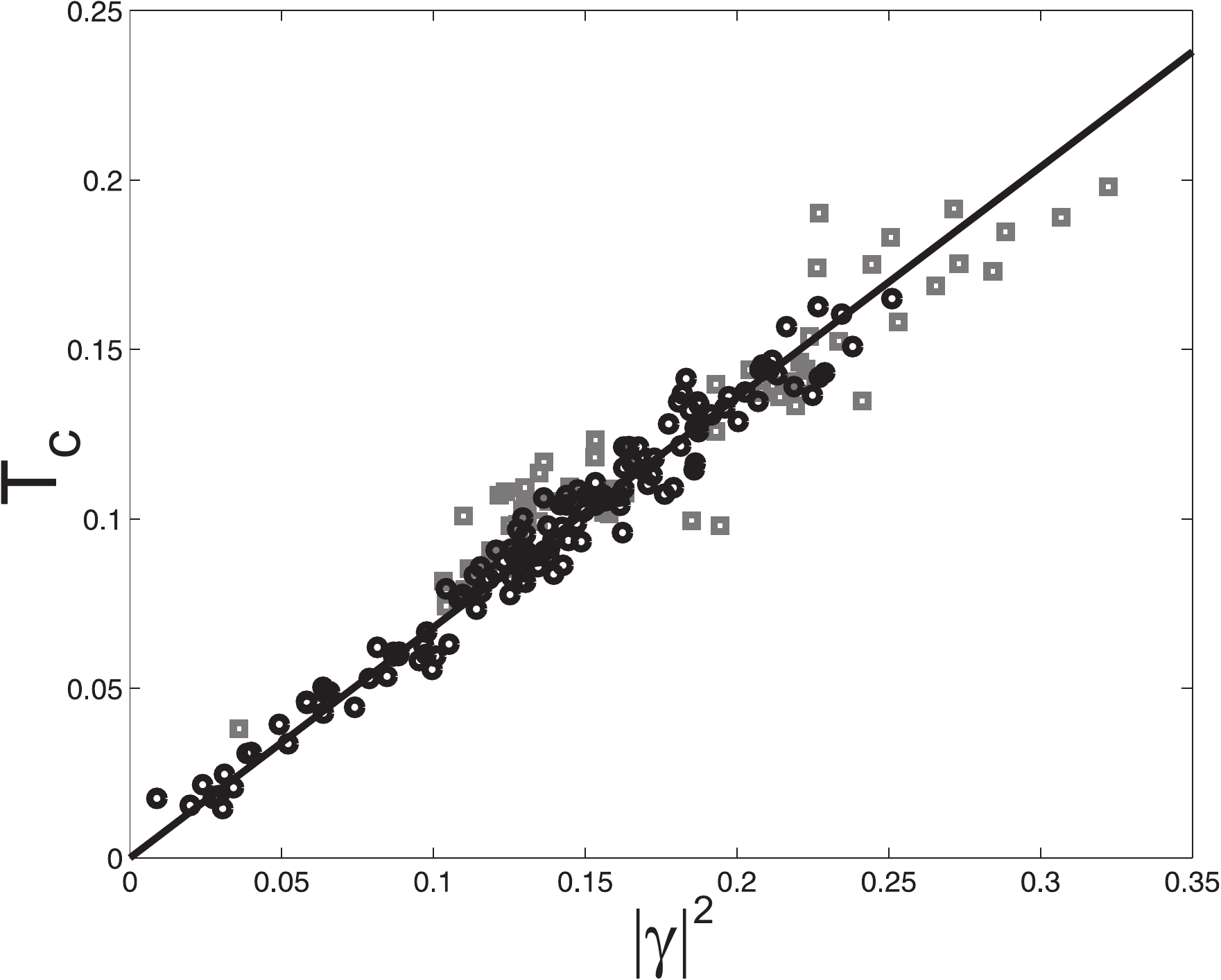}\\
  \caption{Total transmission
  of the controlled fraction of the wavefront.
  Data from seven measurement sequences on two different samples
  collapses to a universal curve.
  There are no free parameters in the data or the theory.
  Circles, results for $\mum{11.3}$-thick sample;
  squares, results for $\mum{5.7}$-thick sample;
  solid line, linear regression with slope of $0.68$.
  The theoretical curve for $\bucket=2/3$ is not shown
  as it overlaps with the fit.}\label{fig:results-corrected-transmission}
\end{figure}

\noindent From Eq.~\eqref{eq:Ttot-opt-experiment} we expect that $\Tcontrol=|\gamma|^2\bucket$. As a
result, all measurements should follow a universal curve with a slope of $\bucket$. In
Fig.~\ref{fig:results-corrected-transmission} we plotted the controlled transmission for seven measurement
sequences performed on two different samples. All data points collapse to a single line, regardless of the
thickness of the sample. This is a clear indication that the transmission of the ideal wavefront
($\bucket$) is a universal parameter.

\enlargethispage{-\baselineskip}

For the $\mum{11.3}$-thick sample the linear regression gives a
slope $0.68\pm0.01$ and for the $\mum{5.7}$-thick sample we find a
slope of $0.68\pm0.02$, where the uncertainties represent the 0.95\%
confidence interval of the regression. Combining the data from both
samples, we find $\bucket=0.68\pm0.07$ where the uncertainty is
determined by the worst case systematical errors due to the
extrapolation of the intensity profile (see
Section~\ref{sec:dorokhov-details}).

\section{Details of the data analysis\label{sec:dorokhov-details}}
\enlargethispage{-\baselineskip}

\noindent In this section we give experimental details about the data analysis. In
Section~\ref{sec:difftrans-camera} we explain how the totally transmitted power was measured using a CCD
camera. This measurement method introduces a small but significant error in the determination of
$\bucket$. Other possible sources of systematical measurement error are discussed in
Section~\ref{sec:systematic-error}. In Section~\ref{sec:Pin-measurement} we explain how the incident
intensity was measured accurately.

\subsection{Diffuse transmission measurement with a camera\label{sec:difftrans-camera}}
\noindent We measured the totally transmitted power with the CCD cameras behind the sample. The cameras
collect most of the transmitted light, but the diffuse spot has such a wide profile that its tails are not
imaged onto the CCD chip. Therefore, the camera image is extrapolated to obtain the total intensity.

It turned out to be hard to reliably fit a theoretical curve to the
complete intensity profile, as the profile depends on the incident
wavefront and many other parameters. According to diffusion theory,
the transmission profile asymptotically decays exponentially with a
decay length of $(L+z_{e1}+z_{e2})/\sqrt{6}$ (see
Section~\ref{sec:intro-diffusion}). It, therefore, makes sense to
extrapolate the camera image exponentially.

\begin{figure}
\centering
  \includegraphics[width=\smallimage]{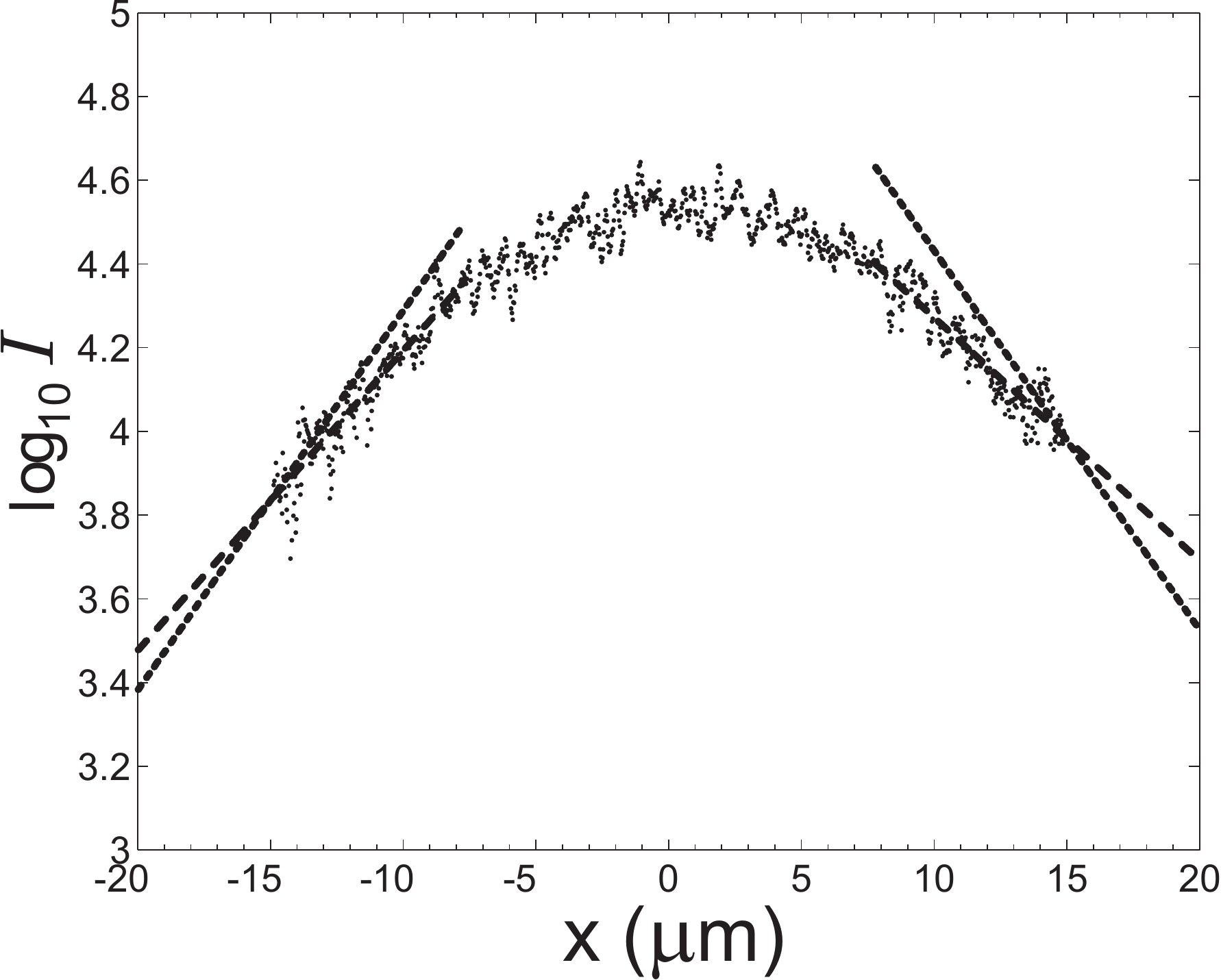}\\
  \caption{Logarithmic plot of the transmitted intensity profile for a
  reference measurement. The intensity is integrated per column.
  Dots, measurements; Dashed line, exponential extrapolation;
  Dotted line, steepest slope that is consistent with diffusion theory.}\label{fig:camera-extrapolation}
\end{figure}

A typical transmission profile of a reference measurement is shown
in Fig.~\ref{fig:camera-extrapolation}. It can be seen that the
exponential extrapolation gives a slight overestimation of the
intensity since the measured profile does not fully reach into the
exponential asymptotic limit. Alternatively, we extrapolated the
profile with the steepest exponential decay that is in agreement
with diffusion theory. For this, we took the lower limit for the
sample thickness ($L=\mum{10.8}$ and $L=\mum{5.2}$), the minimal
mean free path ($\ell=0.7$) and the minimal extrapolation lengths
($z_{e1}=z_{e2}=2/3\ell$). This alternative extrapolation method
(dotted lines in Fig.~\ref{fig:camera-extrapolation}) gives a lower
limit for the total intensity. By averaging the results of the two
methods we find a total intensity of $(2.92\pm0.08)\cdot 10^7$
counts for the profile in Fig.~\ref{fig:camera-extrapolation}, where
the error margin is the difference between the results of the two
extrapolation methods. A fraction of 82\% of the total intensity
fell on the CCD.

The typical uncertainty in this experiment is about $4\%$. The
profiles for the reference intensity $\Ptot$ and the optimized
intensity $\tPtot$ are almost equal at the edge of the image (see
Fig.~\ref{fig:results-before-after-dorokhov}). Therefore, it is
likely that the error caused by the extrapolation is either positive
for both profiles, or negative for both profiles. In this case, the
extrapolation only causes an uncertainty of $\pm 2\%$ in $\bucket$.
By calculating $\bucket$ for all combinations of error margins a
worst case error margin of $\pm 9\%$ is found.

\subsection{Possible causes of systematical error\label{sec:systematic-error}}
It is possible that small systematical errors are introduced by the
imaging system (microscope objective, lenses and camera) that record
the transmitted light. We measured the intensity distribution coming
from a sample that is illuminated with a wide ($>\mum{500}$) spot.
In this situation, the intensity distribution of the transmitted
field should be flat (except for speckle). We used this flat-field
reference image to compensate for the position dependent sensitivity
of the imaging system.

The angular dependency of the imaging system was not calibrated.
However, the width of the optimized spot was equal to the
diffraction limit of $\lambda/(2n_2)$ to within an experimental
uncertainty of 6\%. Therefore, we expect the effect of angular
dependency to be below $6\%$.

\subsection{Measurement of the incident power\label{sec:Pin-measurement}}
\noindent With the cameras we measure the total power that is
transmitted through a sample before and after an optimization. We
also measure $\Pin$, which is a reference for the incident power.
This reference value is equal to the total power that is incident on
the sample times the detection efficiency of the transmitted light.
When $\Pin$ is known, we can calculate absolute values for the
transmission coefficients. In measuring $\Pin$, great care was taken
not to introduce systematical errors. A complicating factor is that
the optimized wavefront is different for every measurement, which
causes variations in the power that reaches the sample. To
compensate for these variations, we determined $\Pin$ using
\begin{equation}
\Pin =  \Ptot\frac{P_1}{P_2} \frac{1}{1-R}
\end{equation}
where $P_1$ is the detected transmitted power with an empty
substrate in place of the sample. The substrate was illuminated with
an unshaped wavefront: all pixels of the modulator were set to the
same phase, so that the incident light forms a focused wave. $P_2$
was measured using the same unshaped wavefront, with the sample in
place. The sample was at exactly the same position where the
optimization was performed. $\Ptot$ is the reference transmission of
the shaped wavefront. It was measured by translating the sample
after each optimization (see
Fig.~\ref{fig:results-sample-translation}). The factor $1/(1-R)$
compensates for the Fresnel reflection at the front surface of the
substrate. For a uniformly illuminated microscope objective with
$\NA=0.95$, the average Fresnel reflection coefficient is $R=0.11$.
We determined the transmission coefficient of the unshaped wavefront
using $\Ttot=(1-R)P_2/P_1$.

\section{Conclusion}
\noindent We have found experimental evidence of the existence of
open transport channels for light in opaque, strongly scattering
materials. A wavefront shaping algorithm was used to focus light
through opaque objects. We observed that the transmitted intensity
in the target focus increased by almost three orders of magnitude.
Simultaneously, the intensity outside the focus increased, even
though the algorithm only optimized the intensity in the focus. This
result cannot be explained by diffusion theory or by an uncorrelated
matrix model. We conclude that the elements of the transmission
matrix are correlated as predicted by random matrix theory.
Optimizing the transmission at a single point creates a wavefront
that selectively couples to open transport channels in the medium.

The total transmission has a linear relation to the intensity in the
target focus. This relation confirms the theoretical results from
Section~\ref{sec:imperfect}. The total transmission through a
disordered sample was increased by a factor of up to $1.4$.

We used the intensity in the target focus to derive the degree of
control over the incident wavefront. By considering only the
transmission of the controlled fraction of the wavefront, we were
able to compare the results for different samples and different
realizations of disorder. All results showed a universal behavior
that is in excellent quantitative agreement with the distribution of
channel transmission coefficients as is predicted by random matrix
theory.

\bibliography{../../bibliography}
\bibliographystyle{Ivo_sty}

\setcounter{chapter}{9}
\chapter{Summary and outlook\label{cha:summary}}

\noindent In this thesis, we explored the use of wavefront shaping to steer light through strongly
scattering materials. We found that scattering does not irreversibly scramble the incident wave. By
shaping the incident wavefront, an opaque object can form any desired wavefront and focus light as sharply
as an aberration free lens. We used feedback from a target behind, or in, an opaque object to shape the
incident wave. This way, light was focused through, or inside, an opaque object for the first time ever.

Wavefront shaping provides a new way to perform experiments on strongly scattering samples. By studying
the total intensity transmission of a shaped wavefront, we found experimental evidence of open transport
channels for light. This is an example of how wavefront shaping can be used to study fundamental
properties of scattered waves.

Our results lay both an experimental and a theoretical basis for
wavefront shaping experiments. The concepts that were developed are
generally applicable to any linear complex system. Therefore, our
methods can be directly translated to other disordered systems such
as scattering of microwaves, sound, seismic waves, electrons or
neutrons. Even though we have only taken some first steps in this
field, the outline of future applications and exciting research
subjects becomes visible. We now discuss some of the possible
applications of our work.

Opaque object can be made to focus light with an excellent resolution. Therefore, thin layers of paint
might be used as cheap disposable lenses. Scattering can also be used to improve the resolution and,
thereby, the data transfer rate of a free space communication system. Instead of creating a focus,
multiply scattered light can also be made to form a beam. This could be useful when performing laser
spectroscopy of opaque materials. The scattered coherent light could then be collected with a high
efficiency, thus greatly improving the sensitivity. Even better, incoherent background light does not form
into a beam and, therefore, would not be detected. The wavefront that focuses light through an opaque
object is unique for each object. This property could be used in security applications. When, for example,
a small photodiode is covered with a layer of white paint, it is only possible to focus light on it when
the matching wavefront is known.

Light can also be focused inside an opaque object. The applications of this method are endless. One could,
for example, focus light on a single fluorescently labelled cell below the human skin or inside bone. This
application would allow one to study, track or destroy individual (malignant) cells. To make this
futuristic application possible, the wavefront needs to be constructed faster than the cells in the tissue
move. We demonstrated that wavefront shaping in a living flower petal is possible. For wavefront shaping
in human tissue, especially in perfused tissue, the speed of the wavefront shaping process needs to
improve dramatically. This speed improvement is possible with the use of extremely fast light modulators
that have recently been developed.

Our method can also be used to focus light inside non-random complex systems. This way, one can focus
light on a quantum dot in a three-dimensional photonic crystal cavity, or excite a desired mode in a
complex photonic metamaterial or plasmonic device. All in all, wavefront shaping can be used to guide
light (or any other wave) to any desired position in an arbitrary complex environment.


\setcounter{chapter}{10}
\selectlanguage{dutch}
\renewcommand{\thefigure}{\arabic{figure}}

\chapter*{Nederlandse samenvatting}
\addcontentsline{toc}{chapter}{Nederlandse samenvatting}
\chaptermark{Nederlandse samenvatting}
Dit proefschrift gaat over de voortplanting van licht in
ondoorzichtige witte dingen zoals velletjes papier, laagjes verf,
eierschillen, bloemblaadjes en tanden. Witte materialen verstoren de
voortplanting van licht: als je bijvoorbeeld met een zaklantaarn op
een velletje papier schijnt, gaat het licht niet als een bundel door
het papier heen, maar in plaats daarvan gloeit het hele papier op
met een vaag schijnsel. Het licht is diffuus geworden: het straalt
gelijkmatig naar alle richtingen uit en er is niet meer te zien waar
het licht oorspronkelijk vandaan kwam. Het is daarom ook onmogelijk
om door het materiaal heen te kijken.

Tot nu toe was het niet mogelijk om te be\"invloeden hoe licht zich
door een ondoorzichtig materiaal voortplant: het licht werd altijd
diffuus. Wij hebben een techniek bedacht waarmee we het licht w\'el
kunnen sturen. Met deze techniek kun je licht door een
ondoorzichtige laag heen projecteren of zelfs concentreren op een
punt diep binnenin een ondoorzichtig object. Hiermee wordt het
misschien ooit mogelijk om licht door huid of bot heen op \'e\'en
enkele gemarkeerde kankercel te focusseren, of om door de
ingewikkelde structuur van een microchip heen \'e\'en enkele
transistor te belichten. Ook bij andere toepassingen, zoals
telecommunicatie, optische experimenten aan microstructuren en
spectroscopie aan poeders kan onze techniek uitkomst bieden.
Uiteindelijk zou dit ook kunnen leiden tot een verbetering van
bestaande medische beeldvormende technieken.

\section*{Interferentie en verstrooiing}
\begin{figure}
\centering
  \includegraphics[width=\wideimage]{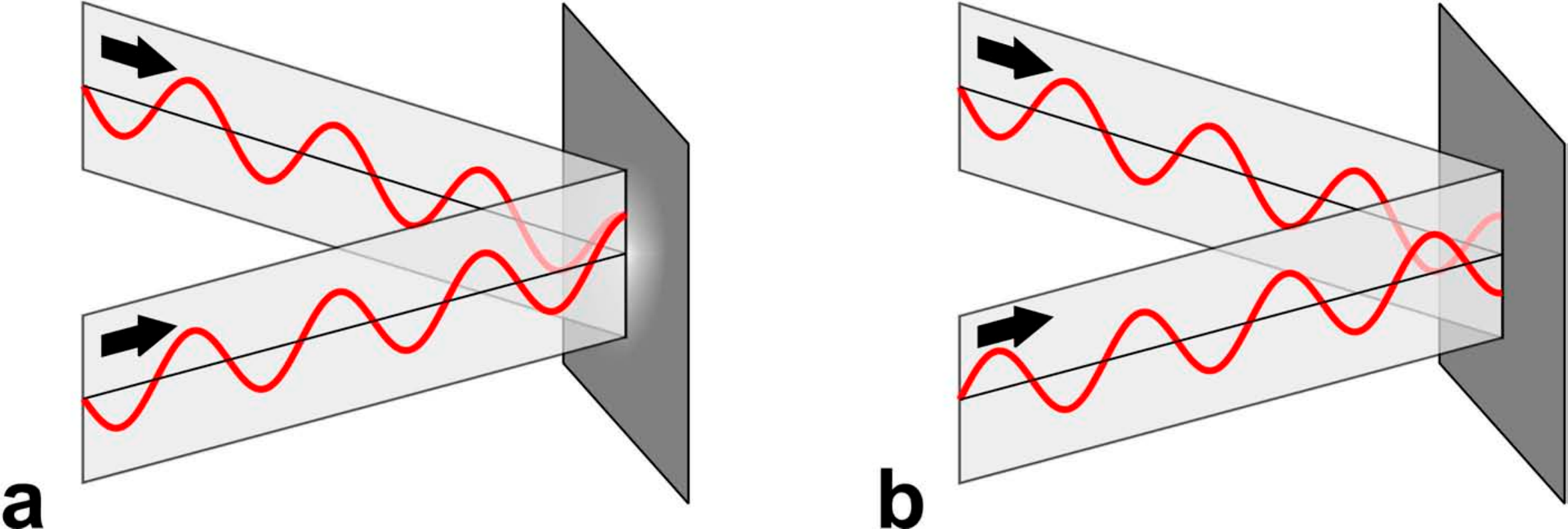}\\
  \caption{Interferentie. \subfig{a} De golven komen tegelijk aan. Ze zijn in fase en versterken elkaar. \subfig{b} De golven zijn uit fase en doven elkaar uit.}\label{fig:interferentie}
\end{figure}

Licht is een golf. Zoals alle golven kunnen lichtgolven met elkaar
interfereren. Interferentie is niets anders dan dat twee golven bij
elkaar opgeteld worden. Als twee golven tegelijkertijd aankomen zijn
ze in fase, zoals te zien is in Figuur~\ref{fig:interferentie}a. De
piek van de ene golf telt dan op bij de piek van de andere golf en
het resultaat is een golf met grote pieken en dalen (dit heet
constructieve interferentie). Het kan ook zijn dat twee golven uit
fase zijn. De ene golf komt dan wat later aan dan de andere zodat de
pieken van die golf samenvallen met de dalen van de andere golf; het
resultaat is dan dat de twee golven elkaar opheffen. Deze situatie
is afgebeeld in Figuur~\ref{fig:interferentie}b.

In een wit materiaal zitten allemaal kleine deeltjes waar het licht op botst. Bij iedere botsing verandert
de voortplantingsrichting van het licht. Het licht `stuitert' door het materiaal, een beetje zoals het
balletje in een flipperkast. Dit proces heet verstrooiing van licht. Al het verstrooide licht interfereert
met elkaar, zodat er een complex wan\-or\-de\-lijk interferentiepatroon ontstaat. Dit patroon staat bekend
als laserspikkel (Engels: speckle). Een voorbeeld van een spikkelpatroon is te zien in
Figuur~\ref{fig:speckle}a in Hoofdstuk~\ref{cha:focusing-through}. Het hangt af van de fase van alle
interfererende golven of ergens een heldere of juist een donkere plek ontstaat.

\begin{figure}
\centering
  \includegraphics[width=\mediumimage]{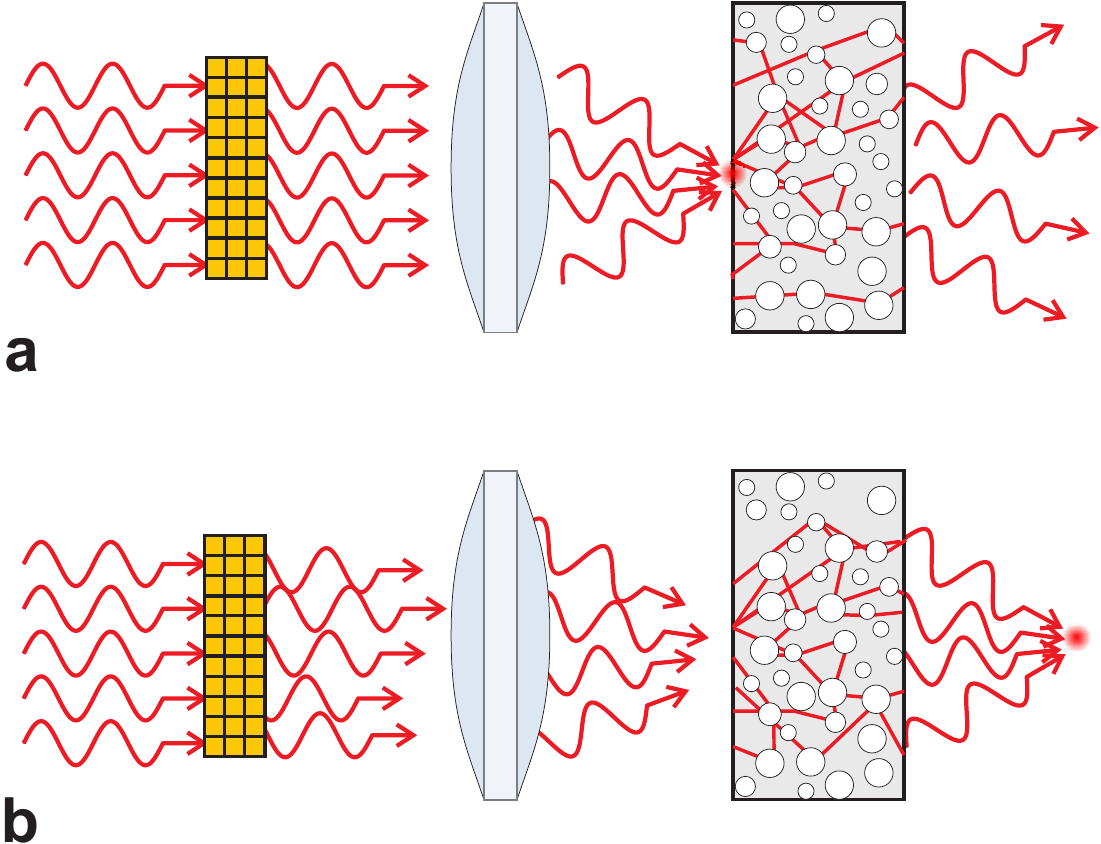}\\
  \caption{Principe van het experiment. \subfig{a} een vlakke golf wordt gefocusseerd op een ondoorzichtig wan\-or\-de\-lijk
materiaal. Door verstrooiing ten gevolge van microscopische
oneffenheden legt het licht een zeer complexe route af. Na honderden
keren verstrooien bereikt een klein deel van het licht de achterkant
van het materiaal en vormt een diffuse vlek. \subfig{b} Door de fase
van het binnenkomende licht op precies de goede manier aan te passen
is het mogelijk om de bijdragen van alle vlakjes constructief te
laten interfereren op een punt achter het medium. Het medium werkt
in dit geval als een wan\-or\-de\-lijk\-e lens en focusseert het
licht.}\label{fig:principe}
\end{figure}

\section*{Licht focusseren met een eierschil}
We hebben een opstelling ontworpen die een bundel laserlicht in
duizenden vlakjes verdeelt en dan de fase van elk van deze vlakjes
afzonderlijk kan aanpassen. Het licht van elk vlakje zal op een
andere manier verstrooid worden (zie Figuur~\ref{fig:principe}a).
Door de fase van de vlakjes op de juiste manier in te stellen kunnen
we nu zorgen dat het verstrooide licht op een bepaalde plek
constructief interfeert (Figuur~\ref{fig:principe}b). De
lichtintensiteit op die plek neemt daardoor enorm toe, soms wel met
een factor duizend. Het resultaat is dat een ondoorzichtig object,
zoals een eierschil, licht focusseert tot een scherp punt. Eigenlijk
werkt de eierschil dus als een `ondoorzichtige lens'.

Er zijn ongelooflijk veel mogelijke combinaties van fases te maken.
Om de juiste combinatie te vinden gebruiken we een computer die de
hoeveelheid licht in het gewenste focus meet en de fases aanpast. In
het begin is de lichtintensiteit heel laag, maar door de fase van
een vlakje aan te passen kunnen we de hoeveelheid licht laten
toenemen. De computer herhaalt dit proces voor alle vlakjes.
Uiteindelijk hebben we dan een heel ingewikkeld gevormde bundel
gemaakt die precies compenseert voor alle botsingen aan de deeltjes
in het materiaal. In hoofdstuk \ref{cha:focusing-through} laten we
zien dat deze methode de optimale combinatie van fases vindt. Omdat
deze combinatie afhangt van de positie van alle deeltjes in het
materiaal, werkt onze methode nog niet voor materialen waarin de
deeltjes bewegen, zoals melk of doorbloed weefsel. Met recent
ontwikkelde technologie is het echter mogelijk om het proces zo te
versnellen dat we in de toekomst `real-time' kunnen compenseren voor
de bewegende deeltjes.

\section*{Belangrijkste resultaten}
Het belangrijkste resultaat van dit proefschrift is dat we in staat zijn om licht door een ondoorzichtig
materiaal heen te focusseren. Deze doorbraak heeft nationaal en internationaal veel aandacht gekregen (zie
bijvoorbeeld \emph{`Witte verf of eierschaal als brandglas voor bundel laserlicht'} in het NRC Handelsblad
van 14 juli 2007). Het is niet alleen gelukt om verstrooid licht in een scherp focus samen te laten komen,
we begrijpen de resultaten ook kwantitatief. We hebben uitgerekend hoe intens het focus is (Hoofdstuk
\ref{cha:focusing-through}), wat er gebeurt als de deeltjes in het materiaal bewegen (Hoofdstuk
\ref{cha:algorithms}) en hoe groot het focus is (Hoofdstuk \ref{cha:diffraction-limit}). Het blijkt dat
het focus van, bijvoorbeeld, een eierschil precies even scherp is als het focus dat je met de best
mogelijke lens zou kunnen maken. In Hoofdstuk \ref{cha:communication} laten we verder nog zien dat
wan\-or\-de\-lijk\-e verstrooiing gebruikt kan worden om meer optische informatie te versturen dan normaal
mogelijk is.

We zijn ook in staat om licht binnenin een materiaal te focusseren,
iets wat zeer interessant is voor bijvoorbeeld medische
toepassingen. Dit experiment is technisch veel lastiger dan licht
ergens dwars doorheen focusseren. Om de juiste combinatie van fases
te vinden is het namelijk nodig om de lichtintensiteit op de
gewenste plek te meten. Als je, bijvoorbeeld, binnen in de huid wilt
focusseren, is het niet mogelijk om op die plek een detector aan te
brengen. We hebben daarom gebruik gemaakt van zeer kleine bolletjes
met een fluorescerende verf erin. Dit soort bolletjes zijn w\'el in
te brengen. Door nu te meten hoeveel het bolletje fluoresceert,
meten we indirect hoeveel licht er van buiten af op het bolletje
valt. Dit experiment staat beschreven in Hoofdstuk
\ref{cha:focusing-inside}.

Verstrooiing van licht lijkt wiskundig gezien erg veel op het transport van elektronen in een metaaldraad.
We kunnen daarom met licht voorspellingen testen die gedaan zijn voor elektronische systemen. Het voordeel
van onze optische methode is dat we het binnenkomende licht heel precies kunnen vormen, iets wat met
elektronen nu niet mogelijk is. E\'en van de meest verrassende voorspellingen is dat het mogelijk moet
zijn om een golf te maken die voor de volle 100\% wordt doorgelaten door een ondoorzichtig materiaal. In
Hoofdstuk \ref{cha:dorokhov-theory} laten we zien wat dit betekent voor ons experiment: als we de
binnenkomende lichtbundel optimaliseren om op \'e\'en punt licht te focusseren moet de intensiteit rondom
dat punt ook omhoog gaan. Er zal dus in totaal veel meer licht doorgelaten worden (en er zal dus ook veel
minder licht worden gereflecteerd). We hebben in een experiment aangetoond dat de totale transmissie inderdaad significant stijgt (Hoofdstuk
\ref{cha:dorokhov-experiment}). Hiermee bevestigen we een belangrijk theoretisch aspect van het transport
van elektronen, geluid, licht en andere golven.

\section*{Conclusie}
We hebben een methode ontwikkeld om met behulp van interferentie laserlicht door ondoorzichtige materialen
heen te sturen. Het op deze manier sturen van laserlicht is een nieuw onderwerp in de optica. Dit
proefschrift kan daarom het beste gezien worden als een eerste verkenning van dit gebied. Het is nu al
duidelijk dat er nog heel veel verder te experimenteren valt en dat er interessante toepassingen verwacht
kunnen worden. We hopen dan ook dat dit nieuwe vakgebied ook de komende jaren nog tot veel spectaculaire
resultaten zal leiden.


\setcounter{chapter}{11}
\selectlanguage{dutch}%

\chapter*{Dankwoord}
\addcontentsline{toc}{chapter}{Dankwoord}
\chaptermark{Dankwoord}

\vskip -20pt
\begin{tabular}{ll}
\begin{minipage}[l]{0.45\textwidth}
\textit{``Niet in de afzondering zullen we onszelf ontdekken, maar onderweg, in de stad, in de menigte,
als ding onder de dingen, als mens onder de mensen.''}
\\
\end{minipage}
&
\begin{minipage}[l]{0.45\textwidth}
\textit{``Ce n'est pas dans je ne sais quelle retraite que nous nous d\'ecouvrirons, c'est sur la route,
dans la ville, au milieu de la foule, chose parmi les choses, homme parmi les hommes.''}
\end{minipage}
\end{tabular}

\begin{flushright}
- Jean Paul Sartre, Situations (1976)\\
\end{flushright}

\noindent In geen enkel ander hoofdstuk van mijn proefschrift zul je
het woordje `ik' te\-gen\-ko\-men om de simpele reden dat alles wat
ik bereikt heb het resultaat is van een intensieve samenwerking
waaraan iedereen op zijn of haar eigen wijze heeft bijgedragen.
Daarom wil ik hier iedereen bedanken die mij in de afgelopen vier
jaar heeft bij\-ge\-staan en heeft gezorgd voor een leuke sfeer in
de vakgroep en daarbuiten. Een paar mensen wil ik apart noemen.

Allereerst mijn begeleiders Ad en Allard. Allard, het was jouw idee waar dit project allemaal mee begon.
Vanaf het eerste moment was ik enthousiast en je hebt dat enthousiasme steeds aangewakkerd met de vele
(soms wilde) idee\"en die in onze ge\-sprek\-ken ontstonden. Ik vond het heel prettig om met je samen te
werken en soms ook over andere dingen dan natuurkunde te kunnen praten.

Ad, het is mede dankzij jouw flexibele instelling dat we uiteindelijk met zijn allen dit mooie resultaat
hebben behaald. Ik bewonder je scherpe inzicht in de na\-tuur\-kun\-de en je visie op hoe goede wetenschap
hoort te werken. Ik denk met plezier terug aan mijn bezoeken aan het Amolf en de momenten dat we over een
zin gebogen zaten om n\'et dat ene woord te vinden dat precies de goede lading had.

Na de verhuizing bleef in Enschede een kleine, maar gezellige groep
over. Bas, bedankt voor je snelle hulp met computers. Cock, dank je
voor al het werk dat je hebt verricht aan mijn opstelling. Je hebt
altijd voor een goede sfeer in de groep gezorgd en ik ben blij je
als kamergenoot gehad te hebben. Hannie, mijn meest nieuwschierige
kamergenoot en groepspsychologe. Karen Munnink stond altijd klaar
voor hulp met formulieren en andere dingen waar ik niets van snap.
L\'eon de masterchief van het chemisch lab en daarbuiten. Phil,
blijf zwemmen! Raymond, het was altijd een prettige verrassing als
je weer wat leven in de brouwerij kwam brengen. Willem Tjerkstra, de
ideale schrijver om mee naar de film te gaan. En natuurlijk Willem
Vos die als een soort vliegende keep zowel in Amsterdam als in
Enschede de vakgroepen be\-stuurt. Willem, dank je voor je
ondersteuning, inbreng en enthousiasme.

Wouter en Elbert, het was erg leuk en leerzaam om jullie te mogen begeleiden. Jullie grote inzet en
creatieve idee\"en hebben veel voor mij betekend. Elbert heeft in grote mate bijgedragen aan Hoofdstukken
\ref{cha:setup} en \ref{cha:focusing-inside}, hartelijk dank hiervoor.

Het was altijd gezellig om op het AMOLF langs te komen. In het bijzonder herinner ik me de vrolijke kamer
met Bart, Merel en Ramy; Alex, die mij heeft ge\"inspireerd om weer Sartre gaan lezen en Bernard, het is
altijd een plezier met je te praten, ook al kan ik je af en toe niet helemaal bijhouden. Otto, Paulo,
Patrick, Sanli bedankt voor de inspirerende discussies. Verder wil ik iedereen in de Photon Scattering en
Photonic Bandgaps groepen bedanken voor de prettige samenwerking. Ook veel dank aan de Photon Scattering
groep voor het uitlenen van de TiO$_2$ samples en aan Simone voor het uitlenen van haar melktand.

Ook met sommige oud groepsleden heb ik een hechte band opgebouwd.
Karen en Tom, jullie waren geweldige collega's, zeker ook toen de
dingen tegenzaten. Peter, Boris, Martijn, jullie hebben me wegwijs
gemaakt in de natuurkunde. Tijmen, op de \'e\'en of andere manier
komen we altijd op bijzondere plaatsen terecht... Femius, Ivan en
Nina, Lydia, Professor Valentin Freilikher, en alle andere geweldige
mensen bij COPS, dankjewel!

I would like to thank Professor Sir John Pendry and Professor Vinod
Subramaniam for their enthusiastic response to our research.

Ik heb de afgelopen vier jaar niet alleen maar werkend doorgebracht.
Ik wil al mijn vrienden bedanken die er voor hebben gezorgd dat ik
af en toe het laboratorium uit kwam om de echte wereld te
(her)ontdekken. in het bijzonder Claartje, Kell, Marieke en Maarten,
Mieke en Willem Mees, Moes, Nienke en Pieter, Wietse en alle mensen
van de Parkweg. Frans, met jou hardlopen was altijd erg ontspannend
en gezellig.

Thomas, je woont ver weg, maar je bent dichtbij. Je hebt me enorm
geholpen met alles wat me de afgelopen jaren bezig gehouden heeft.
Dank je. !`Est\'ibaliz, cuida bien de el! Henrik, Marc, ik ben blij
jullie als paranimfen te hebben. Ik zal proberen niet flauw te
vallen, maar het stelt me gerust dat er twee goede vrienden (met
veel toneelervaring) klaar staan om het over te nemen. Iris, je hebt
veel voor me betekend de afgelopen jaren. Ik hoop dat jij ook heel
veel plezier zult beleven aan je onderzoek.

Papa, Mama, mijn hoofd duizelt als ik bedenk wat er in de afgelopen
jaren allemaal is gebeurd. Al die tijd hebben jullie mij gesteund en
waren jullie enthousiast en ge\"interesseerd in mijn onderzoek.
Dankjewel. Ik ben blij jullie te zien genieten van jullie reizen en
van de kleine en grote dingen in het leven. Casper en Karin, het is
goed om familie te hebben die je steunt en die ook weet hoe ze een
goed feest moeten organiseren, bedankt voor de leuke tijden!

Bert, je bent er helaas niet meer, maar jouw aanstekelijke enthousiasme voor de wetenschap heeft mij voor
altijd gevormd. Dank je wel hiervoor.

\selectlanguage{english}

\pagestyle{empty} \newpage\mbox{} \newpage\mbox{}
\end{document}